\documentclass[a4paper,11pt]{article}
\pdfoutput=1 

\usepackage[utf8]{inputenc}

\usepackage[no-natbib-sort]{jheppub}
\usepackage{enumitem}

\usepackage{youngtab}
\usepackage{amsfonts}
\usepackage{amsmath}
\usepackage{amsbsy}
\usepackage{graphicx}
\usepackage{amssymb}
\usepackage{amsthm}
\usepackage{bm}
\usepackage{comment}
\usepackage{epsfig}
\usepackage{latexsym}
\usepackage{pdflscape}
\usepackage{cancel} 
\usepackage{color}
\usepackage{cleveref}
\usepackage{bm,nicefrac}

\makeatletter
\def\@fpheader{\relax}
\makeatother

\DeclareSymbolFont{AMSa}{U}{msa}{m}{n}
\DeclareSymbolFont{AMSb}{U}{msb}{m}{n}
\DeclareMathSymbol{\fieldR}{\mathalpha}{AMSb}{"52}

\newcommand{\beq}{\begin{eqnarray}}
\newcommand{\eeq}{\end{eqnarray}}
\newcommand{\bea}{\begin{eqnarray}}
\newcommand{\eea}{\end{eqnarray}}
\newcommand{\be}{\begin{equation}}
\newcommand{\ee}{\end{equation}}
\newcommand{\bq}{\begin{equation}}
\newcommand{\eq}{\end{equation}}

\def\ie{{\it \emph{i.e.}} }

\def\dd{{\rm d}}

\def\6{\partial}

\newcommand{\bb}{\mathbb}

\def\6{\partial}

\usepackage{color}
    \definecolor{darkgreen}{rgb}{0,0.5,0}
    \definecolor{darkblue}{rgb}{0,0,0.6}
    \definecolor{purple}{rgb}{0.4,.2,0.7}

\newcommand{\fig}[1]{Fig.~\ref{#1}}
\newcommand{\figs}[1]{Figs.~\ref{#1}}
\newcommand{\sect}[1]{Sec.~\ref{#1}}
\newcommand{\taut}{T}
\newcommand{\eho}{\hat{\overline{\cal E}}}
\newcommand{\nc}{N}
\title{Crossing a large-$\nc$ phase transition at finite volume}

\author[a,b]{Yago Bea,}
\author[c]{Oscar J.~C.~Dias,}
\author[d]{Thanasis Giannakopoulos,}
\author[b,e]{David Mateos,}
\author[b]{Mikel Sanchez-Garitaonandia,}
\author[f,g]{Jorge E.~Santos,}
\author[d]{Miguel Zilh\~ao}

\affiliation[a]{School of Mathematical Sciences, Queen Mary University of London, Mile End Road, \\ London E1 4NS, United Kingdom.}
\affiliation[b]{Departament de F\'\i sica Qu\`antica i Astrof\'\i sica and Institut de Ci\`encies del Cosmos (ICC),\\  Universitat de Barcelona, Mart\'\i\  i Franqu\`es 1, ES-08028, Barcelona, Spain.}
\affiliation[c]{STAG research centre and Mathematical Sciences, University of Southampton, UK.}
\affiliation[d]{Centro de Astrof\'{\i}sica $E$ Gravita\c c\~ao -- CENTRA,
  Departamento de F\'{\i}sica, Instituto Superior T\'ecnico -- IST, Universidade
  de Lisboa -- UL, Av.\ Rovisco Pais 1, 1049-001 Lisboa, Portugal }
\affiliation[e]{Instituci\'o Catalana de Recerca i Estudis Avan\c cats (ICREA), Passeig Llu\'\i s Companys 23, \\ ES-08010, Barcelona, Spain.}
\affiliation[f]{DAMTP, Centre for Mathematical Sciences, Wilberforce Road, Cambridge, CB3 0WA, UK.}
\affiliation[g]{Institute for Advanced Study, Princeton, NJ 08540, USA.}

\emailAdd{ahw704@qmul.ac.uk}
\emailAdd{jorge.casalderrey@ub.edu}
\emailAdd{ojcd1r13@soton.ac.uk}
\emailAdd{athanasios.giannakopoulos@tecnico.ulisboa.pt}
\emailAdd{dmateos@fqa.ub.edu}
\emailAdd{mikeliccub@icc.ub.edu}
\emailAdd{jss55@cam.ac.uk}
\emailAdd{miguel.zilhao.nogueira@tecnico.ulisboa.pt}

\preprint{ICCUB-20-015}
\begin{abstract}

\end{abstract}

\abstract{The existence of phase-separated states is an essential feature of infinite-volume systems with a thermal, first-order phase transition. At energies between those at which the phase transition takes place, equilibrium homogeneous states are either metastable or suffer from a spinodal instability. In this range the stable states are inhomogeneous, phase-separated states. We use holography to investigate how this picture is modified at finite volume in a strongly coupled, four-dimensional gauge theory. We work in the planar limit, $\nc \to \infty$, which ensures that we remain  in the thermodynamic limit. We uncover a rich set of inhomogeneous states dual to lumpy black branes on the gravity side, as well as first- and second-order phase transitions between them.  We establish their local (in)stability properties and show that fully non-linear time evolution in the bulk takes unstable states to stable ones.}

\begin{document} 
\maketitle
\flushbottom

\section{Introduction}\label{sec:Intro}
\begin{figure}[t]
\centerline{
		\includegraphics[width=.9\textwidth]{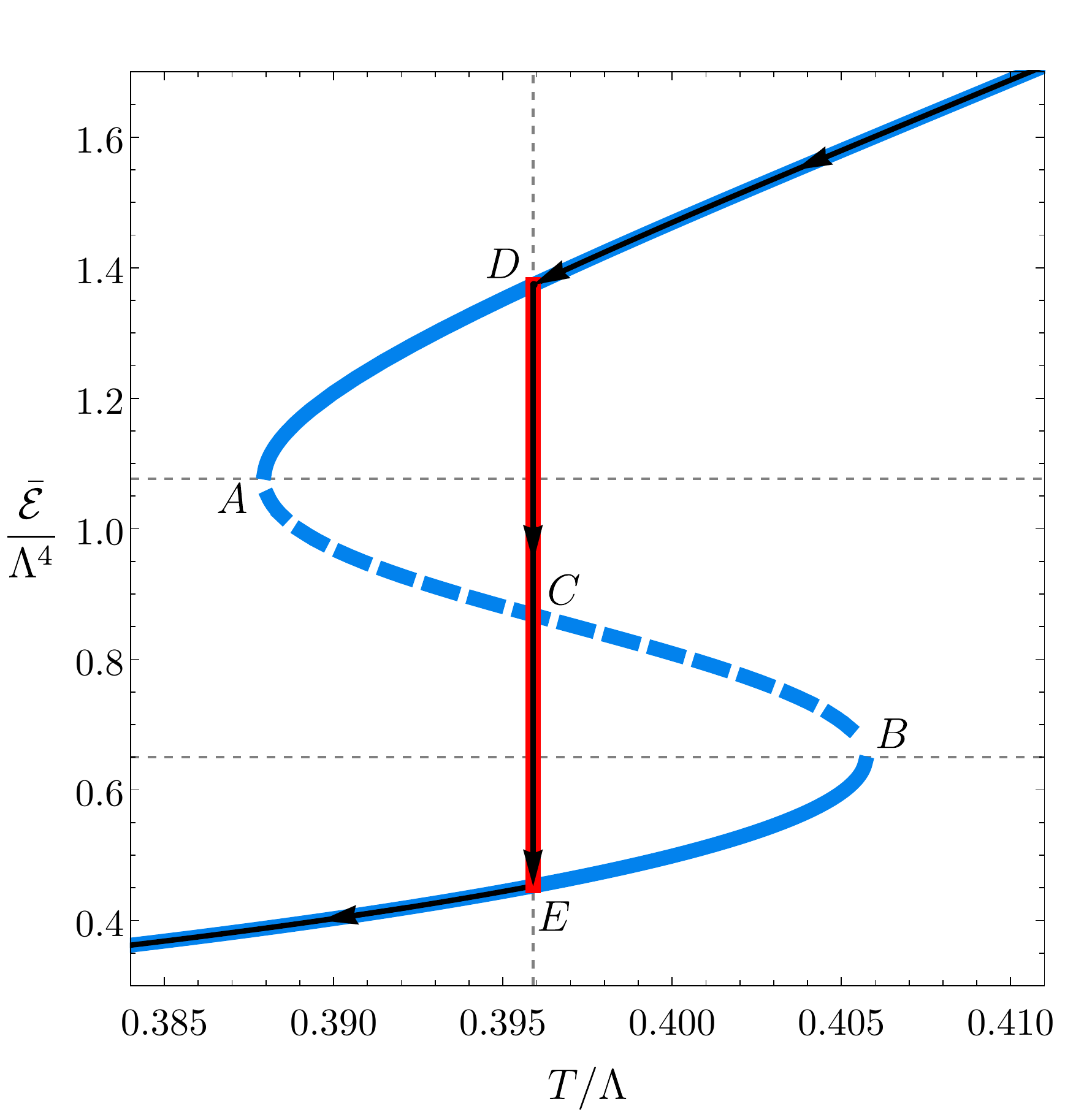}
}
\caption{Phase diagram in the infinite-volume limit. The dashed, vertical line indicates the critical temperature $T=T_c$. The blue curve corresponds to homogeneous states. The red line corresponds to inhomogeneous, phase-separated states. Solid segments  indicate locally dynamically stable states; dashed segments indicate unstable ones.  The black curves with arrows indicate the sequence of maximum-entropy states as the average energy density $\overline{\mathcal{E}}$ decreases. In the canonical ensemble there is one first-order phase transition at which the system jumps betweens points $D$ and $E$. In the microcanonical ensemble there are two second-order phase transitions at points $D$ and $E$ between homogeneous and inhomogeneous states.
}
\label{fig:transition0INTRO}
\end{figure}

Phase coexistence is an essential feature of systems with a first-order phase transition. Consider for example \fig{fig:transition0INTRO}. This shows the energy density as a function of the temperature, in the infinite-volume limit, for the four-dimensional gauge theory that we will study in this paper. 
The blue curve indicates homogeneous states with  energy density $\overline{\mathcal{E}}$, which we measure in units of the microscopic scale in the gauge theory $\Lambda$. 
In the canonical ensemble there is a first-order phase transition at a critical temperature $T_c$ indicated by the dashed, vertical line. The thermodynamically preferred, lowest-free energy  states at $T>T_c$ lie on the upper branch and have energies above that of point $D$. Similarly, at $T<T_c$ the preferred states are  on the lower branch with energies below that of point $E$.  States between points $A$ and $D$, and between $B$ and $E$, are locally but not globally thermodynamically stable. Finally, states between points $A$ and $B$ are locally thermodynamically  unstable. The region between $A$ and $B$ is known as the ``spinodal region".

In the canonical ensemble, setting $T=T_c$ does not select a unique state. For this reason, it is convenient to work in the microcanonical ensemble, in which the control parameter is the energy  instead of the temperature. In this case  the preferred, maximum-entropy  configuration  for energy densities between points $D$ and $E$ is well understood in the infinite-volume limit: it is a phase-separated state in which part of the volume is in the phase associated to point $D$ and the other part is in the phase associated to point $E$ (see Sec.~\ref{sec:PerturbativeO0}).  The fraction of volume occupied by each phase is determined by the average energy density $\overline{\mathcal{E}}$, which lies between $D$ and $E$.  The two phases are separated by a universal interface, i.e.~by an interface whose spatial profile is independent of the way in which the phase-separated configuration is reached. Since the temperature is constant and equal to $T_c$ across the entire volume,  these states lie on the red, vertical segment $DE$ in \fig{fig:transition0INTRO}. We conclude that, at infinite volume in the microcanonical ensemble, the sequence of preferred states as the energy density  decreases  is that indicated by the black arrows in \fig{fig:transition0INTRO}.

The thermodynamic statements above have dynamical counterparts. Since the total energy is conserved under time evolution, it is again convenient  to think of the system  in the microcanonical ensemble. Imagine preparing the system in a homogeneous state. If the energy density lies above point $D$ or below point $E$ then this state is dynamically stable against small or large perturbations. If instead the energy density is between points $A$ and $D$ or between $B$ and $E$ then we expect the system to be dynamically stable against small perturbations but not against large ones. This means that, if subjected to large enough a perturbation, the system will dynamically evolve to a phase-separated configuration. The average energy density in this inhomogeneous configuration will be the same as in the initial, homogenous state, but the  entropy will be higher. Finally, if the initial energy density  is between $A$ and $B$ then the state is dynamically unstable even against small perturbations. This instability, known as ``spinodal instability", implies that  the slightest perturbation will trigger an evolution towards a phase-separated configuration of equal average energy but higher entropy.

If the system of interest is an interacting, four-dimensional quantum field theory then following the real-time evolution from an unstable homogeneous state to a phase-separated configuration can be extremely challenging with conventional methods. For this reason, in \cite{Attems:2017ezz,Attems:2019yqn}  holography was used to study this evolution in the case of a four-dimensional gauge theory with a gravity dual (see also \cite{Janik:2017ykj,Bellantuono:2019wbn} for a case in which the gauge theory is three-dimensional). In order to regularise the problem, Refs.~\cite{Attems:2017ezz,Attems:2019yqn}  considered the gauge theory formulated on $\bb{R}^{1,2}\times S^1$ with periodic boundary conditions on a circle of size $L$. For simplicity, translational invariance along the non-compact spatial directions was imposed, thus effectively reducing the dynamics to a 1+1 dimensional problem along time and the compact direction. The compactness of the circle makes the spectrum of perturbations discrete and simplifies the technical treatment of the problem. Ref.~\cite{Attems:2017ezz} provided a first example of the time evolution from a homogeneous state to an inhomogeneous one. A systematic study was then performed in \cite{Attems:2019yqn}. In this reference the focus was on the infinite-volume limit, understood as the limit in which $L$  is much larger than any other scale in the problem such as the microscopic gauge theory scale $\Lambda$, the size of the interface, etc. It was shown that, if slightly perturbed, an initial homogeneous state with energy density between $A$ and $B$ always evolves towards a phase-separated configuration, and that the latter is dynamically stable. 

In addition to its implications for  gauge theory dynamics, the spinodal  instability of states between $A$ and $B$ is interesting also on the gravity side, where it implies that the corresponding black branes are afflicted by a long-wavelength dynamical instability. Although this is similar \cite{Buchel:2005nt,Emparan:2009cs,Emparan:2009at} to the Gregory-Laflamme  (GL) instability of black strings in spacetimes with vanishing cosmological constant \cite{Gregory:1993vy}, there is an important difference: In the GL case all strings below a certain mass density are unstable, whereas in our case only states between points $A$ and $B$  are unstable. Having clarified this, since the term ``GL-instability" is familiar within part of the gravitational community, in this paper we will  use the terms ``spinodal instability" and ``GL-instability" interchangeably to refer to the dynamical instability between points $A$ and $B$.

The purpose of this paper is to extend the analysis of the equilibrium states summarised in  \fig{fig:transition0INTRO}, as well as the 
systematic analysis of their dynamical stability properties of \cite{Attems:2019yqn}, to the case of finite volume.  In particular, we would like to: (i) classify all possible states, homogeneous or inhomogeneous, available to the system; (ii) determine which ones are thermodynamically preferred;  (iii) establish the local dynamical stability or instability of each state; and (iv) investigate the time evolution from unstable states to stable ones.  For this purpose we will place the system in a box, impose translational invariance along two of its directions and vary the size $L$ of the third direction. We will then see that the results depend on the value of $L$ compared to  a hierarchy of length scales 
\be
L_K < L_{\Sigma_1} < L_{\Sigma_2} \,.
\ee
These three scales are an intrinsic property of the system at finite volume that cannot be determined through an infinite-volume analysis. Depending on the ratio of $L$ to these scales we will uncover: (i)  a large configuration space of inhomogeneous states, both stable and unstable; (ii) a rich set of  first- and second-order thermodynamic phase transitions between them; and (iii) the possible  time evolutions from dynamically  unstable to dynamically stable states. Note that the existence of phase transitions is not in contradiction with the finite volume of the gauge theory  because we work in the planar limit, $\nc \to \infty$, which effectively acts as a thermodynamic limit. Our results are summarised in \sect{disc}. The reader who is only interested  in this summary  can go directly to this section.

\section{Nonconformal lumpy branes: nonlinear static solutions}\label{sec:BVPnonlinearLumpy}

\subsection{Setup of the physical problem and general properties of the system}\label{sec:setup}

\begin{figure}[t]
\centerline{
\includegraphics[width=.48\textwidth]{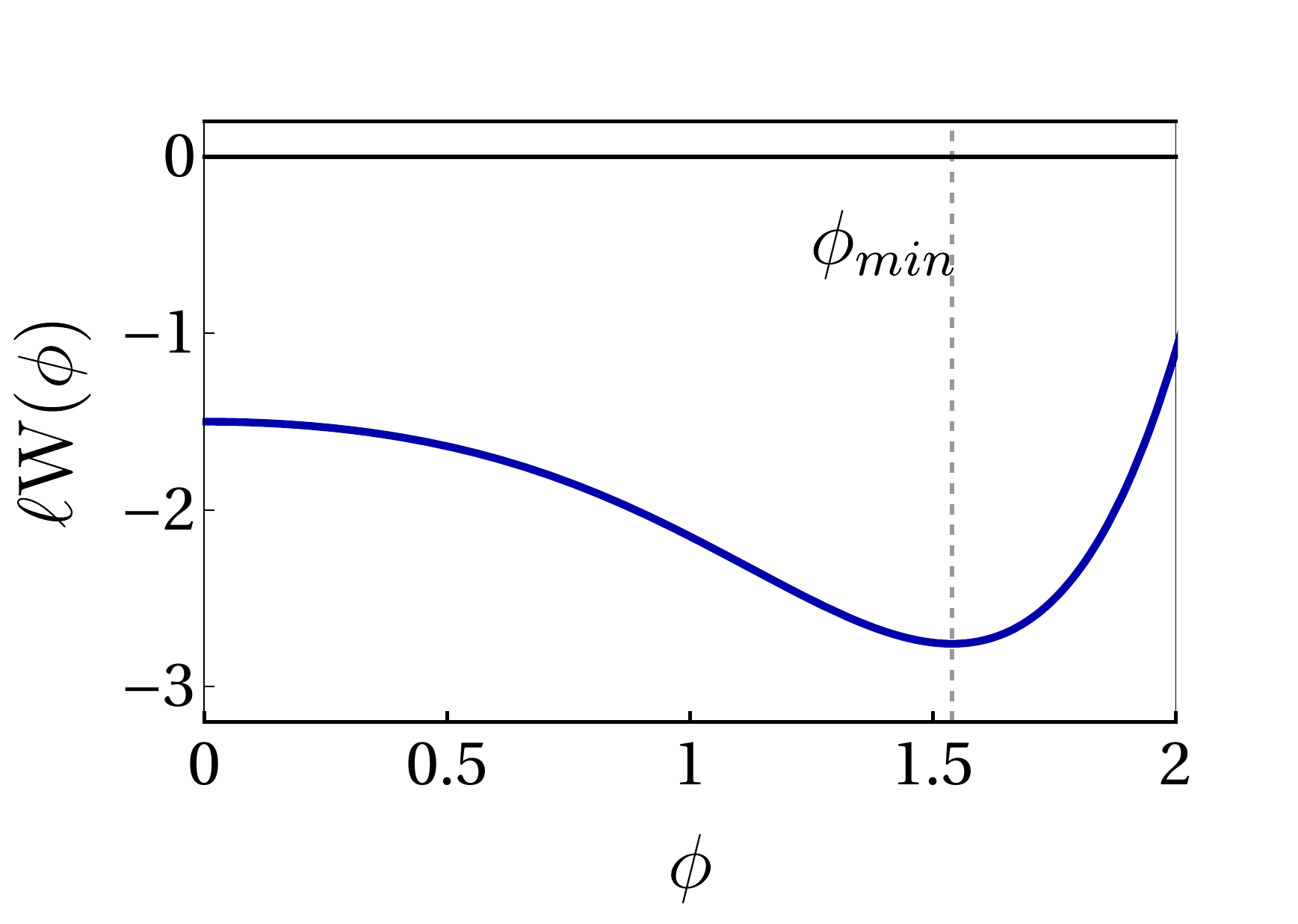}
\hspace{0.2cm}
\includegraphics[width=.48\textwidth]{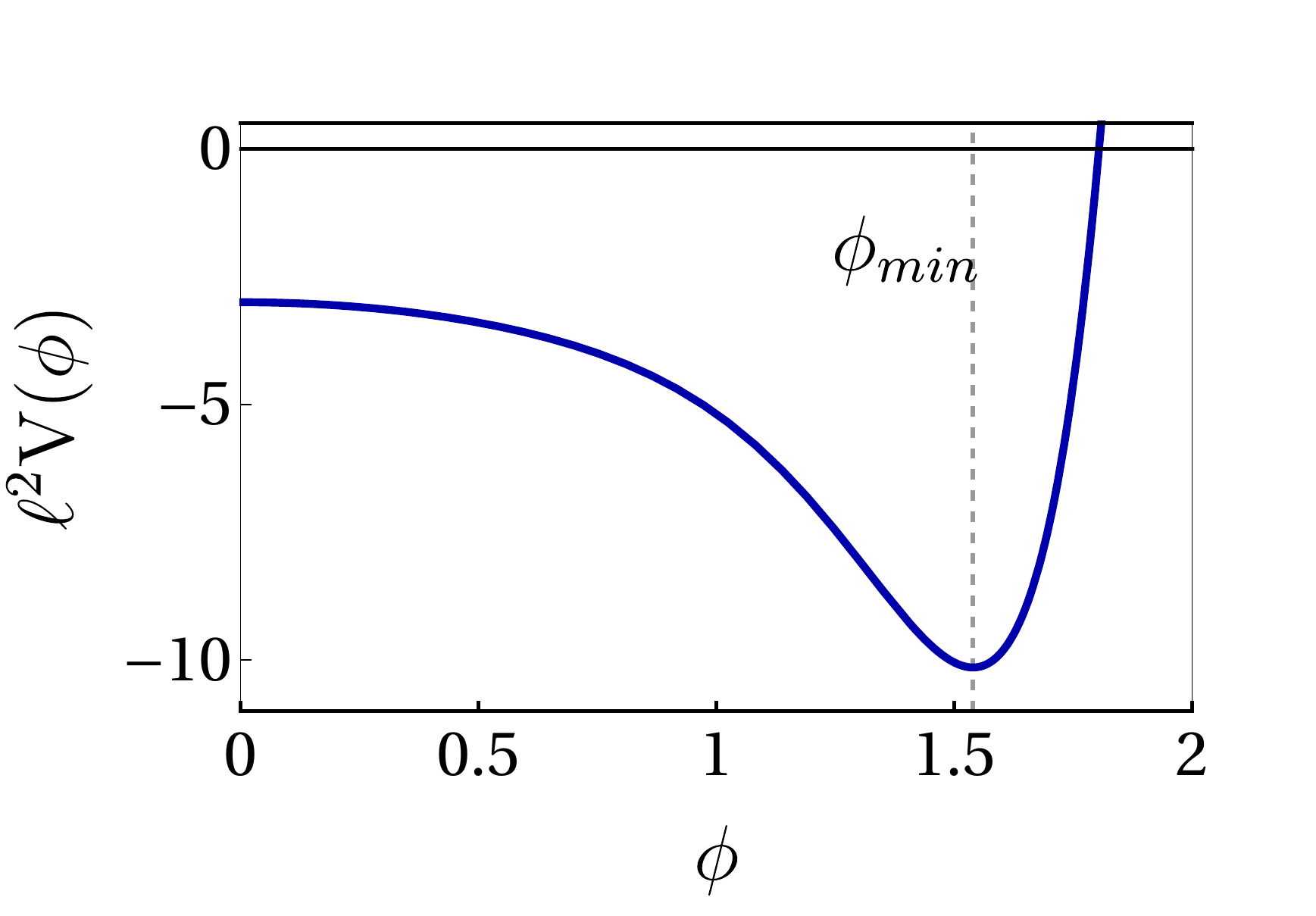}
}
\caption{Superpotential (left) and potential (right) of our model.}
\label{potentialplots}
\end{figure}

We consider the AdS-Einstein-scalar model with action
\begin{equation}
\label{eq:action}
S=\frac{1}{2\kappa^2} \int d^5 x \sqrt{-g} \left[ {\cal R}  - 2 \left( \nabla \phi \right) ^2 - 4 V(\phi) \right ],
\end{equation}
where $\kappa^2=8\pi G_5$ with $G_5$ Newton's constant, 
$g$ is the determinant of the metric $g_{\mu\nu}$, ${\cal R}$ is the associated Ricci scalar and $\phi$ is a real scalar field. The potential $V(\phi)$ can be derived from the superpotential 
\begin{equation} \label{superpotential}
W(\phi)=-\frac{1}{\ell}\left(\frac{3}{2}+\frac{1}{2} \phi^2 + \frac{\phi^4}{4 \phi_M^2} - \frac{\phi^6}{\phi_Q}\right)
\end{equation}
through the usual relation 
\begin{equation}\label{VfromW}
V(\phi)=-\frac{4}{3}W(\phi)^2+\frac{1}{2}W'(\phi)^2 \,.
\end{equation}
The positivity of energy theorem for the AdS-Einstein-Scalar model is subtle \cite{Amsel:2007im}.  For a given potential, one can find up to two superpotentials $W^{\pm}$ that satisfy \eqref{VfromW}. One way of distinguish these two possible solutions is to inspect the small $\phi$ behaviour of the superpotential, namely
\begin{equation}
-\ell\,W^{\pm} = \frac{3}{2}+\frac{\Delta_{\pm}}{2}\phi^2+\mathcal{O}(\phi^4)\,,
\end{equation}
where $\Delta_+=3$ and $\Delta_-=1$ for our potential. It is not a coincidence that $\Delta_{\pm}$ correspond to the conformal dimensions of the operator dual to $\phi$ in standard, and alternative quantisation, respectively. One can show that if $W^-$ exists globally, then so does $W^+$\cite{Amsel:2007im}. Our superpotential \eqref{superpotential} is of the $W^-$ form irrespectively of the value of the dimensionless parameters $\phi_M$ and $\phi_Q$. For a sourced solution such as ours, the existence of $W^-$ ensures that all solutions of our model have positive energy \cite{Amsel:2007im}. For any value of $\phi_M$ and $\phi_Q$ the potential $V(\phi)$ has a maximum at \mbox{$\phi=0$}, corresponding to an ultraviolet (UV) fixed point of the dual gauge theory. We will choose the values $\phi_Q=10$ and $\phi_M=1$, for which  $W$ and $V$  take the form shown in \fig{potentialplots}. 
In this case  both functions  have a minimum at $\phi_{\min}\approx 1.54$, 
corresponding to an infrared (IR) fixed point of the gauge theory. The potential has an additional maximum at $\phi_{\max}\approx 3.65$ and diverges negatively, i.e.~$V(\phi)\to -\infty$, as $\phi\to +\infty$. However, values of $\phi$ larger than $\phi_{min}$ will play no role in our analysis. 

Our motivation to choose this model is simplicity. The superpotential \eqref{superpotential} is the same as in \cite{Attems:2017ezz,Attems:2019yqn,Attems:2018gou} except for the $\phi^6$ term, which was absent in those references but was introduced in \cite{Bea:2018whf}. As in \cite{Attems:2017ezz,Attems:2019yqn,Attems:2018gou,Bea:2018whf}, the dual gauge theory is  a Conformal Field Theory (CFT) deformed by a dimension-three scalar operator with source $\Lambda$. On the gravity side this scale  appears as a boundary condition for the scalar $\phi$. The first two terms in the superpotential are fixed by the asymptotic AdS radius $\ell$ and by the dimension of the dual scalar operator. As in \cite{Attems:2017ezz,Attems:2019yqn,Attems:2018gou}, the present model also possesses a first-order phase transition. In those references this was achieved by choosing the value of $\phi_M$  appropriately, with no need to include a $\phi^6$ term. However, this leads to a phase diagram in which the energy densities at points $D$ and $E$ differ from one another by three orders of magnitude. This huge ratio makes the numerical treatment of the system extremely challenging. In contrast, by including the $\phi^6$ term as in \cite{Bea:2018whf} and choosing the values of $\phi_M$ and $\phi_Q$ as quoted above, this ratio is of order unity, as is clear from \fig{fig:Unif}.

We are interested in finding static, ``lumpy" black brane solutions of \eqref{eq:action} that can break translational invariance along a spatial gauge theory direction $\tilde x$ while being isometric along the remaining two spatial directions $x_2$ and $x_3$ directions. The most general ansatz compatible with such symmetries is
\begin{subequations}
\label{eq:nogauge}
\begin{align}
\mathrm{d}s^2 =& -Q_1(\tilde{x},Z)\mathrm{d}t^2+Q_2(\tilde{x},Z)\mathrm{d}Z^2+2 Q_3(\tilde{x},Z)\mathrm{d}\tilde{x} \mathrm{d}Z+Q_4(\tilde{x},Z)\mathrm{d}\tilde{x}^2
\nonumber \\[2mm]
&+Q_5(\tilde{x},Z)\mathrm{d}x_2^2+Q_6(\tilde{x},Z)\mathrm{d}x_3^2+2\,Q_7(\tilde{x},Z)\mathrm{d}x_2 \mathrm{d}x_3\,,\\[2mm]
\phi =& \: Q_8(\tilde{x},Z)\,, 
\end{align}
\end{subequations}
where $Z$ is the holographic coordinate. We shall be interested in solutions which are isotropic in $x_2$ and $x_3$, so we take $Q_5=Q_6$ and $Q_7=0$\,. To fix the gauge completely we demand $Q_3=0$, and
\begin{subequations}
\begin{align}
&Q_1(\tilde{x},z)= \frac{\ell^2 q_1(\tilde{x},Z)^2}{Z^2} \left(1-\frac{Z^4}{Z_+^4} \right)\,,
\\[2mm]
&Q_2(\tilde{x},Y)= \frac{\ell^2}{Z^2\,q_2(\tilde{x},Z)} \left(1-\frac{Z^4}{Z_+^4}\right)^{-1}\,,
\\[2mm]
&Q_4(\tilde{x},Y)= \frac{\ell^2}{Z^2}q_2(\tilde{x},Z) \, q_3(\tilde{x},Z)\,,
\\[2mm]
&Q_6(\tilde{x},Y)= \frac{\ell^2}{Z^2\,q_1(\tilde{x},Z)}\,,
\\[2mm]
&Q_8(\tilde{x},Y)= \frac{Z}{\ell} \,q_4(\tilde{x},Z)\,,
\end{align}
 together with the condition
\begin{equation}
q_1(\tilde{x},Z_+)^2\,q_2(\tilde{x},Z_+)=\alpha_{\Lambda}\,,
\label{eq:choice}
\end{equation}
\end{subequations}
where $\alpha_{\Lambda}>0$ is a positive constant whose physical significance will be  discussed later; see \eqref{BCs:IR}. The coordinate $\tilde{x}$ is  periodic with period $L$, and we take $\tilde{x}\in[-L/2,L/2]$ and $Z\in[0,Z_+]$. 
The translationally invariant directions $x_{2,3}$ have arbitrary periods $L_{2}$ and $L_{3}$. These will play essentially no role in our discussion, since only densities per unit area in the $23$-plane will matter. Thus we will take them to be the same, i.e.~$L_{2}=L_3\equiv {\cal L}$ and  $x_{2,3}\in [0,{\cal L}]$.

We shall also be interested in solutions which are $\mathbb{Z}_2$-symmetric around $\tilde{x}=0$, which means that we can restrict our domain of integration to $\tilde{x}\in[0,L/2]$, at the expense of imposing $\partial_{ \tilde{x}} q_j\big|_{\tilde{x}=0}=0$, for $j=1,2,3,4$.
In order to vary $L$ in a numerically efficient manner, we further change to a new coordinate 
\be
x=\frac{2\tilde{x}}{L} \,, 
\ee
and take all functions to take values in $x\in [0,1]$.  Note that our ansatz \eqref{ansatzZ} together with our periodicity conditions further imply that $\partial_x q_j\big|_{x=1}=0$ ($j=1,2,3,4$).

Putting everything together brings \eqref{eq:nogauge} to the following simplified form
\begin{eqnarray}\label{ansatzZ}
 \mathrm{d}s^2&=& \frac{\ell^2}{Z^2}\Bigg[ -q_1(x,Z)^2 \left(1-\frac{Z^4}{Z_+^4} \right)\mathrm{d}t^2+\frac{1}{q_2(x,Z)}\left(1-\frac{Z^4}{Z_+^4} \right)^{-1}\mathrm{d}Z^2 \nonumber\\
 &&\hspace{0.7cm} +\left(\frac{ L}{2}\right)^2  q_2(x,Z) \, q_3(x,Z)\mathrm{d}x^2+\frac{1}{q_1(x,Z)}\left(\mathrm{d}x_2^2 + \mathrm{d}x_3^2\right) \Bigg],\\[2mm]
\phi &=&\frac{Z}{\ell} \,q_4(x,Z)\,.\nonumber
\end{eqnarray}  
Our gauge choice is such that the determinant of the metric along the Killing directions $(t,x_2,x_3)$ is fixed and defines the radial (holographic) direction $Z$. The conformal boundary is located at $Z=0$ where we demand
\begin{equation}
q_1=q_2=q_3=1\,.
\end{equation}
In this sense we can denote this gauge choice as the ``double Wick rotation Schwarzschild gauge" and,  as far as we are aware, this is the first time it is introduced. The advantage of this gauge choice (at least in the present system) is that the fields  $q_j$ ($j=1,2,3,4$) have an asymptotic power law decay without irrational powers (nor logarithmic terms; more below), unlike e.g.~the DeTurck gauge.\footnote{\label{footGauge}This feature is particularly important when finding $q_j$ numerically using pseudospectral collocation methods to discretize the numerical grid, as we will do. Due to the absence of the irrational powers near the conformal boundary, the numerical scheme will exhibit exponential convergence when reading asymptotic charges. This is unlike e.g.~the DeTurck gauge that has  power law decays also with irrational powers and therefore does not have exponential convergence in the continuum limit \cite{Donos:2014yya,Marolf:2019wkz}.}

Our gauge choice --- condition \eqref{eq:choice} --- reveals that $Z=Z_+$ is a null hypersurface, where the norm of the Killing vector field $\partial/\partial t$ vanishes.\footnote{Strictly speaking, in order to prove this we need to introduce regular coordinates at the horizon located at $Z=Z_+$. This can be achieved if we use ingoing (or outgoing) Eddington-Finkelstein coordinates of the Schwarzschild brane.} Thus, $Z=Z_+$ is a Killing horizon, and $\alpha_{\Lambda}$ controls its associated surface gravity or, equivalently, temperature. In fact, we find
\begin{equation}
\label{temp}
T=\frac{\sqrt{\alpha_{\Lambda}}}{\pi}\frac{1}{Z_+}\,.
\end{equation}

Our solutions have two important scaling symmetries. The first one  is 
\begin{align}\label{scaling1}
\{t,Z,x_i\}\to \{\lambda_1 t,\lambda_1 Z, \lambda_1 x_i\},\quad \{ q_{1,2,3},q_4 \}\to \{ q_{1,2,3}, \lambda_1^{-1} q_4 \},\quad \{\ell,Z_+\}\to \left\{\lambda_1 \ell,\lambda_1 Z_+ \right\}\,
\end{align}
where \mbox{$x_i=\{x,x_2,x_3\}$}. This  leaves the equations of motion and scalar field invariant and rescales the line element as $\mathrm{d}s^2\to \lambda_1^2 \,\mathrm{d}s^2$, namely $g_{\mu\nu}\to \lambda_1^2 \, g_{\mu\nu}$. It follows that we can use this scaling symmetry to fix the AdS radius to $\ell \equiv 1$. In other words, under the scaling $g_{\mu\nu}\to \lambda_1^2 \, g_{\mu\nu}$, the affine connection $\Gamma^\gamma_{\phantom{\gamma}\mu\nu}$, and the Riemann ($R^\alpha_{\phantom{\alpha}\beta \mu\nu}$) and Ricci ($R_{\mu\nu}$) tensors are left invariant. It follows from the trace-reversed equations of motion that the AdS radius must scale as $\ell \to \lambda_1 \ell$ and we can use this scaling symmetry to set $\ell \equiv 1$.
 
The second scaling symmetry (known as a dilatation transformation, one of the conformal transformations) is  
\begin{align}\label{scaling2}
\{t,Z,x_i\}\to  \{\lambda_2 t,\lambda_2 Z, \lambda_2 x_i\},\quad \{ q_{1,2,3},q_4 \}\to \{ q_{1,2,3}, \lambda_2^{-1}  q_4 \},\quad  \{\ell,Z_+\}\to \{\ell, \lambda_2 Z_+\}\,.
\end{align}
This leaves the metric, scalar field  and equations of motion invariant. It follows that we can use this symmetry to set the horizon radius at 
$Z=Z_+\equiv 1$ or, equivalently, the temperature \eqref{temp} to 
\be
T=\frac{\sqrt{\alpha_{\Lambda}}}{\pi}\,.
\ee
We will see below  that this  is just a convenient choice of units with no effect on the physics. 

Let us turn now our attention to the scalar field. It follows from \eqref{VfromW} and \eqref{superpotential} that the scalar field potential has the  Taylor expansion about $\phi=0$
 \begin{equation}
 V(\phi)\big|_{\phi\sim 0}= -\frac{3}{\ell^2} -\frac{3}{2\ell^2}\,\phi ^2+\mathcal{O}\left(\phi^4 \right) \,,
 \end{equation}
 and thus it describes a scalar field with mass $\mu^2=V''(0)=-3/\ell^2$. According to AdS/CFT, the conformal dimension of the dual operator is simply given by
\begin{equation}
\Delta_\pm=2\pm \sqrt{4+\mu^2\ell^2} \quad \Leftrightarrow  \quad \Delta_-=1 \:\: \hbox{or}  \:\: \Delta_+=3,
\end{equation} 
and these give the two independent asymptotic decays $Z^{\Delta_\pm}$ of the scalar field. Actually, since $\Delta_\pm$ are integers, the nonlinear equations of motion might also generate logarithmic decays of the form $\sum_{n=3} c_n Z^n\ln Z$ where the coefficients $c_n$ depend exclusively on the amplitude of the two independent terms. When this is the case, the conserved charges depend on $c_3$. However, for the potential we use (only with even powers of $\phi$ and double Wick rotation Schwarzschild gauge choice) it turns out that logarithmic terms are not generated by the equations of motion. 
So, for our system, the scalar field decays asymptotically as 
\begin{equation}
\phi\Big|_{Z\to 0}\sim \Lambda Z^{\Delta_-} + \phi_2 Z^{\Delta_+} +\cdots = Z\left( \Lambda + \phi_2 Z^2 \right)+\cdots 
\end{equation}
where $\Lambda$ and $\phi_2$ are two arbitrary integration constants and $\cdots$ represent higher order powers of $Z$ (with no logarithms) whose coefficients are fixed in terms of $\Lambda$ and $\phi_2$ by the equations of motion. The fact that $\Delta_-=1$ motivates our choice of  ansatz for the scalar field in \eqref{ansatzZ}. The  Breitenl\"ohner-Freedman (BF) bound of the system is $\mu_{\rm BF}^2\ell^2=-4$ and thus $\mu^2 \ell^2=-3$ coincides precisely with the unitarity bound $\mu_{\rm BF}^2\ell^2+1$. It follows that only the mode $Z^{\Delta_+}$ with the faster fall-off  is normalizable. In the AdS/CFT correspondence, the non-normalizable mode $\Lambda$ is the source of a boundary operator $\mathcal{O}_\phi$ since it determines the deformation of the boundary theory action. On the other hand, the normalizable modes $\phi_2$ are identified with states of the theory with $\phi_2$ being proportional to the expectation value $\langle \mathcal{O}_\phi \rangle$ of the boundary operator (in the presence of the source $\Lambda$). $\Delta_+=3$ is then the (mass) conformal dimension of the boundary operator $\mathcal{O}_\phi$ dual to $\phi$. Under the  scaling symmetries \eqref{scaling1}-\eqref{scaling2} the scalar source transforms as $\Lambda\to \lambda_1 \lambda_2 \Lambda$.  As a consequence, the ratio $T/\Lambda$ is left invariant by these scalings. Since the physics only depends on this ratio,  setting  $T=\sqrt{\alpha_{\Lambda}}/\pi$ as we did above is just a convenient choice of units  with no effect on the physics. In general, throughout this paper we will measure all dimensionful physical quantities in units of $\Lambda$. 

The undeformed boundary theory --- a CFT --- corresponds thus to the Dirichlet boundary condition $\Lambda=0$, and we have a pure normalizable solution. For this reason the planar AdS$_5$ Schwarzschild solution  \eqref{ansatzZ} with $q_{1,2,3}=1, \,q_4=0$ is often denoted as the (uniform) ``conformal" brane of the theory \eqref{eq:action}. In contrast, if we turn-on the source the  dual gauge theory is no longer conformal. In particular, there are such solutions  \eqref{ansatzZ} with $q_{1,2,3,4}(Z,x)=q_{1,2,3,4}(Z)$ (and $q_4(0)=\Lambda$)  that are translationally invariant along $x,x_2,x_3$. These are often denoted as the ``uniform nonconformal" branes of the theory. This is one family of solutions that we will construct in this manuscript. Still with $\Lambda\neq 0$, we can then have solutions that break translational invariance along $x$ (while keeping the isometries along the other two planar directions). Our main aim is to construct these nonuniform solutions, which we denote as ``lumpy nonconformal branes", and study their thermal competition with the uniform nonconformal branes in a phase diagram of static solutions of  \eqref{eq:action}, both in the micro-canonical and canonical emsembles. Of course there are also nonconformal branes that break translation invariance  along the other two directions $x_{2,3}$. These are cohomogeneity-4 solutions that we will not attempt to construct. Fortunately, the cohomogeneity-2 lumpy branes  that we  will find seem to already allow us to understand the key properties of the most general system.

\subsection{Setup of the boundary-value problem}\label{sec:setupBVP}

Finding the nonconformal brane solutions necessarily requires resorting to numerical methods. 
For that, we find it convenient to change our radial coordinate into  
\begin{equation}\label{coordtransf}
y=\frac{Z^2}{Z_+}\,, \quad \hbox{and define} \quad y_+\equiv \frac{1}{Z_+}\,,
\end{equation}
so that the ansatz \eqref{ansatzZ} now reads 
\begin{eqnarray}\label{ansatz}
 \mathrm{d}s^2&=& \frac{1}{y}\Bigg[-y_+^2 \,q_1(x,y)^2(1-y^2) \mathrm{d}t^2+\frac{1}{q_2(x,y)} \frac{ \mathrm{d}y^2}{4y(1-y^2)} \nonumber\\
 &&\hspace{0.7cm} +\left(\frac{ L}{2}\right)^2 y_+^2  \, q_2(x,y) \, q_3(x,y) \mathrm{d}x^2+\frac{y_+^2}{q_1(x,y)}\left( \mathrm{d}x_2^2 +  \mathrm{d}x_3^2\right)\Bigg],\\[2mm]
\phi &=&\frac{\sqrt{y}}{y_+} \,q_4(x,y)\,, \nonumber
\end{eqnarray}
with compact coordinates  $y\in [0,1]$ and $x\in [0,1]$. The horizon is located at $y=1$ and the asymptotic boundary at $y=0$. Note, that we used the two scaling symmetries \eqref{scaling1}-\eqref{scaling2} to set $\ell\equiv 1$ and $y_+\equiv 1$. 

In these conditions we now need to find the minimal set of Einstein-scalar equations --- the equations of motion (EoM) --- that allows us to solve for all $q_j(x,y)$  while closing the full system of equations, $g_{\mu\nu}=T_{ab}$ and $\Box \phi=0$, of the action \eqref{eq:action}. This is a nested structure of PDEs. This structure motivates also in part our original choice of gauge in the ans\"atze \eqref{ansatzZ} and \eqref{ansatz}. For this reason, rather than presenting the final EoM, it is instructive to explain their origin and nature. Prior to any gauge choice and symmetry assumptions, the differential equations that solve \eqref{eq:action}  are second order for all the fields. The symmetry requirements we made fix some of these fields. Additionally, the fact that we have chosen to fix the gauge freedom of the system using the `double Wick rotation Schwarzschild' gauge means that our system of equations should have the structure of an ADM-like system but with the spacelike coordinate $x$ playing the role of ``time". Of course, our problem is ultimately an  elliptic problem. However, it is instructive to analyse our EoM adopting the above ``time-dependent" viewpoint. In doing so, one expects that the equations of motion take a nested structure of PDEs that include a subset of ``evolution'' equations (to be understood as evolution in the $x$-direction) but also a subset of non-dynamical (``slicing'' and ``constraint'') equations. We now describe in detail this nested structure. 

We have a total of five equations of motion. 
Two of these EoM are dynamical {\it evolution} equations for $q_1$ and $q_4$. The main building block of these equations is the Laplacian operator $\partial_x^2+\partial_y^2$  (acting either on $q_1$ or $q_4$) that describes the spatial dynamics as the system evolves in $x$. With respect to the familiar ADM time evolution in the Schwarzschild gauge, this Laplacian replaces the wave operator $\partial_t^2-\partial_y^2$. Additionally, we have two (first-order) {\it slicing} EoM for $\partial_y q_2$ and $\partial_y q_3$ that can be solved at each constant-$x$ spatial slice for $q_2$ and $q_3$. Besides depending on $\partial_y q_{2,3}$ and $q_{2,3}$ --- but, quite importantly, not on $\partial_x q_{2,3}$ --- these equations also depend on $q_{1,4}$ (that are determined ``previously'' by the evolution equations) and on their first derivatives (both along $x$ and $y$). This means that we can integrate the slicing equations of motion along the radial direction to find $q_{2,3}$ at a particular constant-$x$ slice. Finally, the EoM still includes a fifth PDE that expresses $\partial_x q_2$ as a function of $\left(q_{1,2,3,4},\partial_x q_{1,4}, \partial_y q_{1,4}\right)$. Let us schematically denote it as ${\cal C}(x,y)=0$. This is a {\it constraint} equation. To see this note that, after using the evolution and slicing EoM and their derivatives, the Bianchi identity $\nabla^{\mu}(R_{\mu\nu}-g_{\mu\nu}R/2)=0$ implies the constraint evolution relation\footnote{Note that in standard ADM {\it time}-evolution problems the constraint relation that must vanish involves the {\it time} derivative, i.e.~it is schematically of the form  $\partial_t \left(\sqrt{-g}  \,\tilde{\cal C} \right)+\tilde{F}(t,y) \sqrt{-g}  \,\tilde{\cal C}=0$, where $y$ is a radial coordinate.  Interestingly, in our `double Wick rotation of the ADM gauge' --- where $x$ is the evolution coordinate --- it is the radial derivative $\partial_y$ (and {\it not} $\partial_x$) that appears in the vanishing constraint relation.} 
\begin{equation}\label{ConstraintRelation}
\partial_y \left(\sqrt{-g}  \,{\cal C} \right)+F(x,y) \sqrt{-g} \, {\cal C}=0 \,,
\end{equation}
where $F(x,y)$ is a function that is regular at the horizon whose further details are not relevant.\footnote{Let $\hat{\cal C}\equiv\sqrt{-g} \,{\cal C}$. Solving  $\partial_y \hat{\cal C}+F(x,y) \,\hat{\cal C}=0$ yields $\,\hat{\cal C}(x,y)=\hat{\cal C}(x,1)\exp\left( \int_1^y F(x,Y) dY \right)$ which converges if $F(x,y)$ is regular at the horizon $y=1$.} It follows from this  constraint evolution relation that if the constraint equation is obeyed at a given $y$, say at the horizon ${\cal C}(x,y)\big|_{y=1}=0$, then it is obeyed at any other $y \in [0,1]$. In practice, this means that we just need to impose the constraint as a {\it boundary condition} at $y=1$, say. It is then preserved into the rest of the domain.

Altogether, the strategy to solve the EoM is thus the following. There are effectively four EoM, two second order PDEs for $q_{1,4}(x,y)$ and two second order PDEs for $q_{2,3}(x,y)$. We need to solve these equations  for $q_{1,2,3,4}(x,y)$ as a boundary-value problem. 
One of the boundary conditions is imposed at the horizon and takes the form ${\cal C}(x,y)\big|_{y=1}=0$, whereas the others are the physically motivated boundary conditions discussed next.

Our integration domain is a square bounded in the radial direction by $y=0$ (the asymptotic boundary) and $y=1$ (the horizon). Along the 
$x$-direction the boundaries are at $x=0$ and $x=1$.
The angular coordinate $x$ is periodic in the interval  $[0,1]$. We can thus use this symmetry to  impose Neumann boundary conditions for all $q_j$ at $x=0$  and  $x=1$:
\begin{eqnarray}\label{BCs:x}
&& \partial_x q_j(x,y)\big|_{x=0}=0\,, \quad \hbox{for} \:\: j=1,2,3,4\,, \\
&&  \partial_x q_j(x,y)\big|_{x=1}=0\,, \quad \hbox{for} \:\: j=1,2,3,4.
\end{eqnarray}
Consider now the asymptotic UV boundary at $y=0$. The invariance of the EoM under dilatations
\eqref{scaling2} guarantees that asymptotically our PDEs are of the Euler type and thus $y=0$ is a regular singular point. The order of our PDE system (two second-order and two first-order PDEs) is 6. It follows that we have a total of 6 free UV independent parameters. We have explicitly checked that, for our gauge choice and scalar potential, all our functions $q_j$ admit a Taylor expansion in integer powers of $y$ (in particular, without logarithmic terms) that contains precisely 6 independent parameters. In this Taylor expansion, at a certain order (as expected due to the fact that we fixed the gauge and our system is cohomogeneity-2) we need to use a differential relation that is ultimately enforced by the Bianchi identity:
\begin{equation}\label{Bianchi}
q_2^{(1,2)}(x,0)=\frac{8}{3}\,\Lambda\,q_4^{(1,1)}(x,0).
\end{equation}
The requirement that our nonconformal branes asymptote to AdS fixes two of the six UV integration constants to unity, namely $q_{1}(x,0)$ and $q_{3}(x,0)$ at the boundary (it then follows directly from the EoM that $q_{2}(x,0)=1$). We complement these boundary conditions with a Dirichlet boundary condition for the scalar field function $q_4$ which introduces the source $\Lambda$. This will be a running parameter in our search for solutions. Altogether we thus impose the boundary conditions at the UV boundary:
 \begin{equation}\label{BCs:UV}
q_j(x,y)\big|_{y=0}=1\,, \quad \hbox{if} \:\: j=1,2,3;\qquad q_4(x,y)\big|_{y=0}=\Lambda.
\end{equation}
Finally, we discuss the boundary conditions imposed at the horizon, $y=1$. It follows directly from the four equations of motion that $q_{1,2,3,4}(x,1)$ are free independent parameters and $q_{1,2,3,4}$ must obey a set of four mixed conditions that fix their first radial derivative as a function of $q_{1,2,3,4}(x,1)$ and $\partial_x q_{1,2,3,4}(x,1)$ that is not enlightening to display.
When assuming a power-law Taylor expansion for 
\be
q_j(x,y)\big|_{y\sim 1}=\underset{k=0}{\sum}  c_{j(k)}(x)(1-y)^k
\ee
 we are  already imposing  boundary conditions that discard two integration constants that would describe contributions that diverge at the horizon.

Having imposed the boundary conditions, we must certify that we have a well-defined elliptic (boundary-value) problem. For that, we need to confirm that the number of free parameters at the UV boundary matches the IR number of free parameters. Recall that, before imposing boundary conditions, we have 6 integration constants at the UV and another 6 in the IR. A power-law Taylor expansion about the asymptotic boundary, 
\be
q_j(x,y)\big|_{y\sim 0}=\underset{k=0}{\sum} a_{j(k)}(x) y^k \,, 
\ee
concludes that, after imposing the boundary conditions \eqref{BCs:UV} that fix $a_{1(0)}=1, a_{3(0)}=1$ and $a_{4(0)}=\Lambda$ ($a_{2(0)}$ is not a free parameter since it is fixed by the equations of motion), we are left with three free UV parameters, namely $a_{1(2)}(x), a_{2(2)}(x), a_{4(1)}(x)$. Note that we will give $\Lambda$ as an input parameter so the boundary-value problem will not have to determine it. As we shall find later, the energy density depends on these three parameters, whereas the expectation value  of the dual operator sourced by $\Lambda$ is proportional to $a_{4(1)}(x)$. On the other hand, at the horizon, after imposing the aforementioned boundary conditions that eliminate two integration constants, one finds that there are 4 free IR parameters.

So we have 3 UV free parameters but 4 IR free parameters. In order  to have a well-defined boundary value problem (BVP) the number of free UV parameters must match the IR number. Note however that in our discussion of the boundary conditions we have not yet imposed the constraint equation ${\cal C}(x,y)=0$. As described above we just need to impose it at the horizon $y=1$, where ${\cal C}(x,1)=0$ simply reads 
\be
2 q_2(x,1) \partial_x q_1(x,1)^2+ q_1(x,1)\partial_x q_2(x,1)=0 \,.
\ee
This is solved by $q_1(x,1)=\sqrt{\alpha_{\Lambda}}/\sqrt{q_2(x,1)}$ where $\alpha_{\Lambda}$ is a constant to be fixed below. It follows that, if we impose this Dirichlet condition together with the three aforementioned mixed conditions for $q_{2,3,4}$ at the  horizon
 \begin{equation}\label{BCs:IR}
q_1(x,y)\big|_{y=1}=\frac{\sqrt{\alpha_{\Lambda}}}{\sqrt{q_2(x,1)}}\,; \qquad \partial_y q_j\big|_{y=1}=  \partial_y q_j\left(q_{1,2,3,4},\partial_x q_{1,2,3,4}\right)\big|_{y=1} \quad \hbox{if} \:\: j=2,3,4,
\end{equation}
then we have just three free IR parameters, namely $q_{2,3,4}(x,1)$. We fix the value of $\alpha_{\Lambda}$ as follows. For a given source $\Lambda$, the lumpy nonconformal branes that we seek merge with the {\it uniform} nonconformal branes at the onset of the Gregory-Laflamme-type instability of the latter.  We use this merger (where the fields of the two solutions must match) to fix the constant $\alpha_{\Lambda}$ to be the value of $q_1(y)^2q_2(y)\big|_{y=1}$ (no $x$-dependence) of the nonuniform solution when it merges with the uniform brane. 

This discussion can be complemented as follows (which also allows us to set the IR boundary condition we impose to search for the uniform branes). Note that when looking for {\it uniform} branes, the PDE system of EoM reduces to an ODE system without any $x$-dependence. In particular, the constraint equation ${\cal C}(x,1)=0$ reduces to ${\cal C}(1)=0$ and is trivially obeyed, since all of its terms involve terms with partial derivatives in $x$. So when searching for uniform branes we use the IR boundary conditions \eqref{BCs:IR} but with the first condition replaced by the Dirichlet condition $q_1(y)\big|_{y=1}=1$:
 \begin{equation}\label{BCs:IRunif}
q_1(y)\big|_{y=1}=1\,; \qquad \partial_y q_j\big|_{y=1}=  \partial_y q_j\left(q_{2,4}\right)\big|_{y=1} \quad \hbox{if} \:\: j=2,3,4; \qquad\hbox{(if uniform branes).}
\end{equation}
This is a choice of normalization that does not change physical thermodynamic quantities.
 Solving the EoM for a given source $\Lambda$ we then find, in particular, the value of $q_2(y)$ at $y=1$. We can then read the constant $\alpha_{\Lambda}=q_1(y)^2q_2(y)\big|_{y=1}=q_2(y)\big|_{y=1}$ of the uniform brane with source $\Lambda$. This value of $\alpha_{\Lambda}(\Lambda)$ is then the one we plug in the boundary condition \eqref{BCs:IR} to find the {\it non}-uniform branes with the same source $\Lambda$. The UV boundary conditions for the uniform branes is still given by \eqref{BCs:UV} (with the replacement $q_j(x,y)|_{y=0}\to q_j(y)|_{y=0}$).

\subsection{Thermodynamic quantities}\label{sec:setupThermo}

Having found the nonconformal solutions \eqref{ansatz} that obey the boundary conditions \eqref{BCs:x}-\eqref{BCs:IR}, we will now implement the holographic renormalization procedure in order to obtain the relevant 
thermodynamic quantities (recall that $\ell\equiv 1$ and $y_+\equiv 1$).
 Our (non)uniform branes are asymptotically $AdS_5$ solutions with a scalar field with mass $\mu^2=-3$. In these conditions, the holographic renormalization procedure to find the holographic stress tensor ${\cal T}_{ab}$ and expectation value $\langle {\cal O}_{\phi} \rangle$ of the operator dual to the scalar field $\phi$ was developed in \cite{Bianchi:2001de,Bianchi:2001kw}. We apply it to our system. We first need to introduce the Fefferman-Graham (FG) coordinates $(z,\chi)$ that are such that the asymptotic boundary is at $z=0$ and $g_{zz}=1/z^2$ and $g_{za}=0$ (with $a=t,\chi,w_{2,3})$ at all orders in a Taylor expansion about $z=0$. In terms of the radial and planar coordinates $(y,x)$ of \eqref{ansatz} the FG coordinates are
 \begin{align}  \label{FGcoord}
&y= z^2 + \frac{\Lambda ^2}{3}z^4 + \frac{1}{12} z^6 \left(\Lambda^4-3+\frac{3}{2} q_2^{(0,2)}(\chi,0)\right) +\mathcal{O}(z^8);  \nonumber \\
&x=\chi-\frac{q_2^{(1,2)}(\chi,0)}{96 L^2}\,z^6 +\mathcal{O}(z^8).
\end{align}    
The expansion of the gravitational and scalar fields around the boundary up to the order that contributes to the thermodynamic quantities is then
\be
ds^2= \frac{1}{z^2}\left[ dz^2 + ds^2_{\partial}+z^2 \,ds^2_{(2)}+z^4\, ds^2_{(4)}+\mathcal{O}(z^6)\right] 
\label{FGexpansion1}
\ee
where
\begin{eqnarray}\label{FGexpansion2}
&& ds^2_{\partial}=g_{ab}^{(0)}{\rm d}x^a {\rm d}x^b=-dt^2+L^2 d\chi^2+ dx_2^2+ dx_3^2,  \nonumber\\[2mm]
&& ds^2_{(2)}=g_{ab}^{(2)}{\rm d}x^a {\rm d}x^b=-\frac{\Lambda^2}{3}\,ds^2_{\partial}\,,
\nonumber\\[2mm]
&&  ds^2_{(4)}=g_{ab}^{(4)}{\rm d}x^a {\rm d}x^b= \frac{1}{36}\left( 27-\Lambda^4-36  \, q_1^{(0,2)}(\chi,0)+\frac{9}{2}\,q_2^{(0,2)}(\chi,0) \right)dt^2\nonumber\\
&& \hspace{1.3cm}+ \frac{L^2}{36}\left( 9-7 \Lambda^4+\frac{27}{2} \,q_2^{(0,2)}(\chi,0)-72  \,\Lambda\,q_4^{(0,1)}(\chi,0)  \right) d\chi^2  \nonumber\\
&& \hspace{1.3cm}+ \frac{1}{36}\left( 9+ \Lambda^4-18  \,q_1^{(0,2)}(\chi,0)-\frac{9}{2} \,q_2^{(0,2)}(\chi,0)  \right) \Big( dx_2^2+ dx_3^2 \Big)\,;  \\[2mm]
&& \phi = \Lambda\,z+\phi_2 \,z^3+ \mathcal{O}(z^5) \,,
\label{FGexpansion3} \\[2mm] 
&& \phi_2= \left( \frac{\Lambda^3}{6}+q_4^{(0,1)}(\chi,0) \right)\,.
\label{FGexpansion4}
\end{eqnarray}
The holographic quantities can now be computed using the holographic renormalization procedure of Bianchi-Freedman-Skenderis \cite{Bianchi:2001de,Bianchi:2001kw}.\footnote{Note however that we use different conventions for the Riemann curvature, that is to say, with respect to \cite{Bianchi:2001de,Bianchi:2001kw} our action \eqref{eq:action} has the opposite relative sign between the Ricci scalar ${\cal R}$ and the scalar field kinetic term  $\left( \nabla \phi \right)^2$.}
At the end of the day, for our system, the expectation value of the holographic stress tensor is given by 
\begin{equation}\label{holoT}
 \langle {\cal T}_{ab} \rangle
=\frac{2\ell^3}{\kappa^2} \Bigg[ g_{ab}^{(4)}+ g_{ab}^{(0)}\Bigg( \Lambda\, \phi_2-\frac{\Lambda^4}{18} -\frac{\Lambda^4}{4\phi_M^2} \Bigg)  \Bigg]\,,
\end{equation}
where the metric components $g_{ab}^{(0)}$, $g_{ab}^{(4)}$ and the scalar decay $\phi_2$ can be read directly from \eqref{FGexpansion1}-\eqref{FGexpansion4},  and we recall that $\phi_M$ is a parameter of the superpotential \eqref{superpotential} that we will eventually set to  $\phi_M=1$.
Similarly, the expectation value of the dual operator sourced by $\Lambda$ is
\begin{equation}\label{holoVEV}
 \langle {\cal O}_{\phi} \rangle
=\frac{2\ell^3}{\kappa^2} \Bigg( \frac{\Lambda^3}{\phi_M^2} -2\phi_2 \Bigg).
\end{equation}
The trace of the  expectation value yields the expected Ward identity associated to the conformal anomaly
\be
\langle {\cal T}_{a}^{\:a} \rangle=-\Lambda  \langle {\cal O}_{\phi} \rangle \,,
\ee
 which reflects the fact that our branes are not conformal.\footnote{Note that the  holographic gravitational conformal anomaly contribution ${\cal A}_{\rm grav}$ \cite{Bianchi:2001de,Bianchi:2001kw} and the scalar conformal anomaly contribution ${\cal A}_{\rm scalar}$ vanish for our system.} 
Furthermore,  after using the Bianchi relation \eqref{Bianchi}, we confirm that the expectation value of the holographic stress tensor is conserved, i.e.
\be
\nabla^a\langle {\cal T}_{ab} \rangle=- \langle {\cal O}_{\phi} \rangle\nabla_b \Lambda=0 \,.
\ee
In \eqref{holoT} and \eqref{holoVEV} we have reinstated the appropriate power of $\ell$ in order to remind the reader that, for an $SU(\nc)$ gauge theory, the prefactor in these expressions typically scales as 
\be
\frac{2\ell^3}{\kappa^2} \propto \nc^2 \,.
\ee
In the rest of the paper we will work with rescaled quantities obtained by multiplying the stress tensor and the scalar operator by the inverse of this factor. 

Evaluating \eqref{holoT} explicitly we find the following expressions for the energy density $\mathcal{E}$, the longitudinal pressure $P_L$ (along the inhomogenous direction $x$) and the transverse pressure $P_T$ (along the homogenous directions $x_2$ and $x_3$):
\begin{align}  
\mathcal{E} (\chi)&=
\frac{\Lambda^4}{4}\left(\frac{1}{\phi_M^2}-\frac{5}{9}\right)
+\frac{1}{4}\left(3-4\, q_1^{(0,2)}(\chi,0) +\frac{1}{2}\,q_2^{(0,2)}(\chi,0)\right)-\Lambda\,q_4^{(0,1)}(\chi,0)\label{E} \,, \\[2mm]
P_L&=
 \frac{\Lambda^4}{4}\left(\frac{1}{\phi_M^2}+\frac{1}{3}\right)
-\frac{1}{4}\left(1+3\, q_2^{(0,2)}(\chi,0) \right)+
\Lambda\,q_4^{(0,1)}(\chi,0) \,, \label{PL} \\[2mm]
P_T(\chi)&= 
 \frac{\Lambda^4}{4}\left(\frac{1}{\phi_M^2}-\frac{5}{9}\right)
-\frac{1}{4}\left(1-4\, q_1^{(0,2)}(\chi,0) -q_2^{(0,2)}(\chi,0)\right)-\Lambda\,q_4^{(0,1)}(\chi,0) \,. \label{PT}
\end{align}
Note that $P_L$ is the pressure conjugate to the dimensionful coordinate $\tilde{x}$, not to the dimensionless coordinate $x$. Moreover, for static configurations, conservation of the stress tensor implies that $P_L$ is constant along the inhomogeneous direction, i.e.~independent of $\chi$. 
The temperature $T$ and the entropy density $s$ of the nonconformal branes can be read simply from the surface gravity and the horizon area density  of the solutions \eqref{ansatz}, respectively:
\begin{eqnarray}\label{calTS}
&& T=\frac{\sqrt{\alpha_{\Lambda}}}{\pi}\,, \nonumber\\
&& s= \pi\sqrt{\alpha_{\Lambda}}\,\sqrt{q_3(\chi,1)} \Big(q_1(\chi,1)\Big)^{-2}\,,
\end{eqnarray}
where have already used the boundary condition \eqref{BCs:IR} that introduces the constant $\alpha_{\Lambda}$ (that we read from the uniform solutions; see discussion below \eqref{BCs:IR}). The  Helmoltz free energy density is  ${\cal F}={\cal E}-T s$. The total energy $E$, entropy $S$ and free energy $F$ are obtained 
by integrating over the total volume:
\be
\label{totalcharges}
E = \mathcal{L}^2 \, L \, \int_0^1 d\chi \, \mathcal{E}(\chi) \,, \qquad 
S = \mathcal{L}^2 \, L \, \int_0^1 d\chi \, s(\chi) \,, \qquad 
F = \mathcal{L}^2 \, L \, \int_0^1 d\chi \, \mathcal{F}(\chi) \,,  
\ee
where we have made use of the fact that the system is homogeneous in the transverse directions. It will also be useful to define average densities  by dividing the integrated quantities by the total volume: 
\be
\overline {\cal E} =  \frac{E}{L \mathcal{L}^2} \,, \qquad 
\overline s =  \frac{S}{L \mathcal{L}^2} \,, \qquad 
\overline f =  \frac{F}{L \mathcal{L}^2} \,.
\ee
For uniform branes these averages coincide with the corresponding densities, since the latter are constant, but for nonuniform branes they do not. 
A quantity that will play a role below is an analogous integral for the expectation value of the scalar operator:
\be
\mathcal{O} = \mathcal{L}^2 \, L \, \int_0^1 d\chi \, 
\langle \mathcal{O}_\phi \rangle (\chi) \,.
\ee
Because of the translational invariance in the transverse directions it will also be convenient to work with densities in the transverse plane, namely with quantities that are only integrated along the inhomogeneous direction. Thus we define the energy, the entropy, the free energy  and the expectation value densities per unit area in the transverse plane as 
\be
\rho =  \frac{E}{\mathcal{L}^2} \,, \qquad 
\sigma =  \frac{S}{\mathcal{L}^2} \,, \qquad 
f =  \frac{F}{\mathcal{L}^2} \,, \qquad 
\vartheta = \frac{\mathcal{O}}{\mathcal{L}^2} \,.
\ee
We will refer to these type of quantities as ``area densities" or ``Killing densities". 
In order to write the first law we will also need the integral of the transverse pressure along the inhomogeneous direction. We therefore define
\be
\frak{p}_L= P_L \,,\qquad \frak{p}_T = L \, \int_0^1 d\chi \, P_T (\chi) \,.
\ee
Note that $\frak{p}_L$ and $\frak{p}_T$ have mass dimension 4 and 3, respectively. 
Finally, we will choose to measure all dimensionful quantities in units of the only gauge theory microscopic scale $\Lambda$. We will  use a ``$\hat{\,\,\,\,}$" symbol to denote the corresponding dimensionelss quantity obtained by multiplying or dividing a dimensionful quantity by the appropriate power of $\Lambda$, thus:
\be
\hat{L}= \Lambda L\,,\qquad  
\hat{T}= \frac{T}{\Lambda} \,,\qquad\hat{\mathcal{E}}=\frac{\mathcal{E}}{\Lambda^4} \,,\qquad  
\hat{\overline{\cal E}}= \frac{\overline{\cal E}}{\Lambda^4} \,,\qquad  
\hat{\rho}=\frac{\rho}{\Lambda^3} \,,\qquad  \hat{f}=\frac{f}{\Lambda^3} \,,\qquad 
 \mbox{etc.}
\ee

We are now ready to write down the first law. In order to do this, we first note that the extensive thermodynamic variables of the system are the total energy $E$, the total entropy $S$, the scalar source $\Lambda$, and the three lengths $L,\,L_2\equiv {\cal L},L_3\equiv {\cal L}$ of the planar directions. It follows that the first law for the total charges of the system is:
\begin{equation}\label{1stA}
\dd E= T \,\dd S+ {\cal O} \,\dd\Lambda+ 
\frak{p}_{L} \, \mathcal{L}^2 \, \dd L 
+ 2 \, \frak{p}_{T} \, \mathcal{L} \, \dd {\cal L} \,.
\end{equation}
We see that $T$, ${\cal O}$, $\frak{p}_L \mathcal{L}^2$ and 
$\frak{p}_T \mathcal{L}$ are the  potentials (intensive variables) conjugate to $S,\Lambda, L$ and ${\cal L}$, respectively. Since the system is translationally invariant along the $x_2$ and $x_3$ directions, under the associated scale transformation $x_{2,3}\to \lambda_0 \,x_{2,3}$ the energy transforms as $E(x_{2,3})\to \lambda_0^2 E(x_{2,3})$ and thus it is a homogeneous function of $\lambda_0$ of degree 2. This means that  for any value of $\lambda_0$ one has:
\begin{equation}\label{EulerA1}
E\left(\lambda_0^2 \,S,\Lambda, L, \lambda_0 \,L_2, \lambda_0 \,L_3\right)= \lambda_0^2 \,E\left(S,\Lambda, L, L_2,  L_3\right).
\end{equation}
We can now apply Euler's theorem for homogeneous functions to write the energy as a function of its partial derivatives:\footnote{Essentially, in the present case, Euler's theorem amounts to take a derivative of the homogeneous relation \eqref{EulerA1} with respect to $\lambda_0$ and then sending $\lambda_0\to 1$.}
\begin{equation}\label{EulerA2}
2 S \frac{\partial E}{\partial S} + L_2 \frac{\partial E}{\partial L_2}+ L_3 \frac{\partial E}{\partial L_3}
=2 E\left(S,\Lambda, L, L_2,  L_3\right).
\end{equation}
The relevant partial derivatives in \eqref{EulerA2} can be read from \eqref{1stA} and, recalling that we are taking $L_2=L_3\equiv {\cal L}$, this yields the Smarr relation for the total charges of the system
\begin{equation}\label{smarrA}
 E=T S +  \frak{p}_T {\cal L}^2\,.
\end{equation}
Dividing by ${\cal L}^2$ we obtain a Smarr relation for the area densities along the transverse plane:
\begin{equation}\label{smarrB}
 \rho=T \sigma +  \frak{p}_T\,.
\end{equation}
Rewriting the first law \eqref{1stA} in terms of these densities and using  \eqref{smarrB} we find the first law for the area densities:
\begin{equation}\label{1stB}
\dd \rho= T \,\dd \sigma+ \frak{p}_{L} \,\dd L+ \vartheta \,\dd\Lambda \,.
\end{equation}
Since we will measure all dimensionful quantities in units of $\Lambda$, it will be useful to find a first law and a Smarr relation for the dimensionless  densities $\hat{\rho}, \hat{\sigma}$, etc. In order to do this  we first use the dilatation transformation \eqref{scaling2}. Under this scale transformation $x^\mu\to \lambda_2 \,x^\mu$ the  energy density $\rho$ transforms as $\rho(x^{\mu})\to \lambda_2^3 \,\rho(x^{\mu})$ and thus it is a  homogeneous function of $\lambda_2$ of degree 3, i.e.
\begin{equation}\label{1stE}
\rho\left(\lambda_2^2 \,\sigma,\lambda_2\, \Lambda, L/\lambda_2 \right)= \lambda_2^3 \,\rho\left(\sigma,\Lambda, L \right).
\end{equation}
Applying Euler's theorem for homogeneous functions we get
\begin{equation}\label{1stF}
2 \sigma \frac{\partial \rho}{\partial \sigma} + \Lambda \frac{\partial \rho}{\partial \Lambda}-L \frac{\partial \rho}{\partial L}
=3 \rho\left(\sigma,\Lambda, L \right) \,.
\end{equation}
Reading  the associated derivatives from the first law \eqref{1stB} we find (another) Smarr relation for the dimensionful  densities:
\begin{equation}\label{smarrC}
3 \, \rho=2 \,T\, \sigma - \frak{p}_L\, L+ \vartheta \,\Lambda\,.
\end{equation}
Dividing this relation by $\Lambda^3$ we get the Smarr relation for the dimensionless  densities:
\begin{equation}\label{Smarr}
3 \, \hat{\rho}=2\, \hat{\taut}\, \hat{\sigma} - \hat{\frak{p}}_L\, \hat{L}+ \hat{\vartheta} \,.
\end{equation}
Finally, we can now rewrite the first law \eqref{1stB} in terms of the dimensionless area densities and use  \eqref{Smarr}  to find the desired first law
\begin{equation}\label{1stLaw}
\dd \hat{\rho}= \hat{T} \,\dd \hat{\sigma}+ \hat{\frak{p}}_{L} \,\dd \hat{L}
\end{equation}
that nonconformal branes with two Killing planar directions $x_{2,3}$ must obey. In a traditional thermodynamic language the first law \eqref{1stLaw} and the Smarr relation \eqref{Smarr} are also known as the Gibbs-Duhem and Euler relations, respectively. In our case they provide valuable tests of  our numerical results. Moreover, they will be useful  to discuss the dominant thermal phases in the microcanonical and canonical ensembles. 
Indeed, in the microcanonical ensemble the dominant phase will be the one that maximises $\hat{\sigma}$ for fixed values of $\hat{\rho}$ and $\hat{L}$. Similarly, in the canonical ensemble the dominant phase will be the one that minimises $\hat{f}$ for fixed $\hat{\taut}$ and $\hat{L}$.

\subsection{Perturbative construction of lumpy branes}\label{sec:Perturbative}

In the previous sections we have setup the BVP that will allow us to find the uniform and nonuniform nonconformal branes \eqref{ansatz} that obey the boundary conditions \eqref{BCs:x}-\eqref{BCs:IR}. This nonlinear BVP can be solved in full generality  using numerical methods. In the uniform case we have a system of coupled quasilinear ODEs that can be solved without much effort. However, in the nonuniform case the ODEs are replaced by PDEs and it is harder to solve the system. We will do this  numerically in \sect{sec:PhaseDiag}. In the present section we will  complement this full numerical analysis with a perturbative nonlinear analysis that finds lumpy branes in the region of the phase diagram where they merge with the uniform branes. This perturbative analysis will already provide valuable physical properties of the system. Additionally, these perturbative results will also be important to test the numerical results of \sect{sec:PhaseDiag}. We  solve the  BVP in perturbation theory up to an order in the expansion parameter where we can distinguish the thermodynamics of the uniform and nonuniform branes. 

We follow a perturbative approach that was developed in \cite{Dias:2017coo} (to find vacuum lattice branes) and that has its roots in \cite{Gubser:2001ac,Wiseman:2002zc,Sorkin:2004qq} (to explore the existence of vacuum nonuniform black strings). More concretely, our strategy to find perturbatively the lumpy branes has three main steps:
\begin{enumerate}
\item The first step is to construct the uniform branes. 
\item Then, at linear ($n=1$) order in perturbation theory, we find the locus in the  space of uniform branes where a zero-mode, namely a mode that is marginally stable, exists. We will refer to this mode as the GL-mode. In practice, we will identify this locus by finding the critical length $L=L_{\hbox{\tiny GL}}$ (wavenumber $k_{\hbox{\tiny GL}}=2\pi/L_{\hbox{\tiny GL}}$) above (below) which uniform branes become locally unstable (stable). As expected from the discussion in \sect{sec:Intro}, this critical length only exists for energy densities between points $A$ and $B$. 
\item The third step is to extend perturbation theory to higher orders, $n\geq 2$, and construct the nonuniform (lumpy) branes that bifurcate  
(in a phase diagram of solutions) from the GL merger curve of uniform branes. 
\end{enumerate}
We describe in detail and complete these three steps in the next three subsections.

\subsubsection{Uniform branes: ${\cal O}(0)$ solution}\label{sec:PerturbativeO0}

The first step is to construct the uniform branes. We solve the system of four coupled ODEs for $q_j(x,y)\equiv Q_j(y)$ (here and below, $j=1,\ldots ,4$)  subject to the boundary conditions \eqref{BCs:UV} and \eqref{BCs:IRunif}, as described in \sect{sec:setupBVP}. This can be done only numerically: we use the numerical methods detailed in the review \cite{Dias:2015nua}. 

There is a 1-parameter family of uniform nonconformal branes. We can take this parameter to be the scalar field source $\Lambda$. This is actually how we construct these solutions since $\Lambda$ is an injective parameter: we give the source $\Lambda$ via the boundary condition \eqref{BCs:UV} and find the associated brane; then we repeat this for many other values of $\Lambda$. The dimensionless energy density $\hat{\cal E}={\cal E}/\Lambda^4$ decreases monotonically as $\Lambda$ grows, so this procedure maps out all possible uniform branes. 
Recall that, once we have found $q_j(x,y)\equiv Q_j(y)$, the thermodynamic quantities of the solution follow straightforwardly from  \sect{sec:setupBVP}.

\begin{figure}[th]
\centerline{
\includegraphics[width=.48\textwidth]{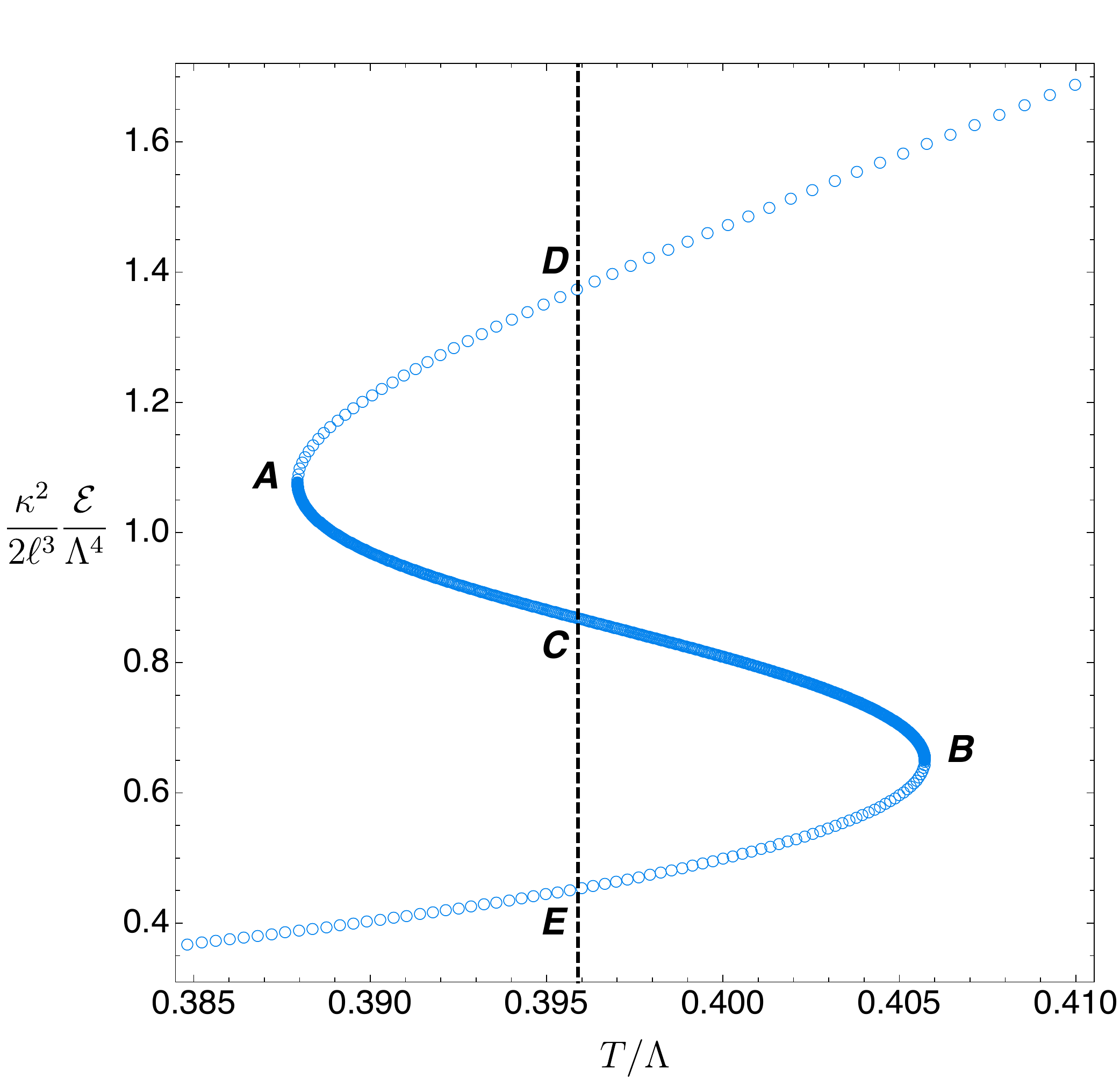}
\hspace{0.1cm}
\includegraphics[width=.495\textwidth]{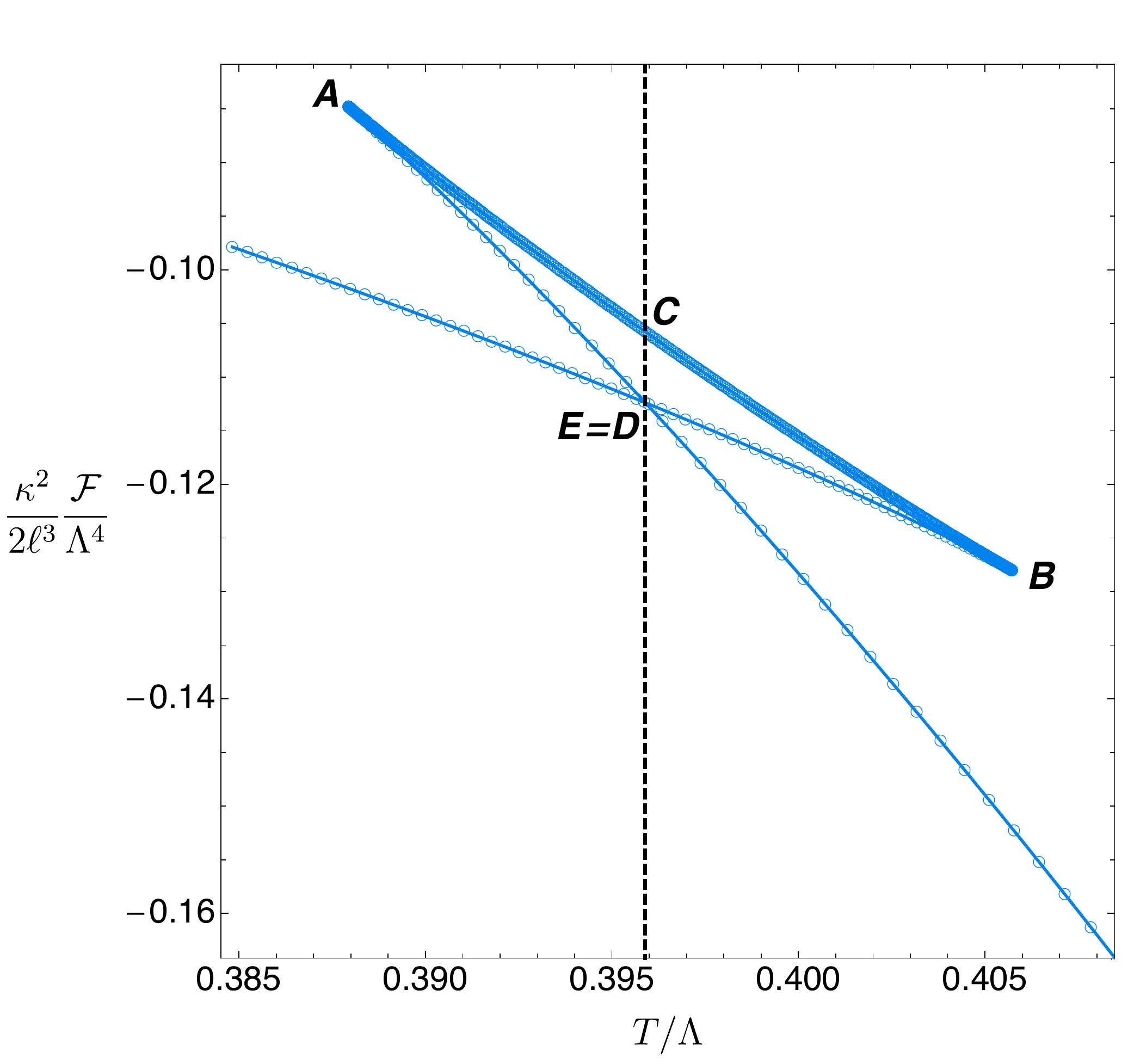}
}
\caption{Dimensionless energy density (left)  and  free energy density (right) as a function of the dimensionless temperature for uniform nonconformal branes in our model. The curve between $A$ and $B$ is the spinodal region and we will refer to it as the ``intermediate branch". At $\hat{\taut}_{\rm c} \simeq 0.3958945$ (vertical dashed line) there is a first order phase transition in the canonical ensemble (see right panel). For reference here and in future plots, 
\mbox{$(\hat{\taut},\hat{\cal E})_A\simeq (0.387944, 1.076417)$}, 
\mbox{$(\hat{\taut},\hat{\cal E})_B \simeq (0.405724,0.650227)$}, 
\mbox{$(\hat{\taut},\hat{\cal E})_C\simeq (0.3958945,0.867956)$}, 
\mbox{$(\hat{\taut},\hat{\cal E})_D\simeq (0.3958945,1.37386)$}  and 
\mbox{$(\hat{\taut},\hat{\cal E})_E\simeq (0.3958945,0.452754)$}.}
\label{fig:Unif}
\end{figure} 

\begin{figure}[th]
\centerline{
\includegraphics[width=.5\textwidth]{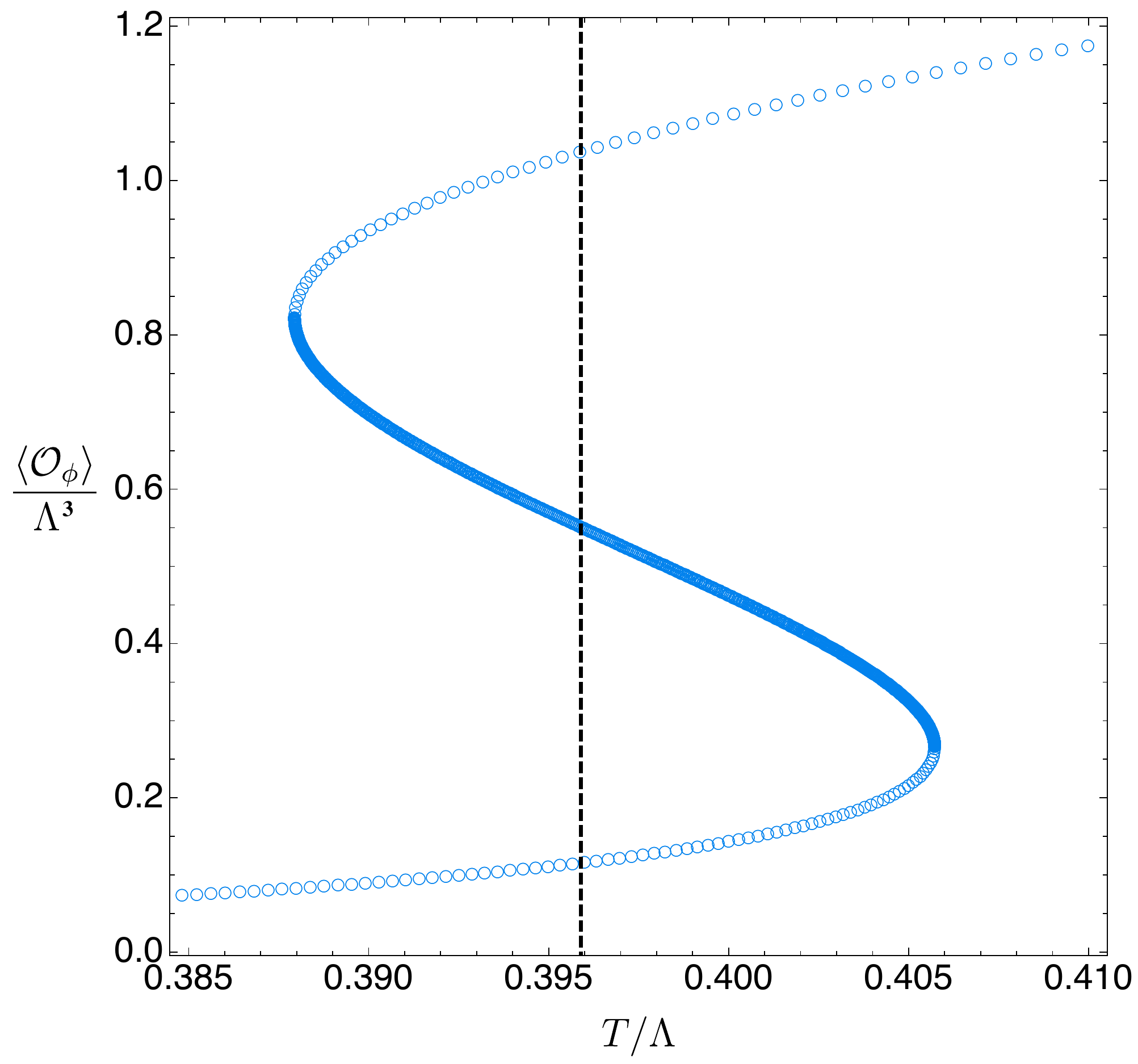}
\hspace{0.1cm}
\includegraphics[width=.475\textwidth]{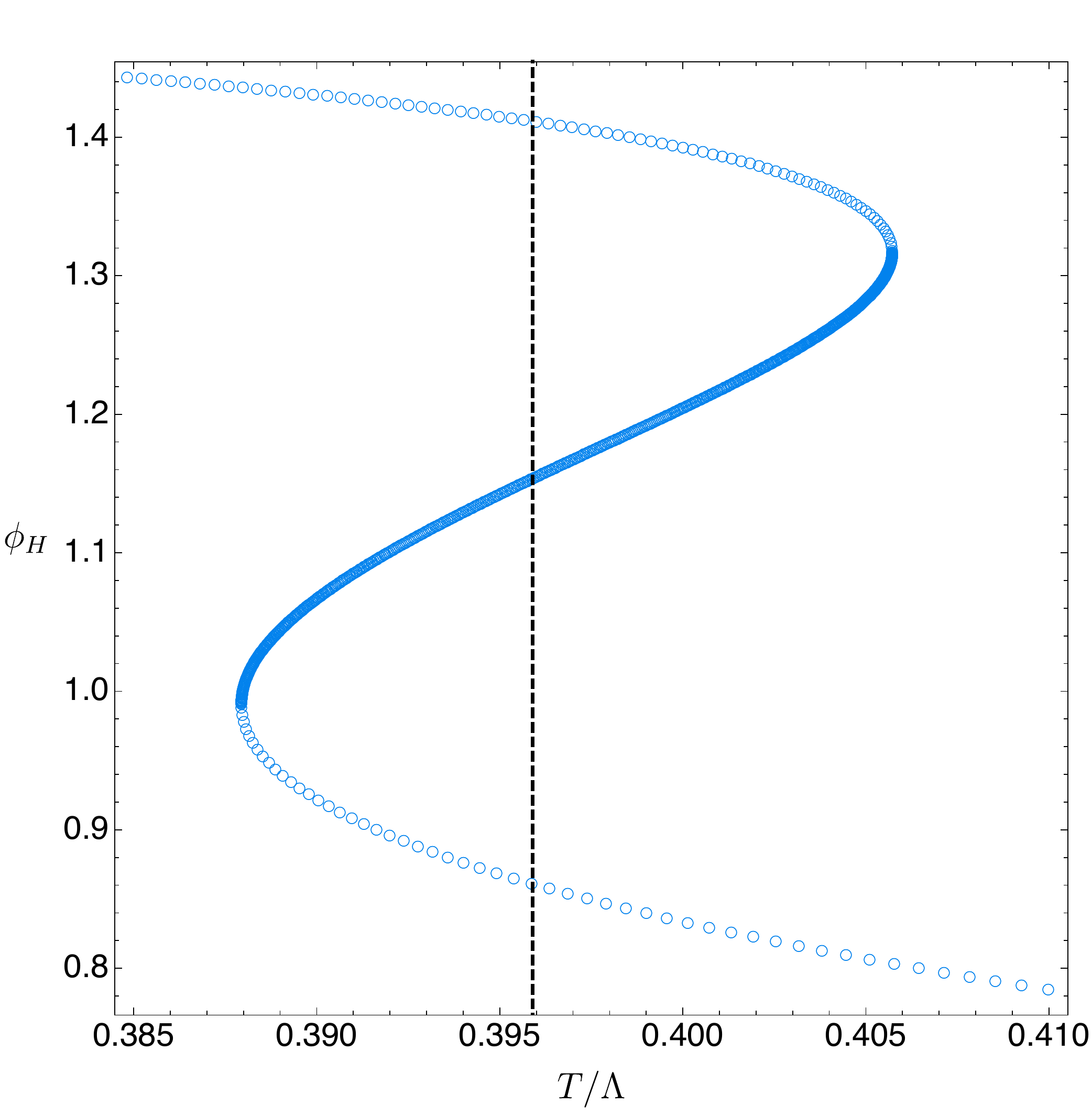}
}
\caption{(Left) Dimensionless expectation value $\langle {\cal O}_{\phi} \rangle/\Lambda^3$ of the operator with source $\Lambda$ as a function of the dimensionless temperature $\hat{\taut}$ of the uniform branes. (Right)  Value of the scalar field of the uniform branes at the horizon $\phi_H$ as a function of $\hat{\taut}$.}
\label{fig:Unif2}
\end{figure} 

The properties of uniform branes are summarized  in \figs{fig:Unif} and \ref{fig:Unif2}. In the left panel of \figs{fig:Unif} we plot the dimensionless energy density $\hat{\cal E}\equiv{\cal E}/\Lambda^4$ as a function of the dimensionless temperature $\hat{\taut}\equiv T/\Lambda$. We see the familiar S-shape associated to the multivaluedness of a first-order phase transition.  Specifically, for a given temperature $\hat{\taut}$ in the window of  temperatures $\hat{\taut}_A\leq T/\Lambda \leq\hat{\taut}_B$  there are three distinct families or branches of uniform branes with  different values of $\hat{\cal E}$. We will refer to these families as the ``heavy", ``intermediate" and ``light" branches. The heavy branch (with higher energy density) starts in the conformal $T/\Lambda\to \infty$ limit and then extends through point $D$ all the way down to point $A$ as the temperature $\hat{\taut}$ decreases. The intermediate branch extends from point $A$, passes though point $C$, towards point $B$. This branch has negative specific heat and is both thermodynamically and dynamically locally unstable. A general discussion of these features can be found in Sec.~2 of 
Ref.~\cite{Attems:2019yqn}. In the present paper we will analyse the zero-mode properties of this instability in \sect{sec:PerturbativeO1}, and its timescale in \sect{sec:GLtimescale}. Finally, the light branch (with lower energy density) starts at point $B$, passes thought point $E$ and extends all the way down towards $T/\Lambda\to 0$. We do not show the plots of $\hat{s}(\hat{\taut})$ and $\hat{\frak{p}}_{L,T}(\hat{\taut})$ because they are qualitatively similar to the plot of $\hat{\cal E}(\hat{\taut})$.

The relevant phase diagram for the canonical ensemble, namely the dimensionless free energy $\hat{\cal F}\equiv {\cal F}/\Lambda^4$ as a function of the dimensionless temperature $\hat{\taut}$, is displayed in the right panel of \fig{fig:Unif}, where we see the expected swallow-tail shape. For a given $\hat{\taut}$, the solution with lowest $\hat{\cal F}$ is the preferred thermal phase. So, as anticipated above, there is a first-order phase transition at $\hat{\taut}=\hat{\taut}_{\rm c}\approx 0.3958945$. This critical temperature is indicated  with a vertical dashed line in the plots of \figs{fig:Unif} and \ref{fig:Unif2}, as well as in subsequent ones  whenever appropriate. For $\hat{\taut}<\hat{\taut}_{\rm c}$ the light uniform branch (the lower branch in the left panel of \fig{fig:Unif}) is the preferred thermal phase, while for fixed $\hat{\taut}>\hat{\taut}_{\rm c}$ the heavy uniform branch (the upper branch in the left panel) dominates the canonical ensemble. In particular, the intermediate uniform branch (between $A$ and $B$) is never the preferred thermal phase. 

For completeness, in \fig{fig:Unif2} we show how the dimensionless expectation value $\langle {\cal O}_{\phi} \rangle/\Lambda^3$ of the operator with source $\Lambda$ changes with the dimensionless temperature $\hat{\taut}$ (left panel) and how the value of the scalar field at the horizon $\phi_H$ varies with $\hat{\taut}$ (right panel). 

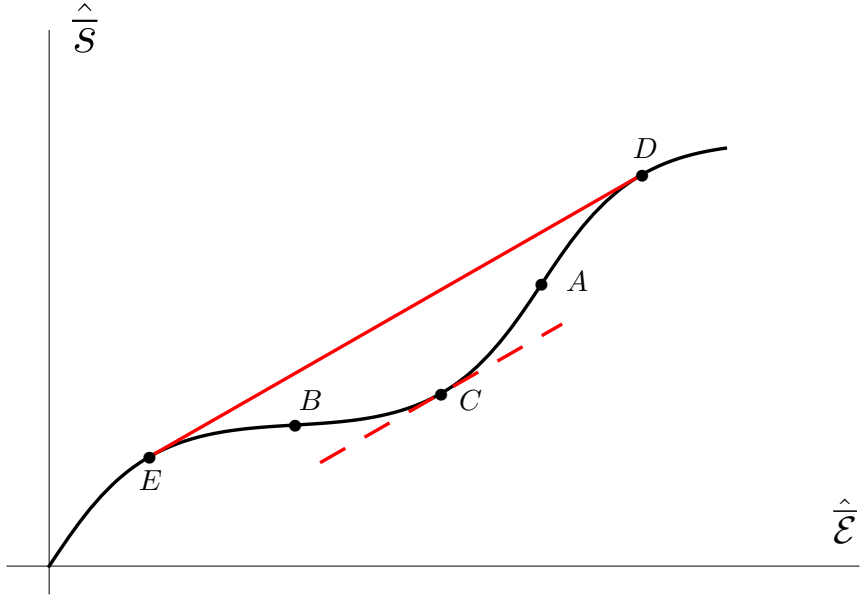
\begin{figure}[t]
	\begin{center}
			\includegraphics[width=.50\textheight]{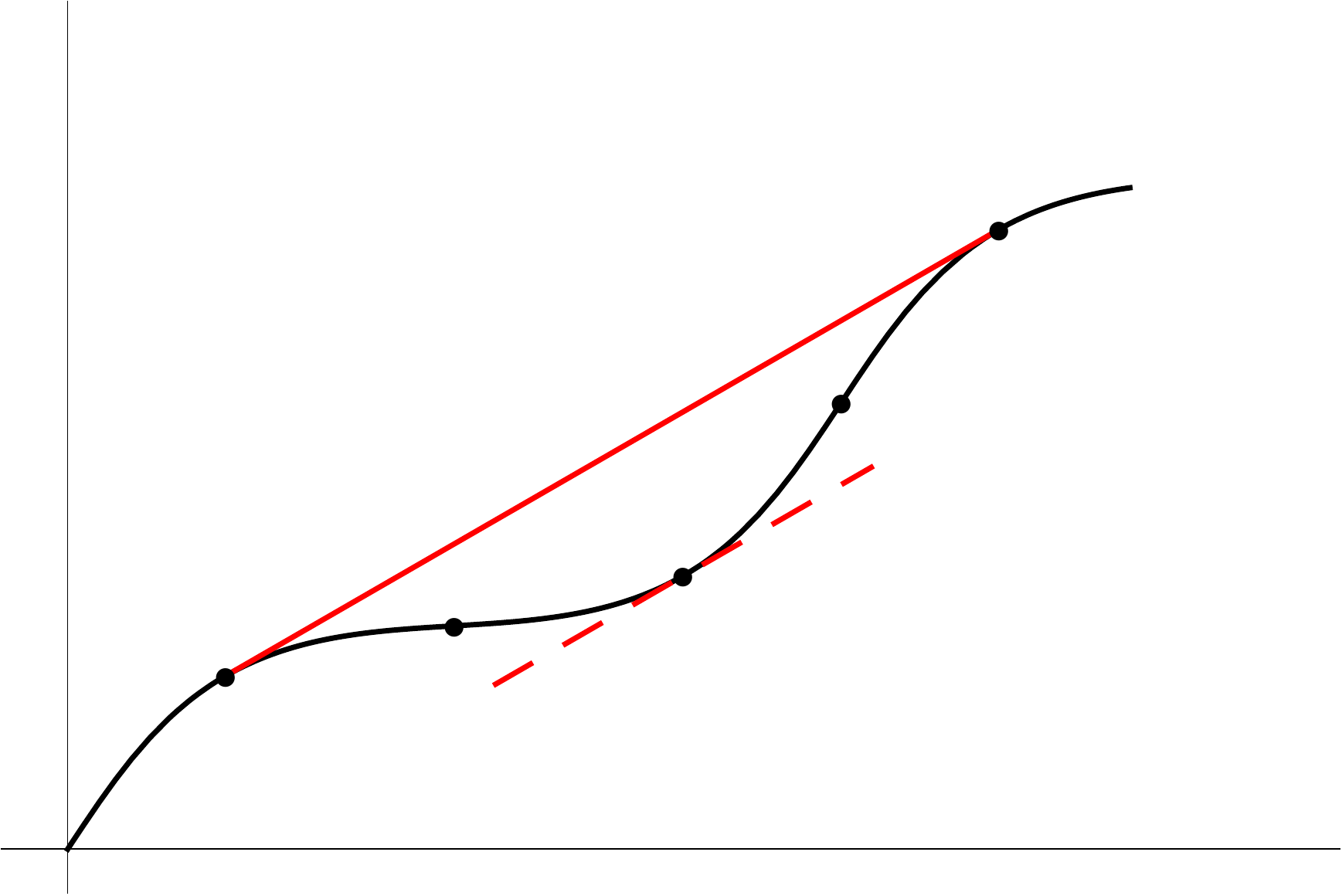} 
				\put(-295,205){\huge $\hat{\overline s}$}
				\put(-10,20){\Large $\hat{\overline{\cal E}}$}
				\put(-210,70){$B$}			
				\put(-110,115){$A$}			
				\put(-150,70){$C$}			
				\put(-85,165){$D$}			
				\put(-270,40){$E$}			
			\caption{\small  Qualitative form of the entropy density in the microcanonical ensemble. The solid red segment corresponds to the average entropy density of phase-separated configurations, as explained in the text. The dashed red segment indicates that the slope of the tangent at point $C$ is the same as that of the solid red segment. 
			This follows from the fact that the temperature at point $C$ is precisely $T_c$.}
			\label{entropyQ}
	\end{center}
\end{figure}

\begin{figure}[th]
\centerline{
\includegraphics[width=.47\textwidth]{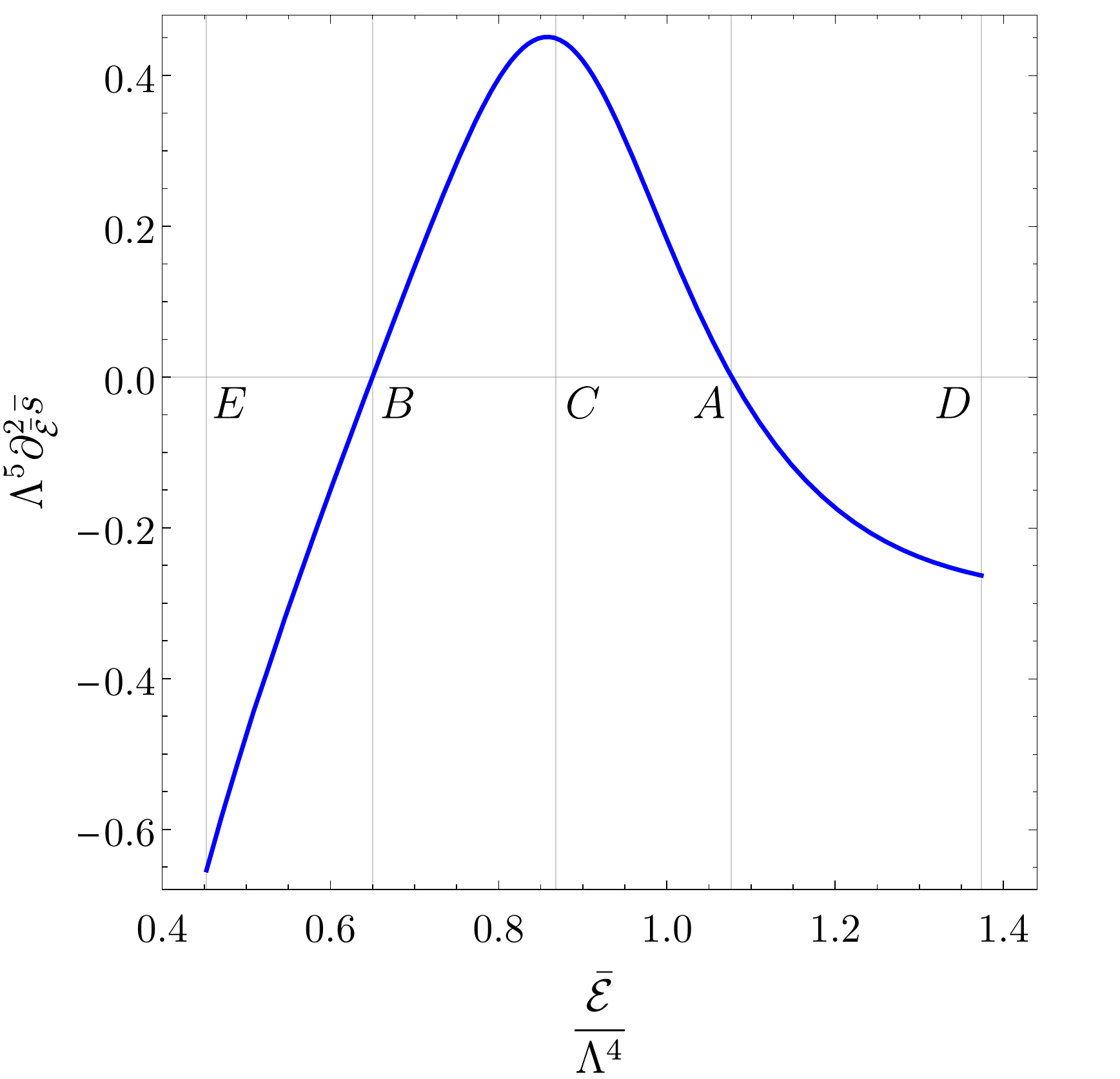}
\hspace{0.1cm}
\includegraphics[width=.5\textwidth]{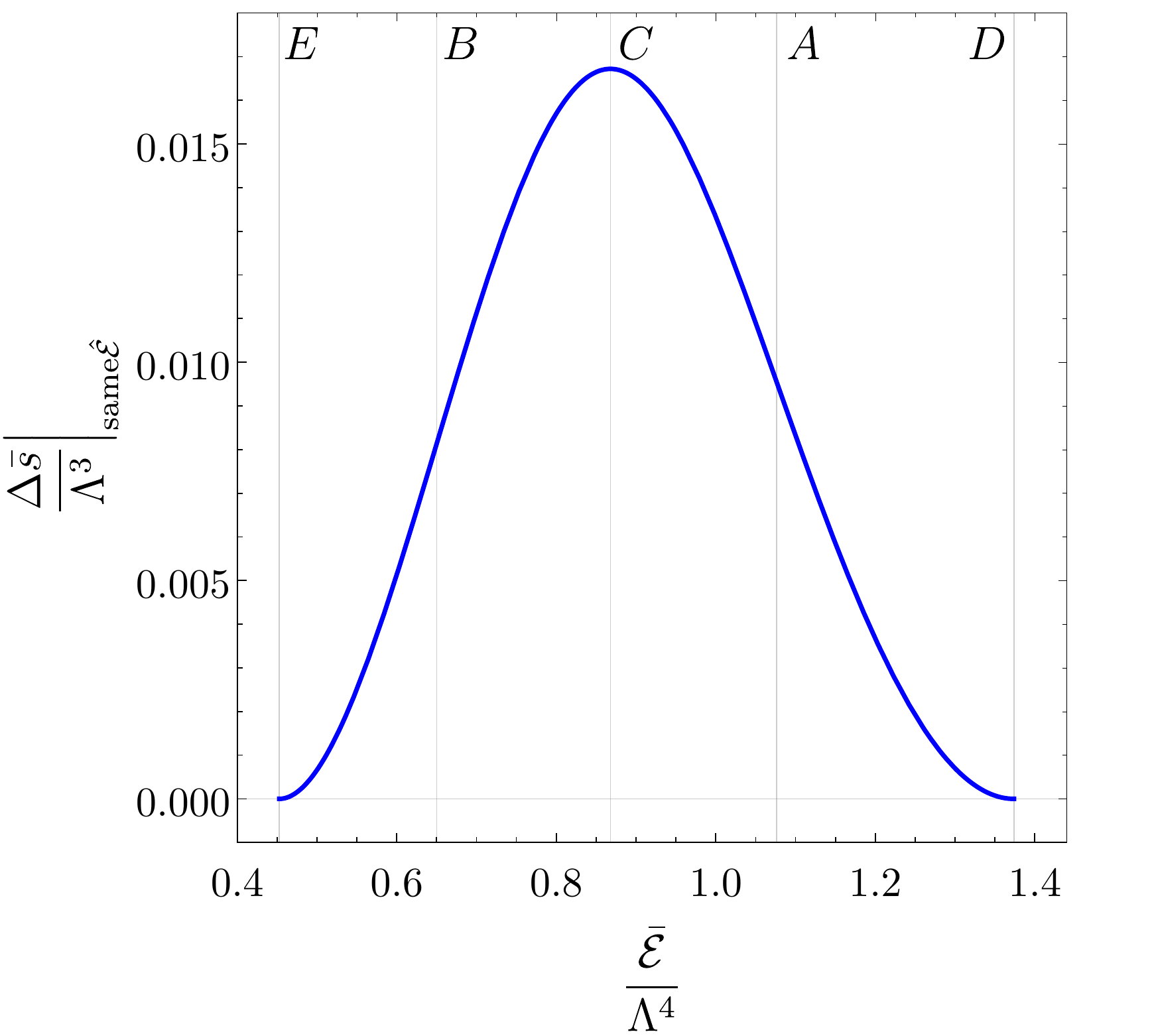}
}
\caption{(Left) Second derivative $\hat{\overline s}''(\hat{\overline{\cal E}})$ of the entropy density with respect to the energy density, showing the convexity/concavity properties discussed in the text. (Right) Difference between the average entropy density of the phase-separated configurations and the entropy density of the homogeneous solutions, showing that the former are preferred in the region between $A$ and $B$.}
\label{fig:convexconcave}
\end{figure}

In the microcanonical ensemble, the relevant phase diagram is the average entropy density $\hat{\overline s}\equiv \overline s/\Lambda^3$ as a function of the average energy density $\hat{\overline{\cal E}}\equiv {\overline{\cal E}}/\Lambda^4$. It is important to consider averaged quantities (which involve integration along the $x$ direction) because inhomogeneous state will play a role. The qualitative form of the function $\hat{\overline s}(\hat{\overline{\cal E}})$ is shown in Fig.~\ref{entropyQ}. 
The key features are as follows. $\hat{\overline s}$ is convex ($\hat{\overline s}''>0$) in the region between $A$ and $B$. This indicates local thermodynamical instability, since the system can increase its total entropy by rising the energy slightly in part of its volume and lowering in another so as to keep the total energy fixed. In the regions $EB$ and $AD$ the entropy function is concave ($\hat{\overline s}''<0$) but there are states with the same total energy and higher total entropy, namely phase-separated configurations in which the phases $E$ and $D$ coexist at the critical temperature. These states are characterised by the fractions $0\leq \nu, (1-\nu) \leq1$ of the total volume occupied by each phase, so their total entropy is of the form $\hat{s}_E + (\hat{s}_D-\hat{s}_E) \nu$, as indicated by the red segment in Fig.~\ref{entropyQ}. Therefore the regions $EB$ and $AD$ are locally but not globally thermodynamically stable. Finally, all states outside the region $ED$ are globally stable. For our system, these qualitative features are difficult to appreciate directly on a plot of $\hat{\overline s}$ versus $\hat{\overline{\cal E}}$ because the curve $\hat{\overline s}(\hat{\overline{\cal E}})$ is very close to a straight line. For this reason we show the convexity/concavity property (the second derivative) in Fig.~\ref{fig:convexconcave}(left) and the difference between the phase-separated configurations and the homogeneous solutions in Fig.~\ref{fig:convexconcave}(right). 

\subsubsection{Gregory-Laflamme physics: $\mathcal{O}(1)$ solution and the spinodal zero-mode}\label{sec:PerturbativeO1}

The intermediate uniform branes with $\hat{\cal E}_B<\hat{\cal E}<\hat{\cal E}_A$ (see left panel of \fig{fig:Unif}), and only these, can  be Gregory-Laflamme (GL) unstable. Roughly speaking, we expect this to happen if their dimensionless length $L\,\Lambda$ (along the $x$ direction) is bigger than the dimensionless thermal scale $\Lambda/T$ of the system. This linear instability  is ultimately responsible for the nonlinear existence of the lumpy solutions. Therefore, our second step is to consider {\it static} perturbations  about the uniform branes, $q_j(x,y)=Q_j(y)+\epsilon \, q_j^{(1)}(x,y)$, that break the $U(1)$ symmetry along $x$ (see \sect{sec:GLtimescale} for time-dependent perturbations). Here, $\epsilon\ll 1$ is the amplitude of the linear perturbation and, ultimately, it will be the expansion parameter of our perturbation theory to higher order. 

We adopt a perturbation scheme that is consistent with our nonlinear ansatz \eqref{ansatz} --- where we recall that $x\in [0,1]$ --- since we want to simply linearize the nonlinear equations of motion that we already have (\sect{sec:setupBVP}) to get the perturbative EoM. In this perturbation scheme we assume an ansatz for the perturbation of the form\footnote{The superscript $\! ^{(n)}\!$ here and henceforth always denotes the order $n$ of the perturbation theory, not order of derivatives.}  
\begin{equation}\label{GLstaticPert}
q_j^{(1)}(x,y)=\frak{q}_j^{(1)}(y)\cos(\pi\, x).
\end{equation}
 This means that the length $L$ of the periodic coordinate $x$ is given in terms of the wavenumber $k$ of the perturbation by $L=2\pi/k$, and it will {\it change} as we climb the perturbation ladder (this is because $k$, and thus $L$, will be corrected at each order; see \sect{sec:PerturbativeOn}). Since the EoM depend on $L$, this relation $L=2\pi/k$ introduces the zero mode wavenumber $k$ in the problem.\footnote{\label{footScheme}We have some freedom in the choice of the perturbation scheme. For example, an {\it alternative} perturbation scheme would be to keep the length $L$ {\it fixed} by absorbing the $L$ factor in the metric component $g_{xx}$ of the ansatz \eqref{ansatz} into a new coordinate $\tilde{x}$. That is to say, we would change the $x\in [0,1]$ coordinate of \eqref{ansatz} into $\tilde{x}=x \frac{L}{2}\in [0, \frac{L}{2}]$. In this case, the $U(1)$ dependence of the perturbation would be $\cos(k \,\tilde{x})$ which would introduce the wavenumber $k=\frac{2\pi}{L}$ in the problem. These two schemes are equivalent. This follows from the observation that the two sets of Fourier modes are equivalent: $\cos(\eta \,k \,\tilde{x})=\cos\left(\eta \frac{2\pi}{L}  \frac{L}{2} x\right)=\cos(\eta\, \pi \,x)$. Further recall from the discussion above \eqref{ansatzZ} that our solutions have  ${\bb Z}_2$ symmetry: the solution in $\tilde{x} \in [-L/2,0[$ can be obtained by simply flipping our solution over the $x=0$ axis (computationally this is useful/efficient since we deploy a given number of grid points to study the range $[0,L/2]$ instead of $[-L/2,L/2$]). This is why we have just a factor of $\pi$ and not $2\pi$ in the arguments of our Fourier cosines.} 

Under these circumstances  the linearized EoM become a simple eigenvalue problem in $k^2$ of four coupled ODEs. Henceforth, we denote this {\it leading-order} wavenumber by $k_{\hbox{\tiny GL}}$.
So we  need to solve our eigenvalue problem to find the eigenvalue $k_{\hbox{\tiny GL}}$ as well as the associated four eigenfunctions $q_j^{(1)}(y)$. Note however that we ``just" need to solve an ODE system of four coupled equations (not PDEs) subject to the {\it linearized versions} of the boundary conditions \eqref{BCs:UV}-\eqref{BCs:IR}. For example, when we linearize \eqref{BCs:UV} using $q_j|_{y=0}=Q_j|_{y=0}+\epsilon \, q_j^{(1)}|_{y=0}$ we find that the linear perturbations $\frak{q}_j^{(1)}(y)$ must obey the UV Dirichlet boundary conditions  $\frak{q}_j^{(1)}\big|_{y=0}=0$. On the other hand, linearizing the IR boundary conditions \eqref{BCs:IR} we find that the linear perturbations $\frak{q}_j^{(1)}(y)$ must obey the condition $\frak{q}_1^{(1)}\big|_{y=1}=\frac{1}{2 \alpha_{\Lambda}}\frak{q}_2^{(1)}\big|_{y=1}$ and mixed boundary conditions for $\frak{q}_{2,3,4}^{(1)}\big|_{y=1}$. Of course, in this linearization procedure about the uniform brane, we insert the boundary conditions \eqref{BCs:UV} and \eqref{BCs:IRunif} of the leading solution; in particular, we impose $Q_1\big|_{y=1}=1$ and  $Q_2\big|_{y=1}=\alpha_{\Lambda}$. 

\begin{figure}[t!]
\centerline{
\includegraphics[width=.65\textwidth]{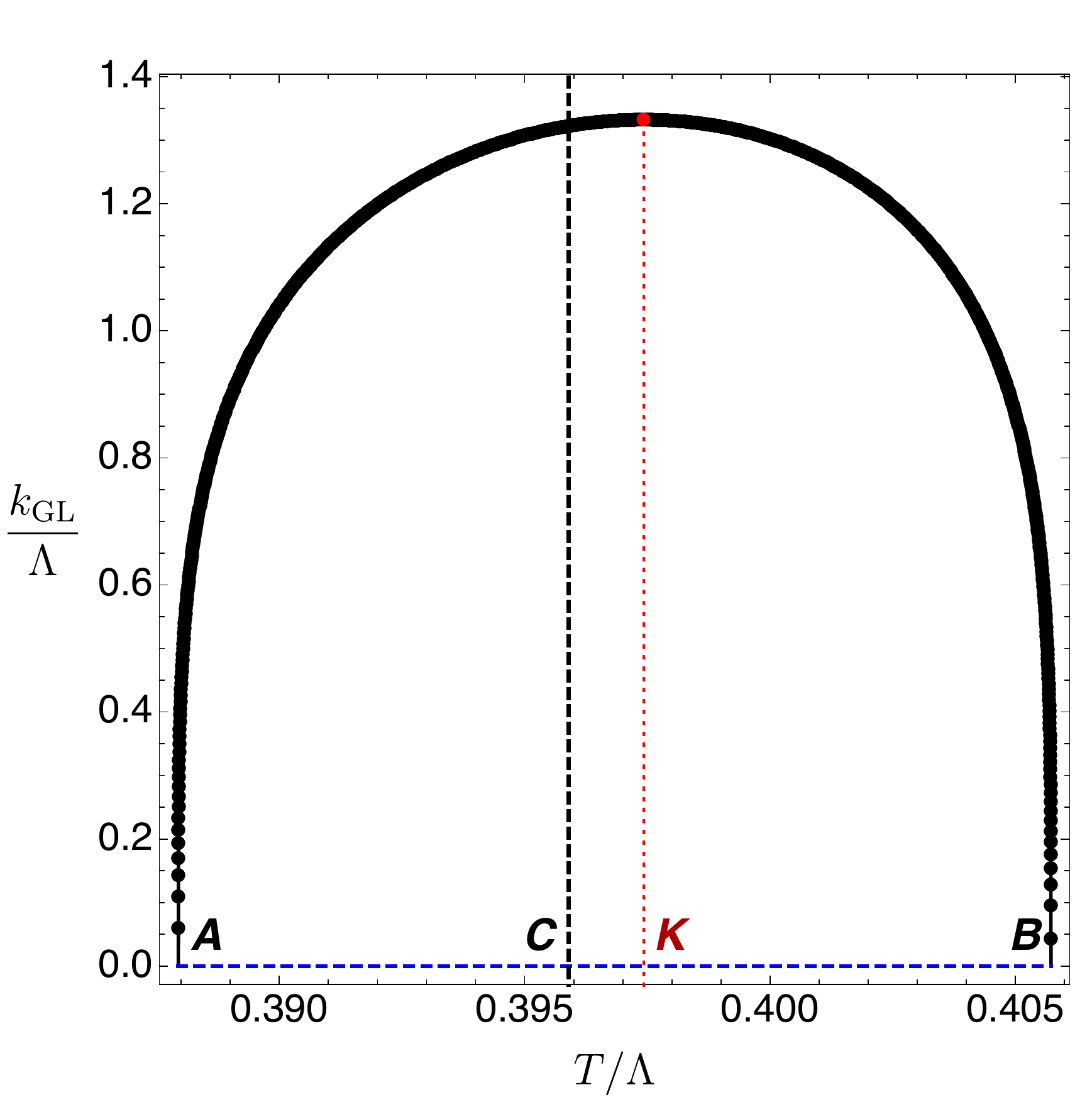}
}
\caption{Zero mode of the GL or spinodal instability, i.e.~its onset wavenumber $\hat{k}_{\hbox{\tiny GL}}$ as a function of the temperature $\hat{\taut}$. For reference \mbox{$(\hat{\taut},\hat{k}_{\hbox{\tiny GL}},\hat{L}_{\hbox{\tiny GL}})_{C}\simeq (0.3958945,1.322508,4.750961)$}, and the maximum of the instability occurs for 
\mbox{$(\hat{\taut},\hat{k}_{\hbox{\tiny GL}},\hat{L}_{\hbox{\tiny GL}})_{K}\simeq (0.397427,1.332306,4.716021)$}.}
\label{fig:GLk0}
\end{figure}

Summarizing this second step, the above perturbation procedure at  $\mathcal{O}(\epsilon)$ finds the critical zero mode of the Gregory-Laflamme (GL) instability of uniform branes with energy densities  $\hat{\cal E}_B<\hat{\cal E}<\hat{\cal E}_A$. That is to say, it finds the dimensionless critical wavenumber \mbox{$\hat{k}_{\hbox{\tiny GL}}=k_{\hbox{\tiny GL}}/\Lambda$} for the onset of the GL instability, and thus the minimum length \mbox{$L_{\hbox{\tiny GL}}\Lambda=2\pi/\hat{k}_{\hbox{\tiny GL}}$} above which the uniform brane is unstable. This critical value $\hat{k}_{\hbox{\tiny GL}}=\hat{k}_{\hbox{\tiny GL}}(\hat{\taut})$ is only a function of the dimensionless temperature $\hat{\taut}=T/\Lambda$ and is plotted in \fig{fig:GLk0}. We see that $\hat{k}_{\hbox{\tiny GL}}=0$ at the endpoints $A$ and $B$ of the {\it intermediate} uniform branch where $\hat{\taut}=\hat{\taut}_A$ and $\hat{\taut}=\hat{\taut}_B$. These two branes are effectively stable since $\hat{L}_{\hbox{\tiny GL}}\to \infty$ at these two temperatures. However, intermediate branes with $\hat{\taut}_A\leq \hat{\taut} \leq\hat{\taut}_B$ are unstable if their length satisfies  $\hat{L}>\hat{L}_{\hbox{\tiny GL}}=2\pi/\hat{k}_{\hbox{\tiny GL}}$.

\begin{figure}[t!]
\centerline{
\includegraphics[width=.90\textwidth]{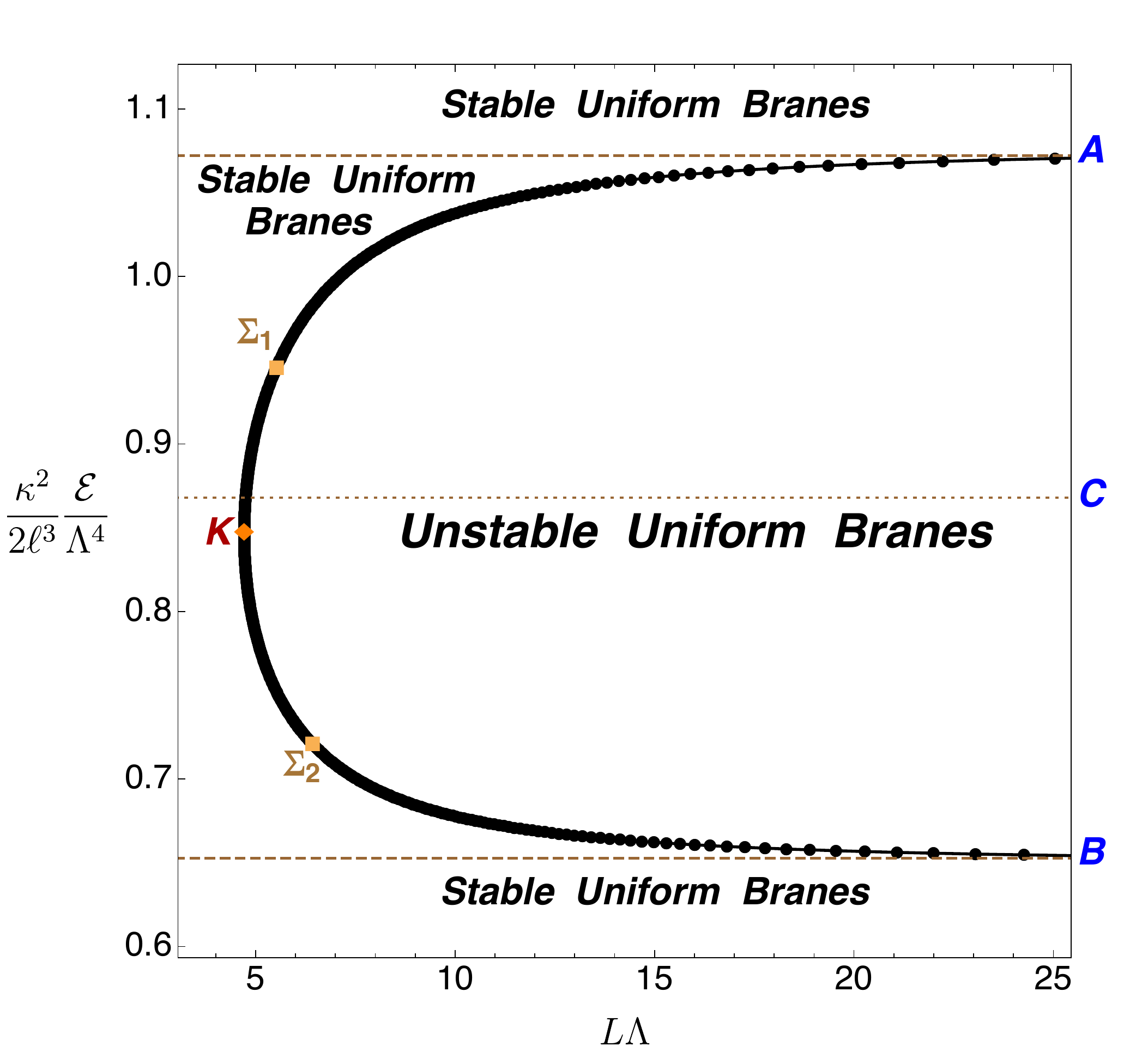}
}
\caption{Stability diagram for uniform nonconformal branes. The interpretation of the two yellow square points $\Sigma_1$ and $\Sigma_2$ will be given when discussing Fig.~\ref{fig:DeltaSigma}. For reference, $(\hat{\taut},\hat{L}_{\hbox{\tiny GL}},\hat{\cal E})_{\Sigma_1}\simeq (0.390817,5.618133,0.950579)$ and  $(\hat{\taut},\hat{L}_{\hbox{\tiny GL}},\hat{\cal E})_{\Sigma_2}\simeq (0.404645,6.592316,0.717060)$ and $(\hat{\taut},\hat{L}_{\hbox{\tiny GL}},\hat{\cal E})_{K}\simeq (0.397427,4.716021,0.846337)$.}
\label{fig:GLstabilityDiag}
\end{figure}

So $\hat{L}_{\hbox{\tiny GL}}$ is parametrized by $\hat{\taut}$, and the energy density of uniform branes is also only a function of the temperature, $\hat{\cal E}=\hat{\cal E}(\hat{\taut})$. It follows that we can identify the onset GL curve of uniform branes in a plot $\hat{\cal E}$ {\it vs} ${\hat{L}}$. This is done in  \fig{fig:GLstabilityDiag}. This plot is effectively a stability phase diagram for the uniform branes since the  black dotted GL onset curve separates the region where the uniform branes are unstable --- namely, the parabola-like shaped interior region $\hat{\cal E}_B<\hat{\cal E}<\hat{\cal E}_A$ with $\hat{L}>\hat{L}_{\hbox{\tiny GL}}$ --- from its complementary region where branes are stable against the spinodal instability. In this figure note that the energy density $\hat{\cal E}=\hat{\cal E}_A$ and $\hat{\cal E}=\hat{\cal E}_B$ corresponds to the energy densities of the uniform solutions $A$ and $B$ in \fig{fig:Unif} and note that $\hat{L}_{\hbox{\tiny GL}}\to \infty$ when the energy density of the black dashed GL onset curve approaches $\hat{\cal E}_A$ or $\hat{\cal E}_B$. 

To summarize,  \fig{fig:GLstabilityDiag} shows that {\it intermediate} uniform branes with a given energy density $\hat{\cal E}_B<\hat{\cal E}<\hat{\cal E}_A$ are unstable if their dimensionless length is higher that the GL critical length, $\hat{L}>\hat{L}_{\hbox{\tiny GL}}$.
Not less importantly, in a phase diagram of solutions, the GL onset curve also signals a bifurcation to a new family of solutions that describes nonuniform or lumpy branes. That is to say, the GL onset curve is a {\it merger line} between the uniform and lumpy nonconformal branes. Perturbation theory at order $\mathcal{O}(\epsilon)$  identifies this merger or intersection line (see \fig{fig:GLstabilityDiag}) of two distinct surfaces in a 3D phase diagram but it cannot describe the properties of the lumpy brane surface as we move away from the merger line (roughly speaking, it cannot describe the ``slope of the lumpy surface" in a 3D phase diagram). For that, we need to proceed to higher order $\mathcal{O}(\epsilon^n)$ in the perturbation theory, as we do in in the next subsection. 

\subsubsection{Lumpy branes: perturbative solution at $\mathcal{O}(n)$}\label{sec:PerturbativeOn}

 To find the solution at order $\mathcal{O}(\epsilon^n)$ we expand the metric functions and wavenumber in powers of $\epsilon$:

\begin{subequations} \label{PTexpansion}
\begin{eqnarray}
&& q_j(x,y)=Q_j(y)+\sum_{n=1}^{\infty} \epsilon^n \,q_j^{(n)}(x,y); \label{PTexpansionA} \\
&& k=\sum_{n=1}^{\infty} \epsilon^{n-1} k^{(n-1)}\equiv k_{\hbox{\tiny GL}}+\sum_{n=2}^{\infty} \epsilon^{n-1} k^{(n-1)},\quad \hbox{with} \:\: L=\frac{2\pi}{k}. \label{PTexpansionB}
\end{eqnarray}
\end{subequations}
In this expansion we have made the identification $k^{(0)}\equiv k_{\hbox{\tiny GL}}$ and we have already found the $n=1$ contribution in the previous section. Recall that this $\{ k_{\hbox{\tiny GL}}, q_j^{(1)}\}$ contribution was found by solving a {\it homogeneous} eigenvalue problem for $k_{\hbox{\tiny GL}}$.  The expansion \eqref{PTexpansion} is such that at order $\mathcal{O}(\epsilon^n)$ we solve the BVP to find the coefficients $\{ k^{(n-1)}, q_j^{(n)}\}$.

 Further note that, as explained above, our choice of perturbation scheme is such that the length $L$ is corrected at each  order $n$ (see also footnote \ref{footScheme}). That is, one has 
 \be
 L=L_{\hbox{\tiny GL}}+\sum_{n=2}^{\infty} \epsilon^{n-1} L^{(n-1)}\,,
 \ee
  where the coefficients $L^{(n-1)}$ can be read straightforwardly from \eqref{PTexpansionB}. This also means that in our choice of scheme, the periodicity of the $x$ circle allows us to introduce a separation ansatz for the perturbation coefficients $q_j^{(n)}(x,y)$ whereby they are expressed as a sum of Fourier modes (with harmonic number $\eta$) in the $x$ direction as
\begin{equation} \label{FourierExp}
q_j^{(n)}(x,y)=\sum_{\eta=0}^{n} \frak{q}_j^{(n,\eta)}(y) \cos(\eta \,\pi \,x). 
 \end{equation}
So here and onwards, $0 \leq \eta\leq n$ identifies a particular Fourier mode (harmonic) of our expansion at order  $\mathcal{O}(\epsilon^n)$. 

At order $n\ge 2$, the perturbation EoM are no longer homogeneous. Instead, they describe an {\it inhomogeneous} boundary value  problem with a source  ${\cal S}^{(n,\eta)}$. Not surprisingly, this source  is a function of the lower order solutions $\{k^{(i-1)}, q_j^{(i)} \}$, $i=1,\ldots,n-1$ (and their derivatives): ${\cal S}^{(n,\eta)}(k^{(i-1)}, q_j^{(i)}$). This source can always be written as a sum of Fourier modes of the system. We find that at order $n\geq 2$, the maximum Fourier mode harmonic that is excited in the source is $\eta=n$. This is due to the fact that at linear order we start with the single $\eta=1$ Fourier mode and the $n^{\rm th}$ polynomial power of this linear mode, after using trigonometric identities to eliminate powers of trigonometric functions, can be written as a sum of Fourier modes with the highest harmonic being $\eta=n$.
This property of our source implies that the solution of the $\mathcal{O}(n)$ EoM can only excite harmonics up to $\eta=n$ and this explains why we capped the sum in \eqref{FourierExp} at $\eta=n$.

To proceed, at each order $\mathcal{O}(n)$, we have to distinguish the Fourier modes $\eta=1$ from the other, $\eta\neq 1$. This is because this particular Fourier mode  $\eta=1$ is the only one that is already excited at linear order $n=1$. 

Start with the generic case $\eta \ne 1$. Then the  differential operator --- call it ${\cal L}_H$ --- that describes the associated homogeneous  system of equations, ${\cal L}_H \, \frak{q}_j^{(n,\eta)}=0$, is the same at each order $n$ and for any Fourier mode $\eta$: it only depends on the uniform brane $Q_j(y)$ we expand about and $k_{\hbox{\tiny GL}}$. The ODE system of 4 inhomogeneous equations is thus of the form 
\begin{equation}\label{highEoMgeneral}
{\cal L}_H \, \frak{q}_j^{(n,\eta)}={\cal S}^{(n,\eta)}, \quad \hbox{if $n\geq 2$ and $\eta\neq 1$.}
\end{equation} 
It follows that the complementary functions of the homogeneous system are the same at each order $n\geq 2$ and $\eta$. But, we also need to find the particular integral of the inhomogeneous system and this is different for each pair $(n,\eta)$ since the sources ${\cal S}^{(n,\eta)}$ differ. The general solution $\frak{q}_j^{(n,\eta)}(y)$ is found by solving \eqref{highEoMgeneral} subject to vanishing UV Dirichlet boundary conditions $\frak{q}_j^{(n,\eta)}\big|_{y=0}=0$ --- since the full solution \eqref{PTexpansion} must obey \eqref{BCs:UV} --- and regularity at the horizon $y=1$. This gives mixed boundary conditions for $\frak{q}_{2,3,4}^{(n,\eta)}$ and a Dirichlet condition for $\frak{q}_1^{(n,\eta)}$, all of which follow from \eqref{BCs:IR}.

Consider now the exceptional case $\eta=1$. In this case, at order $n\geq 2$, our BVP becomes a (non-conventional\footnote{It is not a standard eigenvalue problem because the eigenvalue $k^{(n-1)}$ is not multiplying the unknown eigenfunction $\frak{q}_j^{(n,\eta)}$. Instead, it multiplies an eigenfunction that was already determined at previous $n=1$ order.}) eigenvalue problem in $k^{(n-1)}$. That is to say, the ODE system of 4 inhomogeneous equations is now of the form 
\begin{equation}\label{highEoMspecial}
{\cal L}_H \, \frak{q}_j^{(n,1)}=k^{(n-1)}\frac{k_{\hbox{\tiny GL}} K_{j m}\frak{q}_m^{(1)} }{2y(1-y^2)Q_2^2Q_3} + {\cal S}^{(n,1)}, \quad \hbox{if $n\geq 2$ and $\eta = 1$.}
\end{equation} 
where $K_{j m}$ is a diagonal matrix whose only non-vanishing components are $K_{11}=1=K_{44}$. Recall that ${\cal L}_H$ is an operator that describes two second-order ODEs for $\frak{q}_1$, $\frak{q}_4$ and two first-order ODEs for $\frak{q}_2$, $\frak{q}_3$ and this justifies the presence of this particular $K_{j m}$ in our eigenvalue term. We now have to solve 
\eqref{highEoMspecial} (subject to boundary conditions that are motivated as in the $\eta\neq 1$ case) to find the eigenvalue $k^{(n-1)}$ and the eigenfunctions $\frak{q}_j^{(n,1)}(y)$.

To have a full understanding of the EoM of our perturbation problem one last observation is required. As pointed out above, the highest Fourier harmonic that is excited in our system at order $\mathcal{O}(\epsilon^n)$ is $\eta=n$. This is because the $n^{\rm th}$ polynomial power of the single Fourier mode that is present at linear order, after using trigonometric identities to eliminate powers of trigonometric functions, can be written as a sum of Fourier modes with the highest harmonic being $\eta=n$. But this trigonometric operation also indicates (as we explicitly confirmed) that {\it not} all Fourier modes with $\eta \leq n$ are excited. More concretely, for {\it even} $n\geq 2$ we find that {\it only even} $0\leq \eta\leq n$ modes are present in our system. And for any {\it odd} $n\geq 3$,  {\it only odd} $0\leq \eta\leq n$ modes are excited.
Therefore, up to order  $n=5$ we find that the modes that are excited in our system are:
\begin{subequations}\label{excitedFourierModes}
 \begin{align}
& q_j^{(2)}(x,y)=\frak{q}_j^{(2,0)}(y)+\frak{q}_j^{(2,2)}(y)\cos(2\,\pi \,x),  \\
& q_j^{(3)}(x,y)=\frak{q}_j^{(3,1)}(y)\cos(\pi \,x)+\frak{q}_j^{(3,3)}(y)\cos(3\,\pi \,x), \\
& q_j^{(4)}(x,y)=\frak{q}_j^{(4,0)}(y)+\frak{q}_j^{(4,2)}(y)\cos(2\,\pi \,x)+\frak{q}_j^{(4,4)}(y)\cos(4\,\pi \,x),  \\
& q_j^{(5)}(x,y)=\frak{q}_j^{(5,1)}(y)\cos(2\,\pi\, x)+\frak{q}_j^{(5,3)}(y)\cos(3\,\pi \, x)+\frak{q}_j^{(5,5)}(y)\cos(5\,\pi\, x).
 \end{align}
 \end{subequations}
This last property of our system, together with the previous observation --- see the discussion of \eqref{highEoMspecial} --- that Fourier modes with $\eta=1$ are those that give the wavenumber correction $k^{(n-1)}$ at order $n$, immediately allows us to conclude that $k^{(n-1)}=0$ if $n$ is even. At even $n$ order the $\cos{(\pi\, x)}$ Fourier mode is not excited by the source and thus the only solution of \eqref{highEoMspecial} is the trivial solution.

Finally, note that the $\eta=0$ harmonics are of particular special interest. Indeed note that modes with $\eta\neq 0$ do not contribute (since the integral of a cosine vanishes) to the total thermodynamic quantities of the solution such as the energy $E$, the entropy $S$, etc. It follows from the discussion of \eqref{excitedFourierModes} that odd order $n$ modes do not contribute to correct these thermodynamic quantities.

We can finally summarize the key aspects of the general flow of our perturbation theory as the order $n$ grows:
\begin{enumerate}
\item  even orders $\mathcal{O}(\epsilon^n)$ introduce perturbative corrections to thermodynamic quantities like  energy, entropy, pressure, etc., but they do not correct the wavenumber, $k^{(n-1)}=0$ (and thus do not correct $L$).
\item  odd orders $\mathcal{O}(\epsilon^n)$ give the wavenumber corrections $k^{(n-1)}$ but do not change the energy, entropy and pressure.
\end{enumerate}
We complete this perturbation scheme up to order $\mathcal{O}(\epsilon^5)$: this is the order required to find a deviation between the relevant thermodynamics of the lumpy branes and the uniform phase. 

Once we have found all the Fourier coefficients $\frak{q}_j^{(n,\eta)}(y)$ and wavenumber corrections $k^{(n-1)}$ up to $n=5$, we can reconstruct the four fields $q_j(x,y)$ using \eqref{PTexpansion}. We can then substitute  these fields in the thermodynamic formulas of  
\sect{sec:setupBVP} to obtain all the thermodynamic quantities 
of the system up to $\mathcal{O}(\epsilon^5)$. We find that all of them, as well as the wavenumber, have an even expansion in $\epsilon^n$, with the only exception of the temperature that is simply given by \eqref{calTS}. 

Now that we have the thermodynamic description of lumpy branes up to $\mathcal{O}(\epsilon^5)$, we can compare it against the thermodynamics of uniform branes and find which of these two families is the preferred phase. We are particularly interested in the microcanonical ensemble, so the dominant  phase is the one that has the highest  $\hat{\sigma}$ for a given pair $(\hat{L},\hat{\rho})$. Let $Q_{\rm u}$ and $Q_{\rm nu}$ denote thermodynamic quantities $Q$ for the uniform and nonuniform branes, respectively. 
When comparing these two solutions in the microcanonical ensemble, one must have
\begin{equation}\label{microCondition}
\hat{L}_{\rm nu}=\hat{L}_{\rm u}, \quad \hbox{and} \quad \hat{\rho}_{\rm nu}=\hat{\rho}_{\rm u}.
\end{equation}
Given a lumpy brane with $(\hat{L}_{\rm nu},\hat{\rho}_{\rm nu})$ we must thus identify a uniform brane whose Killing density $\hat{\rho}_{\rm u}$ satisfies \eqref{microCondition}. Equivalently, we can impose that the energy density of the uniform brane obeys 
\be
\label{impose}
{\cal E}_{\rm u}=\frac{\hat{\rho}_{\rm nu}}{\hat{L}_{\rm nu}} \,.
\ee
Both sides of this equation are known as a perturbative expansion in $\epsilon$. This is because the energy density ${\cal E}_{\rm u}$ is a function of the dimensionless temperature $\hat{\taut}_{\rm u}$ which is corrected at each order as $\hat{\taut}_{\rm u}=\hat{\taut}_0+\epsilon^2\, \hat{\taut}_{(2)}+\epsilon^4\, \hat{\taut}_{(4)}+\mathcal{O}(\epsilon^6)$   in our perturbation expansion. Similarly, the Killing energy density $\hat{\rho}_{\rm nu}(\epsilon)$ and the length $\hat{L}_{\rm nu}(\epsilon)$ of lumpy branes are also known as a Taylor expansion in $\epsilon$. Therefore, in practice equation \eqref{impose} becomes
\be
{\cal E}_{\rm u}(\hat{\tau}_0)+\epsilon^2 \, \hat{\tau}_{(2)} \, {\cal E}_{\rm u}'(\tau_0) +\epsilon^4  \left( \hat{\tau}_{(4)} \, {\cal E}_{\rm u}'(\tau_0) +\frac{1}{2}\,\hat{\tau}_{(2)}^2 \,{\cal E}_{\rm u}''(\tau_0) \right) +\mathcal{O}(\epsilon^6) =\frac{\hat{\rho}_{\rm nu}(\epsilon)}{\hat{L}_{\rm nu}(\epsilon)}
\,.
\ee

Taking the Taylor expansion of ${\rho}_{\rm u}(\hat{\taut}_{\rm u})$ we must impose
\be
 \label{unifSameEL}
 {\rho}_{\rm nu} = 
{\rho}_{\rm u}(\hat{\taut}_0)+\epsilon^2 \, \hat{\taut}_{(2)} \, {\rho}_{\rm u}'(\taut_0) +\epsilon^4  \left( \hat{\taut}_{(4)} \, {\rho}_{\rm u}'(\taut_0) +\frac{1}{2}\,\hat{\taut}_{(2)}^2 \,{\rho}_{\rm u}''(\taut_0) \right) +\mathcal{O}(\epsilon^6) \,, 
\ee
 Given a lumpy brane with known $\hat{L}_{\rm nu}(\epsilon)$ and $\hat{\rho}_{\rm nu}(\epsilon)$, equation \eqref{unifSameEL} allows us to find the temperature coefficients $\taut_{(i)}$ of the uniform brane that has the same length and Killing energy density as the lumpy solution, i.e.~the temperature of the uniform brane $\hat{\taut}_{\rm u}$ up to $\mathcal{O}(\epsilon^n)$ that satisfies \eqref{microCondition}.

Having this $\hat{\taut}_{\rm u}$ we can now compute the  entropy density of the uniform brane $\hat{s}_{\rm u}(\taut_{\rm u})$ and the Killing entropy density $\hat{\sigma}_{\rm u}(\taut_{\rm u})=\hat{L}_{\rm u} \, \hat{s}_{\rm u}(\taut_{\rm u})$. More concretely, a Taylor expansion in $\epsilon$ of this equality  yields
 \begin{eqnarray}
 \label{unifEntropySameEL}
&& \hat{\sigma}_{\rm u}^{(0)}+\epsilon^2\, \hat{\sigma}_{\rm u}^{(2)} +\epsilon^4\, \hat{\sigma}_{\rm u}^{(4)} +\mathcal{O}(\epsilon^6) =\left[ \hat{L}_{\rm nu(0)}+\epsilon^2\, \hat{L}_{\rm nu(2)} +\epsilon^4\, \hat{L}_{\rm nu(4)} +\mathcal{O}(\epsilon^6) \right]   \\ 
&& \hspace{3cm} 
\times \left[  \hat{s}_{\rm u}(\hat{\taut}_0)+\epsilon^2 \, \hat{\taut}_{(2)} \, \hat{s}_{\rm u}'(\taut_0) +\epsilon^4  \left( \hat{\taut}_{(4)} \, \hat{s}_{\rm u}'(\taut_0) +\frac{1}{2}\,\hat{\taut}_{(2)}^2 \,\hat{s}_{\rm u}''(\taut_0) \right) +\mathcal{O}(\epsilon^6) \right], \nonumber
\end{eqnarray}
which allows us to find the entropy correction coefficients $\hat{\sigma}_{\rm u}^{(i)}$ and thus the Killing entropy density $ \hat{\sigma}_{\rm u}(\hat{\taut}_{\rm u})$ up to order $\mathcal{O}(\epsilon^6)$ of the uniform brane that has the same $(\hat{L},\hat{\rho})$ as the particular lumpy brane we selected. This procedure \eqref{microCondition}-\eqref{unifSameEL} can now be repeated for all lumpy branes.

\begin{figure}[t]
\centerline{
\includegraphics[width=.48\textwidth]{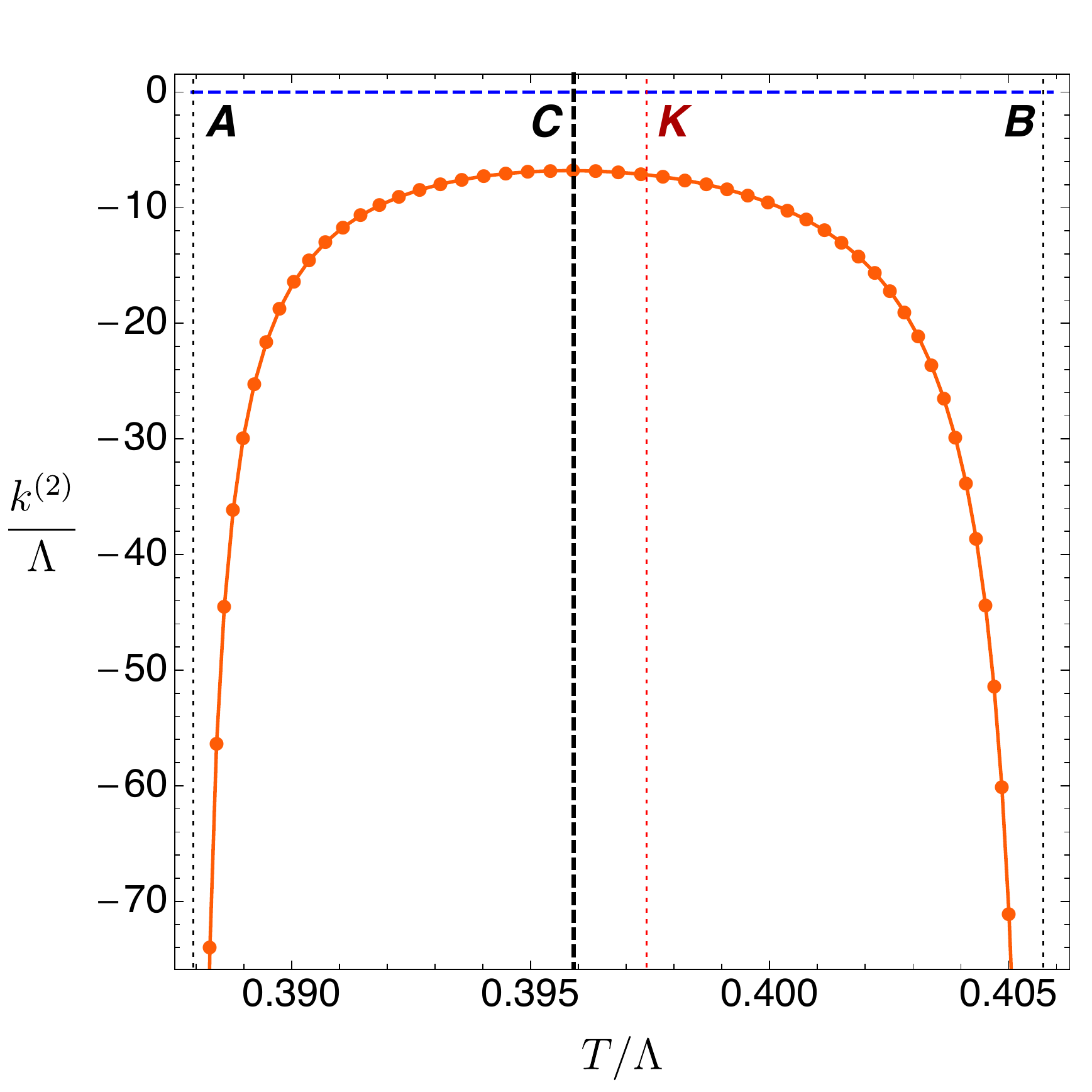}
\hspace{0.5cm}
\includegraphics[width=.49\textwidth]{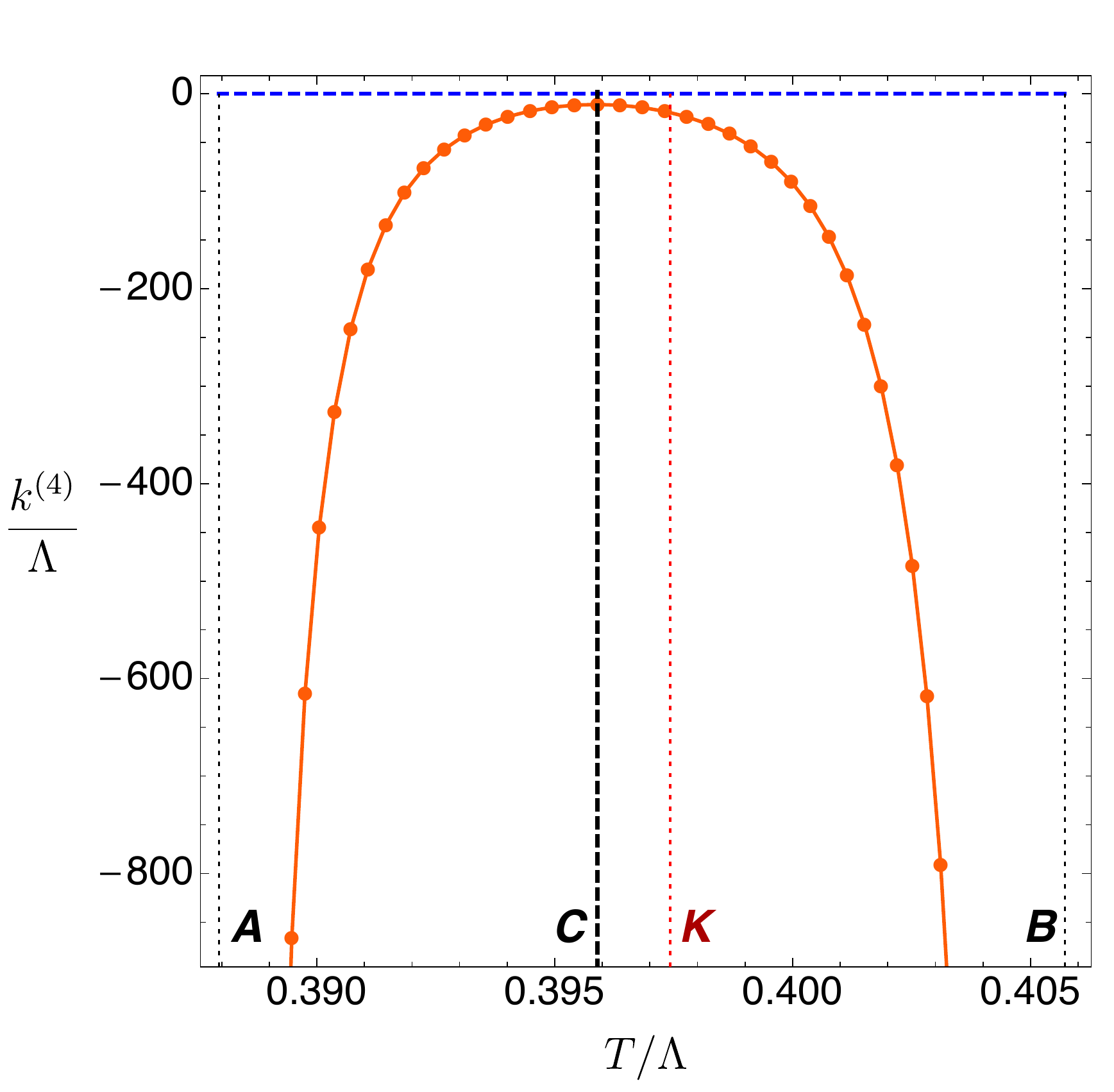}
}
\caption{Wavenumber corrections $k^{(2)}$ (left panel) and $k^{(4)}$ (right panel), as defined in \eqref{PTexpansionB}, as a function of the uniform brane temperature. For reference, $\hat{\taut}_{C}\simeq 0.3958945$ and the maximum of the instability occurs for $\hat{\taut}_{K}\simeq 0.397427$.}
\label{fig:k2k4}
\end{figure} 

We are now ready to discuss our higher-order perturbative findings.
First, in Fig.~\ref{fig:k2k4} we plot the wavenumber corrections $k^{(2)}$ (left panel) and $k^{(4)}$ (right panel), as defined in \eqref{PTexpansionB}. The fact that these higher order quantities grow large as one approaches $\hat{\taut}_A$ and $\hat{\taut}_B$ tells us that our perturbation theory breaks down in these regions. We will come back to this below.

Second, in order to determine the dominant phase, we are interested in the entropy difference between a nonuniform and a uniform brane when the two have the same length $\hat{L}$ and Killing energy density $\hat{\rho}$. This is given by  
 \begin{eqnarray}\label{unifEntropySameEL}
 \Delta \hat{\sigma}(\epsilon)\big|_{\hbox{\footnotesize same } \hat{L}, \hat{\rho}} &=&\Big[ \hat{ \sigma}_{\rm nu}(\epsilon)-\hat{\sigma}_{\rm u}(\epsilon) \Big]_{\hbox{\footnotesize same } \hat{L}, \hat{\rho}}
 \nonumber\\
&=& \left(\hat{\sigma}_{\rm nu}^{(0)}-\hat{\sigma}_{\rm u}^{(0)}\right)+\epsilon^2 \left(\hat{\sigma}_{\rm nu}^{(2)}-\hat{\sigma}_{\rm u}^{(2)}\right) +\epsilon^4 \left(\hat{\sigma}_{\rm nu}^{(4)}-\hat{\sigma}_{\rm u}^{(4)}\right)+\mathcal{O}(\epsilon^6) \,\,\,\,\,\,\,
\\
&\equiv& \Delta\hat{\sigma}^{(0)}+\epsilon^2 \Delta\hat{\sigma}^{(2)} +\epsilon^4 \Delta\hat{\sigma}^{(4)} +\mathcal{O}(\epsilon^6) \,.
\nonumber
\end{eqnarray}
By construction $\Delta\hat{\sigma}^{(0)}\equiv 0$ since the leading order of our perturbation theory describes the merger line of lumpy branes with uniform branes. Moreover, the first law for the Killing densities \eqref{1stLaw} can be rewritten, in the perturbative context, as $\partial_\epsilon \hat{\rho}= \hat{\taut} \,\partial_\epsilon \hat{\sigma}+ \hat{\frak{p}}_{L} \,\partial_\epsilon \hat{L}$ and has itself an expansion in $\epsilon$ that must be obeyed at each order. The leading-order term of this expansion implies that $\Delta\hat{\sigma}^{(2)}\equiv 0$, a condition that we actually  use to test our numerical results. Therefore  the first non-trivial contribution to $\Delta \hat{\sigma}(\epsilon)$ occurs at fourth order, namely
\be
 \Delta \hat{\sigma}(\epsilon)\big|_{\hbox{\footnotesize same } \hat{L}, \hat{\rho}} =\Big[ \hat{ \sigma}_{\rm nu}(\epsilon)-\hat{\sigma}_{\rm u}(\epsilon) \Big]_{\hbox{\footnotesize same } \hat{L}, \hat{\rho}} = 
\epsilon^4 \Delta\hat{\sigma}^{(4)}+\mathcal{O}(\epsilon^6) \,.
 \ee
This is the reason why we have to extend our perturbation analysis up to $\mathcal{O}(\epsilon^5)$.

\begin{figure}[th]
\centerline{
\includegraphics[width=.70\textwidth]{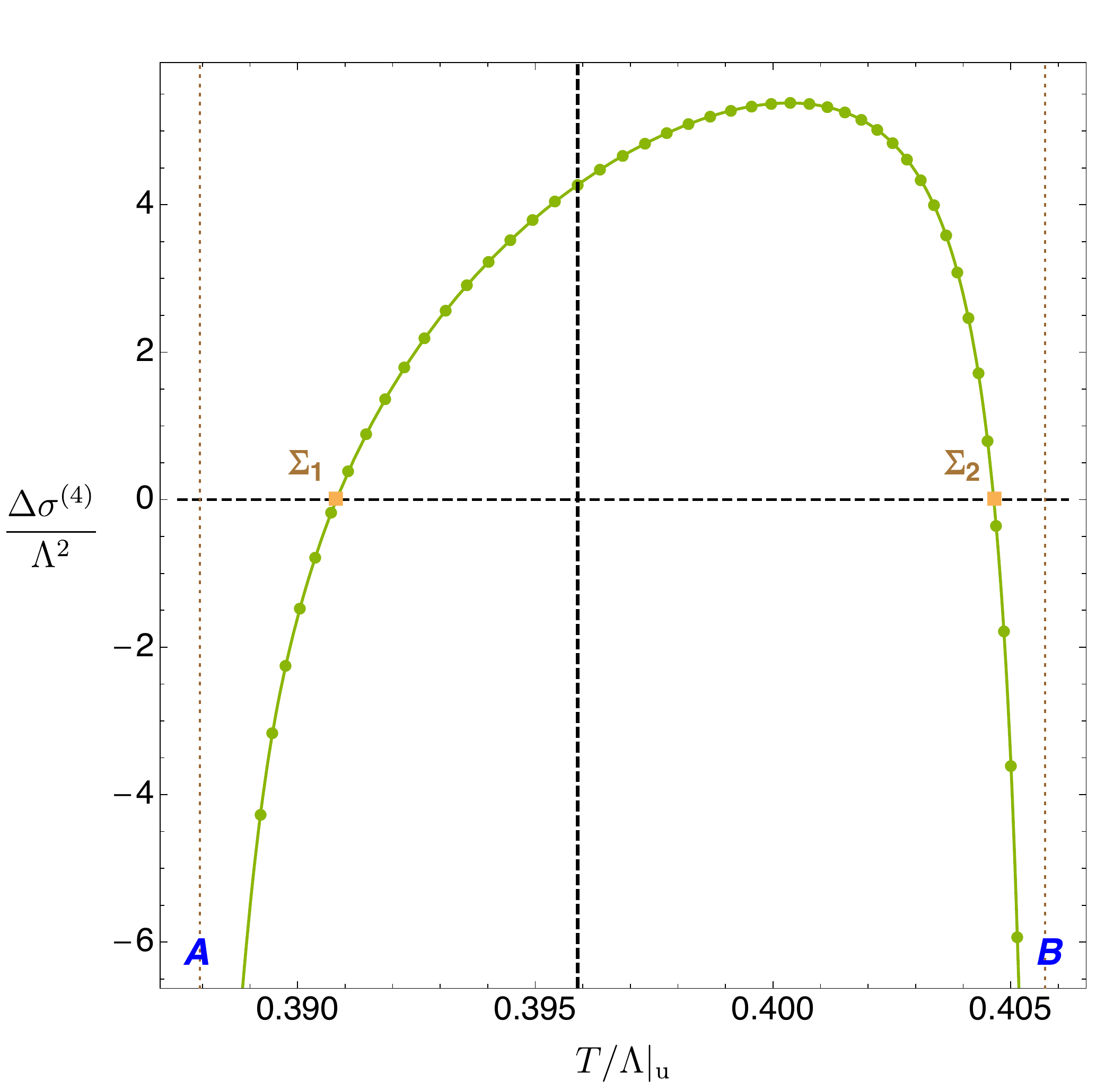}
}
\caption{Perturbative  identification of the dominant microcanonical phase. The horizontal axis shows the temperature $\hat{\taut}_{\rm u}$ of the uniform brane  that has the same $(\hat{L},\hat{\rho})$ as the lumpy brane we compare it with. The vertical axis shows the difference between the Killing  entropy densities of the lumpy and the uniform branes. Thus lumpy branes (uniform branes) dominate if $\Delta\hat{\sigma}^{(4)}> 0$ ($\Delta\hat{\sigma}^{(4)}<0$). For reference, $\hat{\taut}_{\rm c}\simeq 0.3958945$ (vertical black dashed line),  $(\hat{\taut},\hat{\cal E},\hat{L}_{\hbox{\tiny GL}})_{\Sigma_1}\simeq (0.390817,0.950579,5.618133)$ and  $(\hat{\taut},\hat{\cal E},\hat{L}_{\hbox{\tiny GL}})_{\Sigma_2}\simeq (0.404645,0.717060,6.592316)$.}
\label{fig:DeltaSigma}
\end{figure} 

We conclude that, for given $(\hat{L},\hat{\rho})$, if $\Delta\hat{\sigma}^{(4)}>0$ then the lumpy branes are the preferred phase; otherwise the uniform branes are the dominant phase. We should thus plot  the coefficient $\Delta\hat{\sigma}^{(4)}$ of \eqref{unifEntropySameEL} as a function of $\hat{L}$ and $\hat{\rho}$. However, we find it clearer to plot instead  $\Delta\hat{\sigma}^{(4)}$ as a function of the  temperature 
$\hat{\taut}_{\rm u}$ of the uniform brane that has the  same $(\hat{L},\hat{\rho})$ as the lumpy brane we compare it with. This is done in Fig.~\ref{fig:DeltaSigma}. Recall that uniform branes can be GL-unstable only in the range $\hat{\taut}_A\leq \hat{\taut} \leq\hat{\taut}_B$ and $\hat{\cal E}_B<\hat{\cal E}<\hat{\cal E}_A$, see Fig.~\ref{fig:Unif}. It follows that lumpy branes bifurcate from the uniform branch at the GL zero mode for temperatures in the range  $\hat{\taut}_A\leq\hat{\taut} \leq\hat{\taut}_B$. Fig.~\ref{fig:DeltaSigma} plots this range of temperature and shows that for $\hat{\taut}_{\Sigma_1}<\hat{\taut} <\hat{\taut}_{\Sigma_2}$, where the values of $\hat{\taut}_{\Sigma_1}$ and $\hat{\taut}_{\Sigma_2}$ are identified in the caption, the lumpy branes are the preferred thermodynamic phase since $\Delta\hat{\sigma}^{(4)}>0$. However, for $\hat{\taut}_A<\hat{\taut} <\hat{\taut}_{\Sigma_1}$ and $\hat{\taut}_{\Sigma_2}<\hat{\taut} <\hat{\taut}_B$, we have $\Delta\hat{\sigma}^{(4)}<0$ and thus uniform branes dominate over the lumpy phase when they have the same dimensionless length $\hat{L}$ and Killing energy density $\hat{\rho}$. Going back to Fig.~\ref{fig:GLstabilityDiag}, for completeness we have also identified these points $\Sigma_1$ and $\Sigma_2$ in the associated GL merger curve. 

 Figs.~\ref{fig:k2k4} and \ref{fig:DeltaSigma} also illustrate the regime of validity of our perturbative expansion. For example, in Fig.~\ref{fig:DeltaSigma} we see that $\Delta\hat{\sigma}^{(4)}$ grows arbitrarily negative as we approach the endpoints $A$ and $B$ of the {\it intermediate} branes with temperature $\hat{\taut}_A\simeq 0.387944$ and $\hat{\taut}_B \simeq 0.405724$ (see also Fig.~\ref{fig:Unif}). But once the associated entropy correction becomes of the order of our expansion parameter, $\Delta\hat{\sigma}^{(4)}\epsilon^4\sim \epsilon$, perturbation theory breaks down. So we should not trust our perturbative results close to the endpoints $A$ and $B$.
 
Even away from $\hat{\taut}_A$ and $\hat{\taut}_B$, our perturbation theory is certainly valid only for $\epsilon \ll 1$. Therefore we expect it to describe accurately the properties of lumpy branes close to their GL merger line with the uniform branes (where $\epsilon=0$) but not far away from this merger. To learn what happens further away, we need to solve the full nonlinear BVP using  numerical methods. This is what we do in the next subsection.

\subsection{Full nonlinear solutions and phase diagram of nonconformal branes}\label{sec:PhaseDiag}

To find accurately the lumpy branes and thus their thermodynamics in the full phase space where they exist, one needs to resort to numerical methods to solve nonlinearly the associated BVP, which  was set up in \sect{sec:setupBVP}.
It consists of a coupled set of four quasilinear PDEs --- two second-order PDEs for $q_{1,4}(x,y)$ and two second-order PDEs for $q_{2,3}(x,y)$ ---  that allow us to find the brane solutions \eqref{ansatz} that obey the boundary conditions \eqref{BCs:x}-\eqref{BCs:IR}. 

We solve our BVP using a Newton-Raphson algorithm. For the numerical grid discretization we use a pseudospectral collocation with a Chebyshev-Lobatto grid and the Newton-Raphson linear equations are solved by LU decomposition. These methods are reviewed and explained in detail in the review \cite{Dias:2015nua} and used in e.g.~\cite{Dias:2015pda,Dias:2016eto,Dias:2017uyv,Dias:2017opt,Bena:2018vtu}.
As explained in Sects.~\ref{sec:setup} and \ref{sec:setupBVP} (see in particular  footnote \ref{footGauge} and the associated discussion) our gauge was judiciously chosen to guarantee that our solutions have analytical polynomial expansions at all the boundaries of the integration domain. In these conditions the pseudospectral collocation guarantees that our numerical results have exponential convergence with the number of grid points. We further use the first law and the Smarr relations  \eqref{1stLaw}-\eqref{Smarr} to check our numerics. In the worst cases, our solutions satisfy these relations with an error that is smaller than 1\%. As a final check of our  full nonlinear numerical results, we compare them against the perturbative expansion results of \sect{sec:Perturbative}. 

As usual, to initiate the Newton-Raphson algorithm one needs an educated seed. We use the perturbative solutions of \sect{sec:Perturbative} as seeds for the lumpy branes near the GL merger line with the uniform branes. The uniform branes are a 1-parameter family of solutions parametrized by the dimensionless temperature $T/\Lambda$. In contrast, the lumpy branes are a 2-parameter family of solutions that we can take to be $T/\Lambda$ and the dimensionless length $L \Lambda$. This means that we need to scan a 2-dimensional parameter space. Our strategy to do so follows two routes. In one of them we follow lines of constant-temperature lumpy branes as their length $L \Lambda$ changes. The temperature $T$ is given by \eqref{calTS} where the constant $\alpha_{\Lambda}$ and $\Lambda$ (to build the dimensionless ratio $T/\Lambda$) are read from the uniform solution at the GL merger. The minimum length of these branes is the GL length $\hat{L}_{\hbox{\tiny GL}}$ computed in \sect{sec:PerturbativeO1}, and constant-temperature branes exist for arbitrarily large $L \Lambda$.
In a second route, we generate curves of lumpy branes that have fixed dimensionless length $L \Lambda$.  In this path the temperature $T/\Lambda$ of the branes changes but at the GL merger with the uniform branes, see e.g.~Fig.~\ref{fig:GLk0}, we know both the temperature $\hat{\taut}$ and the associated GL length $\hat{L}_{\hbox{\tiny GL}}(\hat{\taut})$. 
 Altogether these two solution-generating procedures  allow us to construct a grid of two ``orthogonal-like" lines of solutions that span the phase space of lumpy branes. Further, recall that once we have the numerical solutions $q_j(x,y)$, the thermodynamic quantities of the lumpy branes are read straightforwardly from the expressions discussed in \sect{sec:setupThermo}.

\begin{figure}[th]
\centerline{
\includegraphics[width=1\textwidth]{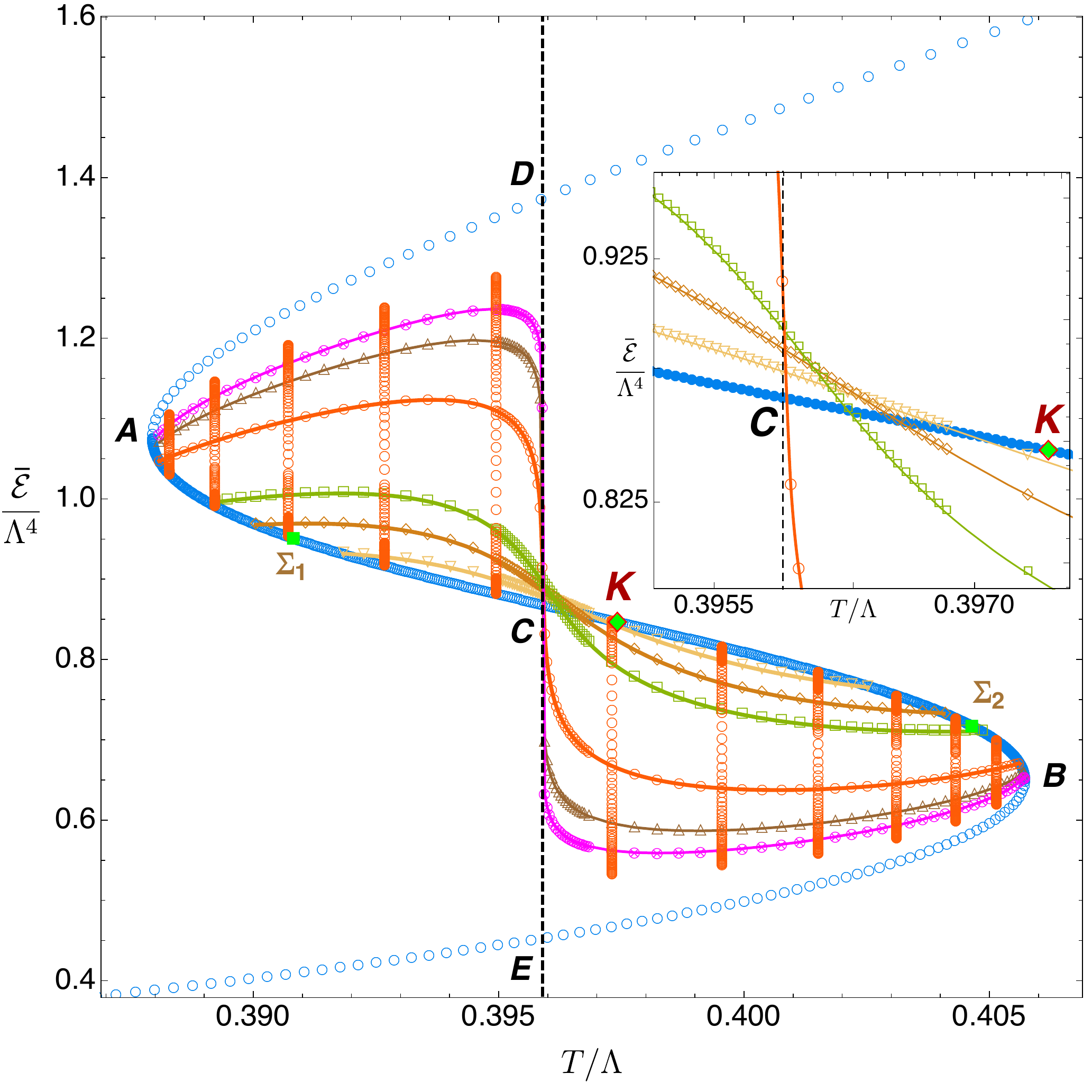}
}
\caption{Phase diagram of Fig.~\ref{fig:Unif} now with both uniform branes  (blue circles) and some nonuniform brane solutions at constant temperature $\hat{\taut}$ (orange circles) or constant length $\hat{L}$.  The eleven constant-temperature vertical lines have (from left to the right): 
\mbox{$\hat{\taut}\simeq \{0.388292,0.389219,0.390711,0.392677,0.394948,0.397308,0.399547,0.401511,0.403112$}, \mbox{$0.404320, 0.405141\}$}.
The six lumpy-brane curves at constant $\hat{L}$ have (from bottom to top on the left): $\hat{L}_{\triangledown} \simeq 5.299674$, $\hat{L}_{\hbox{\large $\diamond$}}\simeq 6.004224$, $\hat{L}_{\hbox{\tiny $\square$}}\simeq 6.900924$,  
$\hat{L}_{\hbox{\tiny$\bigcirc$}} \simeq 11.501849$, $\hat{L}_{\triangle} \simeq 17.906849$, $\hat{L}_{\otimes} \simeq 24.311849$. The inset plot is a zoom in around the region that contain $C$ and $K$ (see also Fig.~\ref{fig:GLk0}) and here we plot the uniform-brane curve and just the four constant-$\hat{L}$ curves $ \{\hat{L}_{\triangledown}, \hat{L}_{\hbox{\large $\diamond$}}, \hat{L}_{\hbox{\tiny $\square$}}, \hat{L}_{\hbox{\tiny$\bigcirc$}} \}$. For reference, $(\hat{\taut},\hat{\cal E})_{\Sigma_1}\simeq (0.390817,0.950579)$, $(\hat{\taut},\hat{\cal E})_{\Sigma_2}\simeq (0.404645,0.717060)$ and $(\hat{\taut},\hat{\cal E})_{K}\simeq (0.397427,0.846337)$.
}
\label{fig:LumpyUnifET}
\end{figure} 

After these preliminaries we are ready to discuss our numerical nonlinear findings. A first important plot is shown in Fig.~\ref{fig:LumpyUnifET}, where we show the dimensionless average energy density 
$\hat{\overline{\mathcal{E}}}=\overline{\mathcal{E}}/\Lambda^4$ as a function of the dimensionless temperature $\hat{\taut}=T/\Lambda$. Recall that for uniform branes $\hat{\overline{\mathcal{E}}}$ coincides with the dimensionless energy density, $\hat{\mathcal{E}}$, which is constant across the entire system. This plot contains again the  uniform-brane spinodal curve (blue circles)  already shown in Fig.~\ref{fig:Unif} but this time we also show some representative examples of lumpy brane solutions (all other lines/curves).

As illustrated in  Fig.~\ref{fig:LumpyUnifET}, a first non-trivial conclusion of our study is that lumpy branes exist {\it only} in the temperature window $\hat{\taut}_A\leq \hat{\taut} \leq\hat{\taut}_B$  where $\hat{\taut}_A\simeq 0.387944$ and $\hat{\taut}_B \simeq 0.405724$. That is, they exist only in the temperature range where the GL-unstable, intermediate branch of uniform solutions  (the curve $ACB$) exists. Of course, we should have anticipated that lumpy branes merge with the uniform branes of the {\it intermediate} branch and thus in the window $\hat{\taut}_A\leq \hat{\taut} \leq\hat{\taut}_B$. However, it was a logical possibility that,  away from this merger, lumpy branes might exist also for temperatures outside the range  $\hat{\taut}\in [\hat{\taut}_A,\hat{\taut}_B]$. We have generated considerably more solutions than those shown in Fig.~\ref{fig:LumpyUnifET} in order to test this possibility and, as stated above, we have found that it is not realised. 

To continue interpreting Fig.~\ref{fig:LumpyUnifET}, it is convenient to discuss separately the regions \mbox{$\hat{\taut}_A \leq \hat{\taut}<\hat{\taut}_{\rm c}$} and \mbox{$\hat{\taut}_{\rm c} < \hat{\taut} \leq \hat{\taut}_B$}, i.e.~the regions to the left and to the right, respectively, of the vertical dashed line $DCE$. Recall that this auxiliary line identifies the critical temperature $\hat{\taut}=\hat{\taut}_{\rm c}$ at which the first-order phase transition for uniform branes takes place (see right panel of  Fig.~\ref{fig:Unif}). 

So consider first lumpy branes that exist in the window $\hat{\taut}_A \leq \hat{\taut}<\hat{\taut}_{\rm c}$:

\begin{enumerate}
\item For a given temperature $\hat{\taut}$ in this range, lumpy branes exist with a dimensionless length that satisfies $\hat{L}_{\hbox{\tiny GL}} \leq \hat L \leq \infty$. In particular, the vertical lines of orange circles  of Fig.~\ref{fig:LumpyUnifET} are lumpy branes at constant $\hat{\taut}$ that have $\hat{L}=\hat{L}_{\hbox{\tiny GL}}(\hat{\taut})$ when they bifurcate from the {\it intermediate} uniform-brane branch $AC$. Then they extend for arbitrarily large $\hat{L}$. More precisely, for $\hat{\taut}<\hat{\taut}_{\rm c}$, constant-$\hat{\taut}$ lumpy branes  extend upwards (i.e.~towards higher $\hat{\overline{\mathcal{E}}}$) as $\hat{L}$ grows. However, we find that for a given step increase in $\hat{L}$, the increase in $\hat{\overline{\mathcal{E}}}$ gets smaller and smaller as $\hat{L}$ grows,  i.e.~$(\partial \hat{\overline{\mathcal{E}}}/\partial \hat{L})\big|_{\hat{\taut}}$ is a monotonically decreasing function of $\hat{L}$. This is explicitly observed in the vertical lines that we display: {\it away} from the merger each two consecutive orange circles are separated by the {\it same} step in $\hat{L}$ but the step increase in $\hat{\cal E}$ is significantly decreasing as we move upwards. Due to the large hierarchy of scales that develops it is difficult to construct lumpy branes with $\hat{L}\to \infty$. But the above behaviour strongly suggests that lumpy branes with $\hat{\taut}<\hat{\taut}_{\rm c}$ are precisely bounded by the {\it heavy} uniform branch segment $AD$ when $\hat{L}\to \infty$, i.e.~we conjecture that 
\be
\underset{\hbox{\tiny $\hat{L}$}\to\infty}{\lim}\,\frac{\partial \hat{\cal E}}{\partial \hat{L}}\big|_{\hat{\taut}}\to 0 \qquad \mbox{and} \qquad
\underset{\hbox{\tiny $\hat{L}$}\to\infty}{\lim}\hat{\cal E} \big|_{\hbox{\tiny const}\: \hat{\taut}}\to \hat{\cal E}_{\rm u}^{AD}(\hat{\taut}) \,.
\ee 

\item The other six curves (with $ {\triangledown}, \hbox{\large $\diamond$}, \hbox{\tiny $\square$}, {\hbox{\tiny$\bigcirc$}}, {\triangle},{\otimes}$) in Fig.~\ref{fig:LumpyUnifET}, that intersect the vertical lines, describe six families of lumpy branes at constant $\hat{L}$. Concretely, the chosen fixed $\hat{L}$ increases as the curves go from the bottom to the top (for $\hat{\taut}<\hat{\taut}_{\rm c}$), i.e.~$\hat{L}_{\triangledown}<\hat{L}_{\hbox{\large $\diamond$}}<\hat{L}_{\hbox{\tiny $\square$}} <\hat{L}_{\hbox{\tiny$\bigcirc$}}<\hat{L}_{\triangle}<\hat{L}_{\otimes}$. We find that constant-$\hat{L}$ lumpy branes {\it always} bifurcate from the {\it intermediate} uniform brane branch $AC$ at a temperature/point  that  matches the temperature already found {\it independently} in Fig.~\ref{fig:GLk0}, $\{\hat{\taut},\hat{L}\}=\{\hat{\taut},\hat{L}_{\hbox{\tiny GL}}(\hat{\taut})\}$. This is thus a test of our numerics. In particular, curves with (constant) higher $\hat{L}$ bifurcate from the {\it intermediate} uniform brane with lower $\hat{\taut}$, i.e.~the merger is closer to the endpoint $A$. In the limit $\hat{L}\to \infty$, this bifurcation occurs exactly at $\{\hat{\taut},\hat{\cal E}\}=\{\hat{\taut}_A,\hat{\cal E}_A\}$ i.e.~at point $A$, {\it in agreement with} the GL linear results of Fig.~\ref{fig:GLk0}. As $\hat{L}$ decreases, the bifurcation occurs at temperatures that are increasingly closer to $\hat{\taut}=\hat{\taut}_{\rm c}$. For $\hat{\taut}<\hat{\taut}_{\rm c}$, constant-$\hat{L}$ curves do not intersect further the uniform branch $AC$.  

\end{enumerate}

Let us now follow these constant-$\hat{L}$ curves as they flow into the second relevant region, namely $\hat{\taut}>\hat{\taut}_{\rm c}$. Fig.~\ref{fig:LumpyUnifET} shows that, if this was not already happening for smaller $\hat{\taut}$, all these curves have a drop in their 
$\hat{\overline{\mathcal{E}}}$ as they approach $\hat{\taut}_{\rm c}$ from the left. For very large $\hat{L}$ this drop  is dramatic with an almost vertical slope (see e.g.~the magenta, $\otimes$ curve). Therefore, as best illustrated in the inset plot of Fig.~\ref{fig:LumpyUnifET}  that zooms in  the region around point $C$, all constant-$\hat{L}$ curves pile up around point $C$ in a way such that: 
\begin{enumerate}
\item As $\hat{\taut}\to \hat{\taut}_{\rm c}^{-}$ (approaching from the left) all curves have $\hat{\cal E}>\hat{\cal E}_C$. In particular,  this means that these curves do not intersect the uniform branch $AC$ near $C$.

\item Once at $\hat{\taut}>\hat{\taut}_{\rm c}$, {\it all} constant-$\hat{L}$ curves that bifurcated from the uniform branes in the trench $AC$ cross the uniform brane branch curve between $C$ and $K$. Recall that $K$ describes the uniform brane solution that has the largest GL wavenumber $k_{\hbox{\tiny GL}}$ or, equivalently, that has the lowest $\hat{L}_{\hbox{\tiny GL}}=2\pi/k_{\hbox{\tiny GL}}$; see Fig.~\ref{fig:GLk0}. After this crossing, the constant-$\hat{L}$ lumpy branes keep extending to higher $\hat{\taut}$ with an energy density lower that the {\it intermediate} uniform brane with the same $\hat{\taut}$. This keeps happening until they merge again with the uniform brane in the trench $KB$ at a critical temperature that is again the one predicted by the GL zero-mode analysis, i.e.~at the highest $\hat{\taut}$ that satisfies the condition $\hat{L}=\hat{L}_{\hbox{\tiny GL}}(\hat{\taut})$, see again Fig.~\ref{fig:GLk0}. 
Lumpy-brane curves with higher constant $\hat{L}$ merge with the uniform branch $KB$ at a point that is closer to B. In the limit where $\hat{L}\to \infty$ this merger occurs precisely at point $B$ in Fig.~\ref{fig:LumpyUnifET},  in agreement with the GL linear results of Fig.~\ref{fig:GLk0}.

\item There are constant-$\hat{L}$ lumpy branes with very small $\hat{L}$ that bifurcate from the uniform brane branch {\it only} in the trench $CK$ ({\it instead} of $AC$). Then they extend to higher $\hat{\taut}$, initially with $\hat{\cal E}$ higher that the uniform branes with same $\hat{\taut}$ before they cross the uniform branch $CK$ at a temperature $\hat{\taut}<\hat{\taut}_K$ and proceed to higher $\hat{\taut}$ below point $K$ until they merge again with the uniform brane branch but this time in the trench $KB$ (at a point very close to $K$). This happens for fixed-$\hat{L}$ branes whenever $\hat{L}_{\hbox{\tiny GL}}(\hat{\taut}_K)<\hat{L}<\hat{L}_{\hbox{\tiny GL}}(\hat{\taut}_C)$.  

\end{enumerate}

The three features of the lumpy branes just listed are compatible with the following interpretation that merges our nonlinear findings, summarized in Fig.~\ref{fig:LumpyUnifET}, with the GL linear results of \sect{sec:PerturbativeO1}, summarized in Fig.~\ref{fig:GLk0}. 
Indeed, let us go back to Fig.~\ref{fig:GLk0} and consider an auxiliary horizontal line at constant $\hat{k}_{\hbox{\tiny GL}}$, i.e.~at constant $\hat{L}_{\hbox{\tiny GL}}$. This line intersects the curve $\hat{k}_{\hbox{\tiny GL}}(\hat{\taut})$ at two points. These are the two merger points of constant $\hat{L}$ lumpy branes with the uniform brane that we identify in Fig.~\ref{fig:LumpyUnifET}. One of the mergers --- let us denote it simply as the ``left" merger --- has $\hat{\taut}_A\leq \hat{\taut}\leq\hat{\taut}_K$ and the other --- the ``right" merger --- has $\hat{\taut}_K\leq \hat{\taut}\leq\hat{\taut}_B$. Since the maximum of the GL wavenumber occurs at a temperature that is higher than the one of the first-order phase transition of the uniform system, $\hat{\taut}_K>\hat{\taut}_{\rm c}$, it follows that the ``left" mergers of lumpy branes with constant $\hat{L}_{\hbox{\tiny GL}}(\hat{\taut}_K)<\hat{L}<\hat{L}_{\hbox{\tiny GL}}(\hat{\taut}_C)$ are in the trench $CK$ of Fig.~\ref{fig:LumpyUnifET}. But, for $\hat{L}>\hat{L}_{\hbox{\tiny GL}}(\hat{\taut}_C)$, the ``left" merger is located in the trench $AC$, with the $\hat{L}\to \infty$ ``left" merger being at $A$. On the other hand, the ``right" merger is always located in the trench $KB$ of Fig.~\ref{fig:LumpyUnifET}, with the ``right" merger of the $\hat{L}\to \infty$ lumpy branes being at $B$. Our nonlinear results summarized in Fig.~\ref{fig:LumpyUnifET} further conclude that there are no lumpy branes with $\hat{L}<\hat{L}_{\hbox{\tiny GL}}(\hat{\taut}_K)$. As $\hat{L}$ approaches $\hat{L}_{\hbox{\tiny GL}}(\hat{\taut}_K)$ from above, lumpy branes exist only in a small neighbourhood around point $K$ in Fig.~\ref{fig:LumpyUnifET}, with the characteristics described in item 3 in the list above.\footnote{Note that for other values of the (super)potential parameters $\phi_M$ and $\phi_Q$ in \eqref{superpotential} (we have picked $\phi_M=1$ and $\phi_Q=10$), or in similar spinodal systems, it might well be the case that $\hat{\taut}_{\rm c}>\hat{\taut}_K$ or, for a fine-tuned choice of potential, even $\hat{\taut}_{\rm c}=\hat{\taut}_K$. If that is the case our conclusions should still apply with the appropriate shift of $K$ to the left of $C$ in Figs.~\ref{fig:GLk0} and~\ref{fig:LumpyUnifET}. Note that for this exercise we only need to find the uniform branes of the system and solve for static linear perturbations of these branes which determine the zero-mode GL wavenumber and thus the location of its maximum $K$ with respect to $\hat{\taut}_{\rm c}$. That is to say, we just need to complete the tasks described in sections~\ref{sec:PerturbativeO0} and \ref{sec:PerturbativeO1}.}   

We stress again that lumpy branes exist only in the temperature range $ \hat{\taut}_A \leq  \hat{\taut}\leq \hat{\taut}_B$. Our nonlinear results of Fig.~\ref{fig:LumpyUnifET} give strong evidence that constant-$\hat{\taut}<\hat{\taut}_{\rm c}$ branes extend to arbitrarily large $\hat{L}$ with 
\be
\underset{\hbox{\tiny $\hat{L}$}\to\infty}{\lim}\hat{\cal E} \big|_{\hbox{\tiny const}\: \hat{\taut}}\to \hat{\cal E}_{\rm u}^{AD}(\hat{\taut})\,,
\ee
see the {\it heavy} uniform brane trench $AD$ in Fig.~\ref{fig:LumpyUnifET}. On the other hand, our results also strongly indicate that  constant-$\hat{\taut}>\hat{\taut}_{\rm c}$ branes extend to arbitrarily large $\hat{L}$ with 
\be
\underset{\hbox{\tiny $\hat{L}$}\to\infty}{\lim}\hat{\cal E} \big|_{\hbox{\tiny const}\: \hat{\taut}}\to \hat{\cal E}_{\rm u}^{EB}(\hat{\taut})\,, 
\ee
see the {\it light} uniform brane trench $ED$ in Fig.~\ref{fig:LumpyUnifET}. Moreover, as $\hat{L}$ grows arbitrarily large, the constant-$\hat{L}$ lumpy branes intersect (without merging with) the {\it intermediate} uniform brane branch $AB$ at a point that is arbitrarily close (from the right) to point $C$ in Fig.~\ref{fig:LumpyUnifET} and with a slope 
$(\partial \hat{\overline{\mathcal{E}}} / \partial \hat{\taut})\big|_{\hat{\taut}_{\rm c}}$ that grows unbounded, that is 
\be
\underset{\hbox{\tiny $\hat{L}$}\to\infty}{\lim}\, \left.  \frac{\partial \hat{\overline{\mathcal{E}}}}{\partial \hat{\taut}}\right|_{\hat{\taut}_{\rm c}}\to \infty\,.
\ee
In the limit $\hat{L}\to \infty$ we thus conjecture that lumpy branes are limited by the curve \mbox{$ADCEB$} with two cusps connected by the vertical $DCE$ line in Fig.~\ref{fig:LumpyUnifET}.
To argue further in favour of this conjecture, it is important to explore better the properties of the system in this  $\hat{L}\to \infty$ limit and the associated limiting curve $ADCEB$. For that it is instructive to look at the energy density profile $\hat{\cal E}(x)$ of the lumpy branes as a function of the inhomogeneous direction $x$. 

\begin{figure}[t]
\centerline{
\includegraphics[width=.35\textwidth]{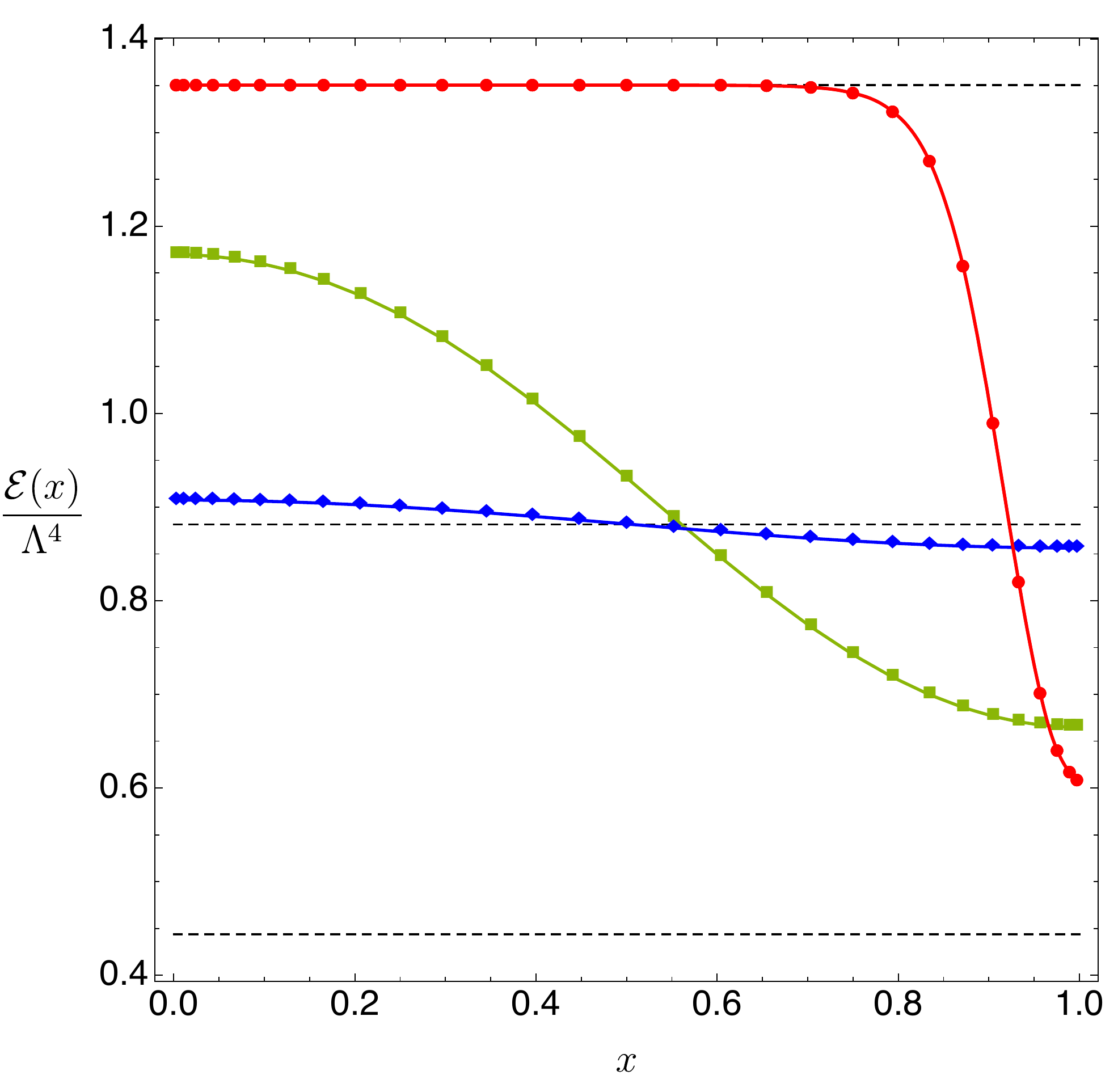}
\includegraphics[width=.35\textwidth]{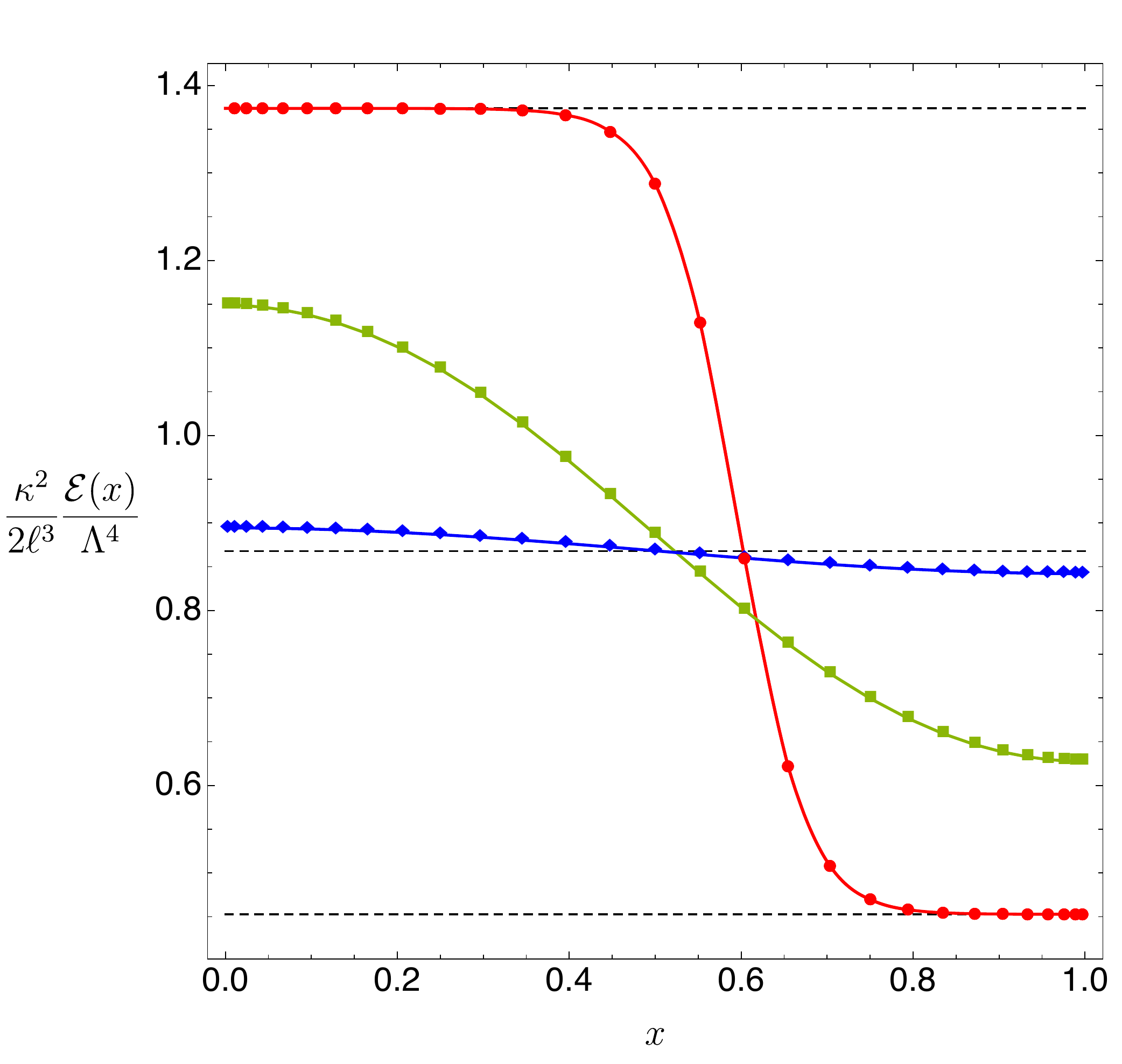}
\includegraphics[width=.35\textwidth]{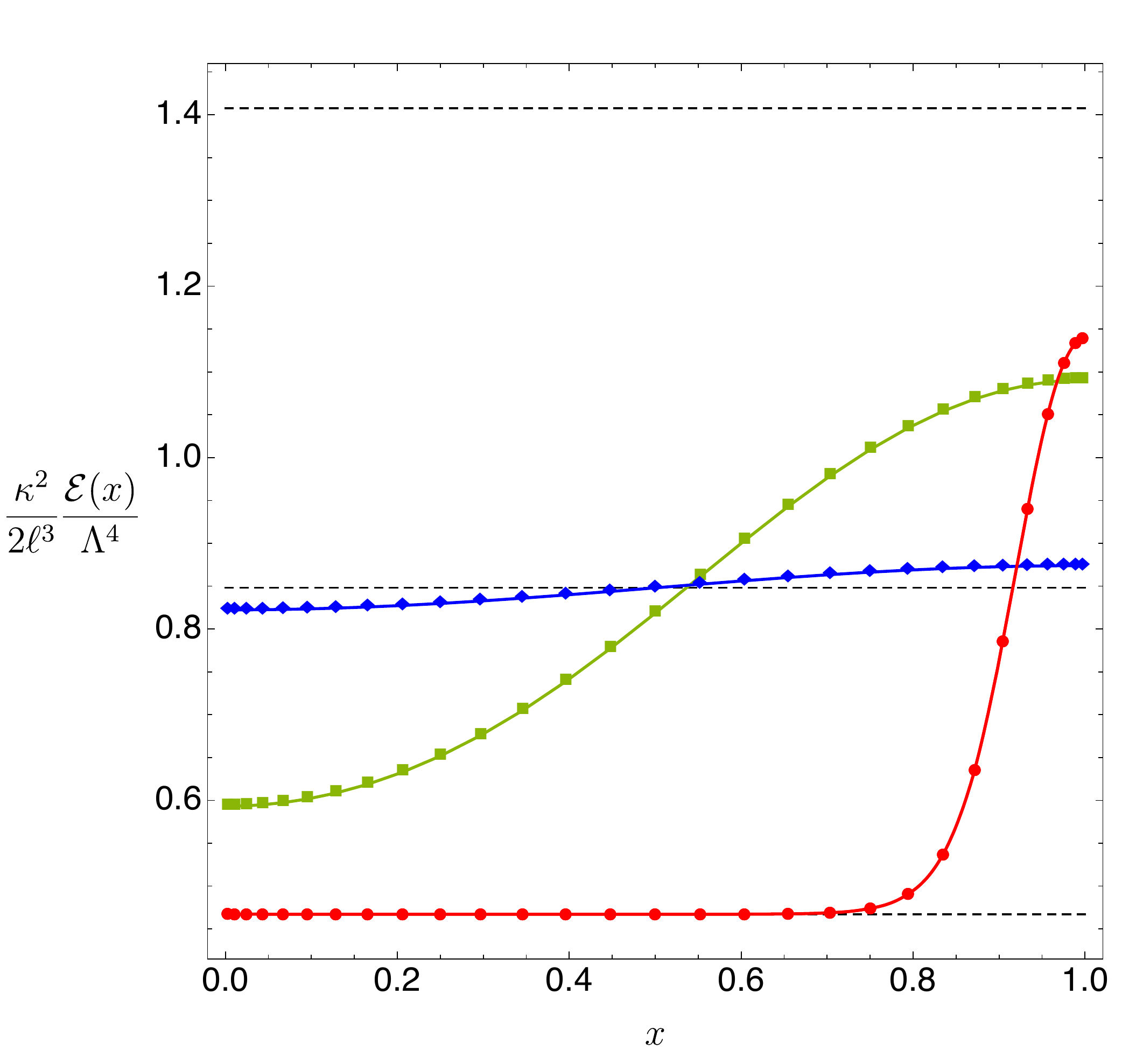}
}
\caption{Energy density profile $\hat{\cal E}(x)$  for three lumpy branes at the same temperature $\hat{\taut}$.$^{\ref{foot13}}$ (Left) Lumpy branes at constant $\hat{\taut}\simeq 0.394948$ (this is the first vertical line of orange circles to the left of $C$ in Fig.~\ref{fig:LumpyUnifET}). For reference, at the merger the lumpy brane with this temperature has \mbox{$\hat{L}=\hat{L}_{\hbox{\tiny GL}}\simeq 4.808993$} and 
$\hat{L}_{\blacklozenge}\simeq 4.819793$, $\hat{L}_{\hbox{\tiny $\blacksquare$}}\simeq 5.996993$, $\hat{L}_{\bullet}\simeq 37.208993$. For this temperature the heavy, intermediate and light uniform branes have energy densities $\hat{\cal E}^{heavy}\simeq 1.350615$, \mbox{$\hat{\cal E}^{inter}\simeq 0.881537$} and $\hat{\cal E}^{light}\simeq 0.443756$, respectively.  These energy densities are indicated by the dashed horizontal  lines in the plot.  
(Middle) Lumpy branes at constant $\hat{\taut}\simeq 0.395894 \lesssim \hat{\taut}_c$ (so, very close to $\hat{\taut}_c\simeq 0.3958945$). For reference, $\hat{L}=\hat{L}_{\hbox{\tiny GL}}\simeq 4.750995$ and $\hat{L}_{\blacklozenge}\simeq 4.761845$, $\hat{L}_{\hbox{\tiny $\blacksquare$}}\simeq 5.944495$,  $\hat{L}_{\bullet}\simeq 46.523495$, and the energy densities of the relevant uniform branes (dashed horizontal lines) are  $\hat{\cal E}^{heavy}\simeq 1.373843$, $\hat{\cal E}^{inter}\simeq 0.867966$, $\hat{\cal E}^{light}\simeq 0.452747$. 
(Right) Lumpy branes at constant $\hat{\taut}\simeq 0.397308$ (this is the first vertical line of orange circles to the right of $C$ in Fig.~\ref{fig:LumpyUnifET}). For reference, at the merger the lumpy brane with this temperature has $\hat{L}=\hat{L}_{\hbox{\tiny GL}}\simeq 4.716232$ and $\hat{L}_{\blacklozenge}\simeq 4.727157$, $\hat{L}_{\hbox{\tiny $\blacksquare$}}\simeq 5.917982$, $\hat{L}_{\bullet}\simeq 34.759982$, and the energy densities of the relevant uniform branes (dashed horizontal lines) are  $\hat{\cal E}^{heavy}\simeq 1.407521$, $\hat{\cal E}^{inter}\simeq 0.848014$, $\hat{\cal E}^{light}\simeq 0.467119$.}
\label{fig:profilesExConsT}
\end{figure}

\begin{figure}[t!]
\centerline{
\includegraphics[width=.48\textwidth]{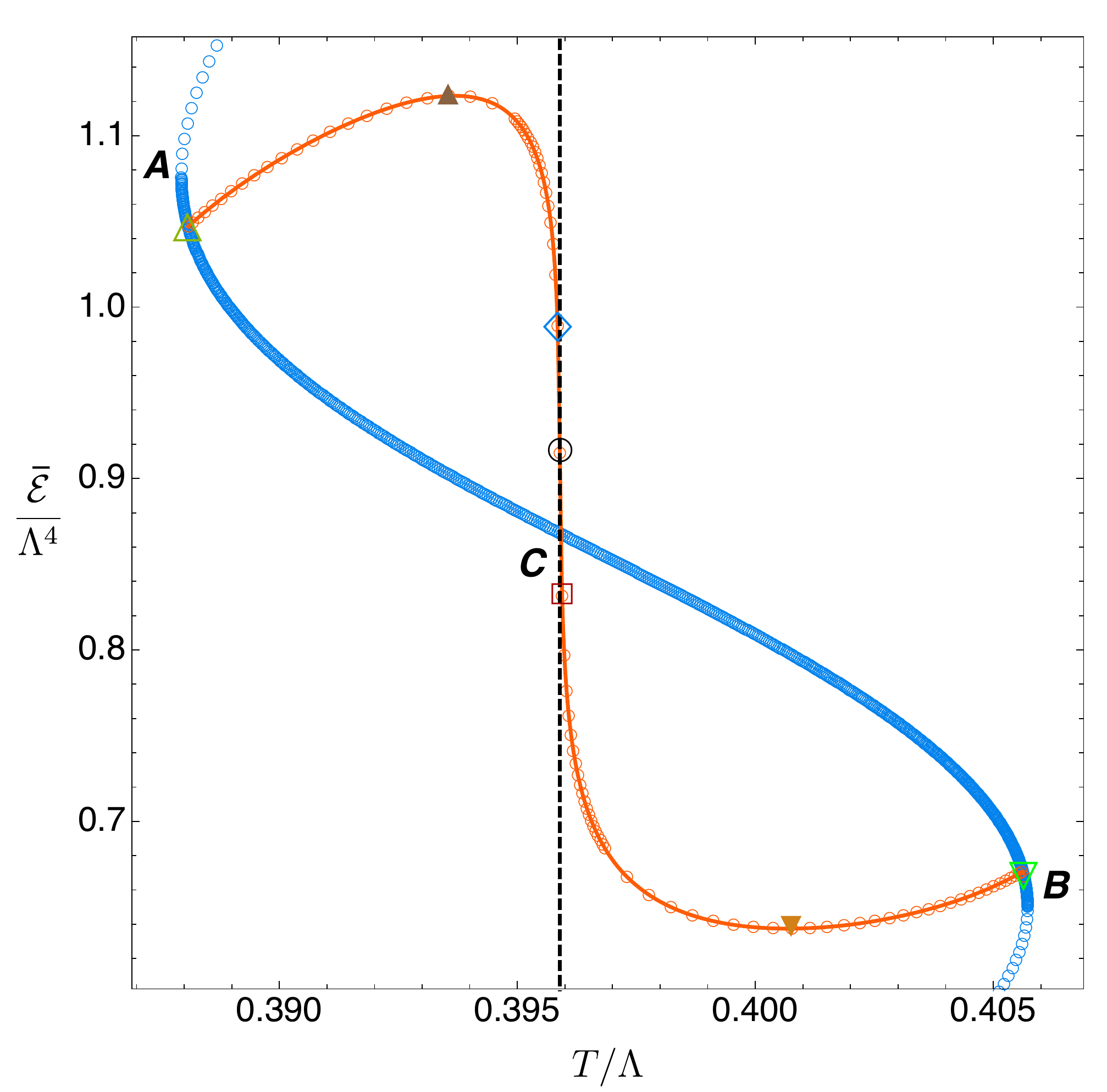}
\hspace{0.1cm}
\includegraphics[width=.51\textwidth]{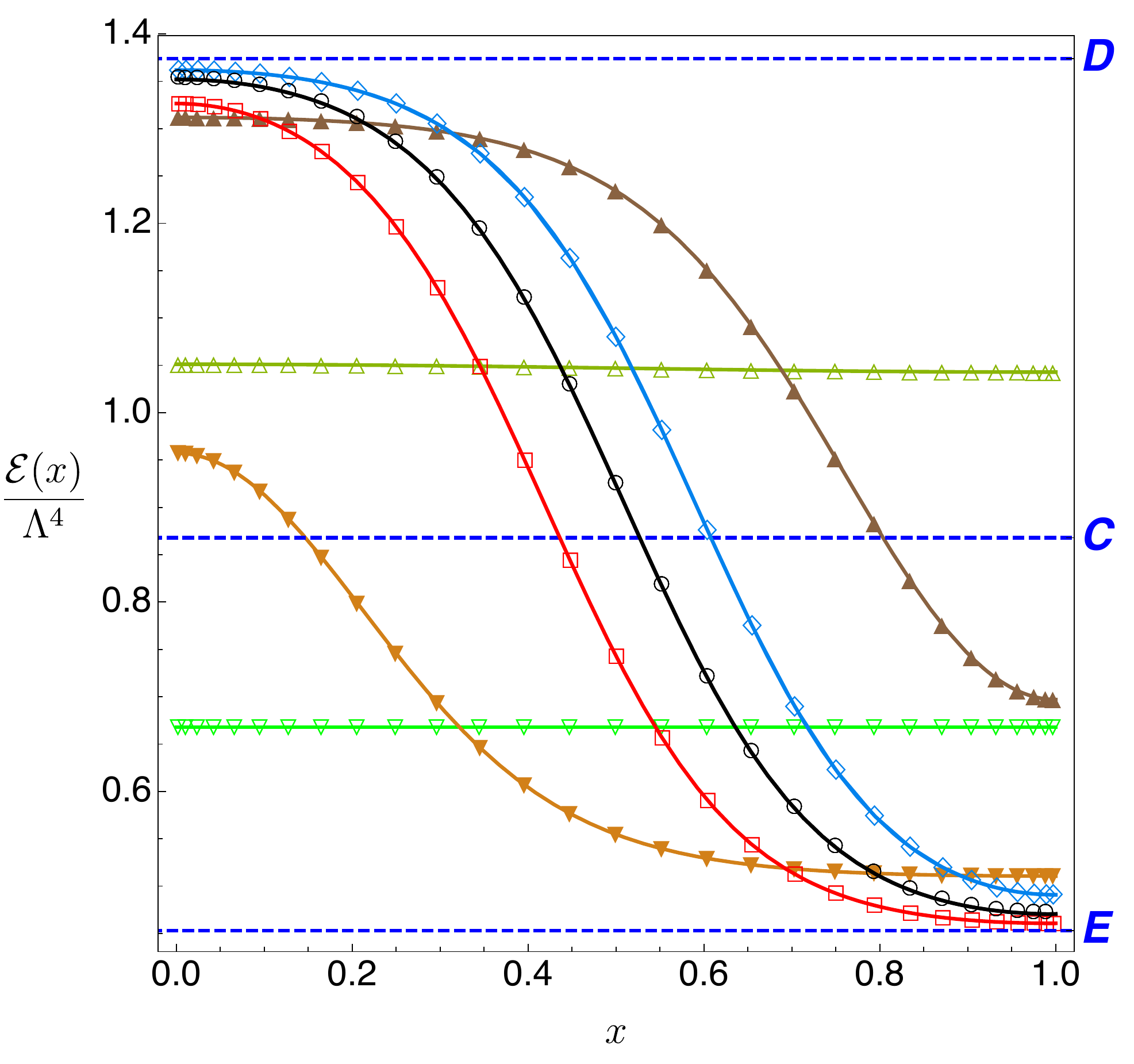}
}
\caption{Lumpy branes at constant  $\hat{L}\simeq 11.501849$. 
(Left) $\hat{\cal E}$ as a function of $\hat{\taut}$. This figure reproduces  Fig.~\ref{fig:LumpyUnifET} but this time it singles out only the relevant lumpy brane with \mbox{$\hat{L}=\hat{L}_{\hbox{\tiny$\bigcirc$}}\simeq 11.501849$} (orange circles) and the uniform branes (blue circles) and zooms in on the relevant region. It also identifies 7 solutions whose energy density profiles are then plotted on the right panel. From left to right these are given by: 
$(\hat{\taut},\hat{\cal E})_{\triangle}\simeq (0.388085, 1.047078),$ 
$(\hat{\taut},\hat{\cal E})_{\blacktriangle}\simeq (0.393560, 1.123216), $ 
$(\hat{\taut},\hat{\cal E})_{\hbox{\large $\diamond$}}\simeq (0.395846, 0.988820), $ 
$(\hat{\taut},\hat{\cal E})_{\hbox{\tiny$\bigcirc$}}\simeq 
(0.3958945, 0.914833), $ 
$(\hat{\taut},\hat{\cal E})_{\square}\simeq (0.395941, 0.831460), $ $(\hat{\taut},\hat{\cal E})_{\blacktriangledown}\simeq  (0.400765, 0.637410), $ 
$(\hat{\taut},\hat{\cal E})_{\triangledown}\simeq (0.405648, 0.667817)$.
(Right) Energy density profile $\hat{\cal E}(x)$  for the 7 lumpy branes pinpointed in the left panel.$^{\ref{foot13}}$ The same shape/colour code is used. }
\label{fig:profilesExConstL}
\end{figure}

In Fig.~\ref{fig:profilesExConsT} 
we first consider lumpy branes with the same $\hat{\taut}$ but different lengths $\hat{L}$.\footnote{\label{foot:profiles}When interpreting these figures recall, from the discussion above \eqref{ansatzZ}, that our solutions have  ${\bb Z}_2$ symmetry: the range $x\in [0,1]$ describes only the brane's half  $\tilde{x}=x L/2 \in [0,L/2]$. To get the other half extension, $\tilde{x} \in [-L/2,0]$, we just need to flip the profiles of Figs.~\ref{fig:profilesExConsT}-\ref{fig:profilesExConstL} along their vertical axis.\label{foot13}}  In the left panel we have the profile of 3 lumpy branes with $\hat{\taut}\simeq 0.394948 <\hat{\taut}_c$; in the  middle panel we have the profile of 3 lumpy branes with $\hat{\taut}\simeq 0.395894 \lesssim \hat{\taut}_c$ (i.e.~almost at $\hat{\taut}_c\simeq 0.3958945$); and, finally, in the right panel we show the profile of 3 lumpy branes with $\hat{\taut}\simeq 0.397308 >\hat{\taut}_c$. In all panels, the blue diamond lines have a length only slightly above $\hat{L}_{\hbox{\tiny GL}}(\hat{\taut})$. Therefore, the profile of these lumpy branes is almost flat and very close to the horizontal dashed line that represents the {\it intermediate} uniform brane with $\hat{\cal E}_{\rm u}^{AC}(\hat{\taut})$ (left/middle panels) or $\hat{\cal E}_{\rm u}^{CB}(\hat{\taut})$ (right panel). Then, the green square curves have a length  of roughly $\hat{L} \sim1.25 \hat{L}_{\hbox{\tiny GL}}(\hat{\taut})$. We see that the profile starts becoming considerably deformed with one of the ``halves" pulling well above (below) the uniform brane profile with the same $\hat{\taut}$. 
Finally, the red disk curves represent lumpy branes that have a length $\hat{L}(\hat{\taut})$  that is considerably higher than $\hat{L}_{\hbox{\tiny GL}}(\hat{\taut}) $ (exact values in the caption). We see that the profile of lumpy branes with $\hat{\taut}\simeq 0.394948 <\hat{\taut}_c$ (left panel) is, in a wide range of $x$ ($x\lesssim 0.7$), very flat with $\hat{\cal E}(x)\sim \hat{\cal E}_{\rm u}^{AD}(\hat{\taut})$, i.e.~with an energy density that is the same as the one of the {\it heavy} uniform brane in the trench $AD$ that has the same $\hat{\taut}$ (upper horizontal dashed line).
Then, for  $x\gtrsim 0.7$, $\hat{\cal E}(x)$ falls considerably towards the  energy density $\hat{\cal E}_{\rm u}^{light}(\hat{\taut})$ of the {\it light} uniform brane that has the same temperature (lower horizontal dashed line). Still in Fig.~\ref{fig:profilesExConsT}, the middle panel shows that as $\hat{\taut}$ approaches $\hat{\taut}_{\rm c}$ and for large $\hat{L}$ (red disks), the profile $\hat{\cal E}(x)$ describes a domain-wall solution that interpolates between $ \hat{\cal E}_{\rm u}^{AD}(\hat{\taut})$ (for small $x$) and   $\hat{\cal E}_{\rm u}^{light}(\hat{\taut})$  (for large $x$). 
On the other hand, for $\hat{\taut}\simeq 0.397308 >\hat{\taut}_c$ (right panel of Fig.~\ref{fig:profilesExConsT}) the roles of the {\it heavy} and {\it light} uniforms get reversed: for $x\lesssim 0.7$ the red disk lumpy curve is almost flat with an energy density close to the one of the {\it light} uniform brane with the same $\hat{\taut}$, $\hat{\cal E}(x)\sim\hat{\cal E}_{\rm u}^{EB}(\hat{\taut})$ (lower horizontal dashed line), while for  $x\gtrsim 0.7$, $\hat{\cal E}(x)$ starts increasing towards the energy density $\hat{\cal E}_{\rm u}^{heavy}(\hat{\taut})$ of the {\it heavy} uniform brane with the same $\hat{\taut}$ (upper dashed horizontal line).

The $\hat{L}\to \infty$ limit of lumpy branes and its association with the limiting curve $ADCEB$ is further revealed when we complement Fig.~\ref{fig:profilesExConsT} with an analysis of the  energy density profile $\hat{\cal E}(x)$ of a constant-$\hat{L}$ family of branes for different values of the temperature. One such analysis  is done in  Fig.~\ref{fig:profilesExConstL} where we fix $\hat{L}\simeq 11.501849$: this picks the fourth constant-$\hat{L}_{\hbox{\tiny$\bigcirc$}}$ curve (from bottom-left) in the plot of Fig.~\ref{fig:LumpyUnifET}. For clarity we single out this curve and reproduce it --- this time only the relevant zoomed in region of Fig.~\ref{fig:LumpyUnifET} --- in the left panel of Fig.~\ref{fig:profilesExConstL}. We pinpoint a total of seven solutions with seven different temperatures (each one with its own distinctive plot marker shape and colour). The first ($\triangle$) and the last ($\triangledown$) solutions are the two mergers with the {\it intermediate} uniform brane, the second ($\blacktriangle$) and sixth ($\blacktriangledown$) solutions are the two extrema of $\hat{\cal E}(\hat{\taut})\big|_{\hat{L}}$, 
and the third ($\hbox{\large $\diamond$}$), fourth ({\Large $\circ$}) and fifth ($\square$) plot markers identify three solutions with $\hat{\taut}$ at or very close to $\hat{\taut}_{\rm c}$. As in  Fig.~\ref{fig:profilesExConsT}, we see that the profile of the two lumpy branes at the merger is flat: they coincide with the uniform branes. As we move to the ``extrema'' solutions with plot markers  $\blacktriangle$ and $\blacktriangledown$ we see, like for similar solutions in Fig.~\ref{fig:profilesExConsT},  that the profile is considerably deformed. More important for our purposes are the solutions with $\hat{\taut}\sim \hat{\taut}_{\rm c}$, e.g.~$\hbox{\large $\diamond$}$, {\Large $\circ$}, $\square$. We see that for such cases the profile reaches its maximum deformation in the sense that the solution clearly interpolates between to regions that are fairly flat. Importantly, the small-$x$ flat region is approaching the energy density $\hat{\cal E}_{\rm u}^{D}(\hat{\taut}_{\rm c})$ of the {\it heavy} uniform brane that has $\hat{\taut}=\hat{\taut}_{\rm c}$ (see the upper, horizontal, dashed, blue line labelled by D). Similarly, the large-$x$ flat region is approaching the energy density $\hat{\cal E}_{\rm u}^{E}(\hat{\taut}_{\rm c})$ of the {\it light} uniform brane that has $\hat{\taut}=\hat{\taut}_{\rm c}$ (see the lower, horizontal, dashed, blue curve labelled by E). We further see that the closer we are to $\hat{\taut}_{\rm c}^-$ ($\hat{\taut}_{\rm c}^+$), the closer we get  to  $\hat{\cal E}_{\rm u}^{D}(\hat{\taut})$ ($\hat{\cal E}_{\rm u}^{E}(\hat{\taut})$). The plot of Fig.~\ref{fig:profilesExConstL} is for a moderate value of $\hat{L}$. Combined with the findings of the discussion of Fig.~\ref{fig:profilesExConsT} we conclude that as $\hat{L}$ grows large and $\hat{\taut}\to \hat{\taut}_{\rm c}$, the flat regions get more extended in $x$ and the domain wall that interpolates between them at $\hat{\cal E}\sim \hat{\cal E}_{\rm u}^{D}(\hat{\taut}_{\rm c})$ and $\hat{\cal E}\sim \hat{\cal E}_{\rm u}^{E}(\hat{\taut}_{\rm c})$ gets  narrower.

Altogether, the findings summarized in Fig.~\ref{fig:profilesExConsT} and Fig.~\ref{fig:profilesExConstL} lead to the following conclusion/conjecture. 
 In the double limit $\hat{L}\to \infty$ and $\hat{\taut}\to \hat{\taut}_{\rm c}$ our results support the conjecture that 
 \be
 \underset{\hbox{\tiny $\hat{L}$}\to\infty}{\lim} \, \left.
 \frac{\partial \hat{\cal E}}{\partial \hat{\taut}}\right|_{\hat{\taut}_{\rm c}} \to \infty \,.
\ee
That is, in this double limit we have a family of lumpy branes that fills up the segment $DCE$ of Fig.~\ref{fig:LumpyUnifET}. All this segment describes infinite-length lumpy branes that are sharp/narrow domain wall solutions interpolating (along $0\leq \tilde{x} \leq \infty$) between two flat regions: one with $\hat{\cal E}(\tilde{x})= \hat{\cal E}_{\rm u}^{D}(\hat{\taut}_{\rm c})$ and the other with $\hat{\cal E}(\tilde{x}) = \hat{\cal E}_{\rm u}^{E}(\hat{\taut}_{\rm c})$.  These are the phase-separated configurations discussed above. As we move up from $C$ to $D$, the region of $\tilde{x}$ with  $\hat{\cal E}(\tilde{x})= \hat{\cal E}_{\rm u}^{D}$ increases while as we move down from $C$ to $E$, the region  of $\tilde{x}$ with  $\hat{\cal E}(\tilde{x})= \hat{\cal E}_{\rm u}^{E}$ increases. We have infinite domain wall solutions that interpolate between the two uniform phases of the system at $\taut=\taut_{\rm c}$. Moreover, keeping the limit $\hat{L}\to \infty$, but relaxing the condition $\hat{\taut}\to \hat{\taut}_{\rm c}$, the results summarized in Figs.~\ref{fig:profilesExConsT} and \ref{fig:profilesExConstL} give evidence to conjecture that infinite-length lumpy branes exist only for $\hat{\taut}_A\leq \hat{\taut} \leq \hat{\taut}_B$ and are exactly at the line $ADCEB$ of Fig.~\ref{fig:LumpyUnifET}. 

We will now discuss the thermal competition between  lumpy and uniform nonconformal branes in the microcanonical ensemble. Recall that  we keep the dimensionless length $\hat{L}$ and the  Killing energy density 
$\hat{\rho}$  fixed and the relevant thermodynamic potential is the Killing entropy density $\hat{\sigma}$. Again, uniform and lumpy branes co-exist for temperatures $\hat{\taut}_A\leq \hat{\taut} \leq \hat{\taut}_B$. So in the microcanonical ensemble, given a lumpy brane with $(\hat{L},\hat{\rho})$, our first task is to find the uniform brane (i.e.~the temperature $\hat{\taut}$ which parametrizes this family) that has the same $(\hat{L},\hat{\rho})$ as the chosen lumpy brane. Once this is done, we can  compare the Killing entropy  densities $\hat{\sigma}$ of the two solutions at the same selected $(\hat{L},\hat{\rho})$ pair. As for the perturbative analysis of \sect{sec:PerturbativeOn} --- see e.g.~the discussion of \eqref{unifEntropySameEL} ---
we compute the entropy difference between the two phases when they have the same $\hat{L}$ and $\hat{\rho}$:  
 \begin{equation}\label{unifEntropySameELnonlinear}
 \Delta \hat{\sigma}\big|_{\hbox{\footnotesize same } \hat{L}, \hat{\rho}} =\Big[ \hat{ \sigma}_{\rm nu}-\hat{\sigma}_{\rm u} \Big]_{\hbox{\footnotesize same } \hat{L}, \hat{\rho}} \,.
\end{equation}
As before, the subscript ``nu'' stands for the nonuniform (lumpy) brane and ``u'' denotes the uniform brane. From our perturbative analysis recall that at the merger curve  (ACB in Fig.~\ref{fig:LumpyUnifET} or the black dotted line in Fig.~\ref{fig:GLstabilityDiag}) between uniform and lumpy branes one must have $ \Delta \hat{\sigma}\big|_{\hbox{\footnotesize same } \hat{L}, \hat{\rho}}=0$. Moreover, in the perturbative analysis leading to Fig.~\ref{fig:DeltaSigma} we found that lumpy branes that bifurcate from uniform branes in the trench $\Sigma_1\Sigma_2$ of Figs.~\ref{fig:LumpyUnifET}~or~\ref{fig:GLstabilityDiag} do so with a positive entropy difference slope. In other words, slightly away from the merger curve we have $ \Delta \hat{\sigma}\big|_{\hbox{\footnotesize same } \hat{L}, \hat{\rho}}>0$. This means that lumpy branes emanating from the GL merger dominate over the uniform branes with the same 
$\hat{L}, \hat{\rho}$, at least ``initially". On the other hand, in the complement of $\Sigma_1\Sigma_2$, i.e.~for lumpy branes bifurcating from $A\Sigma_1$ or $\Sigma_2B$ in Fig.~\ref{fig:LumpyUnifET}, the perturbative analysis of Fig.~\ref{fig:DeltaSigma} shows that $ \Delta \hat{\sigma}\big|_{\hbox{\footnotesize same } \hat{L}, \hat{\rho}}<0$. That is, in this case for a given $(\hat{L}, \hat{\rho})$ lumpy branes have less Killing entropy  density than the uniform solutions and thus the latter are the preferred phase in the microcanonical ensemble. 

\begin{figure}[t]
\centerline{
\includegraphics[width=.51\textwidth]{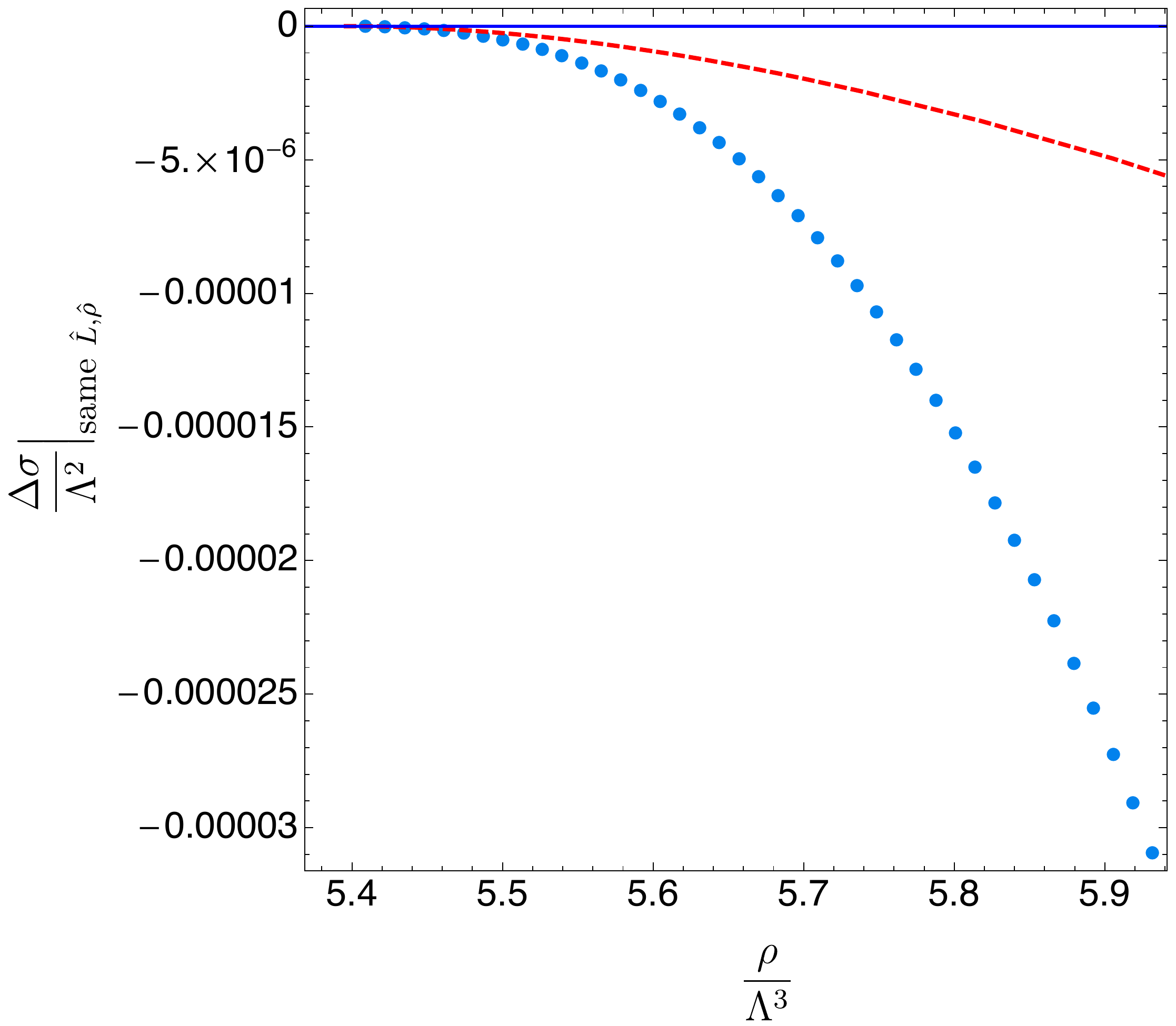}
\hspace{-0.1cm}
\includegraphics[width=.51\textwidth]{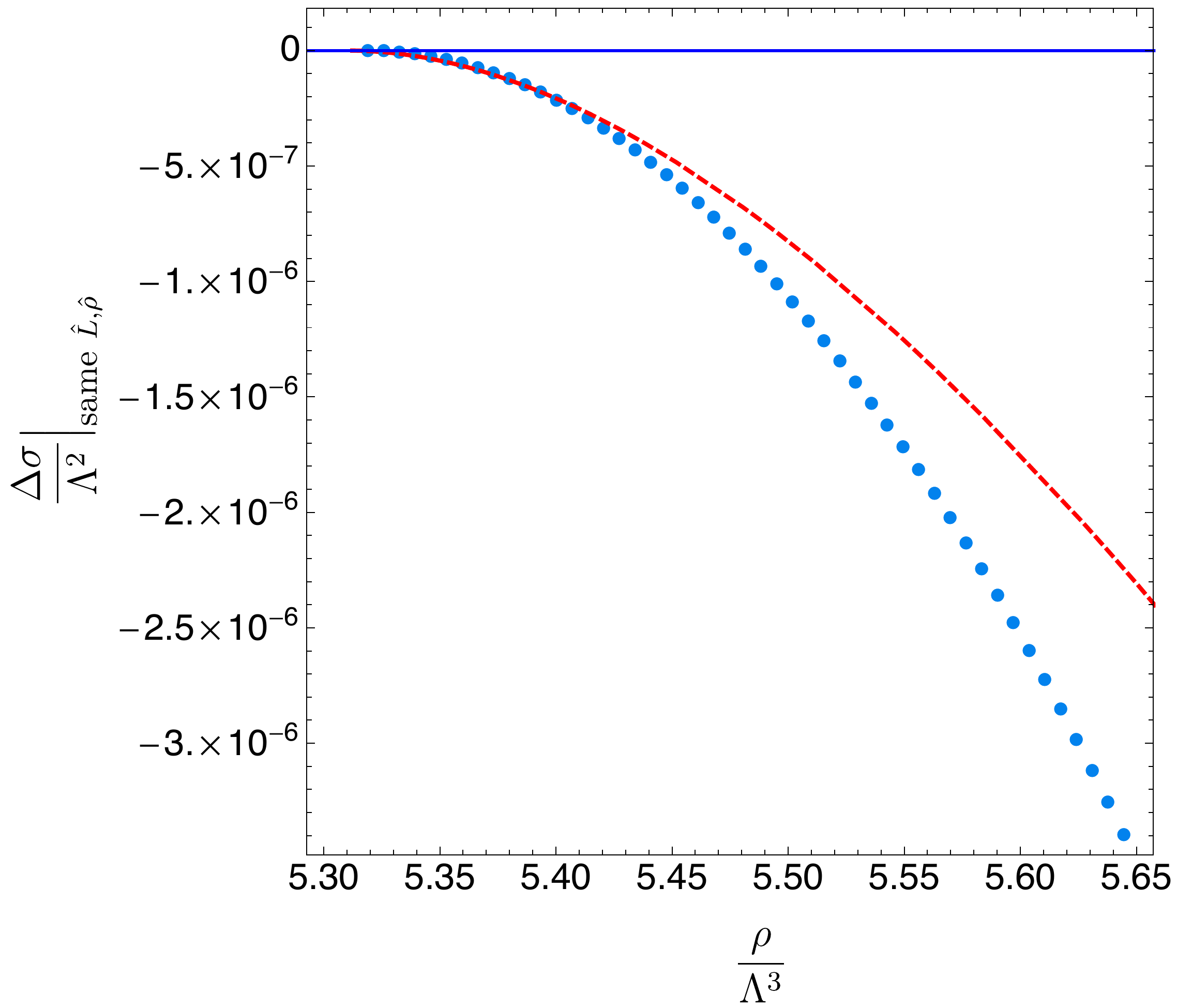}
}
\centerline{
\includegraphics[width=.5\textwidth]{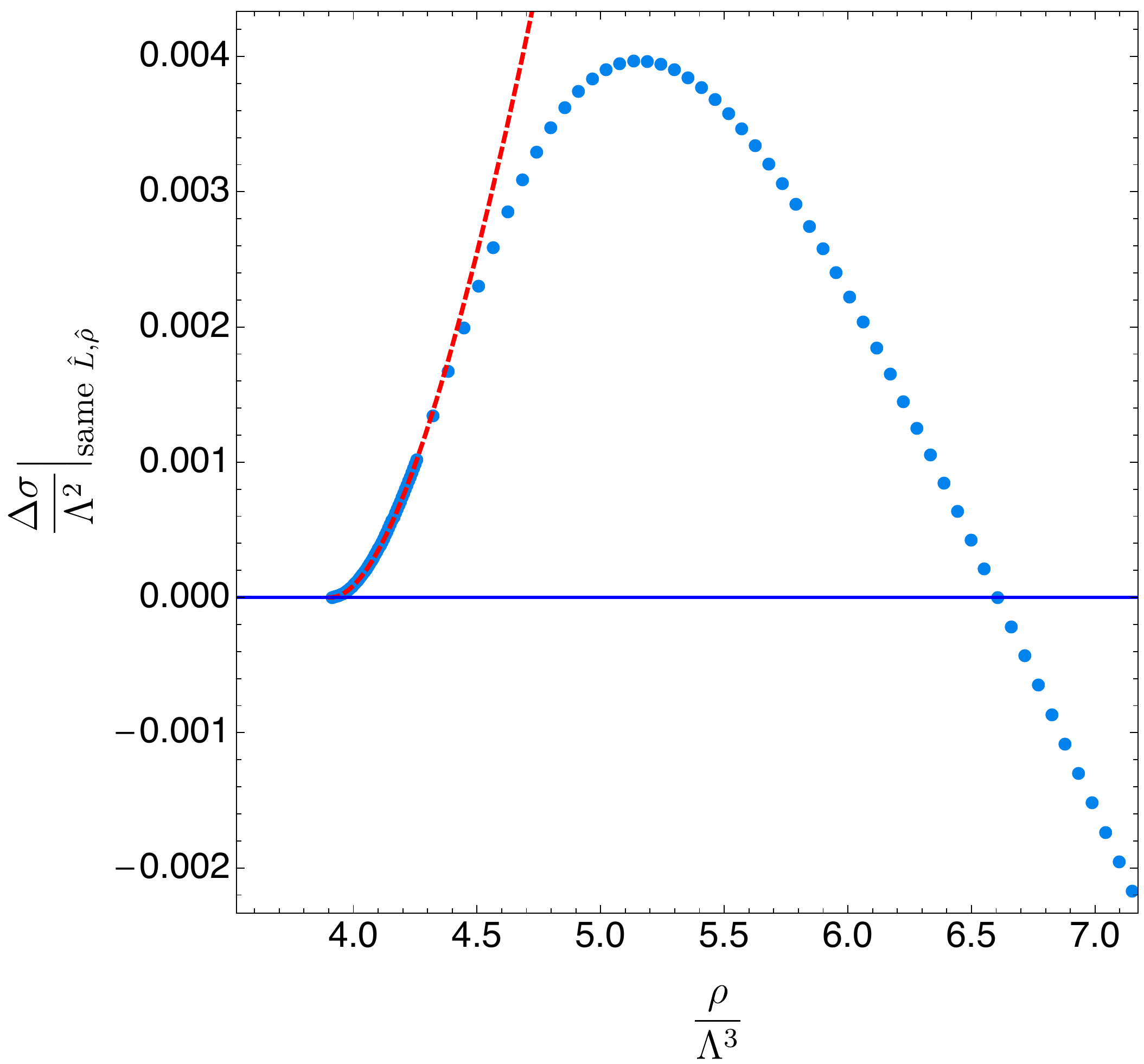}
}
\caption{Difference $\Delta \hat{\sigma}\big|_{\hbox{\footnotesize same } \hat{L}, \hat{\rho}}$ between the Killing entropy densities of  lumpy and uniform branes with the same $(\hat{L}, \hat{\rho})$ as a function of the Killing energy density $\hat{\rho}$ for three constant-$\hat \taut$ families of lumpy branes with $\hat{\taut}\simeq 0.390711<\hat{\taut}_{\Sigma_1}$ (top-left), $\hat{\taut}\simeq 0.405141> \hat{\taut}_{\Sigma_2}$ (top-right) and $\hat{\taut}_{\Sigma_1}< \hat{\taut}\simeq 0.399547< \hat{\taut}_{\Sigma_2}$ (bottom). Recall that $\hat{\taut}_{\Sigma_1}\simeq 0.390817$ and  $\hat{\taut}_{\Sigma_2}\simeq 0.404645$. The blue dots are the numerical results for the lumpy branes. The  dashed, red curves are the perturbative result \eqref{unifEntropySameEL}. The horizontal blue line indicates the 
uniform-brane family.}
\label{fig:CompPertNum}
\end{figure}

\begin{figure}[th]
\begin{center}
\includegraphics[width=.99\textwidth]{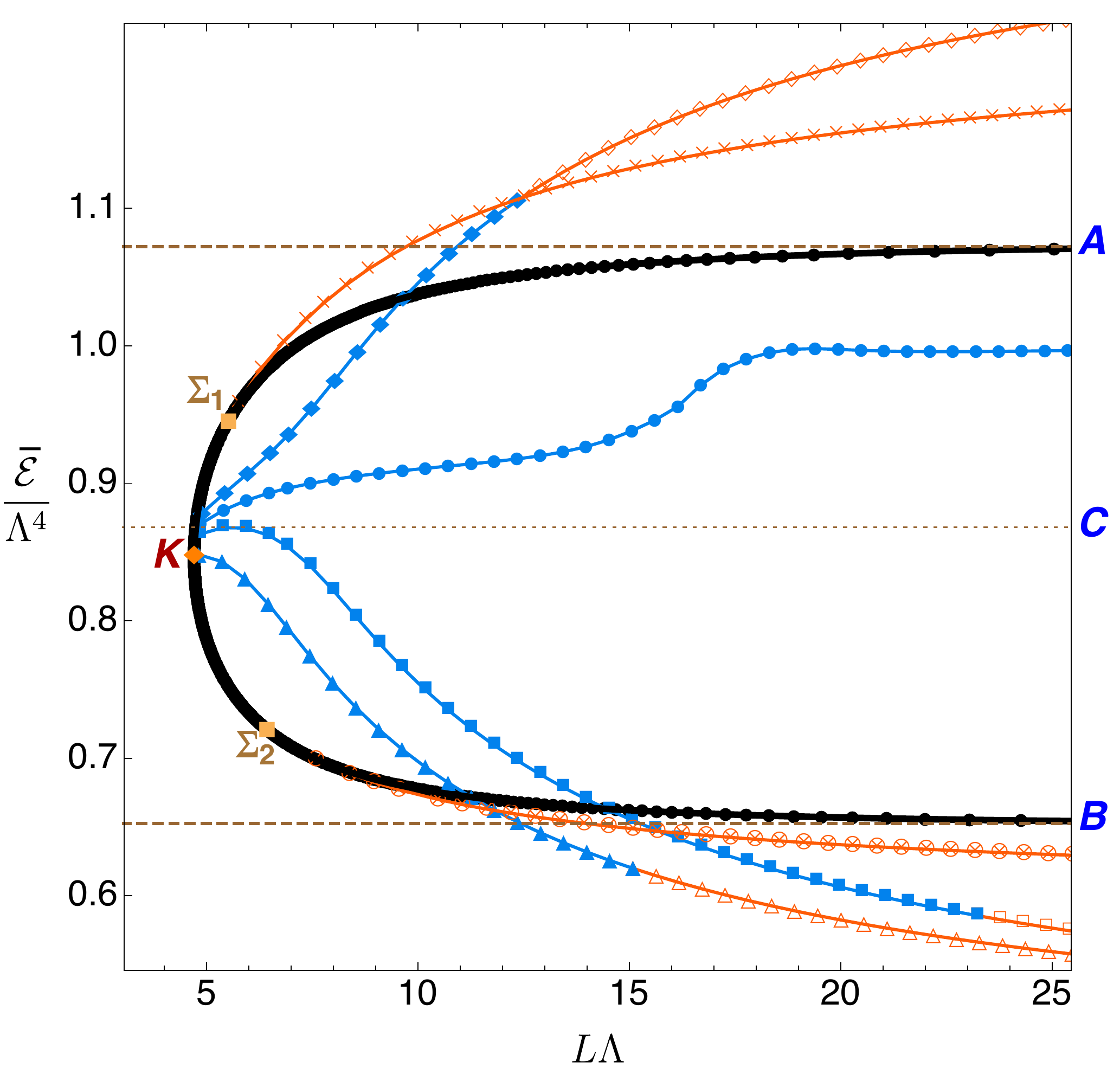}
\end{center}
\vspace{-2mm}
\caption{Same stability diagram as in Fig.~\ref{fig:GLstabilityDiag} with 
the inclusion of some lumpy-brane curves that bifurcate from the GL merger curve. These curves have constant $\hat{\taut}$ given by $\{\hat{\taut}_{\times},\hat{\taut}_{\otimes},\hat{\taut}_{\blacklozenge},\hat{\taut}_{\bullet},\hat{\taut}_{\hbox{\tiny $\blacksquare$}},\hat{\taut}_{\blacktriangle} \}\simeq 
\{ 0.390711,0.405141, 0.395420,0.395894,0.396367,0.397307\}$.  Solid blue markers (empty orange markers), no matter their shape,  indicate positive  (negative) $\Delta \hat{\sigma}\big|_{\hbox{\footnotesize same } \hat{L}, \hat{\rho}}$.} 
\label{fig:GLstabilityDiagLumpyT}
\end{figure}

Now that we have the full nonlinear solutions, our first task is to naturally compare these results with the perturbative results of \sect{sec:PerturbativeOn} that led to Fig.~\ref{fig:DeltaSigma}. On the one hand this will check our numerical results. On the other hand it will  identify the regime of validity of the perturbative analysis, i.e.~how ``far away" from the merger curve it holds. To illustrate this comparison, in Fig.~\ref{fig:CompPertNum} we show $\Delta \hat{\sigma}\big|_{\hbox{\footnotesize same } \hat{L}, \hat{\rho}}$ as a function of $\hat{\rho}$ for a family of lumpy branes that have a fixed temperature $\hat{\taut}$. 
Note that, as $\hat \rho$ changes, so does $\hat L$ in order to keep 
$\hat \taut$ fixed (this effect is better illustrated in \fig{fig:GLstabilityDiagLumpyT}, as we explain below). The plots in the top row of  Fig.~\ref{fig:CompPertNum} illustrate what happens for $\hat{\taut}<\hat{\taut}_{\Sigma_1}$ (left) and $\hat{\taut}>\hat{\taut}_{\Sigma_2}$ (right). In these cases the perturbative analysis summarized in Fig.~\ref{fig:DeltaSigma} predicts that lumpy branes bifurcate from the GL merger with $\Delta \hat{\sigma}\big|_{\hbox{\footnotesize same } \hat{L}, \hat{\rho}}<0$, as indicated by the dashed red curves in Fig.~\ref{fig:CompPertNum}. The numerical nonlinear results, shown as blue dots, indeed confirm this, and they are in excellent agreement with the perturbative results near the merger with the uniform brane. The numerical nonlinear results then show the regime where the perturbative analysis ceases  to be valid and that $\Delta \hat{\sigma}\big|_{\hbox{\footnotesize same } \hat{L}, \hat{\rho}}$ decreases monotonically with $\hat{\rho}$ (we have extended the computation to much higher values of $\rho$ than those shown in the plot). The plot in the bottom row of Fig.~\ref{fig:CompPertNum} illustrates what happens  for a 
constant-$\hat{\taut}$ lumpy brane family that bifurcates from an { intermediate} uniform brane with $\hat{\taut}_{\Sigma_1}<\hat{\taut}<\hat{\taut}_{\Sigma_2}$. In this case the perturbative analysis (see Fig.~\ref{fig:DeltaSigma}) tells us that the bifurcation occurs with $\Delta \hat{\sigma}\big|_{\hbox{\footnotesize same } \hat{L}, \hat{\rho}}>0$. Again the full nonlinear analysis confirms this is the case and is in excellent agreement with the perturbative results near the merger. 
{ However}, in this case the nonlinear analysis provides new crucial information away from the merger: it shows that, although $\Delta \hat{\sigma}\big|_{\hbox{\footnotesize same } \hat{L}, \hat{\rho}}$ initially grows away from the GL merger, at a certain point it reaches a maximum and then it starts to decrease  until it becomes negative. We have extended the computation to much larger values of $\hat \rho$ than those shown in the plot and we have found that, beyond this point, $\Delta \hat{\sigma}\big|_{\hbox{\footnotesize same } \hat{L}, \hat{\rho}}$ becomes  more and more negative as $\hat{\rho}$ becomes larger and larger. Since both $ \hat{ \sigma}_{\rm nu}$ and $ \hat{ \sigma}_{\rm u}$ are non-negative, the reason for this is clearly that $ \hat{ \sigma}_{\rm u}$ becomes arbitrarily large. 
In turn, this is due to the fact that, on a constant-$\taut$ curve, $\hat L$ becomes larger and larger as $\hat \rho$ increases (see \fig{fig:GLstabilityDiagLumpyT}), which causes the integral over the $x$-direction of the entropy density $\hat s$ to diverge. At the value of $(\hat{\rho}, \hat{L})$ where $\Delta \hat{\sigma}$ crosses zero, there is a  phase transition between lumpy and uniform branes.  This is a first-order phase transition since, for example, the temperature changes discontinuously.  We emphasize that, at the  qualitative level, this behaviour is the same for all constant-$\hat{\taut}$ families of lumpy branes that bifurcate from uniform branes in between points $\Sigma_1$ and $\Sigma_2$ in Fig.~\ref{fig:LumpyUnifET}.

\begin{figure}[th]
\begin{center}
\includegraphics[width=.99\textwidth]{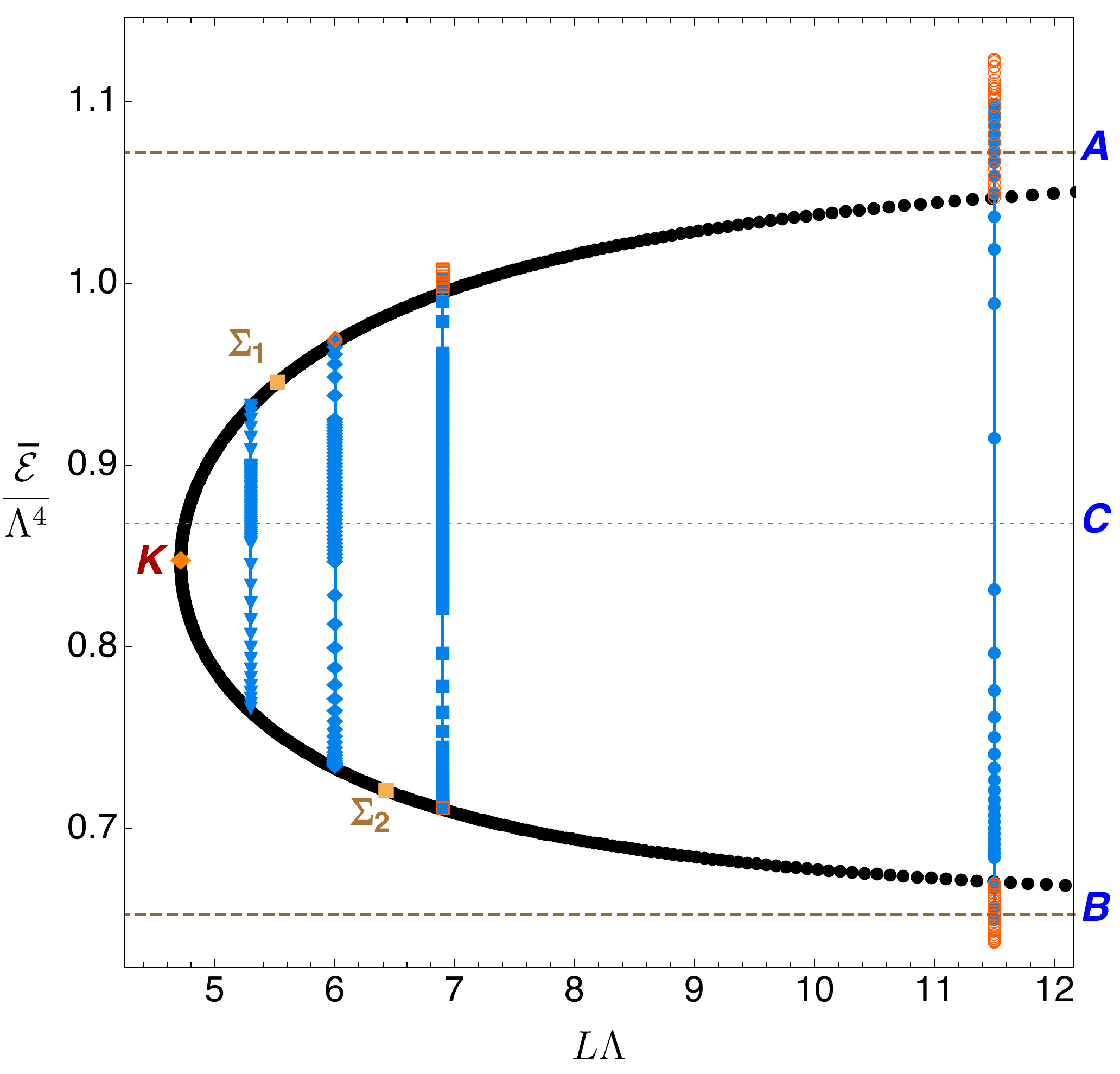}
\end{center}
\vspace{-4mm}
\caption{Same stability diagram as in Fig.~\ref{fig:GLstabilityDiag} with 
the inclusion of some lumpy-brane lines that bifurcate from the GL merger curve. These lines have constant $\hat{L}$ given by $\{\hat{L}_{\blacktriangledown}, \hat{L}_{\blacklozenge},\hat{L}_{\hbox{\tiny $\blacksquare$}}, \hat{L}_{\bullet}\}\simeq \{5.299674, 6.004224, 6.900924, 11.501849 \}$. Solid blue markers (empty orange markers), no matter their shape,  indicate positive  (negative) $\Delta \hat{\sigma}\big|_{\hbox{\footnotesize same } \hat{L}, \hat{\rho}}$. 
Note that some orange circles are on top of some blue disks. This describes the region around the cusps of Fig.~\ref{fig:DsigmaRho} and \ref{fig:DsigmaCalE}.} 
\label{fig:GLstabilityDiagLumpyL}
\end{figure}

To further  understand  this phase  transition, in Figs.~\ref{fig:GLstabilityDiagLumpyT} and \ref{fig:GLstabilityDiagLumpyL} we reproduce again the stability diagram of Fig.~\ref{fig:GLstabilityDiag}, but this time we also plot a few constant-$\hat{\taut}$ or constant-$\hat L$ lumpy branes that depart from the GL merger curve. 
We use two plot marker codes: The { solid blue} markers, no matter their shape, represent the trench where $\Delta \hat{\sigma}\big|_{\hbox{\footnotesize same } \hat{L}, \hat{\rho}}>0$, while the { empty orange} markers, no matter their shape,  describe the region where $\Delta \hat{\sigma}\big|_{\hbox{\footnotesize same } \hat{L}, \hat{\rho}}<0$. For reference, recall  that 
 \begin{eqnarray}
 (\hat{\taut},\hat{L}_{\hbox{\tiny GL}},\hat{\cal E})_{\Sigma_1}
 &\simeq& (0.390817,5.618133,0.950579) \,, \nonumber \\  
 (\hat{\taut},\hat{L}_{\hbox{\tiny GL}},\hat{\cal E})_{\Sigma_2}
 &\simeq &(0.404645,6.592316,0.717060) \,,\nonumber \\
 (\hat{\taut},\hat{L}_{\hbox{\tiny GL}},\hat{\cal E})_{K}
 &\simeq &(0.397427,4.716021,0.846337) \,.
 \label{hierarchy}
\end{eqnarray}
For brevity, henceforth we will use the notation 
\be
\hat{L}_{\Sigma_1} \equiv \hat{L}_{\hbox{\tiny GL}}\big|_{\Sigma_1}  \,,\qquad
\hat{L}_{\Sigma_2} \equiv \hat{L}_{\hbox{\tiny GL}}\big|_{\Sigma_2} \,,\qquad
\hat{L}_{K} \equiv \hat{L}_{\hbox{\tiny GL}}\big|_{K} \,.
\ee
The main conclusions from Figs.~\ref{fig:GLstabilityDiagLumpyT} and \ref{fig:GLstabilityDiagLumpyL} are as follows:
\begin{enumerate}

\item Recall that lumpy branes that bifurcate from the GL merger line 
 at a temperature $\hat{\taut}<\hat{\taut}_{\Sigma_1}$ (i.e.~above $\Sigma_1$ in the figures) or at a temperature $\hat{\taut}>\hat{\taut}_{\Sigma_2 }$ (i.e.~below $\Sigma_2$ in the figures)  have $\Delta \hat{\sigma}\big|_{\hbox{\footnotesize same } \hat{L}, \hat{\rho}}<0$ no matter how large $\hat{L}$ is, as pointed out when discussing Fig.~\ref{fig:CompPertNum} (recall that $\Sigma_1$ and $\Sigma_2$ were introduced in Fig.~\ref{fig:DeltaSigma}). In Fig.~\ref{fig:GLstabilityDiagLumpyT} we display one family of lumpy branes in each of these classes. One has constant  $\hat{\taut}_{\times}\simeq 0.390711 < \hat{\taut}_{\Sigma_1}$ and is always described by empty  orange colour markers, which means that the solutions indeed have  $\Delta \hat{\sigma}\big|_{\hbox{\footnotesize same } \hat{L}, \hat{\rho}}<0$. All the curves that bifurcate from the GL merger above  $\Sigma_1$  have this feature and they always bifurcate towards higher
 $\hat{\overline{\cal E}}$ and higher $\hat{L}$ with respect to the merger point 
 $(\hat{\cal E}_{\hbox{\tiny GL}}, \hat{L}_{\hbox{\tiny GL}})$.  The other family has constant  $\hat{\taut}_{\otimes}\simeq 0.405141 > \hat{\taut}_{\Sigma_2}$ and is again always described by empty orange markers,  which means that the solutions indeed have $\Delta \hat{\sigma}\big|_{\hbox{\footnotesize same } \hat{L}, \hat{\rho}}<0$. All the curves that bifurcate from the GL merger below  $\Sigma_2$  have this feature and they always bifurcate towards lower $\hat{\cal E}$ and higher $\hat{L}$ with respect to the merger point 
 $(\hat{\cal E}_{\hbox{\tiny GL}}, \hat{L}_{\hbox{\tiny GL}})$.

\item The situation is less monotonous for lumpy branes that bifurcate from a point on the GL merger curve that lies between  $\Sigma_1$ and $\Sigma_2$. Recall that these have, close to the merger, $\Delta \hat{\sigma}\big|_{\hbox{\footnotesize same } \hat{L}, \hat{\rho}}>0$. In Fig.~\ref{fig:GLstabilityDiagLumpyT} we display four curves in this class, namely:   the family with constant $\hat{\taut}_{\blacklozenge}\simeq 0.395420 < \hat{\taut}_c$; 
 the family with constant $\hat{\taut}_{\bullet}\simeq 0.395894 \lesssim \hat{\taut}_c$ (so, very close to $\hat{\taut}_c\simeq 0.3958945$);  the family with constant $\hat{\taut}_{\hbox{\tiny $\blacksquare$}}\simeq 0.396367 > \hat{\taut}_c$; and  the family with constant $\hat{\taut}_{\blacktriangle}\simeq 0.397307 > \hat{\taut}_c$ (slightly below  $\hat{\taut}_K\simeq 0.397427$). These curves with $\hat{\taut}_{\Sigma_2}<\hat{\taut}<\hat{\taut}_{\Sigma_2}$ bifurcate
towards  $\hat{L}>\hat{L}_{\hbox{\tiny GL}}$ with $\Delta \hat{\sigma}\big|_{\hbox{\footnotesize same } \hat{L}, \hat{\rho}}>0$. Then, if $\hat{\taut}_{\Sigma_1}<\hat{\taut}<\hat{\taut}_{\rm c}$ (e.g.~the curve with diamond plot markers $\blacklozenge$) they typically move to higher 
$\hat{\overline{\cal E}}$ as $\hat{L}$ increases and $\Delta \hat{\sigma}\big|_{\hbox{\footnotesize same } \hat{L}, \hat{\rho}}$ changes from positive into negative when the  plot markers  change from solid blue $\blacklozenge$ into empty orange $\hbox{\large $\diamond$}$. On the other hand, if $\hat{\taut}_{\rm c}<\hat{\taut}<\hat{\taut}_{\Sigma_2}$ (e.g.~the curves initially with {\hbox{\tiny $\blacksquare$}} and $\blacktriangle$), the constant $\hat{\taut}$-curves typically plunge into lower $\hat{\overline{\cal E}}$
 as $\hat{L}$ increases and $\Delta \hat{\sigma}\big|_{\hbox{\footnotesize same } \hat{L}, \hat{\rho}}$ changes from positive into negative when the  plot markers  change from solid blue  into empty orange (i.e.~\hbox{\tiny $\blacksquare$} $\to$ \hbox{\tiny $\square$} or $\blacktriangle \to \triangle$).

\item When $\hat{\taut}\sim\hat{\taut}_c$ the properties described in the two previous points hold but the constant-$\hat{\taut}$ curves do not escape to large $\hat{\overline{\cal E}}$  (if $\hat{\taut}\lesssim \hat{\taut}$) or small $\hat{\overline{\cal E}}$ (if $\hat{\taut}\gtrsim \hat{\taut}$) so quickly as $\hat{L}$ grows. A good example is given by the dotted ($\bullet$) curve with $\hat{\taut}_{\bullet}\simeq 0.395894 \lesssim \hat{\taut}_c$. The closer one is of $\hat{\taut}_{\rm c}$ the longer $\hat{L}$ must be  for the constant-$\hat{\taut}$ curve to cross the GL merger line again and then acquire $\Delta \hat{\sigma}\big|_{\hbox{\footnotesize same } \hat{L}, \hat{\rho}}<0$. Our results suggest  that in the exact limit $\hat{\taut}\to \hat{\taut}_{\rm c}$ the curve extends to $\hat{L}\to \infty$ without ever leaving the window of energy densities $[\hat{\cal E}_B,\hat{\cal E}_A]$.
 
\end{enumerate}

\begin{figure}[th]
\centerline{
\includegraphics[width=.515\textwidth]{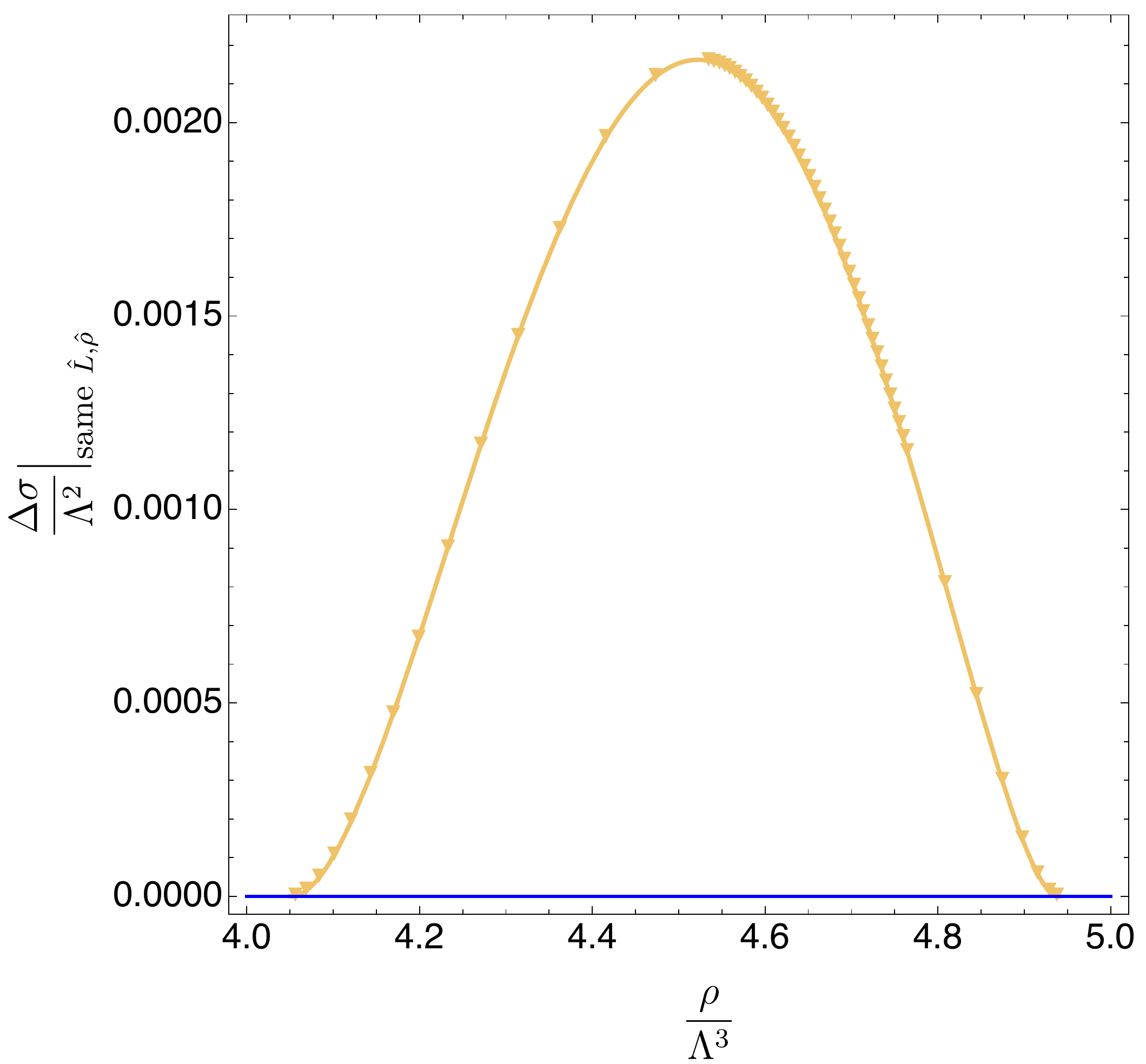}
\hspace{0.1cm}
\includegraphics[width=.50\textwidth]{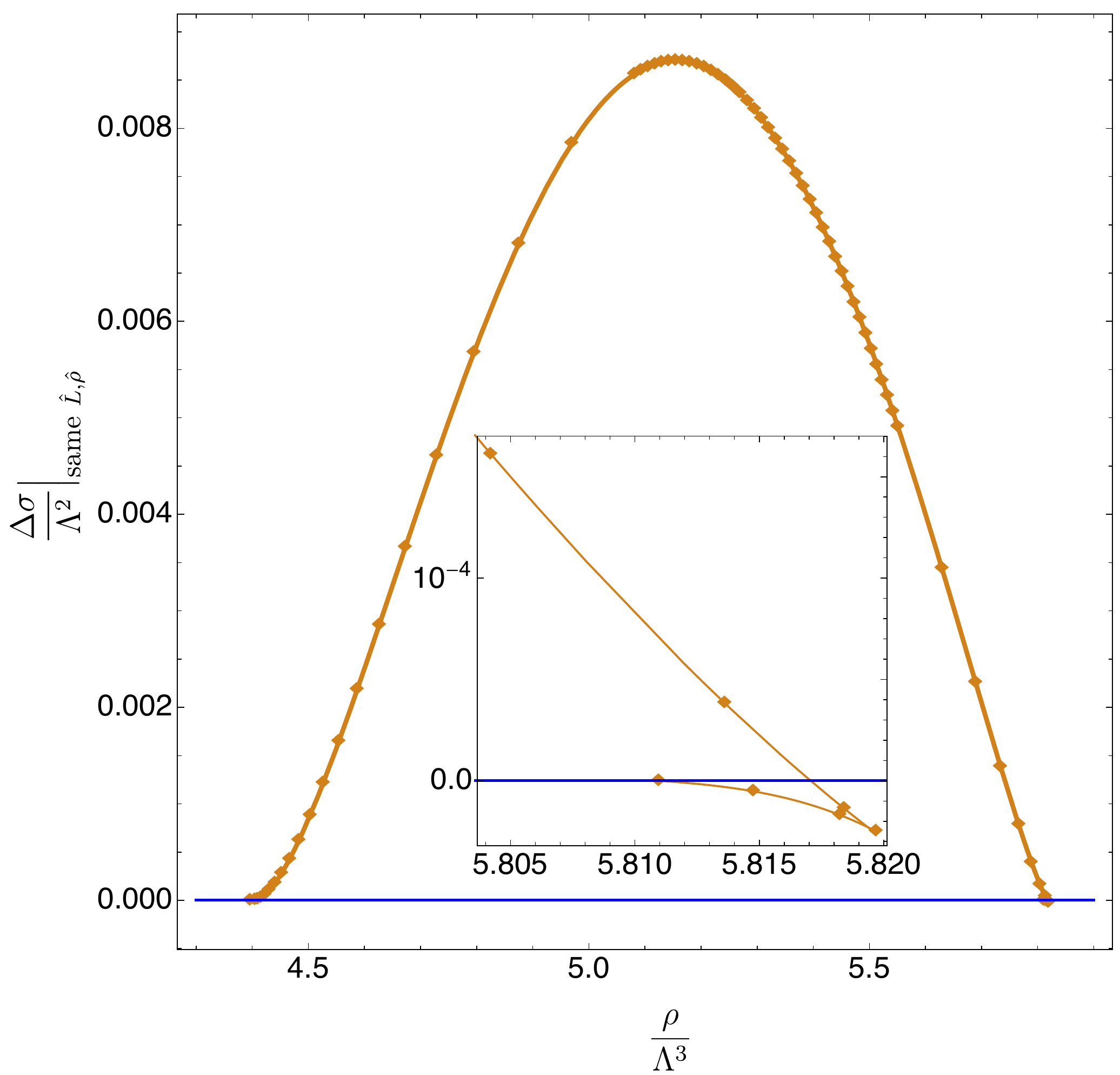}
}
\vspace{2mm}
\centerline{
\includegraphics[width=.50\textwidth]{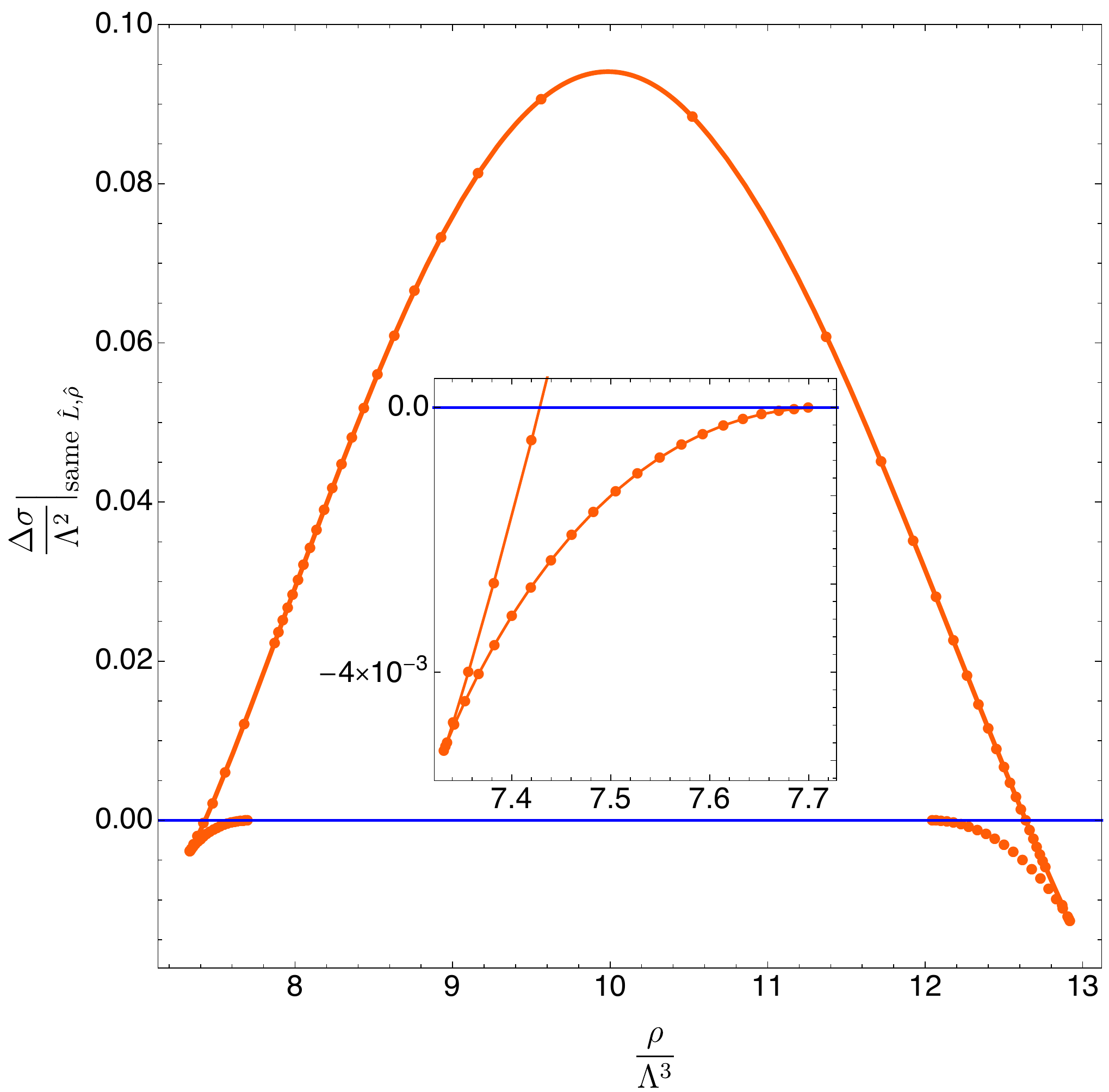}
}
\caption{Microcanonical phase diagram: Killing entropy density difference $\Delta \hat{\sigma}\big|_{\hbox{\footnotesize same } \hat{L}, \hat{\rho}}$ between lumpy and uniform branes with the same $(\hat{L}, \hat{\rho})$ as a function of the Killing energy density $\hat{\rho}$ for 3 families of lumpy branes that have constant $\hat{L}$ given by $\hat{L} \simeq 5.299674 < \hat{L}_{\Sigma_1} $ (top-left), 
$\hat{L}\simeq 6.004224$ which is in the range $\hat{L}_{\Sigma_1} <\hat{L} < \hat{L}_{\Sigma_2}$ (top-right) and $\hat{L}\simeq 11.501849 > \hat{L}_{\Sigma_2}$ (bottom). For reference, $\hat{L}_{\Sigma_1} \simeq 5.618133$ and  $\hat{L}_{\Sigma_2}\simeq 6.592316$.
The three families are those with constant $\{\hat{L}_{\blacktriangledown}, \hat{L}_{\blacklozenge}, \hat{L}_{\bullet}\}$ already displayed in Fig.~\ref{fig:LumpyUnifET}; we use the same shape/colour coding for the markers execpt that here they are all solid. The horizontal blue line with $\Delta \hat{\sigma}\big|_{\hbox{\footnotesize same } \hat{L}, \hat{\rho}}=0$ describes the uniform brane family.  
}
\label{fig:DsigmaRho}
\end{figure}

To complete our  understanding of the microcanonical phase diagram, in Fig.~\ref{fig:DsigmaRho} we plot the Killing entropy density difference $\Delta \hat{\sigma}\big|_{\hbox{\footnotesize same } \hat{L}, \hat{\rho}}$ between lumpy and uniform branes with the same $(\hat{L}, \hat{\rho})$ as a function of the Killing energy density $\hat{\rho}$ for three families of lumpy branes that have {\it constant} $\hat{L}$. 
Note that  
$\hat{L}_K< \hat{L}_{\Sigma_1}< \hat{L}_{\Sigma_2}$ (see e.g.~Fig.~\ref{fig:GLstabilityDiagLumpyL}). The three panels of Fig.~\ref{fig:DsigmaRho} describe representative examples of the following three possible cases: (1) $\hat{L}_K<\hat{L}< \hat{L}_{\Sigma_1}$ (top-left panel), (2) $  \hat{L}_{\Sigma_1}<\hat{L}< \hat{L}_{\Sigma_2}$ (top-right panel), and (3) $\hat{L}>\hat{L}_{\Sigma_2}$ (bottom panel). Together with those in Fig.~\ref{fig:CompPertNum}, the plots in Fig.~\ref{fig:DsigmaRho} are the most important ones in our analysis of the phase diagram. The three panels of Fig.~\ref{fig:DsigmaRho} encode the following conclusions:
\begin{enumerate}
\item The top-left panel is for constant $\hat{L} \simeq 5.299674$ solutions and  illustrates what happens in the three-dimensional microcanonical phase diagram $\Delta \hat{\sigma}\big|_{\hbox{\footnotesize same } \hat{L}, \hat{\rho}}$ versus $( \hat{\rho}, \hat{L})$ when the lumpy branes have $  \hat{L}_K<\hat{L}< \hat{L}_{\Sigma_1}$. We see that in this range of $\hat{L}$, lumpy branes (yellow inverted triangles) bifurcate from the uniform brane (blue line) at low $\hat{\rho}$ with $\Delta \hat{\sigma}\big|_{\hbox{\footnotesize same } \hat{L}, \hat{\rho}}>0$ and, as $\hat{\rho}$ increases, the entropy difference grows until it reaches a maximum and then it decreases monotonically until the lumpy brane merges again with the uniform brane at higher $\hat{\rho}$. Since in this range of $(\hat{L}, \hat{\rho})$ one always has $\Delta \hat{\sigma}\big|_{\hbox{\footnotesize same } \hat{L}, \hat{\rho}}>0$, lumpy branes are the preferred phase in the microcanonical ensemble.  

\item The top-right panel is for constant $\hat{L} \simeq 6.004224$ solutions and illustrates what happens in the 3-dimensional microcanonical phase diagram $\Delta \hat{\sigma}\big|_{\hbox{\footnotesize same } \hat{L}, \hat{\rho}}$ versus 
$( \hat{\rho}, \hat{L})$ when the lumpy branes have $\hat{L}_{\Sigma_1} <\hat{L} < \hat{L}_{\Sigma_2}$. As in the previous case, in this range of $\hat{L}$, lumpy branes (brown diamonds) also bifurcate from the uniform brane (blue line) at low $\hat{\rho}$ with $\Delta \hat{\sigma}\big|_{\hbox{\footnotesize same } \hat{L}, \hat{\rho}}>0$ and, as $\hat{\rho}$ increases, the entropy difference grows until it reaches a maximum. Then it again decreases monotonically but, this time, 
$\Delta \hat{\sigma}\big|_{\hbox{\footnotesize same } \hat{L}, \hat{\rho}}$ becomes negative at a certain $\hat{\rho}$. This first-order phase transition point is best seen in the inset plot that zooms into this region. The entropy difference keeps decreasing as $\rho$ grows until it reaches a cusp. Then, as $\hat{\rho}$ decreases, $\Delta \hat{\sigma}\big|_{\hbox{\footnotesize same } \hat{L}, \hat{\rho}}$ becomes less negative until the lumpy brane with constant $\hat{L}$ merges again with the uniform brane. 

\item Finally, the bottom panel is for constant $\hat{L} \simeq 11.501849$ solutions (whose energy density profile was discussed in Fig.~\ref{fig:profilesExConstL}). It illustrates how the 3-dimensional microcanonical phase diagram $\Delta \hat{\sigma}\big|_{\hbox{\footnotesize same } \hat{L}, \hat{\rho}}$ versus 
$(\hat{\rho}, \hat{L})$ looks like when the lumpy branes have $\hat{L} > \hat{L}_{\Sigma_2}$. In this range of $\hat{L}$, at { both} GL mergers with the uniform brane  (blue line),  lumpy branes (orange disks) bifurcate with $\Delta \hat{\sigma}\big|_{\hbox{\footnotesize same } \hat{L}, \hat{\rho}}<0$. Then, as we move along the constant-$\hat{L}$ line away from the merger points,  there are first two cusps (the left one is shown in more detail in the inset plot) and {two first-order phase transition points} where $\Delta \hat{\sigma}\big|_{\hbox{\footnotesize same } \hat{L}, \hat{\rho}}$  changes sign and becomes positive. For $ \hat{\rho}$ in between these two transition points, one has a lumpy brane with $\Delta \hat{\sigma}\big|_{\hbox{\footnotesize same } \hat{L}, \hat{\rho}}>0$ and thus these lumpy branes are the preferred microcanonical phase. Otherwise, uniform branes dominate the microcanonical ensemble.
\end{enumerate}

To complement this discussion, it is useful to plot the Killing entropy density difference $\Delta \hat{\sigma}\big|_{\hbox{\footnotesize same } \hat{L}, \hat{\rho}}$ between lumpy and uniform branes with the same $(\hat{L}, \hat{\rho})$ as a function of the {average energy density} 
$\eho$ for some families of lumpy branes that have {constant} $\hat{L}$. Recall that the average energy density and the Killing energy density are related through $\eho = \hat \rho / \hat L$. This means that comparing the entropy of uniform and nonuniform brane at the same  $(\hat{L},  \hat{\rho})$ is the same as comparing them at the same 
$(\hat L, \eho)$. 
\begin{figure}[th]
\centerline{
\includegraphics[width=.99\textwidth]{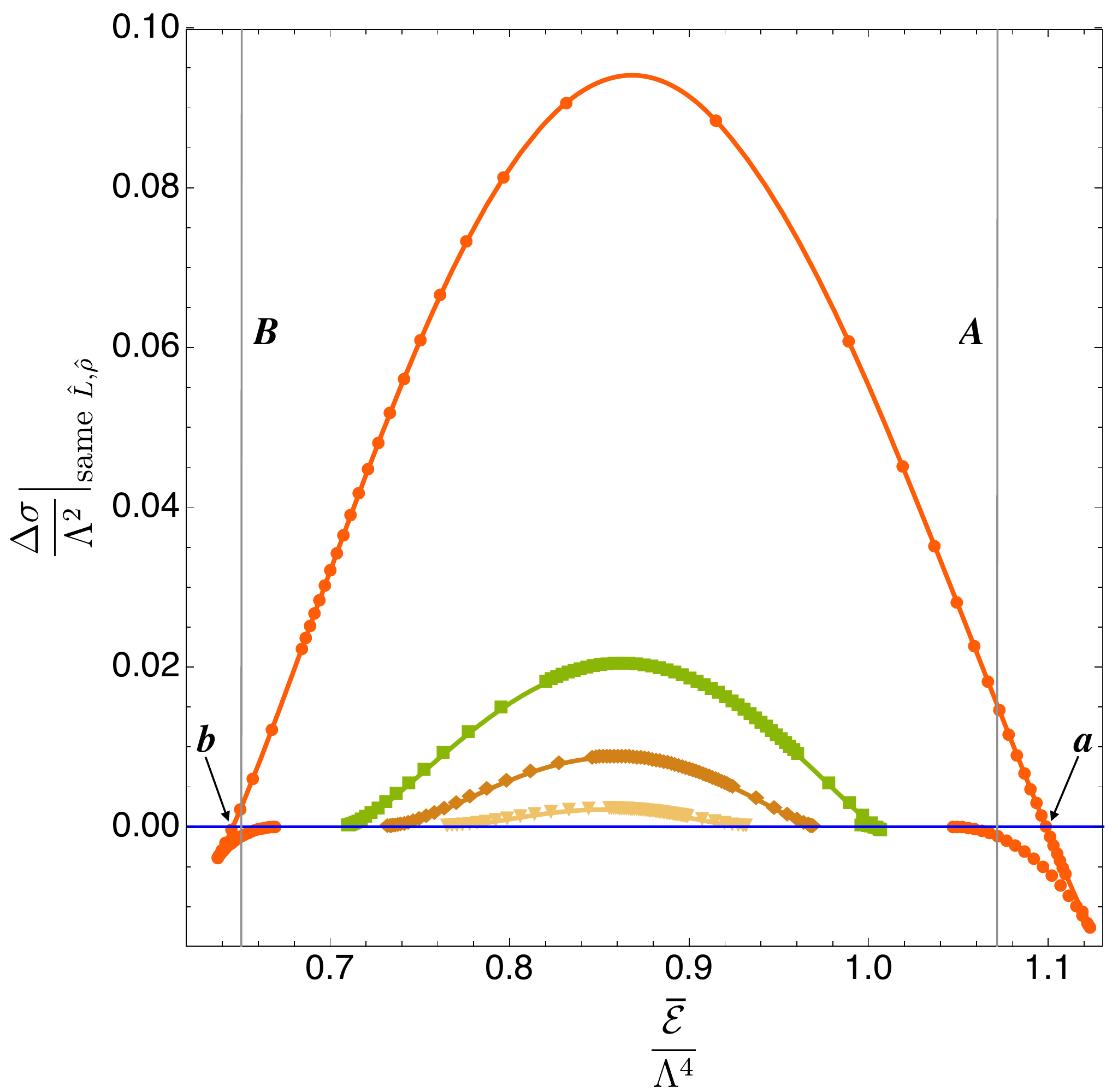}
}
\caption{Killing entropy density difference $\Delta \hat{\sigma}\big|_{\hbox{\footnotesize same } \hat{L}, \hat{\rho}}$ between lumpy and uniform branes with the same $(\hat{L}, \hat{\rho})$ or, equivalently, with the same  $(\hat{L}, \eho)$, as a function of the average energy density 
$\eho$. We show the same four families of lumpy branes with constant 
\mbox{$\{\hat{L}_{\blacktriangledown}, \hat{L}_{\blacklozenge},\hat{L}_{\hbox{\tiny $\blacksquare$}}, \hat{L}_{\bullet}\}\simeq \{5.299674, 6.004224, 6.900924, 11.501849 \}$} already displayed in Fig.~\ref{fig:LumpyUnifET}. We use the same shape/colour coding for the markers as in Fig.~\ref{fig:LumpyUnifET} except that here they are all solid. The families with $\blacktriangledown,\blacklozenge,\bullet$ were also shown in Fig.~\ref{fig:DsigmaRho}, but the family $\hbox{\tiny $\blacksquare$}$ was not.  The horizontal blue line with $\Delta \hat{\sigma}\big|_{\hbox{\footnotesize same } \hat{L}, \hat{\rho}}=0$ describes the uniform brane family. The  grey vertical lines  indicate the turning points $A$ and $B$  in the phase diagram of \fig{fig:Unif}. The labels  ``$a$" and ``$b$" indicate the  lumpy solutions with 
$\hat{L}_{\bullet} \simeq 11.501849$ that lie 
away from the merger curve but have the same entropy density as the corresponding uniform branes. In other words, these are the points away from the merger curve at which $\Delta \hat{\sigma}\big|_{\hbox{\footnotesize same } \hat{L}, \hat{\rho}}$ crosses zero. 
The average energy densities at these points are $\eho_a\simeq 1.09879 > \eho_A$ and $\eho_b\simeq 0.645861< \eho_B$. 
}
\label{fig:DsigmaCalE}
\end{figure} 
Fig.~\ref{fig:DsigmaCalE} shows this comparison for the four constant-$\hat{L}$ families $\{\hat{L}_{\blacktriangledown}, \hat{L}_{\blacklozenge},\hat{L}_{\hbox{\tiny $\blacksquare$}}, \hat{L}_{\bullet}\}$ that were plotted in  Fig.~\ref{fig:GLstabilityDiagLumpyL}. Fig.~\ref{fig:DsigmaCalE}, together with the projections to the $(\hat{L},\eho)$-plane shown in Fig.~\ref{fig:GLstabilityDiagLumpyL}, is the key figure to understand the microcanonical phase diagram because it provides four representative slices of this plot at constant $\hat L$.  Gluing slices of this type together along the $\eho$-axis one obtains the three-dimensional plot of $\Delta \hat{\sigma}\big|_{\hbox{\footnotesize same } \hat{L}, \hat{\rho}}$ versus $(\hat L, \eho)$. 
The  $\hat L \to \infty$ limit of the curves of Fig.~\ref{fig:DsigmaCalE} is the curve in Fig.~\ref{fig:convexconcave}(right). 

So far we have  discussed the lumpy branes only in the microcanonical ensemble. This is the most interesting ensemble because nonuniform branes can dominate this ensemble for certain windows of the parameter space  and in a time evolution we typically fix the length and the average energy density of the solutions (i.e. the latter is conserved). But we may also ask about the role played by the lumpy branes in the canonical ensemble. In this case, we want to fix the length $L\Lambda$ and the temperature $T/\Lambda$ of the solutions and the dominant solution is the one that has the lowest Killing free energy density $f/\Lambda^3$. 

\begin{figure}[th]
\centerline{
\includegraphics[width=.75\textwidth]{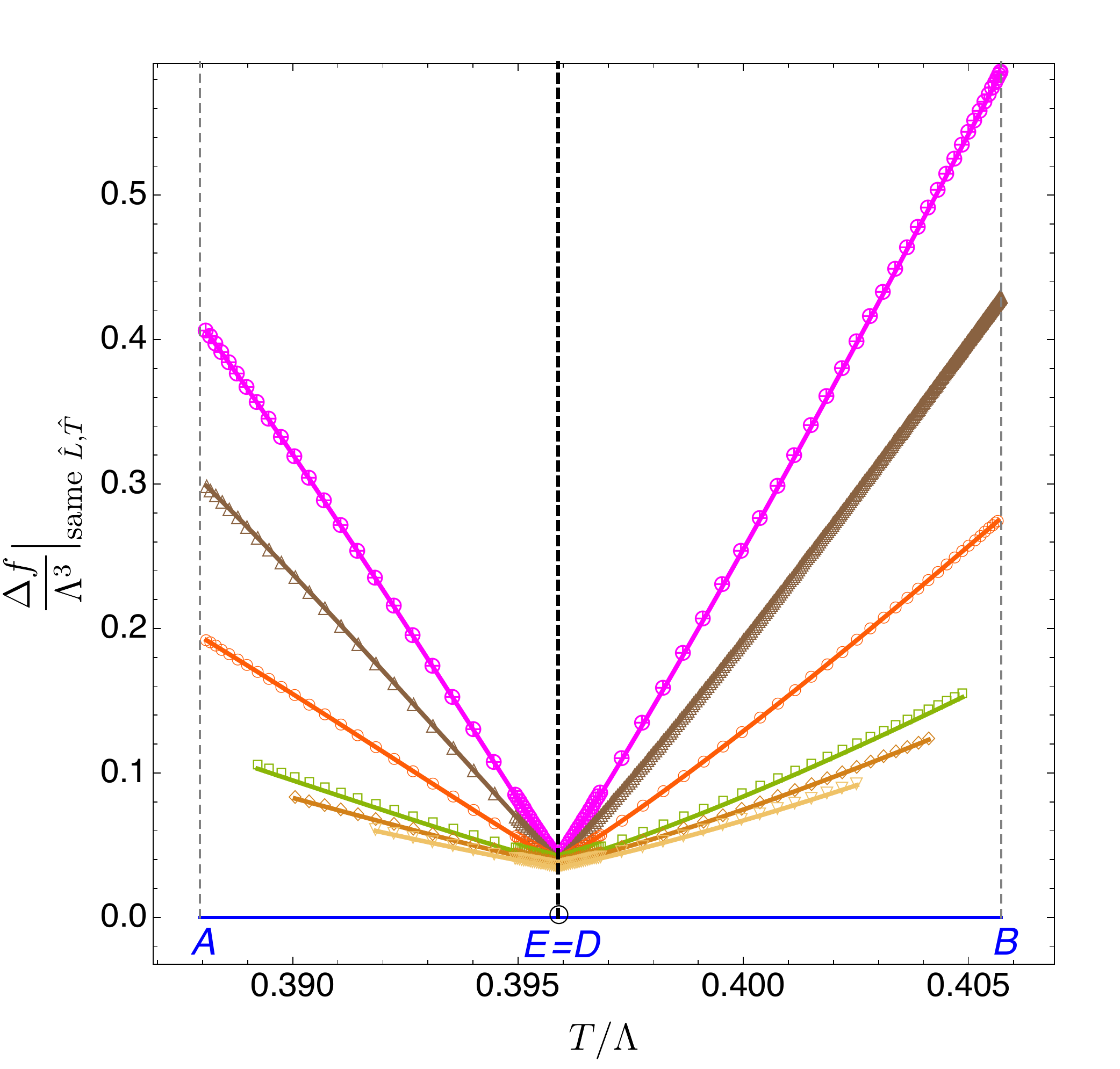}
}
\caption{Canonical phase diagram: Dimensionless Killing free energy density difference $\Delta \hat{f}\big|_{\hbox{\footnotesize same } \hat{L}, \hat{T}}$ as a function of the dimensionless temperature $\hat{T}$ for the six lumpy-brane families at constant $\hat{L}$ already shown in \fig{fig:LumpyUnifET} (with the same colour/shape code). Namely, from the bottom to the top the length of the curves are:  $\hat{L}_{\triangledown} \simeq 5.299674$, $\hat{L}_{\hbox{\large $\diamond$}}\simeq 6.004224$, $\hat{L}_{\hbox{\tiny $\square$}}\simeq 6.900924$,  
$\hat{L}_{\hbox{\tiny$\bigcirc$}} \simeq 11.501849$, $\hat{L}_{\triangle} \simeq 17.906849$, $\hat{L}_{\otimes} \simeq 24.311849$. The black dashed vertical line with $\hat{T}=\hat{T}_D=\hat{T}_E\equiv \hat{\taut}_{\rm c}\sim 0.3958945$ represents the critical temperature first identified in the right panel of \fig{fig:Unif} and the grey vertical dashed lines represent $\hat{T}=\hat{T}_A\sim 0.387944$ and $\hat{T}=\hat{T}_B\sim 0.405724$ between which lumpy branes coexist with uniform branes.
The horizontal blue line with  $\Delta \hat{f}=0$ represents the light uniform brane for $\hat{T}<\hat{T}_c$, and the heavy uniform brane for $\hat{T}>\hat{T}_c$.}
\label{fig:canonical}
\end{figure} 

To address this question it is useful to first recall what happens when we  consider only the uniform brane solutions. In the right panel of \fig{fig:Unif} we have already seen that the light uniform branch (the lower branch in the left panel of \fig{fig:Unif}) is the preferred thermal phase for $\hat{\taut}<\hat{\taut}_{\rm c}$, while for fixed $\hat{\taut}>\hat{\taut}_{\rm c}$ the heavy uniform branch (the upper branch in the left panel of \fig{fig:Unif}) dominates the canonical ensemble. The intermediate uniform branch (between $A$ and $B$ in \fig{fig:Unif}) is never a preferred thermal phase of the canonical ensemble. For this reason it is sometimes  stated in textbooks that, at $\hat{\taut}=\hat{\taut}_{\rm c}\sim 0.3958945$, there is a first-order phase transition at which the system jumps discontinuously  between the light and heavy uniform branes. However, at infinite volume there is actually a degeneracy of states at $\hat{\taut}=\hat{\taut}_{\rm c}$ because the average free energy density of any phase separated state  is the same as that of the homogeneous states at $\hat{\taut}=\hat{\taut}_{\rm c}$. The reason for this is that the interface between the two phases in a phase-separated configuration gives a  volume-independent contribution. This contribution is therefore subleading in the infinite-volume limit with respect to those of the two coexisting phases, whose free energy densities are equal to each other and to those of homogeneous states at $\hat{\taut}=\hat{\taut}_{\rm c}$. 
Therefore, in the infinte-volume limit the system can transition between  points $D$ and $E$ along a sequence of constant-temperature, constant free-energy, phase-separated states. The fact that all these states have the same free energy is the content of Maxwell's construction.

In contrast, at finite volume the inhomogenous, phase-separated states are never thermodynamically favoured. To show this, in \fig{fig:canonical} we compare the free energy of lumpy branes with the light uniform branes if $\hat{\taut}<\hat{\taut}_{\rm c}$, and with the heavy uniform branch if $\hat{\taut}>\hat{\taut}_{\rm c}$. More concretely, we compute the difference between the Killing free energy of the nonuniform brane $\hat{f}_{\rm nu}$ and the light (heavy) uniform Killing free energy $\hat{f}_{\rm u}$ when $\hat{\taut}<\hat{\taut}_{\rm c}$ ($\hat{\taut}>\hat{\taut}_{\rm c}$) that has the same length $L\Lambda$ and  temperature $T/\Lambda$, i.e. $\Delta \hat{f}\big|_{\hbox{\footnotesize same } \hat{L}, \hat{T}} =\big( \hat{f}_{\rm nu}-\hat{f}_{\rm u} \big)_{\hbox{\footnotesize same } \hat{L}, \hat{T}}$. \fig{fig:canonical} shows that for any temperature $\hat{T}_A \leq \hat{T}\leq \hat{T}_B$ where nonuniform branes exist, one always has $\Delta \hat{f}\big|_{\hbox{\footnotesize same } \hat{L}, \hat{T}}>0$. That is to say, the Killing free energy density of the lumpy branes is always  higher than the free energy of the relevant (light or heavy) uniform brane and thus lumpy branes {\it never} dominate the canonical ensemble.\footnote{For completeness, we have verified that the Killing free energy density $\hat{f}$ of lumpy branes is always {\it lower} than the Killing free energy density of the {\it intermediate} branes $AB$, and that they become equal to one another precisely when the merger of these two branches occurs. In any case neither branch is ever preferred at finite volume in the canonical ensemble.}



\subsection{Excited static lumpy branes: beyond the ground state solutions}\label{sec:LumpyCopies}

So far we have discussed only the ``ground state" lumpy branes of our spinodal system. The profile, for example that of the energy density  ${\cal E}(x)$, of these fundamental branes has a single maximum and a single minimum, see Figs.~\ref{fig:profilesExConsT} and \ref{fig:profilesExConstL}.  The phase diagram of the theory also contains  infinitely many more lumpy brane phases whose profiles ${\cal E}(x)$ have $\eta$ maxima and $\eta$ minima for natural integer $\eta$. However, these are  ``excited states" of the theory in the sense that, as we will show below, for given 
$(\hat{L}, \hat{\cal \rho})$ they always have lower Killing entropy density $\hat{\sigma}$ than the ground state lumpy branes that we have constructed above. In other words, lumpy branes with $\eta>1$ are 
subdominant phases of the microcanonical ensemble. In particular, this suggests that they should be dynamically unstable and evolve towards the fundamental lumpy brane if slightly perturbed. In the case of large $\hat L$ this was explicitly verified in \cite{Attems:2019yqn}.

\begin{figure}[th]
\centerline{
\includegraphics[width=.6\textwidth]{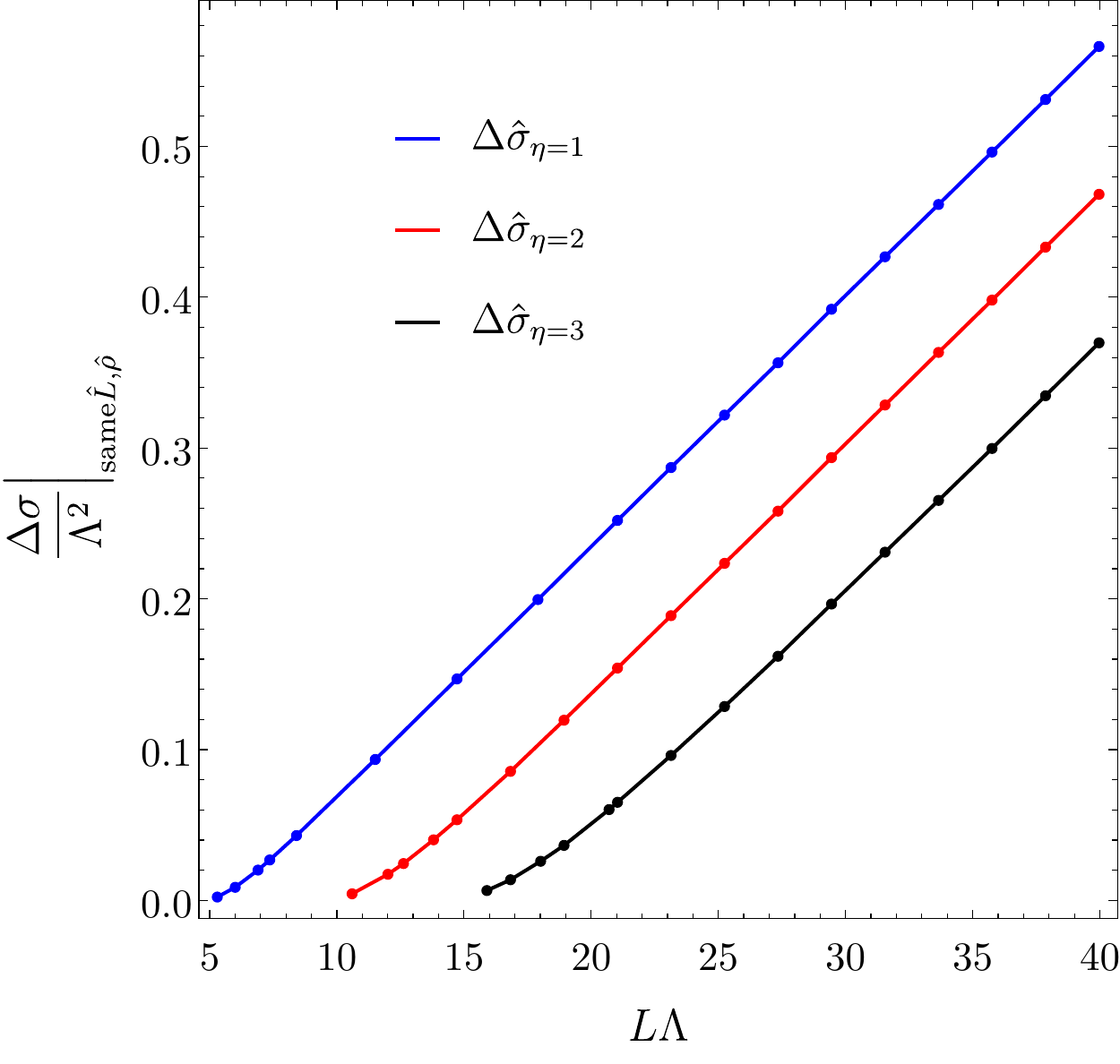}
}
\caption{Killing entropy density difference $\Delta \hat{\sigma}\big|_{\hbox{\footnotesize same } \hat{L}, \hat{\rho}}$ (between lumpy and uniform branes with the same $(\hat{L}, \hat{\rho})$)  as a function of the length $\hat{L}$ for solutions with $\eho=0.85$ for: (1) the fundamental ($\eta=1$) lumpy brane (orange disks), (2) the $\eta=2$ excited lumpy brane (blue diamonds), and (3) $\eta=3$ excited lumpy branes (green squares).}
\label{fig:copies}
\end{figure}

In principle, excited lumpy branes  can be constructed using the perturbative method of Secs.~\ref{sec:PerturbativeO1} and \ref{sec:PerturbativeOn}. At linear order we would have to start with a Fourier mode that describes the $\eta^{\rm th}$ harmonic of the system, namely with
\begin{equation}\label{GLstaticPertCopies}
q_j^{(1)}(x,y)=\frak{q}_j^{(1)}(y)\cos(\eta\, \pi\, x)\,, \quad \hbox{for} \:\: \eta=2,3,4,\cdots
\end{equation}
instead of the $\eta=1$ case of \eqref{GLstaticPert}. However, it is not necessary to perform this construction since the properties of these excited states can be obtained from those of the fundamental ones using extensivity.\footnote{Similar arguments where used to find the thermodynamics of excited nonuniform black strings of the original GL system \cite{Horowitz:2002dc,Harmark:2003eg}). We can also start our linear order analysis with two (or more) harmonics with {\it different}  amplitudes. This allows to construct lumpy branes with two (or more) maxima that have different amplitudes (in the spirit of \cite{Dias:2007hg}).} Indeed, given a solution with $\eta=1$ in a box of size $\hat L$ we can obtain a solution with $\eta > 1$ in a box of size $\eta \hat L$ by taking $\eta$ copies of the initial solution. Once we know all solutions with $\eta=1$ in boxes of any size, as we do, this procedure gives us all possible solutions with $\eta>1$ maxima and minima in all possible boxes. Clearly, if the Killing energy and entropy densities of the initial solution are $\hat \rho$ and $\hat \sigma$, respectively, then those of the new solution are $\eta \hat \rho$ and $\eta \hat \sigma$. In contrast, the average energy density $\eho = \hat \rho/\hat L$ remains invariant. We must now compare the Killing entropy density of the  solution with $\eta$ maxima and minima with that of the corresponding $\eta =1$ brane in a box of size $\eta \hat{L}$. Since the average energy density is invariant when taking copies of the initial solution, this comparison is most easily done by considering $\hat \sigma$ as a function of $\eho$ and $\hat L$. Therefore we must compare the entropy of the excited brane $\hat\sigma_{\eta} (\eho, \eta \hat L)\equiv \eta \times \hat\sigma (\eho, \hat L)$ with that of the fundamental brane 
$\hat \sigma (\eho, \eta \hat{L})$. It follows that if the entropy at fixed $\eho$ grows with $\hat L$ faster than linearly then the fundamental brane always has higher entropy than the excited brane. This is indeed the case, as can be seen by taking constant-$\eho$ slices of \fig{fig:DsigmaCalE}. 
For example, in Fig.~\ref{fig:copies} we do this for $\eho=0.85$ and we compare $\Delta\hat{\sigma}_{\eta=1}$ of the fundamental ($\eta=1$) nonuniform brane (orange $\bullet$) against $\Delta\hat{\sigma}_{\eta=2}$ and $\Delta\hat{\sigma}_{\eta=3}$ of the $\eta=2$ (blue  $\hbox{\tiny $\blacklozenge$}$) and $\eta=3$ (green  $\hbox{\tiny $\blacksquare$}$) excited branes. For a given $(\hat{L}, \hat{\rho})$ or, equivalently, for a fixed  $(\hat{L}, \eho)$, we see that the Killing entropy density decreases as $\eta$ grows: in agreement with the most naive intuition, the fundamental lumpy brane has the highest Killing entropy density and therefore it dominates the microcanonical ensemble over any excited brane. 

The discussion above applies to excited states that can be obtained as copies of a single configuration. Therefore states of this type with $\eta$ maxima and minima have a $\bb{Z}_\eta$ discrete symmetry. There exist more general excited states with maxima and minima of different heights, but we expect these to be subdominant too.  In the case of large $\hat L$ this was explicitly verified in \cite{Attems:2019yqn}.

\subsection{The spinodal (Gregory-Laflamme) timescale}\label{sec:GLtimescale}

In \sect{sec:PerturbativeO1} we saw that intermediate uniform branes with $\hat{\cal E}_B<\hat{\cal E}<\hat{\cal E}_A$ (see left panel of Fig.  \ref{fig:Unif}) can be GL-unstable. To find when the instability appears, we took the uniform branes $Q_j(y)$ of section \ref{sec:PerturbativeO0} and considered static Fourier perturbations of the form \eqref{GLstaticPert} about this background, namely $q_j(x,y)=Q_j(y)+\epsilon \, \frak{q}_j^{(1)}(y)\cos(\pi x)$. This allowed us to find the minimum length $L_{\hbox{\tiny GL}}\Lambda=2\pi/\hat{k}_{\hbox{\tiny GL}}$ (see Figs.~\ref{fig:GLk0} and \ref{fig:GLstabilityDiag}) above which the uniform brane is unstable. This was enough for our purposes of \sect{sec:BVPnonlinearLumpy}, where we were just interested in finding the static lumpy branes.
In particular, we found large regions of the microcanonical phase diagram  where lumpy branes coexist with and are favoured over uniform branes. This suggests that, if we start with initial data that consists of a uniform brane that is GL unstable plus a perturbation, the system should evolve towards a lumpy brane with the same length $\hat{L}$ and  Killing energy density $\hat{\rho}$, and hence also the same  $\eho=\hat{\rho}/\hat{L}$. The initial stages of this time evolution should be well described by the linear GL frequencies. It is thus important to compute the GL timescales of the system.

Consider again the uniform branes constructed in  \sect{sec:PerturbativeO0} in the regime $\hat{\cal E}_B<\hat{\cal E}<\hat{\cal E}_A$ (see left panel of Fig.  \ref{fig:Unif}). Denote the collective fields by $\bar{\psi}(y)=\{\bar{g}_{\mu\nu}(y),\bar{\phi}(y)\}$. We will now allow  for  time-dependent perturbations of this background. 
More concretely, we will use the fact that $\partial_t$ and $\partial_{\tilde{x}}$ are Killing vector fields of the uniform brane background to Fourier decompose the time dependent perturbations as 
\begin{equation}\label{GLtimePert}
\psi(t,x,y)=\bar{\psi}(y)+\epsilon \, \delta\psi^{(1)}(y)e^{i\, k \,\tilde{x}} e^{-i\,\omega \,t} \,.
\end{equation}
This introduces the wavenumber $k$ conjugate to the spatial direction $\tilde{x}=x \frac{L}{2}\in [0,L/2]$ and the frequency $\omega$ of the perturbation. Let $\delta g_{\mu\nu}\equiv h_{\mu\nu}$ be the metric perturbations and  $\delta\phi$ the scalar field perturbation. Perturbations $\delta\psi^{(1)}(y)=\{h_{\mu\nu}(y),\delta\phi(y)\}$ that break the symmetries indicated in \eqref{GLtimePert} excite a total of 8 fields, namely: $\delta\phi$, $h_{tt}$, $h_{ty}$, $h_{t\tilde{x}}$ $h_{yy}$, $h_{\tilde{x}y}$, $h_{\tilde{x}\tilde{x}}$ and $h_{x_2x_2}=h_{x_3x_3}$. 

We have not yet fixed the gauge freedom of the problem. Instead of doing so we construct two gauge invariant-quantities that encode the most general perturbations of the form \eqref{GLtimePert} as described in \cite{Benincasa:2005iv}. The linearized Einstein equations then reduce to (and are closed by) a coupled system of two linear, second-order ODEs for these two gauge-invariant variables. The perturbations must be regular at the horizon in ingoing Eddington-Finkelstein coordinates and preserve the asymptotic AdS structure of the uniform background. This is a non-polynomial eigenvalue problem for the frequency $\omega$ where we give the uniform background and the wavenumber $k$ and find $\omega$. The GL modes of the uniform brane system have purely imaginary frequency.

\begin{figure}[t!]
\centerline{
\includegraphics[width=.55\textwidth]{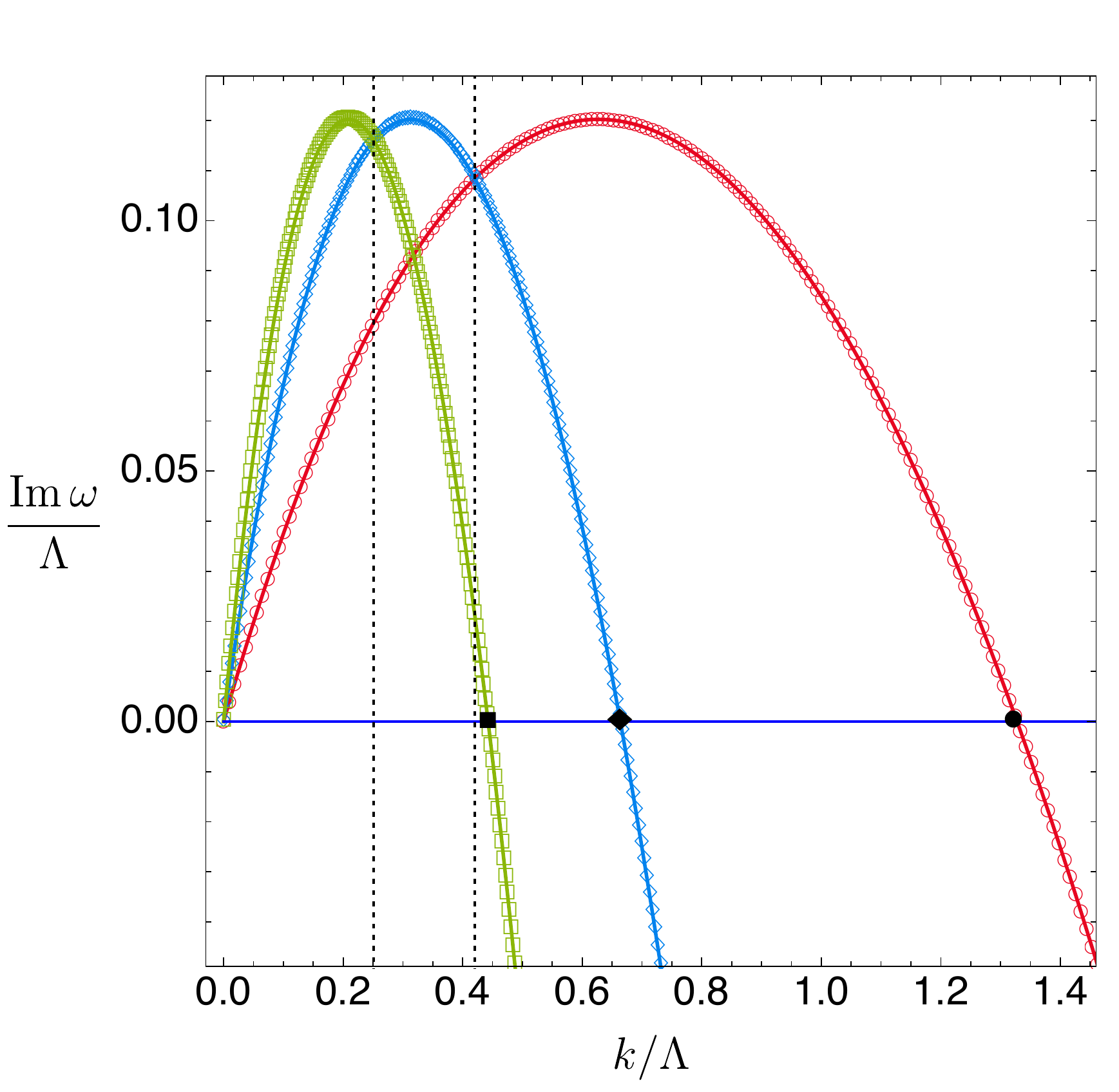}
}
\caption{Dispersion relation of GL modes for a uniform brane with $(\hat{\tau},\hat{\cal E})\simeq (0.395894,0.867966)$. Red circles, blue diamonds and green squares correspond to harmonics with $\eta=1,2$ and 3, respectively.
For reference the maximum of the instability occurs for $(\hat{\tau},\hat{k}_{\hbox{\tiny GL}})_{K}\simeq (0.397427,1.332306)$, \ie $(\hat{\tau},\hat{L}_{\hbox{\tiny GL}})_{K}\simeq (0.397427,4.716021)$. The GL zero mode was identified in \fig{fig:GLk0}.}
\label{fig:GLtimescale}
\end{figure}

In Fig.~\ref{fig:GLtimescale}, 
as an illustrative example, we plot the dimensionless dispersion relation $\hat{\omega}(\hat{k})$ for a particular uniform brane with  $(\hat{\tau},\hat{\cal E})\simeq (0.395894,0.867966)$ that is very close to point $C$ in Fig.~\ref{fig:Unif}, for which $(\hat{\tau},\hat{\cal E})_C\simeq (0.3958945,0.867956)$.\footnote{This corresponds to the same temperature used in the lumpy branes of the top-right panel of Fig.~\ref{fig:profilesExConsT} and in the disk lumpy curve 
$\eho(\hat{L})$ of Fig.~\ref{fig:GLstabilityDiagLumpyT}.}
In  Fig.~\ref{fig:GLtimescale}, the red circle ${\hbox{\tiny$\bigcirc$}}$ curve describes the dispersion relation of the fundamental harmonic $\eta=1$. Not surprisingly, this curve starts at $(\hat{k},\hat{\omega})=(0,0)$ and, as $\hat{k}$ increases,  the dimensionless frequency ${\rm Im}\,\omega/\Lambda$ first grows until it reaches a maximum and then starts decreasing. Precisely at the GL critical wavenumber $\hat{k}=\hat{k}_{\hbox{\tiny GL}}=1.322499$, as computed independently in Fig.~\ref{fig:GLk0}, one has ${\rm Im}\,\omega/\Lambda=0$ and for $\hat{k}>\hat{k}_{\hbox{\tiny GL}}$ the uniform brane is stable. For $0<\hat{k}<\hat{k}_{\hbox{\tiny GL}}$ the uniform brane is GL unstable and the maximum of the instability occurs at $(\hat{k},\hat{\omega})|_{\rm max}\simeq (0.626902,0.120249\, i)$.

Besides the fundamental GL mode, the uniform brane has an infinite tower of integer $\eta$ spatial Fourier harmonics. Uniform branes are also unstable to these higher harmonics but  the minimum unstable GL length $\hat{L}_{\hbox{\tiny GL}, \eta}$  for the $\eta^{\rm th}$ harmonic increases with $\eta$ or, equivalently, the critical GL wavenumber $\hat{k}_{\hbox{\tiny GL}, \eta}$ decreases with $\eta$. As examples, in Fig.~\ref{fig:GLtimescale} we also plot two other curves that describe the dispersion relation of the second ($\eta=2$, blue $\diamond$) and third ($\eta=3$, green ${\hbox{\tiny $\square$}}$) harmonics. Note that the dispersion relation of these higher harmonics can be obtained straightforwardly from that  of the fundamental harmonic. Indeed, note that we can unwrap the $S^1$ and change  the periodicity of its coordinate $\tilde{x}$ from $L$ to $L_\eta= \eta\, L$, for integer $\eta$ \cite{Horowitz:2002dc,Harmark:2003eg}. This also changes the wavenumber from $k=\frac{2\pi}{L}$ into $k_\eta=\frac{k}{\eta}$. Altogether this leaves the phase of the Fourier mode $e^{i k \tilde{x}}$ invariant. But this means that the frequency $\omega_\eta$ of the $\eta^{\rm th}$ harmonic is related to the frequency of the fundamental harmonic simply by $\omega_\eta(k)=\omega(k/\eta)$ and that the critical GL zero mode of the $\eta^{\rm th}$ harmonic is  $\hat{L}_{\hbox{\tiny GL}, \eta}=\eta \,\hat{L}_{\hbox{\tiny GL}}$ or  $\hat{k}_{\hbox{\tiny GL}, \eta}=\hat{k}_{\hbox{\tiny GL}}/\eta$. These properties, namely 
\begin{equation}
\omega_\eta(k)=\omega(k/\eta)\,, \qquad \hat{k}_{\hbox{\tiny GL}, \eta}=\hat{k}_{\hbox{\tiny GL}}/\eta \,,
\end{equation}
are indeed observed in  Fig.~\ref{fig:GLtimescale}.

The linear results of Fig.~\ref{fig:GLtimescale} also provide a guide  to  the full nonlinear time evolution of nonconformal branes. In a microcanonical ensemble experiment,  imagine that we start with a uniform brane in the regime  $\hat{\cal E}_B<\hat{\cal E}<\hat{\cal E}_A$ where it can co-exist with lumpy branes, for example with $\hat{\cal{E}}\simeq 0.867966$. We want to perturb it to drive it towards a lumpy brane with the same $\hat{L}$ and $\hat{\rho}$ and thus same 
$\eho$. What should we do? We certainly have to consider a Fourier perturbation with $\hat{k}<\hat{k}_{\hbox{\tiny GL}}$ as read from Fig.~\ref{fig:GLtimescale} or from  Figs.~\ref{fig:GLk0} and \ref{fig:GLstabilityDiag}. In these circumstances  we still have different options that will result in substantially different time evolutions. Indeed, if we start with a $\hat{k}_{\hbox{\tiny GL}, 2}<\hat{k}<\hat{k}_{\hbox{\tiny GL}}$ where only the fundamental  harmonic is unstable then the system will evolve ``quickly" towards an $\eta=1$ lumpy brane (the quickest evolution  should occur if $\hat{k}\sim \hat{k}|_{\rm max}$). More generically this will still be the case also for a $\hat{k}<\hat{k}_{\hbox{\tiny GL}, 2}$  as long as it is higher than the critical $\hat{k}\sim 0.4205$ where the curves for $\eta=1$ ($\bullet$) and $\eta=2$ ($\diamond$) meet, see the right-most dotted vertical line in Fig.~\ref{fig:GLtimescale}. 
(In this discussion we assume that the initial amplitudes of all modes are similar.) If instead  $0.2513 \lesssim \hat{k}\lesssim 0.4205$, i.e.~in between the two vertical dotted lines of Fig.~\ref{fig:GLtimescale}, then the time evolution of the uniform brane should first approach an $\eta=2$ lumpy  brane before finally moving towards the fundamental $\eta=1$ lumpy brane, which has a higher Killing entropy density. Finally, if the uniform brane is perturbed with a $\hat{k}\lesssim 0.2513$ mode then the system will first evolve towards an $\eta\geq 3$ lumpy brane before being driven towards its fundamental lumpy brane endpoint. These expectations were explicitly verified in the case of large boxes in \cite{Attems:2019yqn}.


\section{Real-time dynamics}

Above we have constructed inhomogeneous static solutions using purely static methods to solve the Einstein equations. We will now examine several aspects of these solutions using real-time dynamical methods. We will first reproduce the static  solutions obtaining excellent agreement. Then we will use the dynamical methods to address two novel aspects not studied above: the local dynamical stability of the inhomogeneous static solutions, and the full time evolution, including the end state, of the  unstable solutions. The reader interested in the numerical methods that we use can consult e.g.~\cite{Attems:2017ezz,Attems:2019yqn,Attems:2018gou,Attems:2016ugt,Attems:2016tby}.


\subsection{Reproducing the static solutions from real-time dynamics}
\label{Reproducing_lumpy_branes}
\begin{figure}[t]
\hspace{-7mm}
\begin{center}
\includegraphics[width=0.9\textwidth]{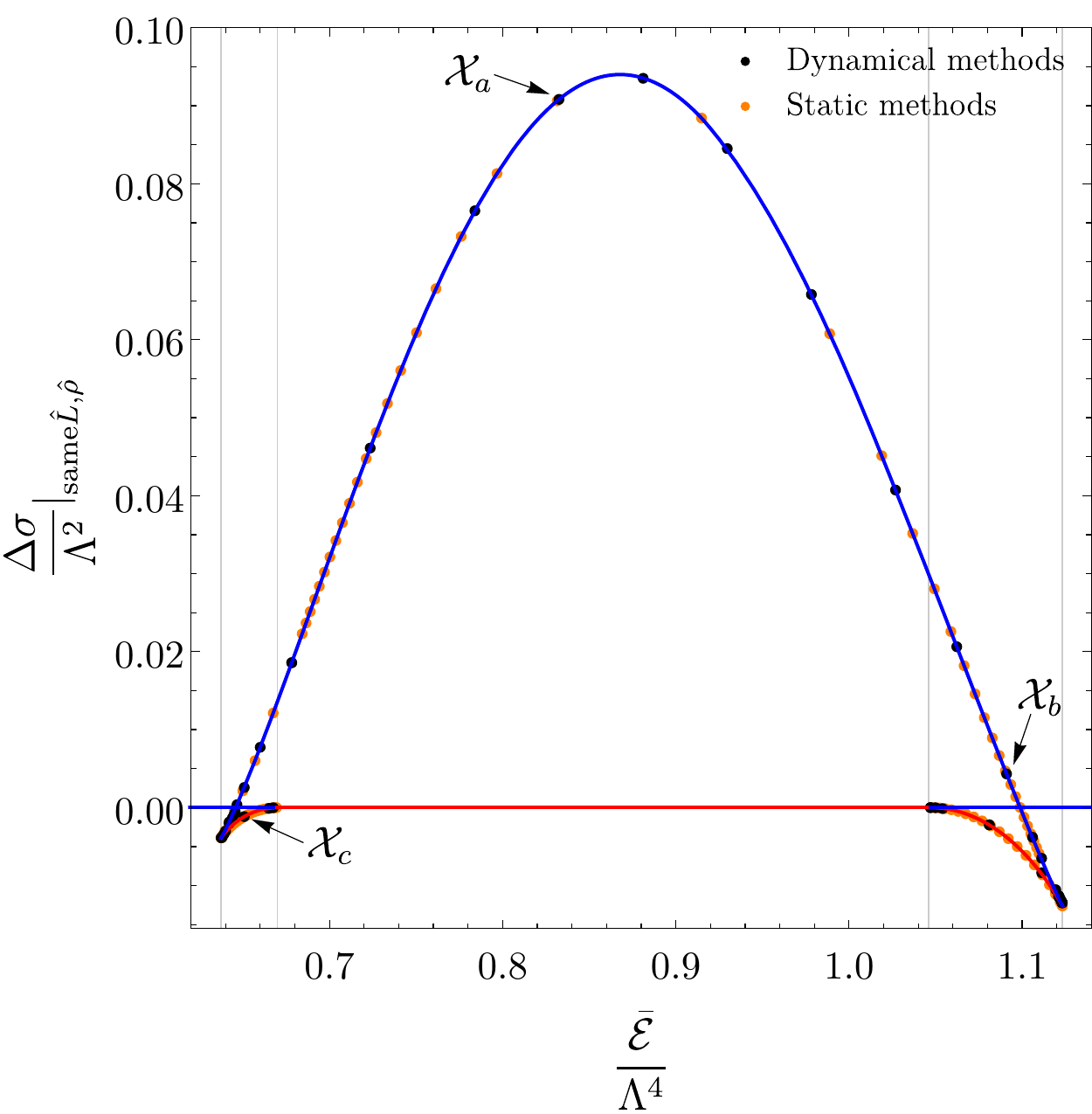}
\end{center}
\caption{Comparison of the entropy density of the static inhomogeneous solutions obtained with dynamical methods (black dots) and with static methods (orange dots) for a system with $\hat{L}\simeq 11.501849$. Blue (red) curves indicate  locally stable (unstable) solutions. 
Orange dots are exactly as in Fig.~\ref{fig:DsigmaRho}(bottom). Grey vertical lines indicate the location of the mergers and the cusps. The representative solutions $\mathcal{X}_a, \mathcal{X}_b$ and $\mathcal{X}_c$ have average energies $\eho_a\simeq 0.831460$, $\eho_b\simeq 1.091$ and $\eho \simeq 0.651$, respectively.
}
\label{fig:Results ComparisonE}
\end{figure}

\begin{figure}[t]
\hspace{-7mm}
\begin{center}
\includegraphics[width=0.9\textwidth]{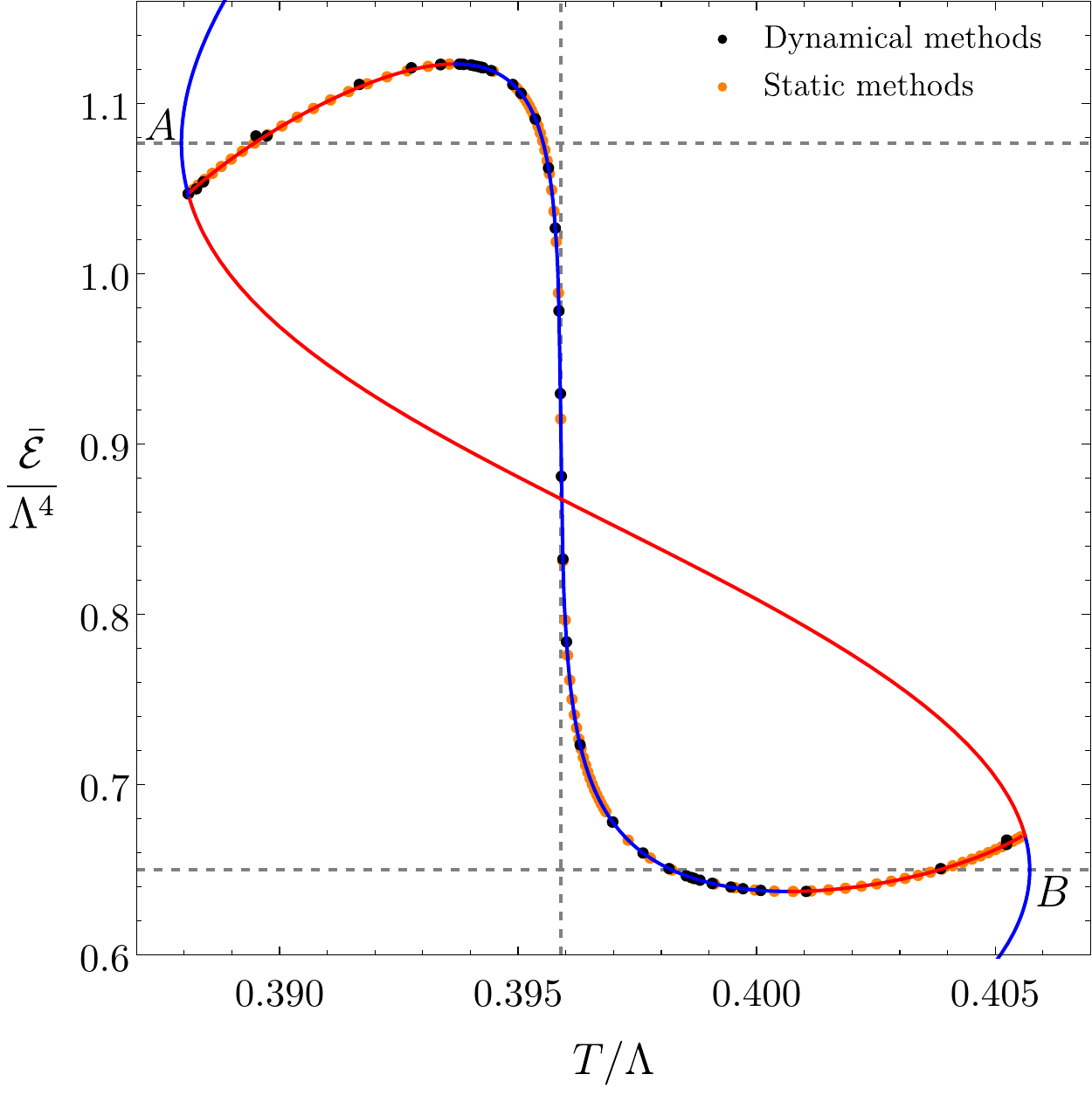}
\end{center}
\caption{Average energy density versus  temperature for static inhomogeneous solutions obtained with dynamical methods (black dots) and with static methods (orange dots) for a system with $\hat{L}\simeq 11.501849$. Blue (red) curves indicate  locally stable (unstable) solutions. 
Orange dots are exactly as in Fig.~\ref{fig:profilesExConstL}(left). 
}
\label{fig:Results ComparisonT}
\end{figure}

In \fig{fig:Results ComparisonE} we compare the Killing entropy density of the static inhomogeneous solutions obtained with dynamical methods (black dots) and with static methods (orange dots) for a system with $\hat{L}\simeq 11.501849$.  In \fig{fig:Results ComparisonT} we compare the average energy density-versus-temperature relation. As is clear from the figures we find excellent agreement. 

The process that we follow to reproduce the static inhomogeneous solutions from real-time dynamical evolution makes it natural to distinguish three cases: 
\begin{enumerate}[label=(\Roman*)]
\item
Lumpy branes whose ($\hat{L}$,$\hat{\rho}$) lies inside the GL merger curve of Figs.~\ref{fig:GLstabilityDiagLumpyT} and \ref{fig:GLstabilityDiagLumpyL}. An example is given by the solution labelled as $\mathcal{X}_a$ in \fig{fig:Results ComparisonE}.
\item
Lumpy branes that are outside the merger curve and have  the largest entropy among lumpy branes with the same ($\hat{L}$,$\hat{\rho}$).   An example is given by the solution labelled as $\mathcal{X}_b$ in \fig{fig:Results ComparisonE}. 
\item
Lumpy branes outside the merger curve with the smallest entropy for a given ($\hat{L}$,$\hat{\rho}$). 
 An example is given by the solution labelled as $\mathcal{X}_c$ in \fig{fig:Results ComparisonE}.
\end{enumerate}

\begin{figure}[h]
\centerline{
\includegraphics[width=.51\textwidth]{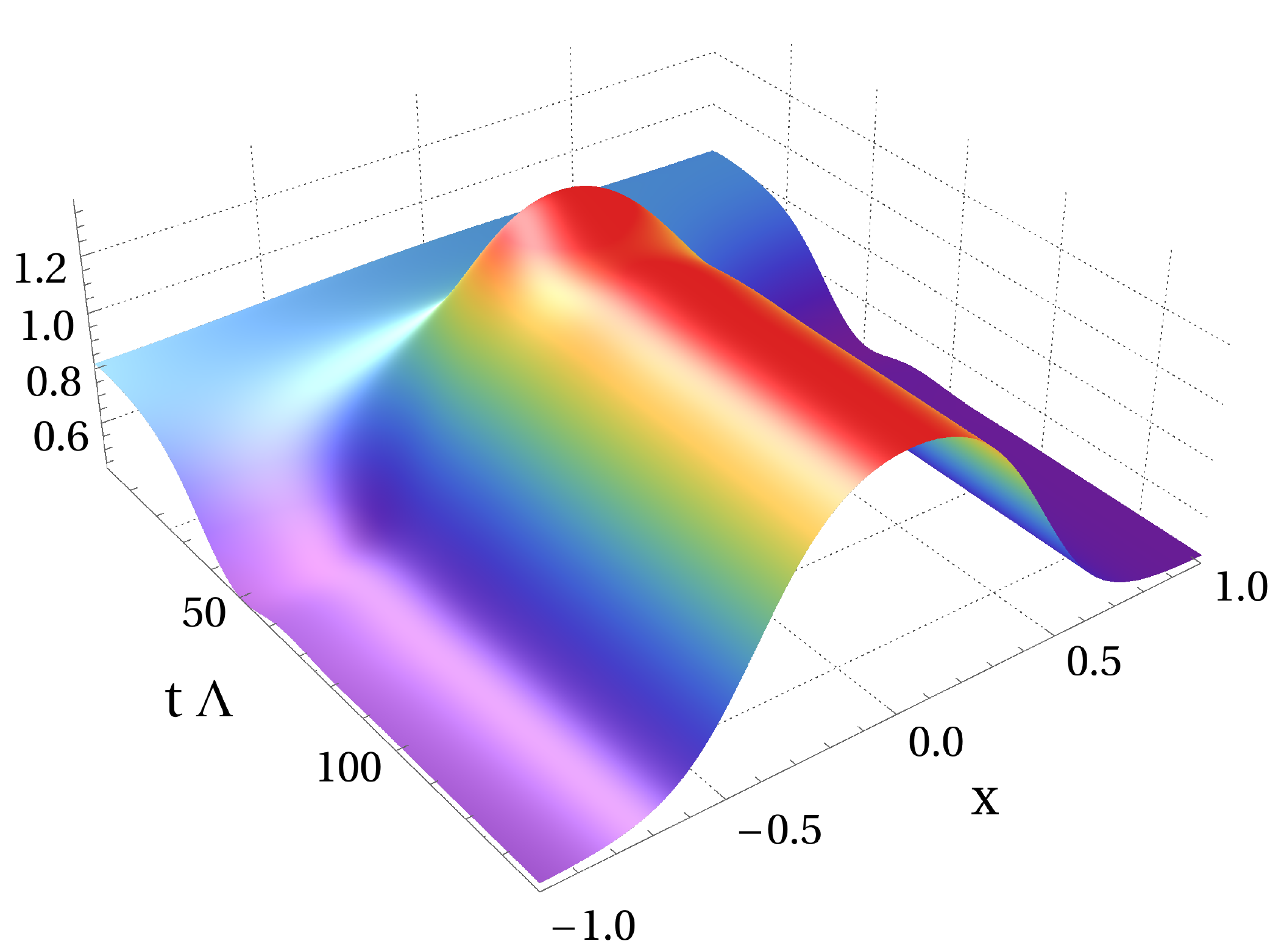}
\put(-225,140){\mbox{{ $\frac{\mathcal{E}}{\Lambda^4}$}}}
\put(-57,24){\fcolorbox{white}{white}{{ $x$}}}
\put(-196,41){\small{\fcolorbox{white}{white}{{$t \Lambda$}}}}
\hspace{0.6cm}
\includegraphics[width=.45\textwidth]{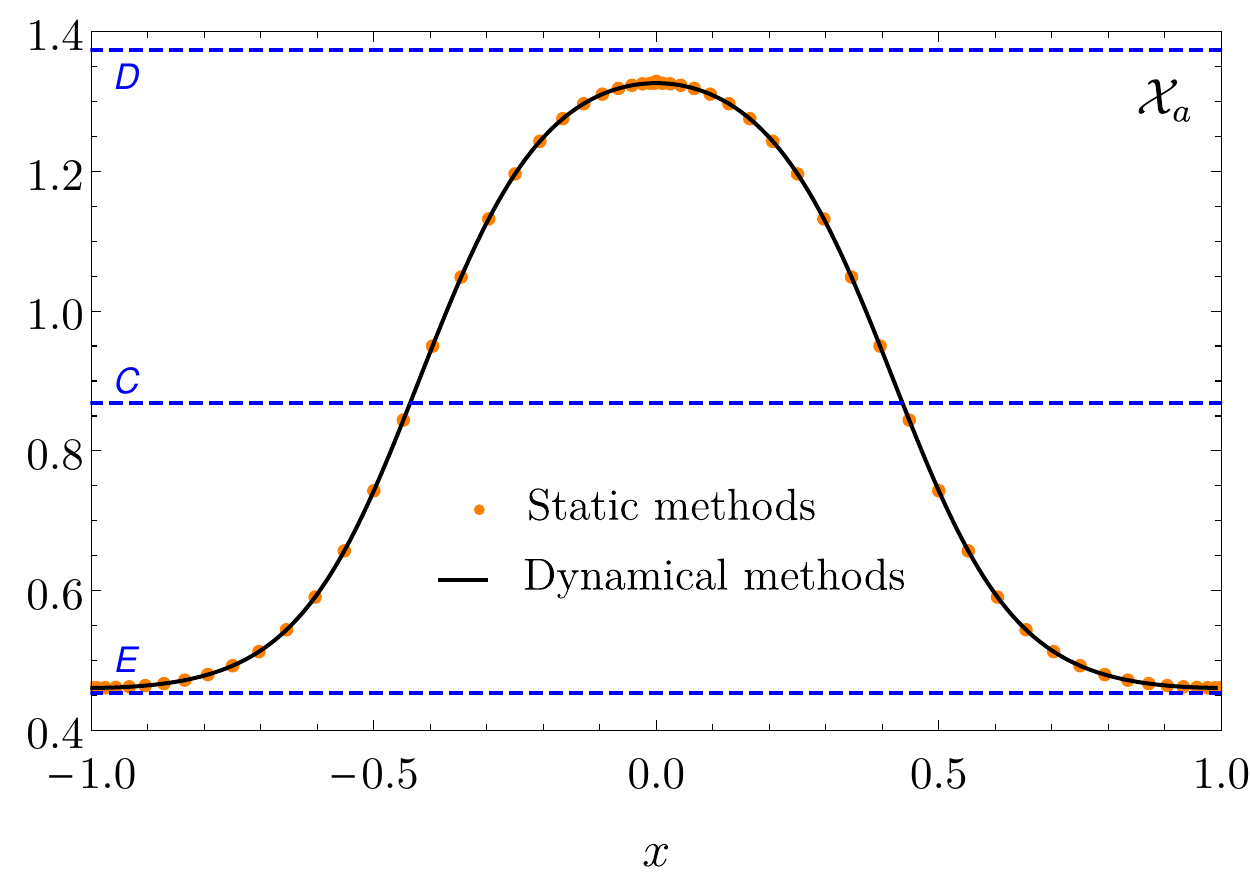}
\put(-208,135){\mbox{{ $\frac{\mathcal{E}}{\Lambda^4}$}}}
}
\centerline{
\includegraphics[width=.51\textwidth]{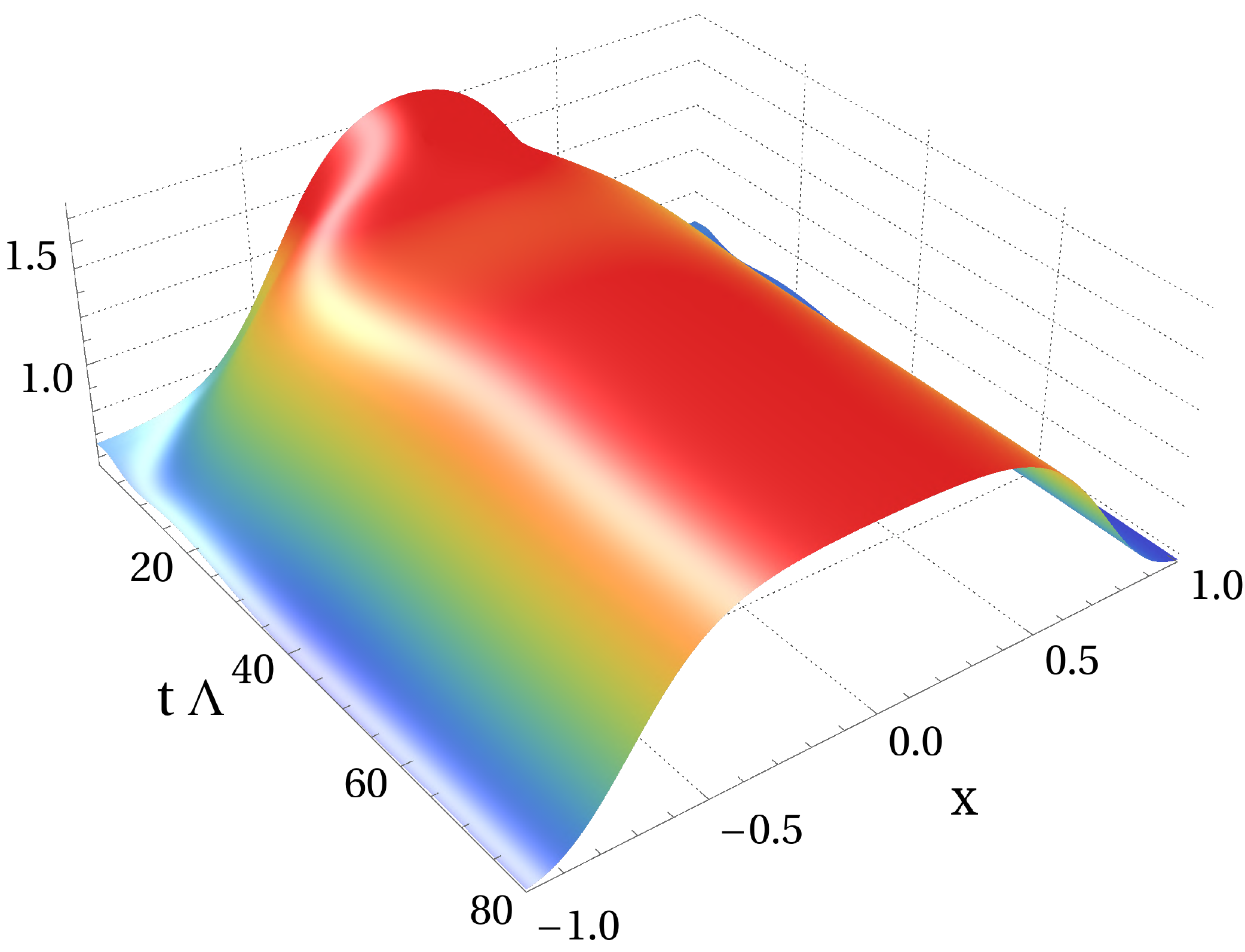}
\put(-226,145){\mbox{{ $ \frac{\mathcal{E}}{\Lambda^4}$}}}
\put(-57,24){\fcolorbox{white}{white}{{ $x$}}}
\put(-198,41){\small{\fcolorbox{white}{white}{{$t \Lambda$}}}}
\hspace{0.6cm}
\includegraphics[width=.45\textwidth]{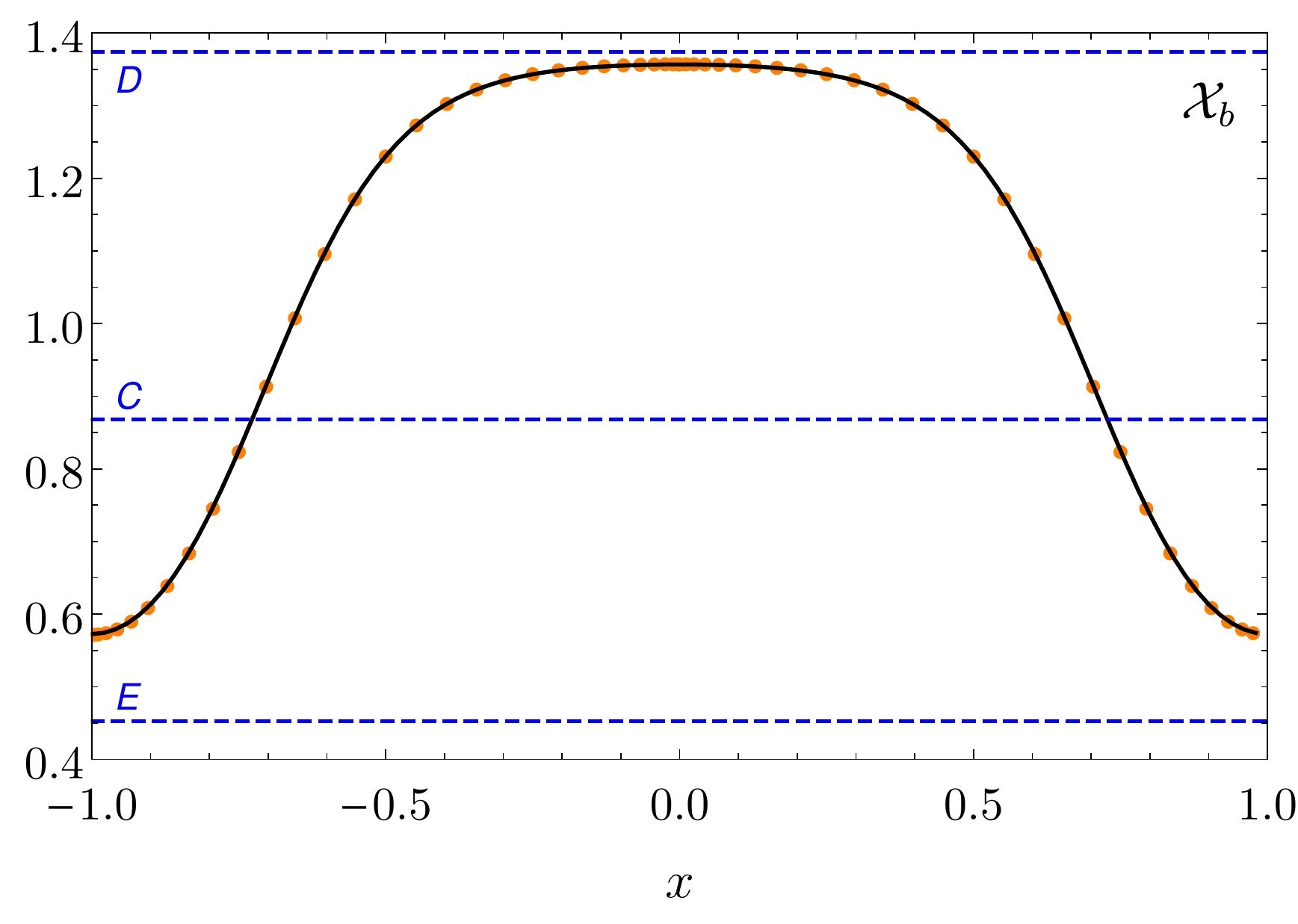}
\put(-208,144){\mbox{{ $\frac{\mathcal{E}}{\Lambda^4}$}}}
}
\centerline{
\includegraphics[width=.51\textwidth]{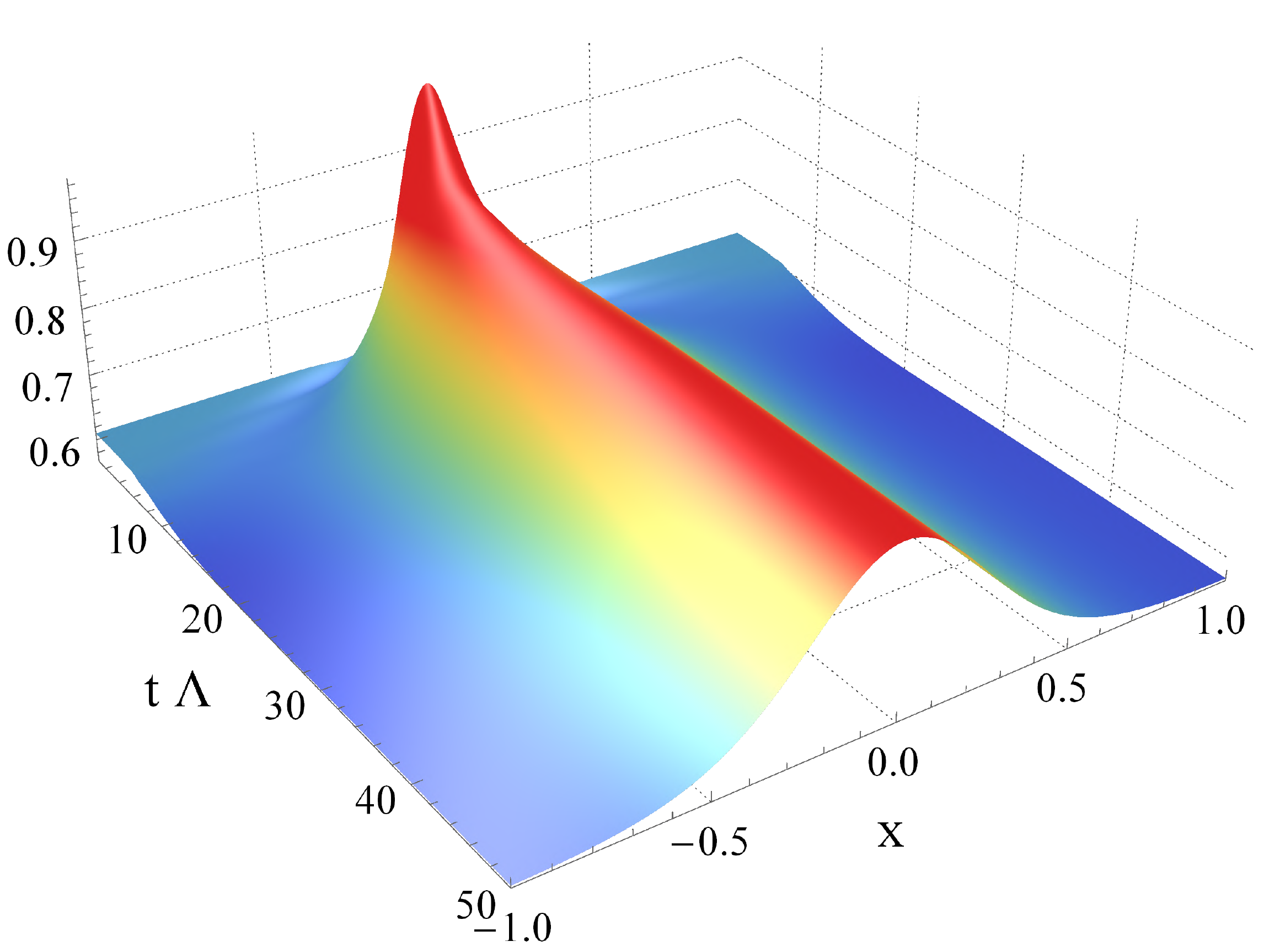}
\put(-225,146){\mbox{{ $\frac{\mathcal{E}}{\Lambda^4}$}}}
\put(-67,17){\fcolorbox{white}{white}{{$x$}}}
\put(-200,41){\small{\fcolorbox{white}{white}{{$t \Lambda$}}}}
\hspace{0.6cm}
\includegraphics[width=.45\textwidth]{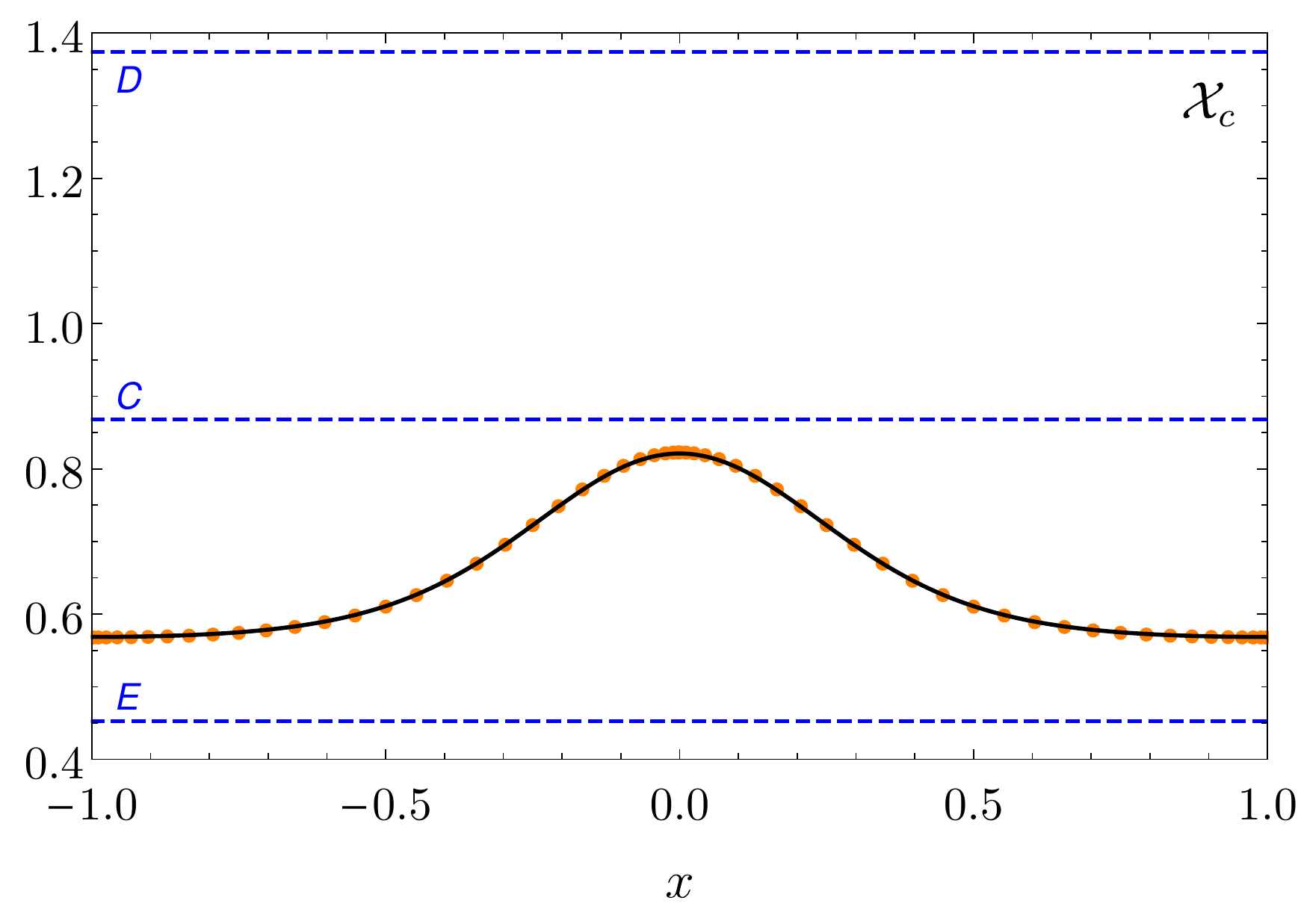}
\put(-208,144){\mbox{{ $\frac{\mathcal{E}}{\Lambda^4}$}}}
}
\caption{Real-time evolution leading to a type I (top-left), a type II (middle-left) and a type III (bottom-left) lumpy brane with \mbox{$\hat{L}\simeq 11.501849$}, labelled $\mathcal{X}_a, \mathcal{X}_b$ and $\mathcal{X}_c$ in \fig{fig:Results ComparisonE}, respectively. On the right panels we compare 
 the energy density profiles  at late times (continuous black lines) with those obtained   by static methods (orange dots). 
}
\label{fig:non linear evolution (1)-(2)}
\end{figure}
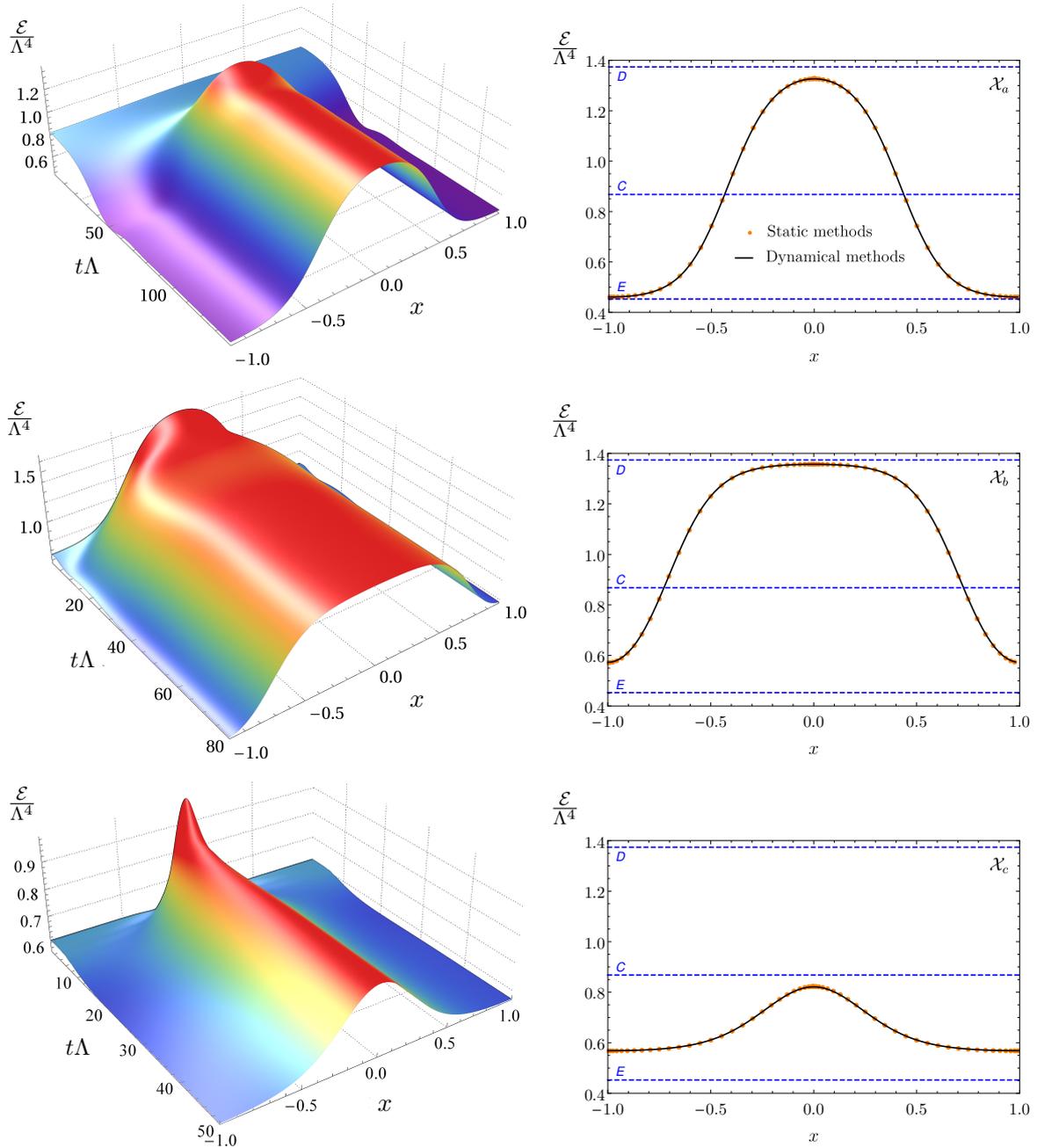

We follow different strategies to find each of these types of solutions.
Solutions of type I are reproduced by following the full evolution of the spinodal instability, as in \cite{Attems:2019yqn, Attems:2017ezz}. 
The initial state is a homogeneous brane with the same  ($\hat{L}$,$\hat{\rho}$) of the lumpy brane that we want to obtain plus a small sinusoidal perturbation corresponding to the lowest Fourier mode that fits in the box. As this solution lies inside the GL merger, this perturbation is unstable and  grows with time.\footnote{There could be other unstable harmonics. However, by considering a sufficiently large amplitude for the first mode the system can always be driven to the fundamental lumpy brane with $\eta=1$.} 
Upon dynamical evolution  the system  eventually enters  the nonlinear regime and finally relaxes to the inhomogeneous solution. In Fig.~\ref{fig:non linear evolution (1)-(2)} we show an example of one of these evolutions (top-left) and the comparison of the solution at asymptotically late times with the solution obtained via static methods (top-right), with excellent agreement.

The previous procedure fails to produce solutions of type II  because the homogeneous system is locally stable, so small perturbations decay in time and the system returns to the initial homogeneous state. Indeed, we consider a uniparametric family of perturbations, not necessarily sinusoidal, with the parameter given by the amplitude $\mathcal{A}$ of the perturbation, and find that if $\mathcal{A}$ is smaller than a certain critical value $\mathcal{A}^*$ then the system evolves back to the homogeneous state.  In order to obtain a lumpy brane as a final state we must start with a homogenous brane plus a perturbation that is so large that the system finds itself directly in the non-linear regime. This is indeed what happens if $\mathcal{A}> \mathcal{A^*}$. In this case the system evolves in time towards the globally preferred state, namely towards a lumpy brane like the one labelled $\mathcal{X}_b$ in \fig{fig:Results ComparisonE}. An example of this evolution is illustrated in  Fig.~\ref{fig:non linear evolution (1)-(2)}(middle-left).  

Finally, if the amplitude of the perturbation is tuned to be exactly  
$\mathcal{A}^*$ then the system evolves in time towards a type III solution like the one labelled as $\mathcal{X}_c$ in \fig{fig:Results ComparisonE}. An example of this evolution is illustrated in  Fig.~\ref{fig:non linear evolution (1)-(2)}(bottom-left). The fact that $\mathcal{A}$ must be precisely tuned in order to reach the type III solution suggests that these solutions are locally dynamically unstable. We will verify this explicitly below. Since numerically it is impossible to tune $\mathcal{A}$ with infinite precision, this means that if we were to evolve the configuration in Fig.~\ref{fig:non linear evolution (1)-(2)}(bottom-left) for sufficiently long times we would see that  either it falls back to the homogeneous state (if $\mathcal{A}$ is slightly smaller than $\mathcal{A}^*$) or it evolves towards a type II configuration (if $\mathcal{A}$ is  slightly larger than $\mathcal{A}^*$). We will confirm this in \fig{fig:non linear evolution (3)}.

\subsection{Local stability}

In this section we study the local stability of the static inhomogeneous solutions by using real-time dynamical methods. We consider an initial state given by the static inhomogeneous solution plus a small perturbation and study its time evolution. The system is said to be locally stable if all possible linear  perturbations decay in time. If at least one of the perturbations grows in time, then the system is said to be locally unstable. 
In order to establish which is the case one must decouple, i.e.~diagonalise, the full set  of linearized equations around the inhomogeneous solution (note that all Fourier modes are indeed coupled to one another because the inhomogenous state breaks translational  invariance). Each eigenmode then evolves in time as $e^{-i \omega t}$, with $\omega$ the corresponding eigenfrequency. If the imaginary part of all the eigenfrequencies is negative the system is locally stable. If at least one of the eigenfrequencies has a positive imaginary part then it is locally unstable.

Rather than performing the exercise above, we will use our numerical code to obtain the  time evolution of a generic small initial perturbation of the inhomogeneous state. Since the perturbation is generic we expect that it will be a linear combination of all the eigenmodes of the system. Thus, after some characteristic time, the eigenmode with the largest imaginary part of omega will dominate the evolution leading to a well defined exponential evolution. We have identified this region of exponential behaviour in all the time evolutions of the perturbed system that we have studied, and we have obtained the real and imaginary parts of omega for the dominant mode by performing fits. Note that this will not result in a mathematical proof of local stability. For example, our generic perturbations may accidentally have a very small projection on some unstable mode, or the positive imaginary part of the frequency of this mode may be exceedingly small and hence go unnoticed, etc. While these possibilities cannot be excluded with absolutely certainty, the detailed searches that we have performed, together with the consistent emergent physical picture, make us confident that they are  highly unlikely.

\begin{figure}[t]
\centerline{
\includegraphics[width=.85\textwidth]{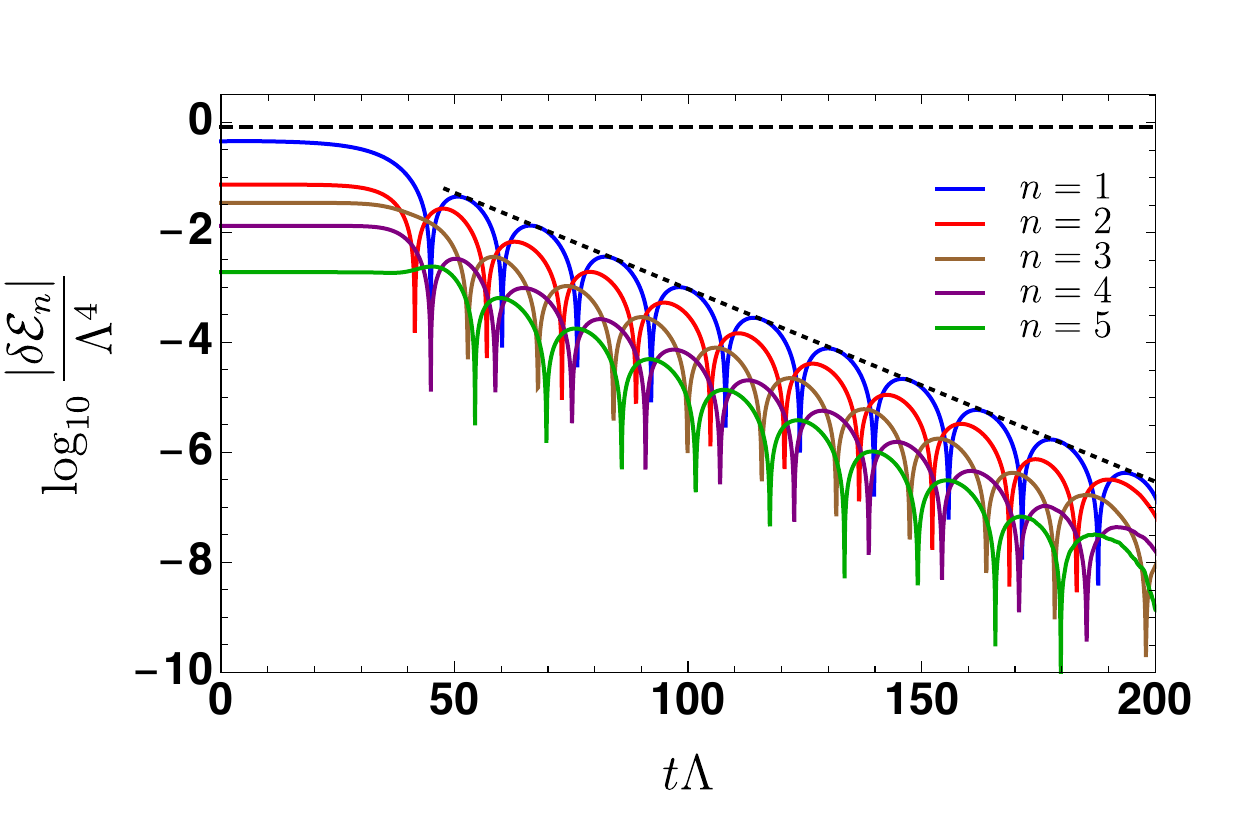}
}
\vspace{-3mm}
\centerline{
\includegraphics[width=.85\textwidth]{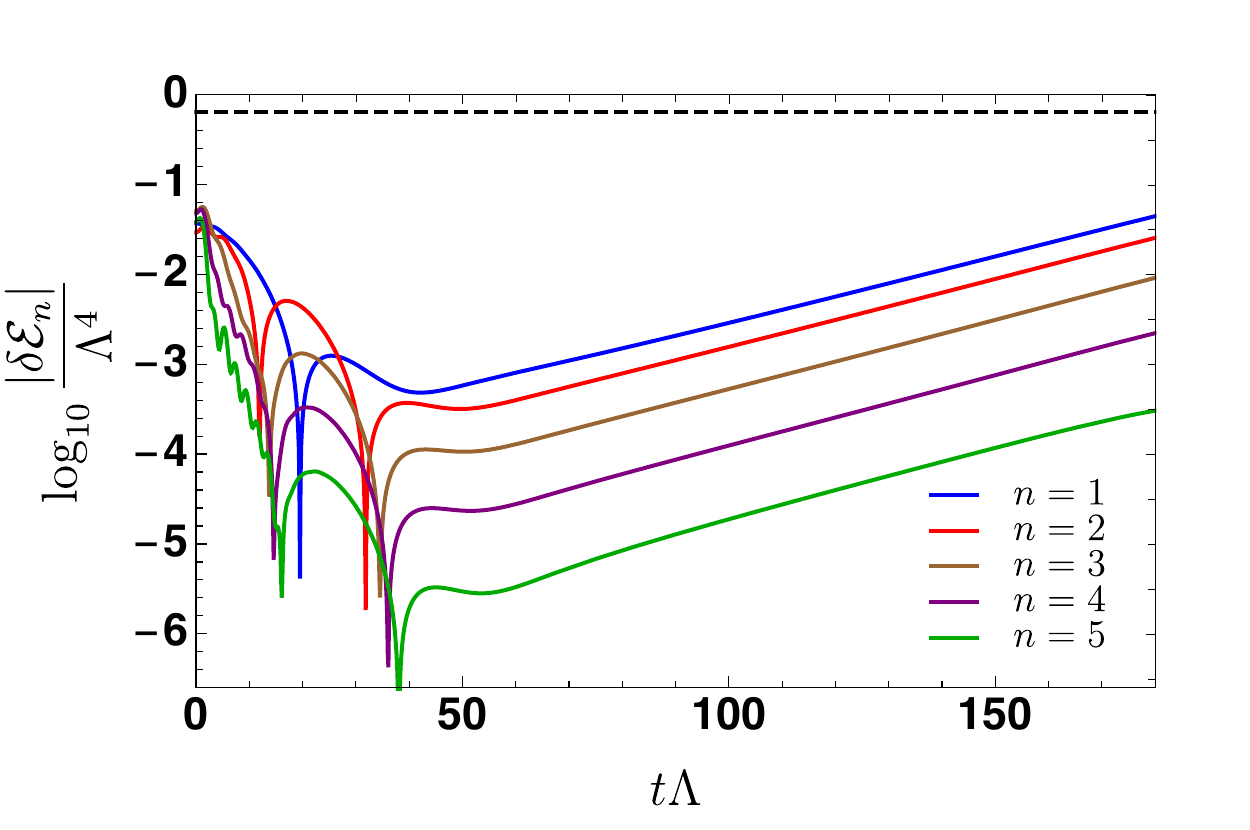}
}
\caption{
(Top) Time evolution of some Fourier modes of a perturbation around the type I, inhomogeneous, static configuration $\mathcal{X}_a$ of Fig.~\ref{fig:non linear evolution (1)-(2)}(top-right) with $\hat{\mathcal{E}}\simeq 0.831460$, $\hat{L}\simeq 11.501849$. The dashed horizontal line indicates the average energy density of the box. 
The region with clear exponentially damped oscillations corresponds to times where the subdominant modes have decayed sufficiently, and the dominant mode has leading amplitude. The dotted line corresponds to a fit to the envelope, from which we extract the imaginary part of omega. The fact that no mode grows in time indicates that $\mathcal{X}_a$ is locally dynamically stable. 
(Bottom) Time evolution of some Fourier modes of a perturbation around 
the type III, inhomogeneous, static configuration $\mathcal{X}_c$ of Fig.~\ref{fig:non linear evolution (1)-(2)}(bottom-right) with 
$\hat{\mathcal{E}}\simeq 0.651$, $\hat{L}\simeq 11.501849$. In this case the leading mode grows exponentially in time, indicating that $\mathcal{X}_c$ is locally dynamically unstable.
}
\label{fig:cos_modes}
\end{figure}

Let us illustrate the procedure with the two examples  in  Fig.~\ref{fig:cos_modes}.
The top panel corresponds to the relaxation to equilibrium at late times of the simulation presented in Fig.~\ref{fig:non linear evolution (1)-(2)}(top-left). Specifically, we take the spatial profile of the energy density at some late time, we subtract from it the profile of the inhomogeneous static solution (which we denoted as $\mathcal{X}_a$ in \fig{fig:Results ComparisonE}), and we decompose this difference into Fourier modes. The time-dependent amplitude of the first few of these Fourier modes is  shown in Fig.~\ref{fig:cos_modes}(top). We see that all of these modes oscillate and decay exponentially in time with the same frequency. This is as expected since the evolution is dominated by the single eigenmode with the slowest decay. The fact that there is no growing mode indicates that the type I, inhomogeneous, static solution to which this configuration asymptotes at late times (namely, $\mathcal{X}_a$ in the current simulation) is locally dynamically stable. 

The second example shown in Fig.~\ref{fig:cos_modes}(bottom) corresponds to the evolution presented in Fig.~\ref{fig:non linear evolution (1)-(2)}(bottom-left), but extended to longer times. Here we plot the Fourier modes corresponding to the difference between the spatial energy profile at a given time and the spatial energy profile of the inhomogeneous, static configuration that we denoted as $\mathcal{X}_c$ in \fig{fig:Results ComparisonE}. We observe a first relaxation in which the stable modes decay but, this time, at later times the system is dominated by an exponential growth of an unstable mode. This confirms that the type III lumpy branes such as $\mathcal{X}_c$ are locally dynamically unstable, as anticipated above. Recall that the initial state in this time evolution is a homogeneous brane plus a large perturbation of amplitude $\mathcal{A}$ that is tuned to be close to a critical value $\mathcal{A}^*$. This tuning is what suppresses the initial amplitude of the unstable mode, hence allowing the time evolution to drive the system close to $\mathcal{X}_c$ for some time. Thus, intuitively, this solution  behaves like a saddle point in configuration space with some stable and some unstable directions (i.e. a metastable configuration). 

\begin{figure}[t]
\centerline{
\includegraphics[width=.75\textwidth]{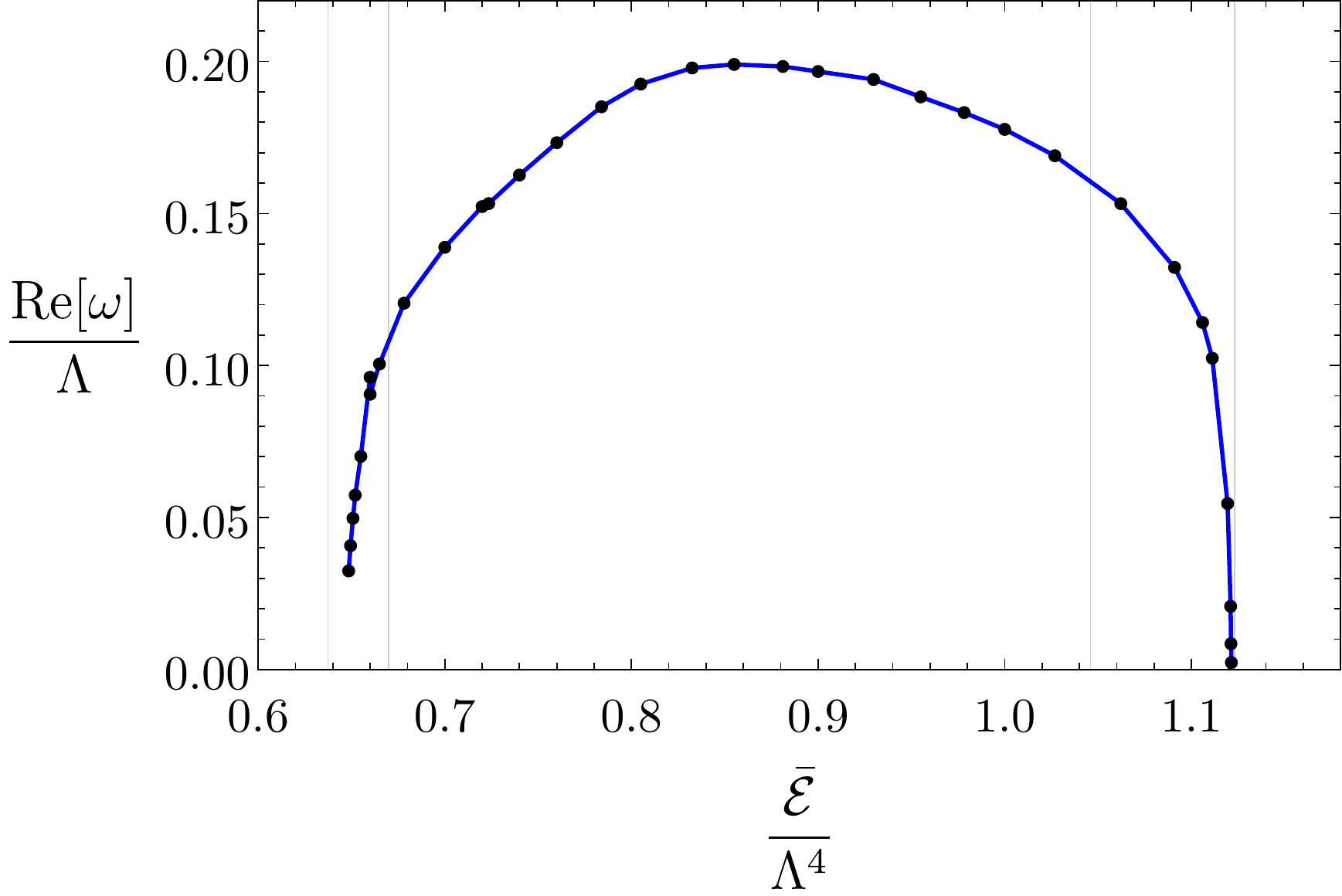}
}
\vspace{8mm}
\centerline{
\includegraphics[width=.78\textwidth]{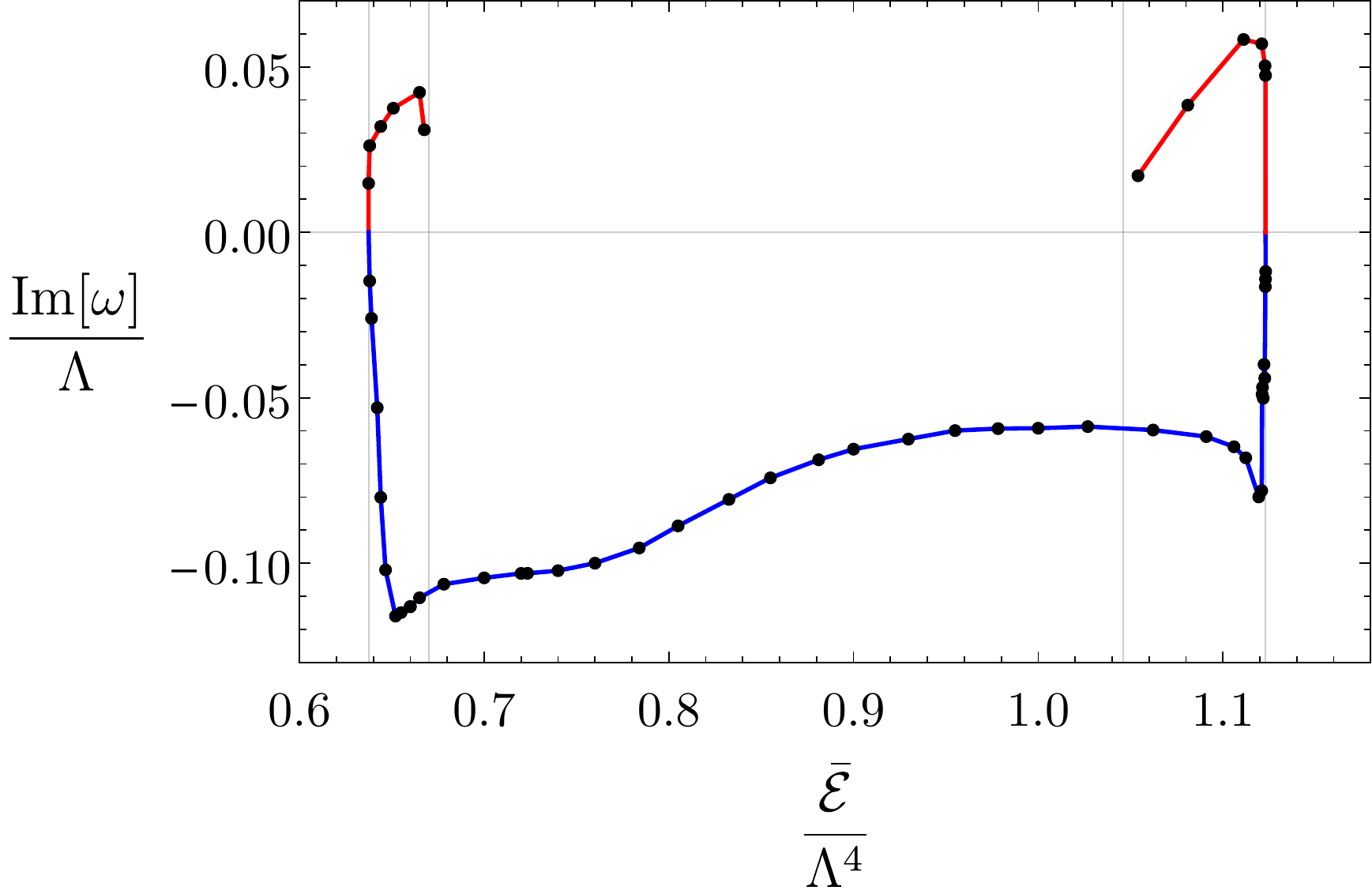}
\hspace{1mm}
}
\caption{Real part (top) and imaginary part (bottom) of the frequency of the dominant linear mode of the perturbations around  static inhomogeneous solutions with $\hat{L}\simeq 11.501849$ and varying energy densities. The real part of omega goes to zero at the cusps and  vanishes for locally unstable solutions. The imaginary part of omega goes to zero at the cusps and at the mergers. Vertical lines indicate the location of the mergers and the cusps, as  in Fig.~\ref{fig:Results ComparisonE}. Blue (red) curves correspond to locally stable (unstable) solutions.
}
\label{fig:Omega}
\end{figure}

We have performed a scan to determine the real and imaginary parts of omega for the dominant mode of  static inhomogeneous solutions with $\hat{L}\simeq 11.501849$ and varying energy densities. The result is shown in Fig.~\ref{fig:Omega}.
We find that type I and II solutions have negative imaginary parts of omega, and so they are locally stable, while the type III solutions have positive imaginary parts of omega, and so they are locally unstable. The imaginary part of omega crosses zero precisely at the ``cusps" of \fig{fig:Results ComparisonE}, that is, at the static solutions lying precisely at the boundary between type II and III solutions. 
The real part of omega is non-vanishing in the locally stable cases, whereas it vanishes in the locally unstable cases, going to zero also at the cusps. In Figs.~\ref{fig:Results ComparisonE}, \ref{fig:Results ComparisonT} and \ref{fig:Omega} we show locally stable solutions in blue and locally unstable solutions in red. 

In this section we have discussed the (in)stability of what we called ``ground-state" or ``fundamental" solutions in \sect{sec:LumpyCopies}, namely of solutions whose spatial energy density profile has a single maximum and a single minimum.  Here we have not explicitly investigated 
the case of ``excited" solutions, namely those with multiple maxima and minima. However, some configurations of this type were studied in  \cite{Attems:2019yqn}, and in all cases they were found to be locally dynamically unstable. As discussed in \sect{sec:LumpyCopies} and \sect{sec:GLtimescale} the reason is that the entropy density can be continuously increased by moving two of these maxima or minima towards each other. Since this seems to be a generic feature, we expect all excited configurations to be locally dynamically unstable.  


\subsection{Full time evolution of the unstable solutions}

In the previous section we studied the local stability properties of the inhomogeneous static solutions, finding some regions of local instability. A natural question  is therefore what is the end state of the evolution if  these locally unstable solutions are perturbed. In this section we perform the full time evolution of the system and determine the end state.

\begin{figure}[t]
\centerline{
\includegraphics[width=.65\textwidth]{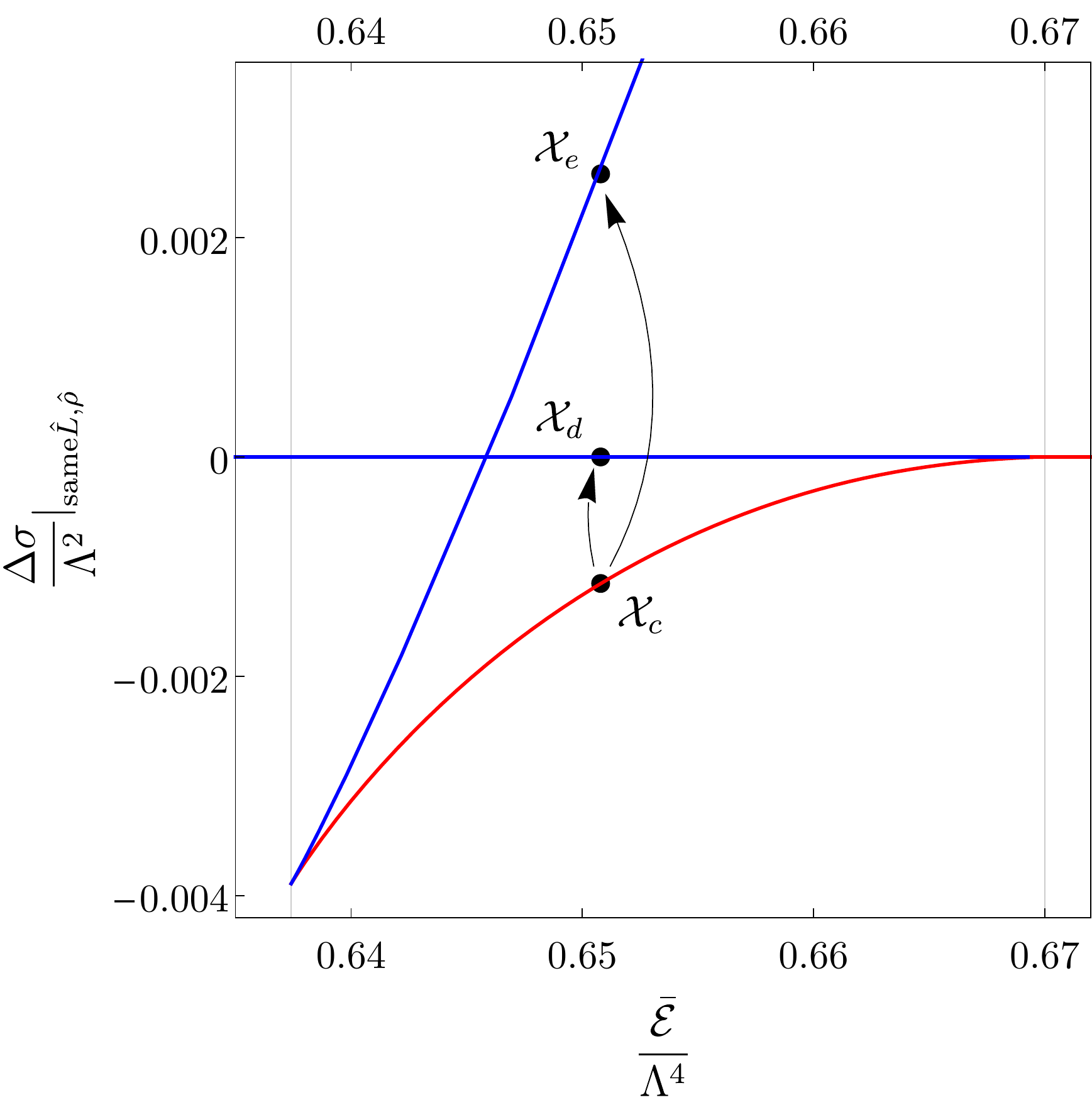}
}
\caption{Zoom-in on the left region of Fig.~\ref{fig:Results ComparisonE}, where $\hat{L}\simeq 11.501849$. $\mathcal{X}_c$, $\mathcal{X}_d$ and $\mathcal{X}_e$ are the three static solutions with the same average energy density 
$\eho \simeq 0.651$. The nonlinear time evolution of Fig.~\ref{fig:non linear evolution (3)}(left) corresponds to an evolution from $\mathcal{X}_c$ to $\mathcal{X}_d$, and Fig.~\ref{fig:non linear evolution (3)}(right) corresponds to an evolution from $\mathcal{X}_c$ to $\mathcal{X}_e$. }
\label{fig:(3) A-B A-C}
\end{figure}

Given a locally unstable solution there are two natural possibilities for the end state of the evolution.  In Fig.~\ref{fig:(3) A-B A-C} we present a concrete example where we show the three static solutions with the same ($\hat{L}$,$\hat{\mathcal{E}})\simeq (11.501849, 0.651$): $\mathcal{X}_c$, $\mathcal{X}_d$ and $\mathcal{X}_e$, where $\mathcal{X}_c$ is the static solution presented in Fig.~\ref{fig:non linear evolution (1)-(2)}(bottom-right). For the locally unstable solution $\mathcal{X}_c$, the two possible candidates for the end state of the evolution are the homogeneous solution $\mathcal{X}_d$ and the inhomogeneous solution $\mathcal{X}_e$, since both of these have larger entropy than $\mathcal{X}_c$. 
By performing full time evolution we confirm that both solutions $\mathcal{X}_d$ and $\mathcal{X}_e$ can be the end state of the evolution, and that which one is reached depends on the initial perturbation. 

\begin{figure}[t]
	\centerline{
		\includegraphics[width=.51\textwidth]{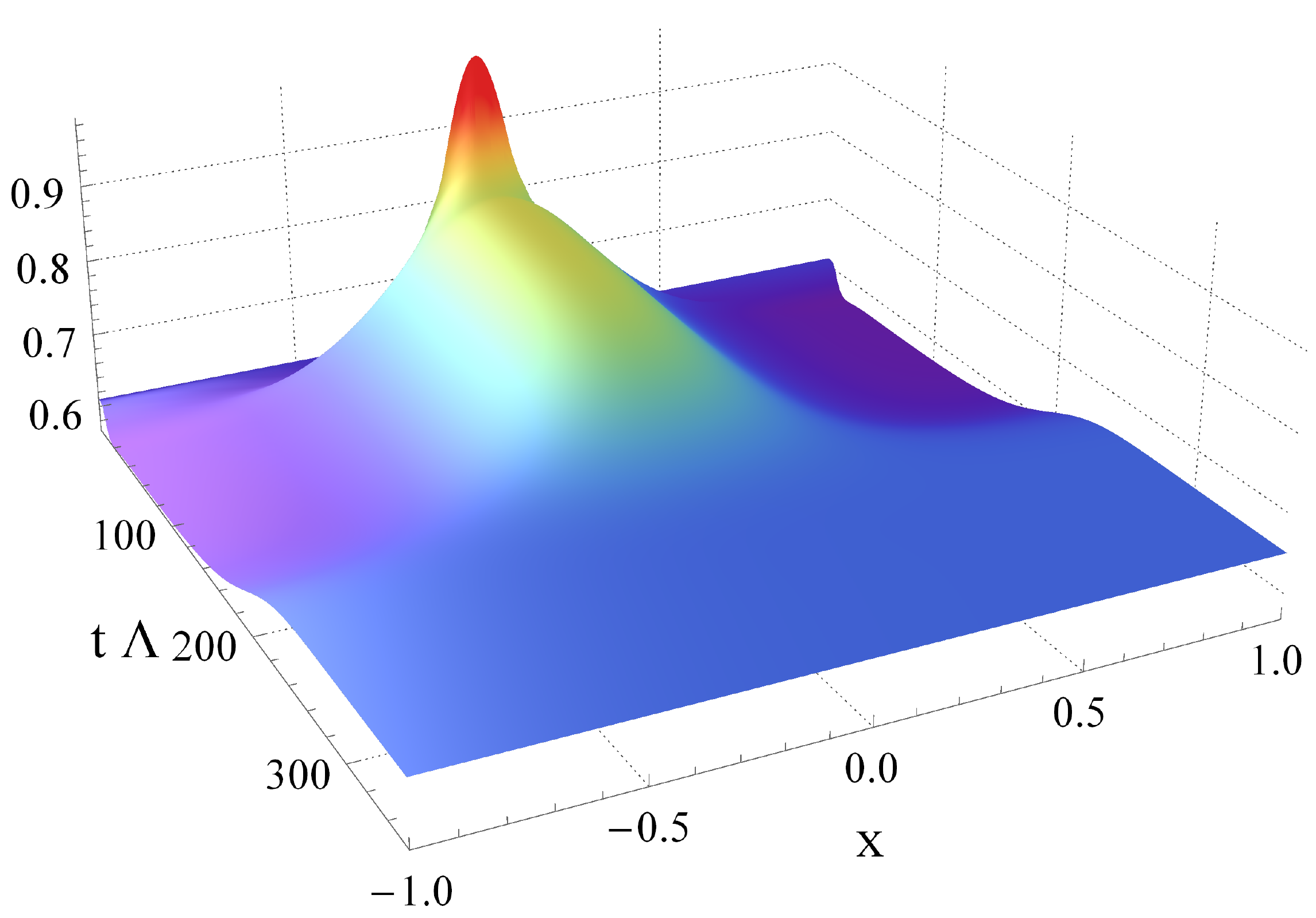}
		\put(-220,143){\mbox{{ $\frac{\mathcal{E}}{\Lambda^4}$}}}
		\put(-78,10){\fcolorbox{white}{white}{{$x$}}}
        \put(-210,42){\fcolorbox{white}{white}{{$t \Lambda$}}}
		\hspace{0.1cm}
		\includegraphics[width=.51\textwidth]{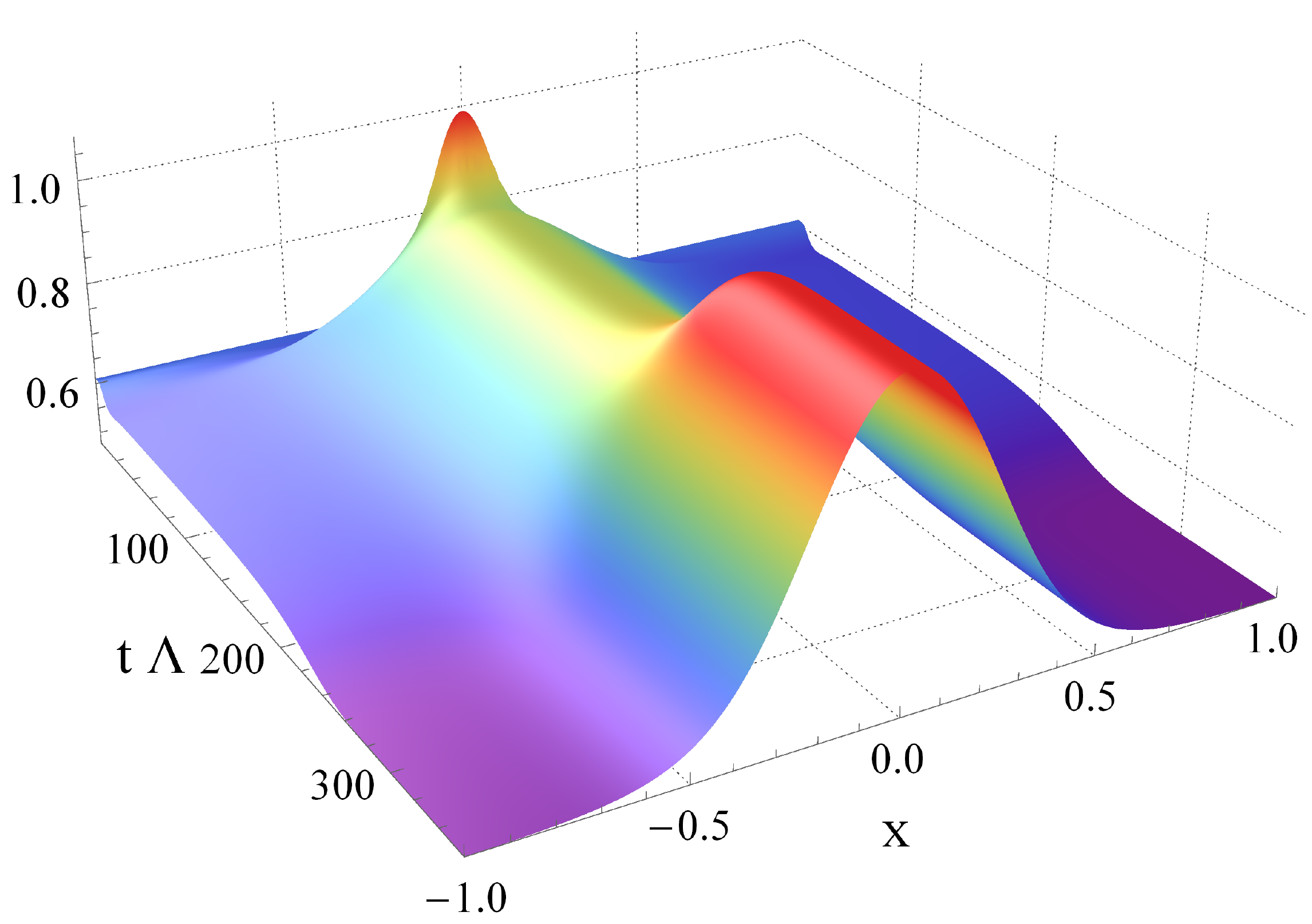}
	    \put(-220,142){\mbox{{ $\frac{\mathcal{E}}{\Lambda^4}$}}}
	    \put(-73,13){\fcolorbox{white}{white}{{$x$}}}
	    \put(-205,40){\fcolorbox{white}{white}{{$t \Lambda$}}}
	}
	\caption{Extension to longer times of the time evolution shown in \fig{fig:non linear evolution (1)-(2)}(bottom-left), whose initial state is a homogeneous configuration plus a large perturbation of amplitude $\mathcal{A}$. At intermediate times this generates the unstable solution  $\mathcal{X}_c$ plus a small perturbation. (Left) If the amplitude of the initial perturbation $\mathcal{A}$ is slightly smaller than the critical value $\mathcal{A}^*$ then the system eventually evolves back to the homogeneous solution labelled  as $\mathcal{X}_d$ in \fig{fig:(3) A-B A-C}. (Right) If the amplitude of the initial perturbation $\mathcal{A}$ is slightly larger than the critical value $\mathcal{A}^*$ then the system eventually evolves towards the stable,  inhomogeneous solution labelled as $\mathcal{X}_e$ in \fig{fig:(3) A-B A-C}.}
	\label{fig:non linear evolution (3)}
\end{figure}
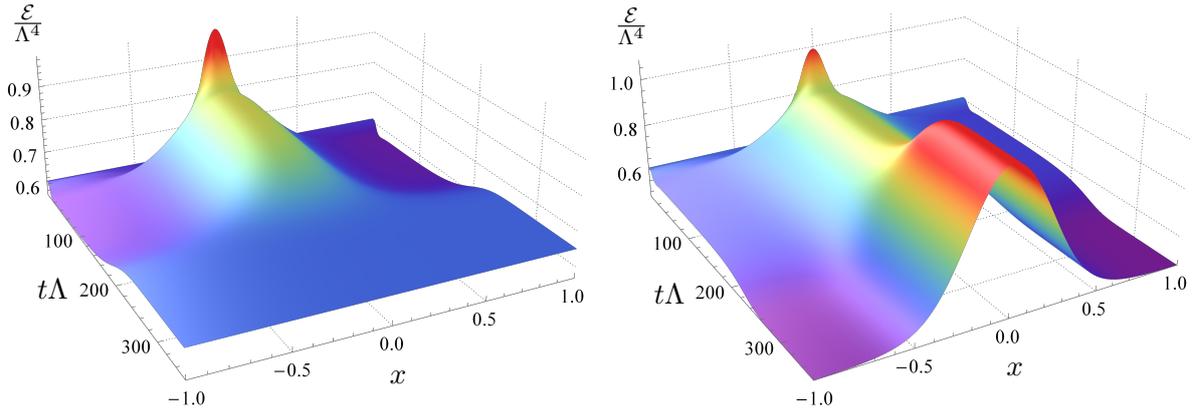

In order to illustrate this  we essentially extend the range of the time evolution shown in \fig{fig:non linear evolution (1)-(2)}(bottom-left). Recall that in that figure we dynamically generated a solution very close to  
$\mathcal{X}_c$ by fine-tuning the amplitude of the initial perturbation to be close to the critical value $\mathcal{A}^*$. Since the amplitude we choose is close but not exactly equal to $\mathcal{A}^*$, the result of this time evolution at intermediate times is not exactly the solution 
$\mathcal{X}_c$ but $\mathcal{X}_c$ plus a small perturbation. Since the perturbation is small the system spends a sizeable amount of time in a very slowly evolving configuration close to $\mathcal{X}_c$, as can be seen from the intermediate-time behaviour in Fig.~\ref{fig:non linear evolution (3)}. However, if the exact amplitude is slightly smaller than the critical one then further time evolution eventually drives the system back to the homogeneous solution labelled as $\mathcal{X}_d$ in \fig{fig:(3) A-B A-C}. This is the case in Fig.~\ref{fig:non linear evolution (3)}(left). If instead the amplitude is slightly larger than the critical one then the system eventually evolves towards the stable, inhomogeneous solution labelled as $\mathcal{X}_e$ in  \fig{fig:(3) A-B A-C}. This is the case in Fig.~\ref{fig:non linear evolution (3)}(right).
Note that the evolution from the unstable to the stable solutions can be viewed as a long, approximately linear regime (when the unstable solution is perturbed), followed by a fast non-linear regime, further followed by another long, approximately linear regime (when the system relaxes to the corresponding stable solution). We have verified that, at the qualitative level, these results apply to all the unstable solutions with $\hat{L}\simeq 11.501849$ and varying energy  densities that we have studied. 


\section{Discussion}
\label{disc}
We have considered a bottom-up, five-dimensional  gravity model that, at infinite volume, possesses a  first-order, thermal phase transition in the canonical ensemble. As usual, we expect this model to be holographically dual to a strongly coupled, large-$\nc$ gauge theory in four dimensions.  We have  placed the system in a box and, for simplicity, we have imposed translational invariance along two sides of the box. We have varied the volume by varying the size  $\hat{L} = \Lambda L$ of the third side, where $\Lambda$ is the microscopic scale of the gauge theory. We have  then constructed what we believe is the complete set of all possible homogeneous or inhomogeneous equilibrium states at finite $\hat L$. 
On the gravity side these correspond to uniform or lumpy branes, respectively. Although we do not have a mathematical proof that this set is indeed complete, we have found no evidence to the contrary in our extensive investigations based both on static and dynamical methods. 

The first effect of the finite volume is that some homogeneous states between points $A$ and $B$ in Fig.~\ref{fig:transition0INTRO} become locally dynamically stable, as illustrated in \fig{fig:GLstabilityDiag}. The reason for this is that the spinodal instability is a long-wavelength instability. If $\hat L$ is below a certain energy-dependent value, then the potentially unstable mode does not fit in the box and the corresponding homogeneous state is actually stable. The unstable states are those in the region inside the parabola in \fig{fig:GLstabilityDiag}. We see that as $\hat L\to \infty$ we recover the fact that all states with energy densities between $A$ and $B$ are unstable, but that at finite $\hat L$ some of them are stable. In particular, there is a value  $\hat L =\hat L_K$ below which all homogeneous states are locally dynamically stable since none of them can accommodate an unstable mode. 

The parabola in \fig{fig:GLstabilityDiag} is  a curve of marginal stability. Therefore we  expect a branch of static, inhomogeneous states to emanate from each point on this curve. One should think of the extra direction in which these branches emanate as the entropy relative to that of the homogeneous state, $\Delta \hat{\sigma}$. The  union of all such inhomogeneous branches  is therefore a surface in the three-dimensional space parametrized by the average energy density in the box $\overline{\mathcal{E}}$, the size of the box $\hat L$, and the entropy $\Delta \hat{\sigma}$. We will refer to this surface as the ``entropy surface". 
The intersection of this surface with the $\hat{\overline{\mathcal{E}}}$-$ \hat{L}$ plane contains the parabola in \fig{fig:GLstabilityDiag} (as well as other points such as the points $a$ and $b$ of Fig.~\ref{fig:DsigmaCalE}). The curves in \fig{fig:GLstabilityDiagLumpyT} are the projections on this plane of constant-$\hat{T}$ slices of the entropy  surface. Similarly, the vertical lines in \fig{fig:GLstabilityDiagLumpyL} are the projections on this plane of constant-$\hat L$ slices. The same slices projected onto the $\hat{\overline{\mathcal{E}}}$-$\hat{T}$ plane are shown in \fig{fig:LumpyUnifET}. This last figure makes it clear that inhomogeneous states only exist in the range of temperatures $\hat{T}_A \leq \hat{T} \leq \hat{T}_B$. 

The structure of the entropy surface is most easily understood by thinking of it as the union of constant-$\hat L$ slices for all $\hat L > \hat L_K$. The shape of each of these slices as a function of the energy density is shown in Figs.~\ref{fig:DsigmaRho} and \ref{fig:DsigmaCalE}. We see that, at the qualitative level, there are three possibilities depending on the value of $\hat L$ in relation to the following hierarchy 
\be
\hat{L}_{K}< \hat{L}_{\Sigma_1}  < \hat{L}_{\Sigma_2} \,.
\ee
These three length scales are an intrinsic property of the theory at finite volume and their  values are given in \eqref{hierarchy}. If $\hat{L}_{K}<\hat L< \hat{L}_{\Sigma_1}< \hat{L}_{\Sigma_2}$, then  $\Delta \hat{\sigma}$ is always positive for all  the values of the energy for which inhomogeneous states exist. This is the case illustrated by \fig{fig:DsigmaRho}(top-left) and by the bottom curve with beige inverted triangles in \fig{fig:DsigmaCalE}. 
If instead $\hat{L}_{K}< \hat{L}_{\Sigma_1}< \hat L <\hat{L}_{\Sigma_2}$ then the $\Delta \hat{\sigma}$ curve  becomes negative and develops a cusp near its endpoint on the right-hand side. This is the case illustrated by \fig{fig:DsigmaRho}(top-right) and by the second-from-the-bottom curve with brown diamonds  in \fig{fig:DsigmaCalE}. Finally, if 
$\hat{L}_{K}< \hat{L}_{\Sigma_1} <\hat{L}_{\Sigma_2}< \hat L$ then the  $\Delta \hat{\sigma}$ curve  becomes negative and develops  cusps near both of its endpoints. This is the case illustrated by \fig{fig:DsigmaRho}(bottom) and by the two top curves with orange circles and green squares, respectively,   in \fig{fig:DsigmaCalE}. 

The shape of the entropy surface that we have just described determines the structure of phase transitions in the microcanonical ensemble. Recall that in the limit $\hat L \to \infty$ the set of globally preferred, maximum-entropy states are those indicated by the black curves (with arrows) in Fig.~\ref{fig:transition0INTRO} 
(see also the discussion around Figs.~\ref{entropyQ}
and \ref{fig:convexconcave}). The direction of the arrows in Fig.~\ref{fig:transition0INTRO} indicates what happens as the energy decreases from an arbitrarily high value. As the energy density decreases towards point $D$ the preferred states are homogeneous branes of decreasing temperature. At $D$ there is a phase transition into inhomogeneous states of constant temperature $T_c$. Since $\hat L \to \infty$ these are phase-separated configurations in which the homogeneous phases $D$ and $E$ coexist. At $E$ there is another  phase transition, in this case  from inhomogeneous to homogeneous states. 
The fact that the fraction of the total volume occupied by each phase varies continuously between 0 and 1 as the energy density varies between $D$ and $E$ suggests that these transitions are continuous in the microcanonical ensemble. Continuity can also be seen more formally as follows. For fixed length and source,  the first law \eqref{1stA} takes the form
\be
\frac{1}{T} = \frac{dS}{dE} \,.
\ee
In the microcanonical  ensemble the total entropy $S$ is the relevant thermodynamic potential, the total energy $E$ is the control parameter and $T$ is a derived quantity. At the points $D$ and $E$ the temperature is continuous, but its derivative $dT/dE$ is not, because this is positive on the homogeneous branch but it vanishes on the inhomogeneous one.

\begin{figure}[t]
\centerline{
		\includegraphics[width=.8\textwidth]
		{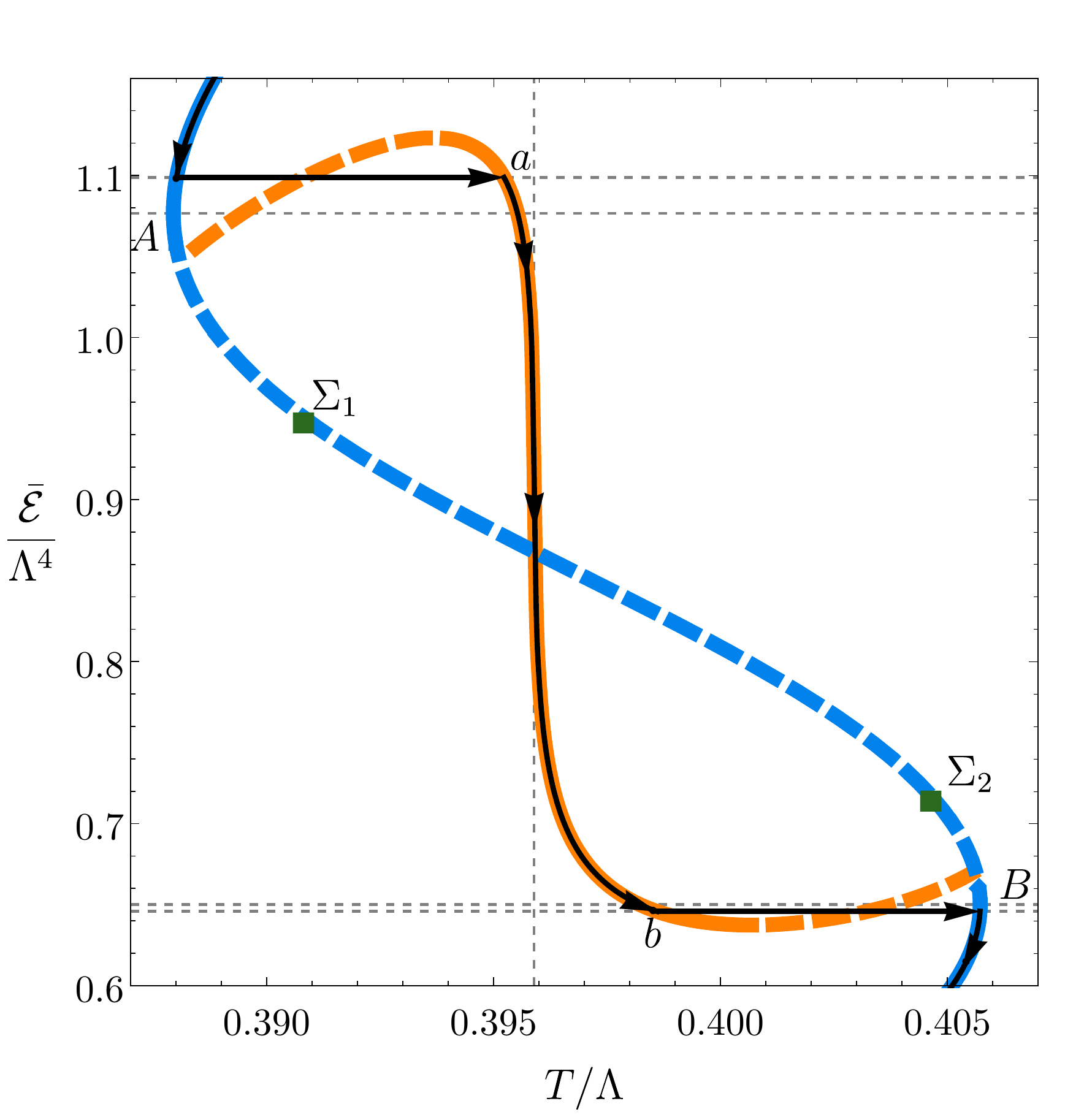}
}
\caption{Phase diagram in the microcanonical ensemble for a length 
$\hat{L} \simeq 11.501849$ such that 
$\hat{L}_K< \hat{L}_{\Sigma_1} <  \hat{L}_{\Sigma_2} < \hat{L}$. The blue and the orange curves correspond to homogeneous and inhomogeneous states, respectively. The orange curve is the same as the curve of orange circles in \fig{fig:LumpyUnifET}.
Solid segments  indicate locally dynamically stable states; dashed segments indicate unstable ones. The black curves with arrows indicate the sequence of globally preferred, maximum-entropy states as the average energy density decreases.  The points $a$ and $b$ corrrespond to those in \fig{fig:DsigmaCalE}. The first and fourth (from top to bottom) dashed horizontal lines indicate the energy densities at these points, whereas the second and third lines indicate the energy densities at the turning points $A$ and $B$. The phase transitions between the homogeneous and the inhomogeneous branches, indicated by the  horizontal arrows, are first-order.
}
\label{fig:transition1}
\end{figure}

This picture is modified at finite $\hat L$.  Note that in this case the system may still exhibit phase transitions since the planar limit that we work in, $\nc \to \infty$, acts effectively as a thermodynamic limit. Consider first
\fig{fig:transition1} which illustrates the structure of phase transitions for a  length $\hat{L} \simeq 11.501849$ such that 
$\hat{L}_K< \hat{L}_{\Sigma_1} <  \hat{L}_{\Sigma_2} < \hat{L}$ (this is the value corresponding to the orange circles in \fig{fig:LumpyUnifET}). 
In this case the preferred states lie on the homogeneous branch until the energy density reaches that of point $a$ in \fig{fig:DsigmaCalE}. At this point a first-order phase transition takes place between the homogeneous state and the state $a$ on the inhomogeneous branch, as indicated by the top horizontal arrow in \fig{fig:transition1}. 
Note that this transition can take place before (as in the case of the orange circles in \fig{fig:LumpyUnifET}) or after (as in the case of the green squares \fig{fig:LumpyUnifET}) the turning point $A$ is reached. The reason that this is a first-order transition is that the temperature changes discontinuously.  As the energy decreases  further,  the preferred states are those on the inhomogeneous branch until the energy density reaches that of point $b$ in \fig{fig:DsigmaCalE}. At this point another first-order phase transition takes place between the inhomogeneous state and a state on the homogeneous branch with the same average energy density, as indicated by the  bottom  horizontal arrow in \fig{fig:transition1}. Note that this state is below the turning point $B$ (i.e. $\hat{\overline{\mathcal{E}}}_b<\hat{\overline{\mathcal{E}}}_B$). As the energy is further decreased the preferred state remains on the homogenous branch. 

\begin{figure}[t]
\centerline{
		\includegraphics[width=.8\textwidth]{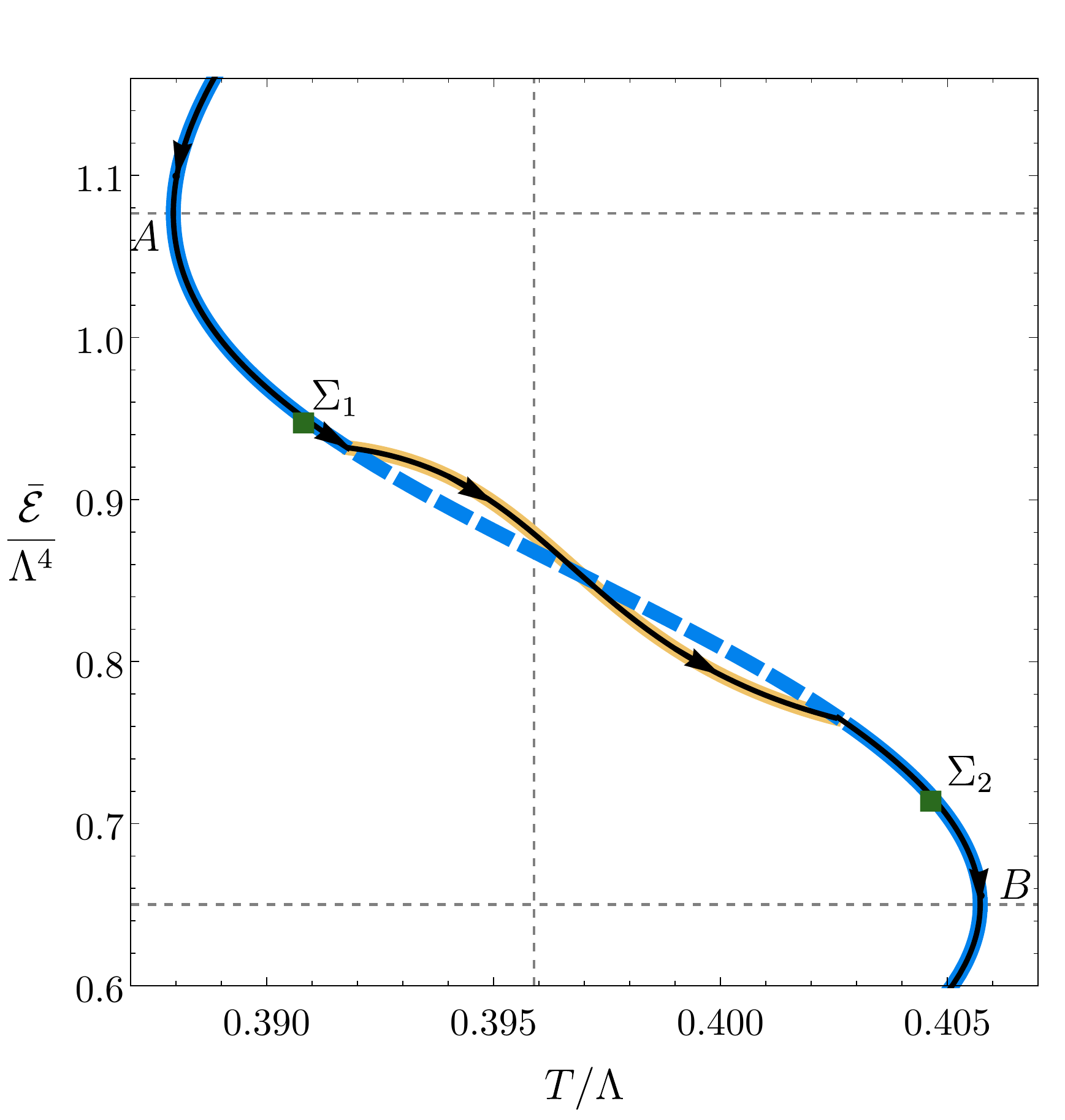}
}
\caption{Phase diagram in the microcanonical ensemble for a length 
$\hat{L}\simeq 5.299674$ such that 
 $\hat{L}_K< \hat{L} <\hat{L}_{\Sigma_1} <  \hat{L}_{\Sigma_2}$. 
 The blue and the beige curves correspond to homogeneous and inhomogeneous states, respectively. The beige curve is the same as the curve of beige inverted triangles  in \fig{fig:LumpyUnifET}.
Solid segments  indicate locally dynamically stable states; dashed segments indicate unstable ones. 
 The black curves with arrows indicate the sequence of globally preferred, maximum-entropy states as the average energy density decreases.  The phase transitions between the homogeneous and the inhomogeneous branches take place at the points where these branches merge and they are second-order. 
}
\label{fig:transition2}
\end{figure} 

Thinking of the infinite-volume case  as the limit $\hat{L}\to \infty$ of the situation described in \fig{fig:transition1} sheds light on the order of the phase transition at infinite volume. As $\hat L$ increases the point $a$ in \fig{fig:transition1} moves up and to the right. This means that the homogeneous and the inhomogeneous states between which the transition takes place become closer to one another. In the limit  $\hat{L}\to \infty$ the point $a$ tends to  the point $D$ (of Figs. \ref{fig:transition0INTRO} or \ref{fig:Unif}) and the transition takes place between two states at the same temperature $\hat{T}=\hat{T}_c$. Since the discontinuity in $\hat{T}$ disappears the phase transition becomes second-order.

Consider now \fig{fig:transition2}, which illustrates the structure of phase transitions for a length $\hat{L} \simeq 5.299674$ such that 
 $\hat{L}_K< \hat{L} <\hat{L}_{\Sigma_1} <  \hat{L}_{\Sigma_2}$  (this is the value corresponding to the inverted beige triangles in \fig{fig:LumpyUnifET}). 
In this case the preferred states lie on the homogeneous branch until the   merger point with the inhomogeneous branch is reached. At this point a transition between the homogeneous and the inhomogeneous branches takes place. Since the transition happens at the merger point, the temperature is continuous and the transition is second-order. As the energy is further decreased the preferred states remain on the inhomogeneous branch until this merges again with the homogeneous branch. At this point another second-order phase transition takes place. Below this point the preferred state lies on the homogeneous branch. 

In the intermediate range of lengths $\hat{L}_K <\hat{L}_{\Sigma_1} <  \hat{L} < \hat{L}_{\Sigma_2}$ the structure of transitions is a hybrid between those described in Figs.~\ref{fig:transition1} and \ref{fig:transition2}. As the energy decreases there is first a first-order phase transition between the homogeneous branch and the inhomogeneous branch. 
In our model the point on the homogeneous branch lies between $A$ and $\Sigma_1$, and the point on the inhomogeneous branch is the analog of point $a$ in  \fig{fig:transition1}. As the energy is further decreased the preferred state remains on the inhomogeneous branch until this merges with the homogeneous one. At this point a second-order phase transition occurs in which the preferred state becomes the one on the homogeneous branch. This second transition is analogous to that in \fig{fig:transition2}. Below this point the preferred state remains on this branch. 

\begin{figure}[t]
\centerline{
		\includegraphics[width=.9\textwidth]{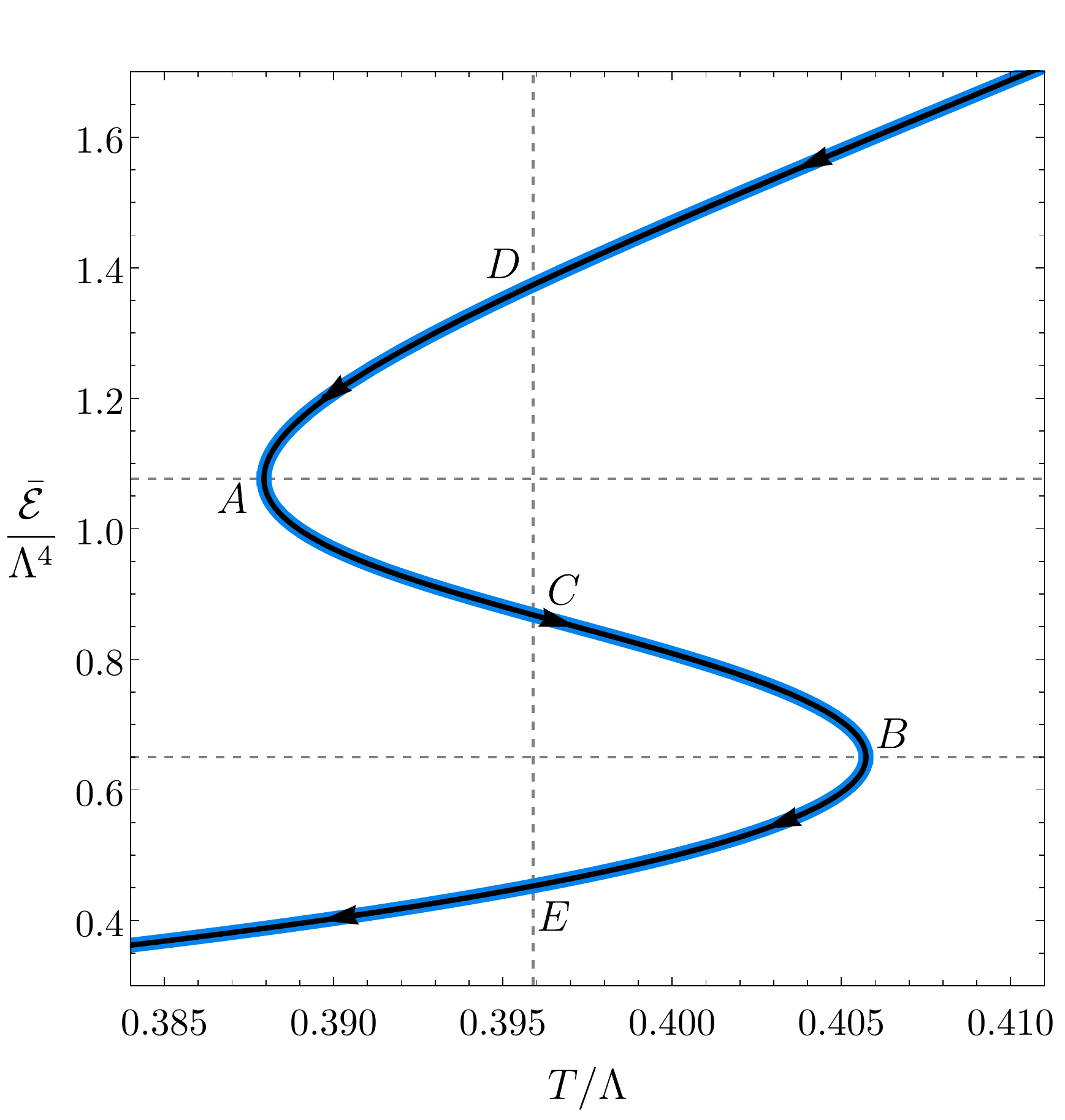}
}
\caption{Phase diagram in the microcanonical ensemble for a length 
$\hat{L}$ such that  $\hat{L} < \hat{L}_K <\hat{L}_{\Sigma_1} <  \hat{L}_{\Sigma_2}$. 
}
\label{fig:transition3}
\end{figure}

Finally, for lengths such that $\hat{L} < \hat{L}_K <\hat{L}_{\Sigma_1} <  \hat{L}_{\Sigma_2}$, no inhomogeneous states exist and all the homogeneous ones are dynamically stable. In this case no phase transitions occur as the energy decreases from infinity to zero, as illustrated in \fig{fig:transition3}.

In the figures above we have used continuous and dashed segments to distinguish between locally dynamically stable and locally dynamically unstable states. In the case of homogeneous states these properties can be established via a perturbative analysis. In the case of inhomogeneous states we  used a numerical code for time evolution to  study the behaviour of small perturbations. The results are shown in  \fig{fig:Results ComparisonE} and can be succinctly summarised as follows: all states on the upper part of the curve, shown in blue, are stable, whereas those on the lower part, shown in red, are unstable (see also a relevant zoom in \fig{fig:(3) A-B A-C}). Once an unstable state is slightly perturbed, its full time evolution and  its end state depend on the perturbation. For example, the locally dynamically unstable state $\mathcal{X}_c$ in  \fig{fig:Results ComparisonE} and \fig{fig:(3) A-B A-C} can  decay to either of the stable states $\mathcal{X}_d$ or $\mathcal{X}_e$. The corresponding time evolutions are shown in Figs.~\ref{fig:non linear evolution (3)}(left) and \ref{fig:non linear evolution (3)}(right), respectively. 

For completeness, note that, unlike in the microcanonical ensemble, at finite volume lumpy branes are never the dominant thermodynamic phase in the canonical ensemble,  as  illustrated by \fig{fig:canonical}.

In our analysis we have benefited from two simplifying assumptions. The first one is that we imposed translational invariance along two of the three spatial directions of the box. Lifting this restriction will generically allow for inhomogeneities to develop in all three  directions. It would be interesting to study this more general setup, in particular the possible interplays between different length scales in different directions. Hopefully, the ``one-dimensional building blocks" that we have investigated will be useful to understand the three-dimensional case. 

The second simplifying assumption is the fact that we have worked in the $\nc \to\infty$ limit, which ensures that the system is in the thermodynamic limit despite the finite volume. In particular, it guarantees that true phase transitions may occur. At finite $\nc$ these transitions will turn into cross-overs. However, the latter can be made arbitrarily rapid by making $\nc$ sufficiently large. This means that our results should be a good approximation to the physics at finite but large $\nc$.

\acknowledgments
We thank Benson Way for discussions.  YB is supported by the European Research Council Grant No.~ERC-2014-StG 639022-NewNGR. OD acknowledges financial support from the STFC Ernest Rutherford grants ST/K005391/1 and ST/M004147/1 and from the STFC ``Particle Physics Grants Panel (PPGP) 2016" Grant No.~ST/P000711/1.
TG acknowledges financial support from FCT/Portugal Grant No.~PD/BD/135425/2017 in the framework
of the Doctoral Programme IDPASC-Portugal.
MSG acknowledges financial support from the APIF program, fellowship APIF\_18\_19/226. JES is supported in part by STFC grants PHY-1504541 and ST/P000681/1. JES also acknowledges support from a J.~Robert~Oppenheimer Visiting~Professorship.
MZ acknowledges financial support provided by FCT/Portugal through the IF
programme, grant IF/00729/2015.
The authors thankfully acknowledge the computer resources, technical expertise and assistance provided by CENTRA/IST. Computations were performed in part at the cluster ``Baltasar-Sete-S\'ois" and supported by the H2020 ERC Consolidator Grant ``Matter and strong field gravity: New frontiers in Einstein's theory" grant agreement No.~MaGRaTh-646597. We also thank the MareNostrum supercomputer at the BSC (activity Id FI-2020-1-0007) for significant computational resources.
We are also supported by grants FPA2016-76005-C2-1-P, FPA2016-76005-C2-2-P, 2014-SGR-104, 2014-SGR-1474, SGR-2017-754, MDM-2014-0369, PID2019-105614GB-C22 and FCT projects UIDB/00099/2020 and CERN/FIS-PAR/0023/2019.





\bibliography{refsSpinodal}{}

\providecommand{\href}[2]{#2}\begingroup\raggedright\begin{thebibliography}{10}

\bibitem{Attems:2017ezz}
M.~Attems, Y.~Bea, J.~Casalderrey-Solana, D.~Mateos, M.~Triana and M.~Zilhao,
  \emph{{Phase Transitions, Inhomogeneous Horizons and Second-Order
  Hydrodynamics}}, \href{http://dx.doi.org/10.1007/JHEP06(2017)129}{\emph{JHEP}
  {\bf 06} (2017) 129}, [\href{https://arxiv.org/abs/1703.02948}{{\tt
  1703.02948}}].

\bibitem{Attems:2019yqn}
M.~Attems, Y.~Bea, J.~Casalderrey-Solana, D.~Mateos and M.~Zilhão,
  \emph{{Dynamics of Phase Separation from Holography}},
  \href{http://dx.doi.org/10.1007/JHEP01(2020)106}{\emph{JHEP} {\bf 01} (2020)
  106}, [\href{https://arxiv.org/abs/1905.12544}{{\tt 1905.12544}}].

\bibitem{Janik:2017ykj}
R.~A. Janik, J.~Jankowski and H.~Soltanpanahi, \emph{{Real-Time dynamics and
  phase separation in a holographic first order phase transition}},
  \href{http://dx.doi.org/10.1103/PhysRevLett.119.261601}{\emph{Phys. Rev.
  Lett.} {\bf 119} (2017) 261601},
  [\href{https://arxiv.org/abs/1704.05387}{{\tt 1704.05387}}].

\bibitem{Bellantuono:2019wbn}
L.~Bellantuono, R.~A. Janik, J.~Jankowski and H.~Soltanpanahi, \emph{{Dynamics
  near a first order phase transition}},
  \href{http://dx.doi.org/10.1007/JHEP10(2019)146}{\emph{JHEP} {\bf 10} (2019)
  146}, [\href{https://arxiv.org/abs/1906.00061}{{\tt 1906.00061}}].

\bibitem{Buchel:2005nt}
A.~Buchel, \emph{{A Holographic perspective on Gubser-Mitra conjecture}},
  \href{http://dx.doi.org/10.1016/j.nuclphysb.2005.10.014}{\emph{Nucl. Phys.}
  {\bf B731} (2005) 109--124},
  [\href{https://arxiv.org/abs/hep-th/0507275}{{\tt hep-th/0507275}}].

\bibitem{Emparan:2009cs}
R.~Emparan, T.~Harmark, V.~Niarchos and N.~A. Obers, \emph{{World-Volume
  Effective Theory for Higher-Dimensional Black Holes}},
  \href{http://dx.doi.org/10.1103/PhysRevLett.102.191301}{\emph{Phys. Rev.
  Lett.} {\bf 102} (2009) 191301}, [\href{https://arxiv.org/abs/0902.0427}{{\tt
  0902.0427}}].

\bibitem{Emparan:2009at}
R.~Emparan, T.~Harmark, V.~Niarchos and N.~A. Obers, \emph{{Essentials of
  Blackfold Dynamics}},
  \href{http://dx.doi.org/10.1007/JHEP03(2010)063}{\emph{JHEP} {\bf 03} (2010)
  063}, [\href{https://arxiv.org/abs/0910.1601}{{\tt 0910.1601}}].

\bibitem{Gregory:1993vy}
R.~Gregory and R.~Laflamme, \emph{{Black strings and p-branes are unstable}},
  \href{http://dx.doi.org/10.1103/PhysRevLett.70.2837}{\emph{Phys. Rev. Lett.}
  {\bf 70} (1993) 2837--2840},
  [\href{https://arxiv.org/abs/hep-th/9301052}{{\tt hep-th/9301052}}].

\bibitem{Amsel:2007im}
A.~J. Amsel, T.~Hertog, S.~Hollands and D.~Marolf, \emph{{A Tale of two
  superpotentials: Stability and instability in designer gravity}},
  \href{http://dx.doi.org/10.1103/PhysRevD.77.049903,
  10.1103/PhysRevD.75.084008}{\emph{Phys. Rev.} {\bf D75} (2007) 084008},
  [\href{https://arxiv.org/abs/hep-th/0701038}{{\tt hep-th/0701038}}].

\bibitem{Attems:2018gou}
M.~Attems, Y.~Bea, J.~Casalderrey-Solana, D.~Mateos, M.~Triana and M.~Zilhão,
  \emph{{Holographic Collisions across a Phase Transition}},
  \href{http://dx.doi.org/10.1103/PhysRevLett.121.261601}{\emph{Phys. Rev.
  Lett.} {\bf 121} (2018) 261601},
  [\href{https://arxiv.org/abs/1807.05175}{{\tt 1807.05175}}].

\bibitem{Bea:2018whf}
Y.~Bea and D.~Mateos, \emph{{Heating up Exotic RG Flows with Holography}},
  \href{http://dx.doi.org/10.1007/JHEP08(2018)034}{\emph{JHEP} {\bf 08} (2018)
  034}, [\href{https://arxiv.org/abs/1805.01806}{{\tt 1805.01806}}].

\bibitem{Donos:2014yya}
A.~Donos and J.~P. Gauntlett, \emph{{The thermoelectric properties of
  inhomogeneous holographic lattices}},
  \href{http://dx.doi.org/10.1007/JHEP01(2015)035}{\emph{JHEP} {\bf 01} (2015)
  035}, [\href{https://arxiv.org/abs/1409.6875}{{\tt 1409.6875}}].

\bibitem{Marolf:2019wkz}
D.~Marolf and J.~E. Santos, \emph{{Phases of Holographic Hawking Radiation on
  spatially compact spacetimes}},
  \href{http://dx.doi.org/10.1007/JHEP10(2019)250}{\emph{JHEP} {\bf 10} (2019)
  250}, [\href{https://arxiv.org/abs/1906.07681}{{\tt 1906.07681}}].

\bibitem{Bianchi:2001de}
M.~Bianchi, D.~Z. Freedman and K.~Skenderis, \emph{{How to go with an RG
  flow}}, \href{http://dx.doi.org/10.1088/1126-6708/2001/08/041}{\emph{JHEP}
  {\bf 08} (2001) 041}, [\href{https://arxiv.org/abs/hep-th/0105276}{{\tt
  hep-th/0105276}}].

\bibitem{Bianchi:2001kw}
M.~Bianchi, D.~Z. Freedman and K.~Skenderis, \emph{{Holographic
  renormalization}},
  \href{http://dx.doi.org/10.1016/S0550-3213(02)00179-7}{\emph{Nucl. Phys.}
  {\bf B631} (2002) 159--194},
  [\href{https://arxiv.org/abs/hep-th/0112119}{{\tt hep-th/0112119}}].

\bibitem{Dias:2017coo}
O.~J. Dias, J.~E. Santos and B.~Way, \emph{{Lattice Black Branes: Sphere
  Packing in General Relativity}},
  \href{http://dx.doi.org/10.1007/JHEP05(2018)111}{\emph{JHEP} {\bf 05} (2018)
  111}, [\href{https://arxiv.org/abs/1712.07663}{{\tt 1712.07663}}].

\bibitem{Gubser:2001ac}
S.~S. Gubser, \emph{{On nonuniform black branes}},
  \href{http://dx.doi.org/10.1088/0264-9381/19/19/303}{\emph{Class. Quant.
  Grav.} {\bf 19} (2002) 4825--4844},
  [\href{https://arxiv.org/abs/hep-th/0110193}{{\tt hep-th/0110193}}].

\bibitem{Wiseman:2002zc}
T.~Wiseman, \emph{{Static axisymmetric vacuum solutions and nonuniform black
  strings}}, \href{http://dx.doi.org/10.1088/0264-9381/20/6/308}{\emph{Class.
  Quant. Grav.} {\bf 20} (2003) 1137--1176},
  [\href{https://arxiv.org/abs/hep-th/0209051}{{\tt hep-th/0209051}}].

\bibitem{Sorkin:2004qq}
E.~Sorkin, \emph{{A Critical dimension in the black string phase transition}},
  \href{http://dx.doi.org/10.1103/PhysRevLett.93.031601}{\emph{Phys. Rev.
  Lett.} {\bf 93} (2004) 031601},
  [\href{https://arxiv.org/abs/hep-th/0402216}{{\tt hep-th/0402216}}].

\bibitem{Dias:2015nua}
O.~J.~C. Dias, J.~E. Santos and B.~Way, \emph{{Numerical Methods for Finding
  Stationary Gravitational Solutions}},
  \href{http://dx.doi.org/10.1088/0264-9381/33/13/133001}{\emph{Class. Quant.
  Grav.} {\bf 33} (2016) 133001}, [\href{https://arxiv.org/abs/1510.02804}{{\tt
  1510.02804}}].

\bibitem{Dias:2015pda}
O.~J. Dias, J.~E. Santos and B.~Way, \emph{{Lumpy AdS$_{5}\times$ S$^{5}$ black
  holes and black belts}},
  \href{http://dx.doi.org/10.1007/JHEP04(2015)060}{\emph{JHEP} {\bf 04} (2015)
  060}, [\href{https://arxiv.org/abs/1501.06574}{{\tt 1501.06574}}].

\bibitem{Dias:2016eto}
O.~J. Dias, J.~E. Santos and B.~Way, \emph{{Localised $AdS_5\times S^5$ Black
  Holes}}, \href{http://dx.doi.org/10.1103/PhysRevLett.117.151101}{\emph{Phys.
  Rev. Lett.} {\bf 117} (2016) 151101},
  [\href{https://arxiv.org/abs/1605.04911}{{\tt 1605.04911}}].

\bibitem{Dias:2017uyv}
O.~J. Dias, J.~E. Santos and B.~Way, \emph{{Localised and nonuniform thermal
  states of super-Yang-Mills on a circle}},
  \href{http://dx.doi.org/10.1007/JHEP06(2017)029}{\emph{JHEP} {\bf 06} (2017)
  029}, [\href{https://arxiv.org/abs/1702.07718}{{\tt 1702.07718}}].

\bibitem{Dias:2017opt}
O.~J. Dias, G.~S. Hartnett, B.~E. Niehoff and J.~E. Santos,
  \emph{{Mass-deformed M2 branes in Stenzel space}},
  \href{http://dx.doi.org/10.1007/JHEP11(2017)105}{\emph{JHEP} {\bf 11} (2017)
  105}, [\href{https://arxiv.org/abs/1704.02323}{{\tt 1704.02323}}].

\bibitem{Bena:2018vtu}
I.~Bena, O.~J. Dias, G.~S. Hartnett, B.~E. Niehoff and J.~E. Santos,
  \emph{{Holographic dual of hot Polchinski-Strassler quark-gluon plasma}},
  \href{http://dx.doi.org/10.1007/JHEP09(2019)033}{\emph{JHEP} {\bf 09} (2019)
  033}, [\href{https://arxiv.org/abs/1805.06463}{{\tt 1805.06463}}].

\bibitem{Horowitz:2002dc}
G.~T. Horowitz, \emph{{Playing with black strings}},  in \emph{{The future of
  theoretical physics and cosmology: Celebrating Stephen Hawking's 60th
  birthday. Proceedings, Workshop and Symposium, Cambridge, UK, January 7-10,
  2002}}, pp.~310--329, 2002.
\newblock \href{https://arxiv.org/abs/hep-th/0205069}{{\tt hep-th/0205069}}.

\bibitem{Harmark:2003eg}
T.~Harmark and N.~A. Obers, \emph{{Phase structure of black holes and strings
  on cylinders}},
  \href{http://dx.doi.org/10.1016/j.nuclphysb.2004.02.022}{\emph{Nucl. Phys.}
  {\bf B684} (2004) 183--208},
  [\href{https://arxiv.org/abs/hep-th/0309230}{{\tt hep-th/0309230}}].

\bibitem{Dias:2007hg}
O.~J.~C. Dias, T.~Harmark, R.~C. Myers and N.~A. Obers, \emph{{Multi-black hole
  configurations on the cylinder}},
  \href{http://dx.doi.org/10.1103/PhysRevD.76.104025}{\emph{Phys. Rev.} {\bf
  D76} (2007) 104025}, [\href{https://arxiv.org/abs/0706.3645}{{\tt
  0706.3645}}].

\bibitem{Benincasa:2005iv}
P.~Benincasa, A.~Buchel and A.~O. Starinets, \emph{{Sound waves in strongly
  coupled non-conformal gauge theory plasma}},
  \href{http://dx.doi.org/10.1016/j.nuclphysb.2005.11.005}{\emph{Nucl. Phys.}
  {\bf B733} (2006) 160--187},
  [\href{https://arxiv.org/abs/hep-th/0507026}{{\tt hep-th/0507026}}].

\bibitem{Attems:2016ugt}
M.~Attems, J.~Casalderrey-Solana, D.~Mateos, I.~Papadimitriou,
  D.~Santos-Oliván, C.~F. Sopuerta et~al., \emph{{Thermodynamics, transport
  and relaxation in non-conformal theories}},
  \href{http://dx.doi.org/10.1007/JHEP10(2016)155}{\emph{JHEP} {\bf 10} (2016)
  155}, [\href{https://arxiv.org/abs/1603.01254}{{\tt 1603.01254}}].

\bibitem{Attems:2016tby}
M.~Attems, J.~Casalderrey-Solana, D.~Mateos, D.~Santos-Oliván, C.~F. Sopuerta,
  M.~Triana et~al., \emph{{Holographic Collisions in Non-conformal Theories}},
  \href{http://dx.doi.org/10.1007/JHEP01(2017)026}{\emph{JHEP} {\bf 01} (2017)
  026}, [\href{https://arxiv.org/abs/1604.06439}{{\tt 1604.06439}}].

\end{thebibliography}\endgroup
\bibliographystyle{JHEP}
\end{document}